\documentclass[prd,aps,tightenlines,superscriptaddress,nofootinbib,showpacs,showkeys,floatfix]{revtex4}
\usepackage{float}
\usepackage{mathrsfs}
\usepackage{amsfonts}
\usepackage{array}
\usepackage{epsfig}
\usepackage{amsmath}    
\usepackage{amssymb}    
\usepackage{graphicx}   
\usepackage{verbatim}   
\usepackage{color}      
\usepackage{subfigure}  
\usepackage{multirow}
\def\beq{\begin{equation}}
\def\b0{\beta_0}

\def\eeq{\end{equation}}
\def\beeq{\begin{eqnarray}}
\def\eeeq{\end{eqnarray}}

\def\to{\rightarrow}

\begin{document}

\title{
New parton distribution functions from a global analysis of quantum chromodynamics}

\author{Sayipjamal Dulat}
\email{sdulat@msu.edu}
\affiliation{
School of Physics Science and Technology, Xinjiang University,\\
 Urumqi, Xinjiang 830046 China }
\affiliation{
Department of Physics and Astronomy, Michigan State University,\\
 East Lansing, MI 48824 U.S.A. }
\author{Tie-Jiun Hou}
\email{tiejiunh@mail.smu.edu}
\affiliation{
Department of Physics, Southern Methodist University,\\
 Dallas, TX 75275-0181, U.S.A. }
\author{Jun Gao}
\email{jgao@anl.gov}
\affiliation{
High Energy Physics Division, Argonne National Laboratory,\\
 Argonne, Illinois 60439, U.S.A.}
\author{Marco Guzzi}
\email{marco.guzzi@manchester.ac.uk}
\affiliation{
School of Physics \& Astronomy, University of Manchester,
Manchester M13 9PL, United Kingdom}
\author{Joey Huston}
\email{huston@pa.msu.edu}
\affiliation{
Department of Physics and Astronomy, Michigan State University,\\
 East Lansing, MI 48824 U.S.A. }
\author{Pavel Nadolsky}
\email{nadolsky@physics.smu.edu}
\affiliation{
Department of Physics, Southern Methodist University,\\
 Dallas, TX 75275-0181, U.S.A. }
\author{Jon Pumplin}
\email{pumplin@pa.msu.edu}
\affiliation{
Department of Physics and Astronomy, Michigan State University,\\
 East Lansing, MI 48824 U.S.A. }
\author{Carl Schmidt}
\email{schmidt@pa.msu.edu}
\affiliation{
Department of Physics and Astronomy, Michigan State University,\\
 East Lansing, MI 48824 U.S.A. }
\author{Daniel Stump}
\email{stump@pa.msu.edu}
\affiliation{
Department of Physics and Astronomy, Michigan State University,\\
 East Lansing, MI 48824 U.S.A. }
\author{ C.--P. Yuan}
\email{yuan@pa.msu.edu}
\affiliation{
Department of Physics and Astronomy, Michigan State University,\\
 East Lansing, MI 48824 U.S.A. }

\begin{abstract}

We present new parton distribution functions (PDFs) at
next-to-next-to-leading order (NNLO)
from the CTEQ-TEA global analysis of quantum chromodynamics.
These differ from previous CT PDFs in several respects, including
the use of data from LHC experiments,
and the new D\O~ charged lepton rapidity asymmetry data, as well as the use
of a more flexible parametrization of PDFs that, in particular,
allows a better fit to different combinations of quark flavors.
Predictions for important LHC processes,
especially Higgs boson production at 13 TeV, are presented.
These CT14 PDFs include a central set and error sets in the Hessian representation.
For completeness, we also present the CT14 PDFs determined at the leading order
(LO) and the next-to-leading order (NLO) in QCD.
Besides these general-purpose PDF sets,
we provide a series of (N)NLO sets with various
$\alpha_s$ values and additional sets in general-mass
variable flavor number (GM-VFN)
schemes, to deal with heavy partons, with up to 3, 4, and 6 active flavors.

\end{abstract}



\maketitle
\newpage
\tableofcontents
\newpage

\section{Introduction \label{sec:Introduction}}

Run-1 at the Large Hadron Collider (LHC) was a great success,
culminating in the discovery of the Higgs boson~\cite{Aad:2012tfa,Chatrchyan:2012ufa}. No physics beyond the
standard model was discovered in this run, however Run-2, with a larger
center-of-mass energy and integrated luminosity, will allow
for an increased discovery potential
for new physics.  Precision measurements
of the Higgs boson and of various electroweak observables will be performed
with extraordinary accuracy in new kinematic
regimes in Run 2. Run-1 achievements, such as the combined ATLAS/CMS
measurement of the Higgs boson mass with 0.2\% accuracy
\cite{CombHiggsMass2015}, will soon be superseded. For both precision measurements and for discovery of possible
new physics, it is important to have the proper tools for the
calculation of the relevant cross sections. These tools include both matrix
element determinations at higher orders in perturbative QCD and
electroweak theory, and precision parton distribution functions
(PDFs).
The need for precision PDFs was driven home
by the recent calculation of the inclusive cross section for
gluon-gluon fusion to a Higgs boson at NNNLO~\cite{Anastasiou:2015ema}. As this
tour-de-force calculation has significantly reduced the
scale dependence of the Higgs cross section, the PDF and $\alpha_s$
uncertainties become the dominant remaining theoretical uncertainty (as of the last
PDF4LHC recommendation).

The CT10 parton distribution functions were published at next-to-leading order (NLO)
in 2010~\cite{Lai:2010vv}, followed by the CT10
next-to-next-to leading order (NNLO) parton distribution
functions in 2013~\cite{Gao:2013xoa}. These PDF ensembles were determined
using diverse experimental data from fixed-target experiments, HERA
and the Tevatron collider, but without data from the LHC. In this
paper, we present a next generation of PDFs, designated as CT14.
The CT14 PDFs include data from the LHC for the first time, as well as
updated data from the Tevatron and from HERA experiments.
Various CT14 PDF sets have been produced
at the leading order (LO), NLO and NNLO
and are available from LHAPDF~\cite{LHAPDF6}.

The CTEQ-TEA philosophy has always been to determine PDFs from
data on inclusive, high-momentum transfer processes, for which
perturbative QCD is expected to be reliable. For example, in the
case of deep inelastic lepton scattering, we only use data with $Q >
2$ GeV and $W > 3.5$ GeV. Data in this region are expected to be
relatively free of non-perturbative effects, such as higher twists or nuclear
corrections.
Thus, there is no need to introduce phenomenological models
for nonperturbative corrections beyond the leading-twist perturbative
contributions.

For the majority of processes in the CT14 global analysis,
theoretical predictions are now included at the NNLO level of
accuracy. In particular, a NNLO treatment \cite{Guzzi:2011ew}
of heavy-quark mass effects in neutral-current DIS is realized
in the  ACOT-$\chi$
scheme~\cite{Aivazis:1993pi,Collins:1998rz,Tung:2001mv}
and is essential for obtaining
correct predictions for LHC electroweak cross
sections~\cite{Tung:2006tb,Nadolsky:2008zw}.
We make two exceptions to this rule, by
including measurements for charged-current DIS and inclusive jet
production at NLO only. In both cases, the complete NNLO contributions are
not yet available, but it can be argued based on our studies that the
expected effect of missing NNLO effects is small relatively to current
experimental errors (cf. Sec.~\ref{sec:Setup}). For both types of processes, the NLO
predictions have undergone various benchmarking tests.
A numerical error was discovered and corrected in the implementation of
the SACOT-$\chi$  scheme for charged-current DIS, resulting in
relatively small changes from CT10 (within the PDF uncertainties).

As in the CT10 global analysis, we use a charm pole mass of 1.3 GeV,
which was shown to be consistent with the CT10 data in
Ref.~\cite{Gao:2013xoa}. The PDFs for $u$, $d$, $s$ (anti-)quarks and the
gluon are parametrized at an initial scale of 1.295 GeV,
and the charm quark PDF is turned on with zero intrinsic charm as
the scale $Q$ reaches the charm pole mass.

The new LHC measurements of $W/Z$ cross sections directly probe the flavor
separation of $u$ and $d$ (anti-)quarks in an $x$-range
around $0.01$ that was not directly
assessed by the previously available experiments.
We also include an updated measurement of
electron charge asymmetry from the D\O~ collaboration~\cite{D0:2014kma}, which
probes the $d$ quark PDF at $x>0.1$.
To better estimate
variations in relevant PDF combinations,
such as $d(x,Q)/u(x,Q)$ and $\bar d(x,Q)/\bar u(x,Q)$,
we increased the number of free PDF parameters to 28,
compared to 25 in CT10 NNLO.
As another important modification, CT14
employs a novel flexible parametrization for the PDFs, based on the use of
Bernstein polynomials (reviewed in the Appendix).
The shape of the Bernstein polynomials is such that a single
polynomial is dominant in each given $x$ range, reducing undesirable
correlations among the PDF parameters that sometimes occurred in CT10.
In the asymptotic limits of $x \rightarrow 0$ or $x\rightarrow 1$, the new
parametrization forms allow for the possibility of arbitrary constant ratios of
$d/u$ or $\bar d/\bar u$, in contrast to the more constrained behavior
assumed in CT10.

The PDF error sets of the CT14 ensemble are obtained using two
techniques, the Hessian method \cite{Pumplin:2001ct} and Monte-Carlo sampling \cite{Watt:2012tq}.
Lagrange multiplier studies~\cite{Stump:2001gu} have also been used to verify the
Hessian uncertainties, especially in regions not well constrained by
data.
This applies at NNLO and NLO; no error sets are provided
at LO due to the difficulty in defining meaningful uncertainties at
that order.

A central value of $\alpha_s(M_Z)$ of 0.118 has been assumed in the
global fits at NLO and NNLO, but PDF sets at
alternative values of $\alpha_s(m_Z)$ are also provided.
CT14 prefers $\alpha_s(M_Z)=0.115^{+0.006}_{-0.004}$ at NNLO
($0.117\pm 0.005$ at NLO) at 90 \% confidence level (C.L.). These
uncertainties from the global QCD fits are larger
than those of the data from LEP and other experiments
included into the world average~\cite{Agashe:2014kda}.
Thus, the central PDF sets are obtained using the value of $0.118$,
which is consistent with the world average value and was recommended by the
PDF4LHC group~\cite{Alekhin:2011sk}.
For the CT14
LO PDFs, we follow the precedent begun in CTEQ6 \cite{Pumplin:2002vw} by supplying two
versions, one with a 1-loop $\alpha_s(M_Z)$ value of 0.130,
and the other with a 2-loop  $\alpha_s(M_Z)$  value of 0.118.

The flavor composition of CT14 PDFs has changed  somewhat compared to
CT10 due to the inclusion of new LHC and Tevatron data sets, to the use of modified
parametrization forms, and to the numerical modifications discussed above.
The new PDFs are largely compatible with CT10 within
the estimated PDF uncertainty. The CT14 NNLO PDFs have a softer
strange quark distribution at low $x$ and a somewhat softer gluon at
high $x$, compared to CT10 NNLO. The $d/u$
ratio has decreased at high $x$ in comparison to CT10,
as a consequence of replacing the 2008 D\O~ electron charge asymmetry
($0.75\mbox{ fb}^{-1}$ \cite{Abazov:2008qv})
measurement by the new $9.7\mbox{ fb}^{-1}$ data set \cite{D0:2014kma}.
The $d/u$ ratio approaches a constant value in the $x\rightarrow 1$ limit
due to the input physics assumption that both $d_{\rm val}$ and
$u_{\rm val}$ behave as $(1-x)^{a_2}$ at $x\rightarrow1$ with the same
value of $a_2$ (reflecting expectations from spectator counting
rules), but allowing for independent normalizations. The $\bar d/\bar
u$ ratio has also changed as a consequence of the new data and the new
parametrization form.

The organization of the paper is as follows. In Sec.~\ref{sec:Setup},
we list the
data sets used in the CT14 fit and discuss further aspects of the
global fits for the central CT14 PDFs and for the error sets. In
Sec.~\ref{sec:OverviewCT14},
we show various aspects of the resultant CT14 PDFs and make
comparisons to CT10 PDFs.
In Sec.~\ref{sec:TheoryVsData},
we show comparisons of NNLO predictions using the
CT14 PDFs to some of the data sets used in the global fits.
Specifically, we compare to experimental measurements of
jet, $W$ and $Z$, $W+c$ cross sections.
In Sec.~\ref{sec:LHCPredictions},
we discuss NNLO predictions using the
CT14 PDFs for Higgs boson production via the gluon-gluon fusion channel and
for top quark and anti-quark pair production.
Our conclusion is given in Sec.~\ref{sec:conclude}.

\section{Setup of the analysis \label{sec:Setup}}

\label{sec:DATA}

\subsection{Overview of the global fit}
The goal of the CT14 global analysis is to provide a new 
generation of PDFs intended for widespread use in high-energy experiments.
As we generate new PDF sets, 
we include newly available experimental data sets and 
theoretical calculations, and redesign the functional forms of PDFs if
new data or new theoretical calculations favor it. 
All changes --- data, theory, and parametrization --- contribute to the 
differences between the old and new generations of PDFs
in ways that are correlated and frequently cannot be separated. 
The most important, but not the only, criterion for the selection of PDFs 
is the minimization of the log-likelihood $\chi^{2}$ that quantifies
agreement of theory and data.
In addition, we make some "prior assumptions" about the forms of the
PDFs. A PDF set that violates them may be rejected 
even if it lowers $\chi^{2}$. 
For example, we assume that the PDFs are smoothly
varying functions of $x$, without abrupt variations or 
short-wavelength oscillations. This is consistent with the
experimental data and sufficient for making new predictions. 
No PDF can be negative at the input scale $Q_{0}$, 
to preclude negative cross sections in the predictions. 
Flavor-dependent ratios or cross section asymmetries must also take physical
values, which limits the range of allowed parametrizations in
extreme kinematical regions with poor experimental constraints. 
For example, in the CT14 parametrization we restricted
the functional forms of the $u$ and $d$ PDFs so that 
$d(x,Q_{0})/u(x,Q_{0})$ would remain finite and nonzero at
$x\rightarrow 1$, cf. the Appendix. 
We now review every input of the CT14 PDF analysis in turn, 
starting with the selection of the new experiments.

\subsection{Selection of experiments}
The experimental data sets that are included in the CT14 global
analysis are listed in Tables~\ref{tab:EXP_1} (lepton scattering)
and \ref{tab:EXP_2}
(production of inclusive lepton pairs and jets).
There are a total of 2947 data points included from 33 experiments,
producing $\chi^2$ value of $3252$ for the best fit (with $\chi^2/N_{pt}=1.10$).
It can be seen from the values of $\chi^{2}$ in Tables \ref{tab:EXP_1}
and \ref{tab:EXP_2} that the data and theory are in reasonable
agreement for most experiments.  The variable $S_n$  in the last
column is an ``effective Gaussian variable'', first introduced in the Sec. 5 of
Ref.~\cite{Lai:2010vv} and defined for the current analysis in
Refs.~\cite{Gao:2013xoa,Dulat:2013hea}. The effective Gaussian variable
quantifies compatibility of any given data set with a particular PDF
fit in a way that is independent of the number of points $N_{pt,n}$ in
the data set.  It maps the $\chi^2_n$ values of
individual experiments, whose probability distributions depend on
$N_{pt,n}$ in each experiment (and thus, are not identical), onto
$S_n$ values that obey a cumulative probability distribution shared by
all experiments, independently of $N_{pt,n}$. Values of
$S_n$ between -1 and +1 correspond to a good fit to the $n$-th
experiment (at the 68\% C.L.). Large positive values ($\gtrsim 2$)
correspond to a poor fit, while large negative values ($\lesssim -2$)
are fit unusually well.

\begin{figure}[b]
\vspace{20pt}
\includegraphics[width=7cm]{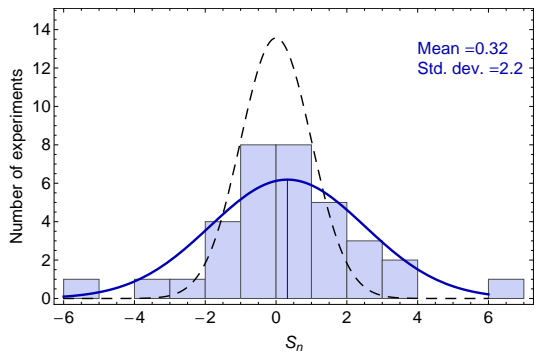}
\caption{Best-fit $S_n$ values of 33 experiments in the CT14
analysis.
\label{fig:Sns}
}
\end{figure}

\begin{table}[tb]
\begin{tabular}{|l|lr|c|c|c|c|}
\hline
\textbf{ID\# }  & \textbf{Experimental data set}  & & $N_{pt,n}$  &$\chi_{n}^2$ & $\chi_{n}^{2}/N_{pt,n}$ & $S_n$ \tabularnewline
\hline
101 & BCDMS $F_{2}^{p}$ & \cite{Benvenuti:1989rh}                & 337 &  384 & 1.14 &  1.74  \tabularnewline
\hline
102 & BCDMS $F_{2}^{d}$ & \cite{Benvenuti:1989fm}                & 250 &  294 & 1.18 &  1.89  \tabularnewline
\hline
104 & NMC $F_{2}^{d}/F_{2}^{p}$ & \cite{Arneodo:1996qe}          & 123 &  133 & 1.08   &  0.68  \tabularnewline
\hline
106 & NMC $\sigma^{p}_{red}$ & \cite{Arneodo:1996qe}             & 201 &  372 & 1.85  &  6.89  \tabularnewline
\hline
108 & CDHSW $F_{2}^{p}$ & \cite{Berge:1989hr}                    & 85  &  72  & 0.85 &  -0.99  \tabularnewline
\hline
109 & CDHSW $F_{3}^{p}$ & \cite{Berge:1989hr}                    & 96  &  80  & 0.83 &  -1.18  \tabularnewline
\hline
110 & CCFR $F_{2}^{p}$ & \cite{Yang:2000ju}                      & 69  &  70  & 1.02 &  0.15  \tabularnewline
\hline
111 & CCFR $xF_{3}^{p}$ & \cite{Seligman:1997mc}                 & 86  &  31  & 0.36 &  -5.73  \tabularnewline
\hline
124 & NuTeV $\nu\mu\mu$ SIDIS & \cite{Mason:2006qa}              & 38  &  24  & 0.62 &  -1.83  \tabularnewline
\hline
125 & NuTeV $\bar\nu \mu\mu$ SIDIS & \cite{Mason:2006qa}         & 33  &  39  & 1.18 &  0.78  \tabularnewline
\hline
126 & CCFR $\nu\mu\mu$ SIDIS & \cite{Goncharov:2001qe}           & 40  &  29  & 0.72 &  -1.32  \tabularnewline
\hline
127& CCFR  $\bar\nu \mu\mu$ SIDIS & \cite{Goncharov:2001qe}      & 38  &  20  & 0.53 &  -2.46  \tabularnewline
\hline
145 & H1 $\sigma_{r}^{b}$ & \cite{Aktas:2004az}                  & 10  &  6.8 & 0.68   &  -0.67  \tabularnewline
\hline
147 & Combined HERA charm production & \cite{Abramowicz:1900rp}  & 47  &  59  & 1.26 &  1.22  \tabularnewline
\hline
159 & HERA1 Combined NC and CC DIS & \cite{Aaron:2009aa}         & 579 &  591 & 1.02 &  0.37  \tabularnewline
\hline
169 & H1 $F_{L}$ & \cite{Collaboration:2010ry}                   & 9   &  17  & 1.92  &  1.7    \tabularnewline
\hline
\hline
\end{tabular}
\caption{Experimental data sets employed in the CT14 analysis.
These are the lepton deep-inelastic scattering experiments.
$N_{pt,n}$, $\chi^2_{n}$ are the number of points and the value of
$\chi^2$ for the $n$-th experiment at the global minimum.
$S_n$ is the effective Gaussian parameter \cite{Lai:2010vv, Gao:2013xoa, Dulat:2013hea} quantifying agreement with each experiment.
\label{tab:EXP_1} }
\end{table}

\begin{table}[tb]
\begin{tabular}{|l|lr|c|c|c|c|}
\hline
\textbf{ID\# }  & \textbf{Experimental data set} &  & $N_{pt,n}$  & $\chi^2_{n}$ & $\chi_{n}^{2}/N_{pt,n}$  & $S_n$ \tabularnewline
\hline
\hline
201 & E605 Drell-Yan process & \cite{Moreno:1990sf}                                    &  119 &  116  & 0.98  &  -0.15 \tabularnewline
\hline
203 & E866 Drell-Yan process, $\sigma_{pd}/(2\sigma_{pp})$ & \cite{Towell:2001nh}                                      &  15  &  13   & 0.87  &  -0.25 \tabularnewline
\hline
204 & E866 Drell-Yan process, $Q^3 d^2\sigma_{pp}/(dQ dx_F)$ & \cite{Webb:2003ps}                                      &  184 &  252  & 1.37  &  3.19 \tabularnewline
\hline
225 & CDF Run-1 electron  $A_{ch}$, $p_{T\ell}>25$ GeV & \cite{Abe:1996us}                                &  11  &  8.9  & 0.81   &  -0.32 \tabularnewline
\hline
227 & CDF Run-2 electron $A_{ch}$, $p_{T\ell}>25$ GeV & \cite{Acosta:2005ud}                               &  11  &  14   & 1.24   &  0.67 \tabularnewline
\hline
234 & D\O~ Run-2 muon $A_{ch}$, $p_{T\ell}>20$ GeV  & \cite{Abazov:2007pm}                                 &  9   &  8.3  & 0.92    &  -0.02 \tabularnewline
\hline
240 & LHCb 7 TeV $35\mbox{ pb}^{-1}$ $W/Z$ $d\sigma/dy_{\ell}$ & \cite{Aaij:2012vn}    &  14  &  9.9  & 0.71     &  -0.73 \tabularnewline
\hline
241 & LHCb 7 TeV $35\mbox{ pb}^{-1}$ $A_{ch}$, $p_{T\ell}>20$ GeV& \cite{Aaij:2012vn}  &  5   &  5.3  & 1.06      &  0.30 \tabularnewline
\hline
260 & D\O~ Run-2 $Z$ rapidity & \cite{Abazov:2006gs}                                   &  28  &  17   & 0.59   &  -1.71 \tabularnewline
\hline
261 & CDF Run-2 $Z$ rapidity & \cite{Aaltonen:2010zza}                                 &  29  &  48   & 1.64   &  2.13 \tabularnewline
\hline
266 & CMS 7 TeV $4.7\mbox{ fb}^{-1}$, muon $A_{ch}$, $p_{T\ell}>35$ GeV& \cite{Chatrchyan:2013mza}         &  11  &  12.1  & 1.10      &  0.37 \tabularnewline
\hline
267 & CMS 7 TeV $840\mbox{ pb}^{-1}$, electron $A_{ch}$, $p_{T\ell}>35$ GeV & \cite{Chatrchyan:2012xt}     &  11  &  10.1  & 0.92     &  -0.06 \tabularnewline
\hline
268 & ATLAS 7 TeV $35\mbox{ pb}^{-1}$ $W/Z$ cross sec., $A_{ch}$ & \cite{Aad:2011dm}   &  41  &  51   & 1.25     &  1.11 \tabularnewline
\hline
281 & D\O~ Run-2 $9.7 \mbox{ fb}^{-1}$ electron $A_{ch}$, $p_{T\ell}>25$ GeV & \cite{D0:2014kma}               &  13  &  35   & 2.67     &  3.11 \tabularnewline
\hline
504 & CDF Run-2 inclusive jet production & \cite{Aaltonen:2008eq}                      &  72  &  105  & 1.45   &  2.45 \tabularnewline
\hline
514 & D\O~ Run-2 inclusive jet production & \cite{Abazov:2008ae}                       &  110 &  120  & 1.09   &  0.67 \tabularnewline
\hline
535 & ATLAS 7 TeV $35\mbox{ pb}^{-1}$ incl. jet production & \cite{Aad:2011fc}         &  90  &  50   & 0.55    &  -3.59 \tabularnewline
\hline
538 & CMS 7 TeV $5\mbox{ fb}^{-1}$ incl. jet production  & \cite{Chatrchyan:2012bja}   &  133 &  177  & 1.33    &  2.51 \tabularnewline
\hline
\hline
\end{tabular}
\caption{Same as Table~\ref{tab:EXP_1}, showing
experimental data sets on Drell-Yan processes and inclusive jet production.
\label{tab:EXP_2} }
\end{table}

The goodness-of-fit for CT14 NNLO is comparable to that of our earlier PDFs, but
the more flexible parametrizations did result in improved agreement
with some data sets. For example, by adding additional parameters
to the $\left\{u,\overline{u}\right\}$ and $\left\{
d, \overline{d} \right\}$ parton distributions,
somewhat better agreement was obtained for the BCDMS and NMC data
at low values of $Q$. The quality of the fit can be also evaluated
based on the distribution of $S_n$ values, which follows a standard
normal distribution (of width 1) in an ideal fit. As in the previous
fits, the actual $S_n$ distribution
(cf. the solid curve in Fig.~\ref{fig:Sns}) is
somewhat wider than the standard normal one (the dashed curve),
indicating the presence of
disagreements, or tensions, between some of the included experiments. The
tensions have been examined before \cite{Lai:2010vv, Collins:2001es,
Pumplin:2009sc, Pumplin:2009nm} and originate largely from experimental
issues, almost independent
of the perturbative QCD order or PDF parametrization form.
A more detailed discussion of the level of
agreement between data and theory will be
provided in Sec.~\ref{sec:TheoryVsData}.

\subsubsection{Experimental data from the LHC}
Much of these data have also been used in previous CT analyses,
such as the one that produced the CT10 NNLO PDFs.
As mentioned, no LHC data were used in the CT10 fits. Nonetheless, the CT10 PDFs have been
in good agreement with  LHC measurements so far.

As the quantity of the LHC data has increased, the time has come to include
the most germane LHC measurements into CT fits.
The LHC has measured a variety of standard model cross sections, yet not all
of them are suitable for determination of PDFs according to the CT
method. For that, we need to select measurements that are
experimentally and theoretically clean and are compatible with the
global set of non-LHC hadronic experiments.

In the CT14 study, we select a few such LHC data sets at $\sqrt{s}=7$\,TeV,
focusing on the measurements that provide novel information
to complement the non-LHC data. From vector boson production
processes, we selected $W/Z$ cross sections
and the charged lepton asymmetry measurement from ATLAS  \cite{Aad:2011dm},
the charged lepton asymmetry in the electron \cite{Chatrchyan:2012xt} and
muon decay channels \cite{Chatrchyan:2013mza} from CMS, and
the $W/Z$ lepton rapidity distributions and charged
lepton asymmetry from LHCb \cite{Aaij:2012vn}. The ATLAS and CMS
measurements primarily impose constraints on the light quark and antiquark PDFs at
$x \gtrsim 0.01$. The LHCb data sets,  while statistically limited,
impose minor constraints on $\bar u$ and $d$ PDFs at $x=0.05-0.1$.

Upon including these measurements, we can relax the parametric constraints on
the sea (anti-)quark PDFs of $u$, $\bar u$, $d$, and $\bar d$.
In the absence of relevant experimental constraints in
the pre-CT14 fits, the PDF parametrizations were chosen so as to
enforce $\bar u/\bar d \rightarrow 1$, $u/d\rightarrow 1$ at
$x\rightarrow 0$ in order to obtain convergent fits. As reviewed in the
Appendix, the CT14 parametrization
form is more flexible, in the sense that only the asymptotic power
$x^{a_1}$ is required to be the same in all light-quark PDFs in the
$x\rightarrow 0$ limit. This choice produces wider uncertainty bands
on $u_{v}$, $d_{v}$, and $\bar u/\bar d$ at $x\rightarrow 0$, with the
spread constrained by the newly included LHC data.

From the other LHC measurements, we now
include single-inclusive jet production at ATLAS \cite{Aad:2011fc}
and CMS \cite{Chatrchyan:2012bja}. These data sets provide
complementary information to Tevatron inclusive jet production cross sections
from CDF Run-2 \cite{Aaltonen:2008eq} and D\O~ Run-2 \cite{Abazov:2008ae}
that are also included. The purpose of jet production cross sections
is primarily to constrain the gluon PDF $g(x,Q)$.
While the uncertainties from the LHC jet cross sections are still
quite large, they probe the gluon PDF across a much wider range of $x$
than the Tevatron jet cross sections.

One way to gauge the sensitivity
of a specific data point to some PDF $f(x,Q)$ at a given $x$ and $Q$
is to compute a correlation cosine between the theoretical
prediction for this point and $f(x,Q)$
\cite{Nadolsky:2008zw, Pumplin:2001ct, Nadolsky:2001yg}.
In the case of CT10 NNLO, the sensitivity of the LHC charge asymmetry data sets
to the valence PDF combinations at $x=0.01-0.1$ was established by this
method in Sec. 7C of \cite{Gao:2013xoa}. However, the somewhat large
strength of correlations at small $x$ that had been observed suggested the
possibility that CT10 light-quark parametrizations were not
sufficiently flexible in the $x$ region probed by the LHC charge
asymmetry.

\begin{figure}[p]
\includegraphics[width=0.49\textwidth]{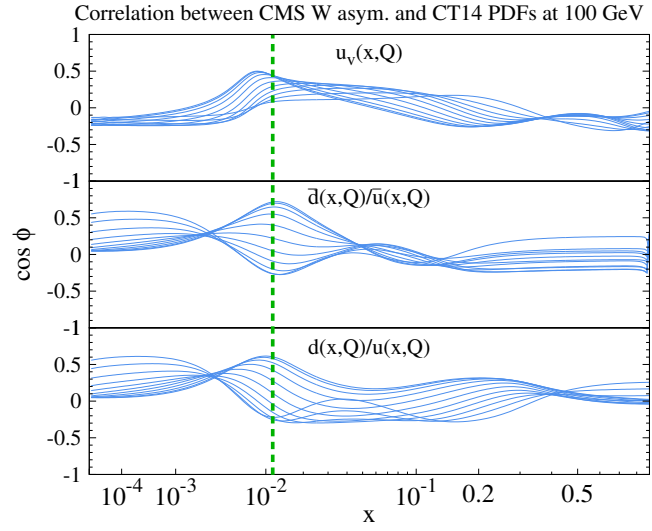}
\caption{The correlation cosine $\cos\phi$ \cite{Nadolsky:2008zw}
between the PDF $f(x,Q=\mbox{100 GeV})$ at the specified
$x$ value on the horizontal axis and
NNLO predictions for muon CMS charge asymmetry
\cite{Chatrchyan:2013mza}.
\label{fig:wasycosphi}}
\end{figure}

Since CT14 has adopted more flexible parametrizations for the
affected quark flavors, the above correlations with $u_v$, $d_v$, and
$d/u$ at small $x$ are now somewhat relaxed, as illustrated by the newly
computed correlations between CT14 NNLO and CMS $A_{ch}$ data
in Fig.~\ref{fig:wasycosphi}. Each line
shows $\cos\phi$ between $f(x,Q)$ and the NNLO prediction
for one of the bins of the data. When the PDF uncertainty
receives a large contribution from $f(x,Q)$, $\cos\phi$ comes out to
be close to $\pm 1$, say, $|\cos \phi| > 0.7$.
With the new parametrization
form, the CMS charge asymmetry is reasonably, but not exceptionally,
correlated with both
$\bar d/\bar u$ and $d/u$ at $x\sim 0.01$ corresponding to
central-rapidity production of weak bosons at $\sqrt{s}=7$ TeV
(indicated by a vertical dashed line in the figure). The correlation
with $u_v$ and $d_v$ is smaller than in CT10.

\begin{figure}[p]
\includegraphics[width=0.49\textwidth]{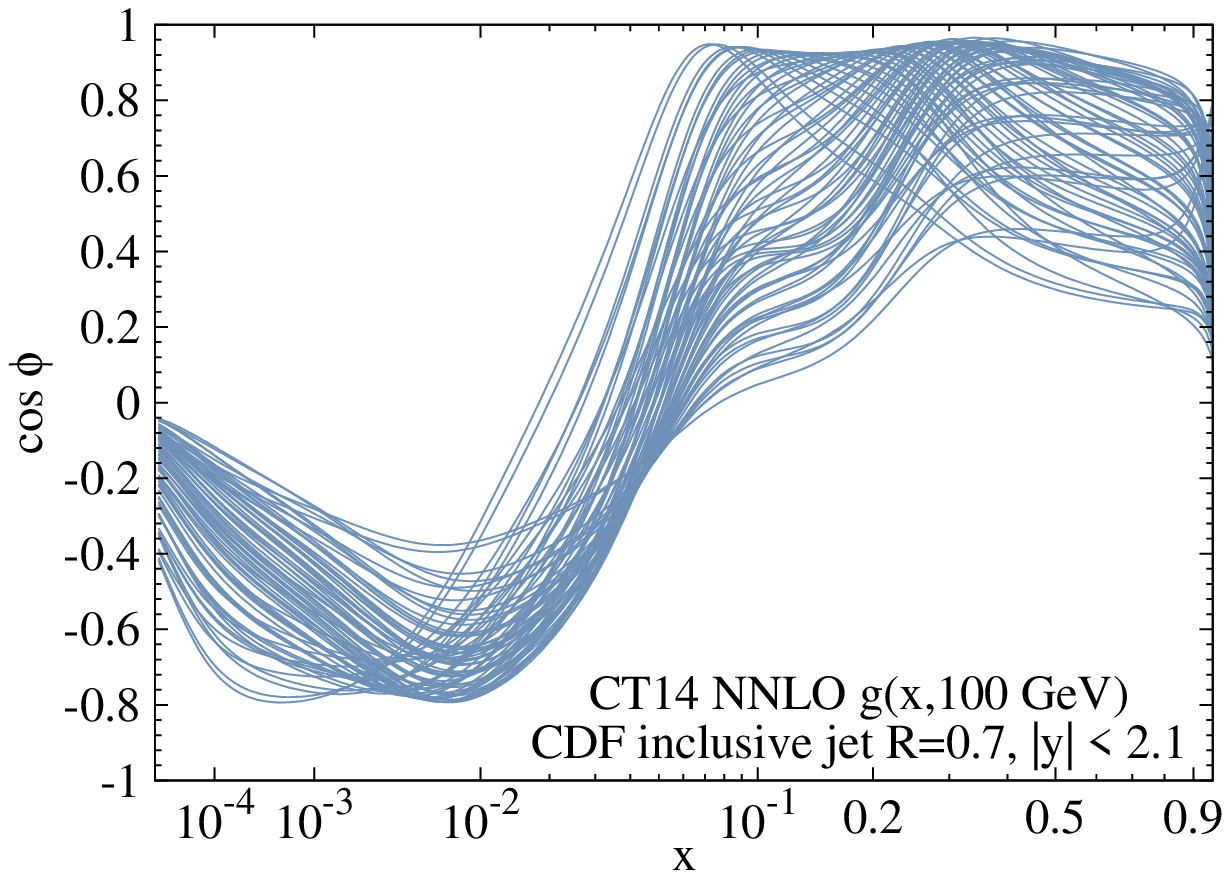}
\includegraphics[width=0.49\textwidth]{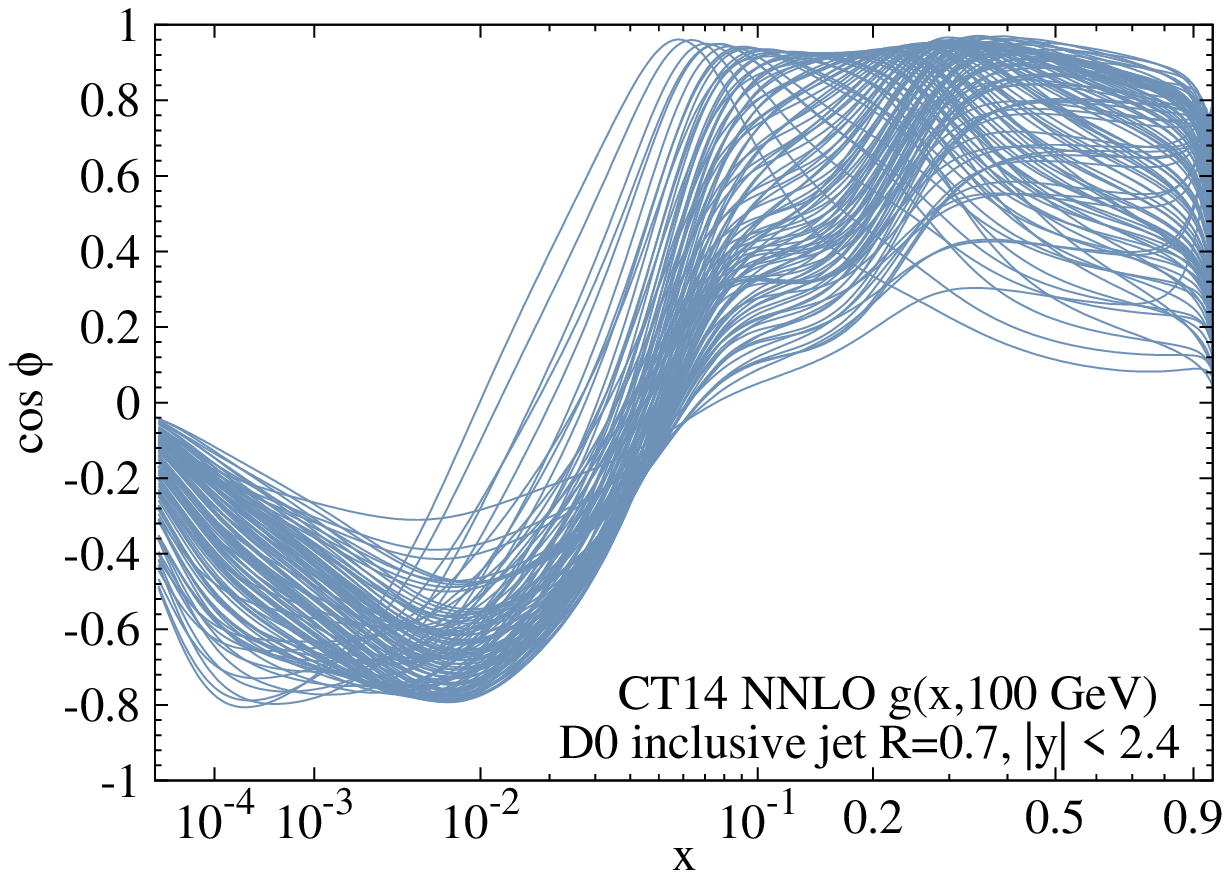}
\includegraphics[width=0.49\textwidth]{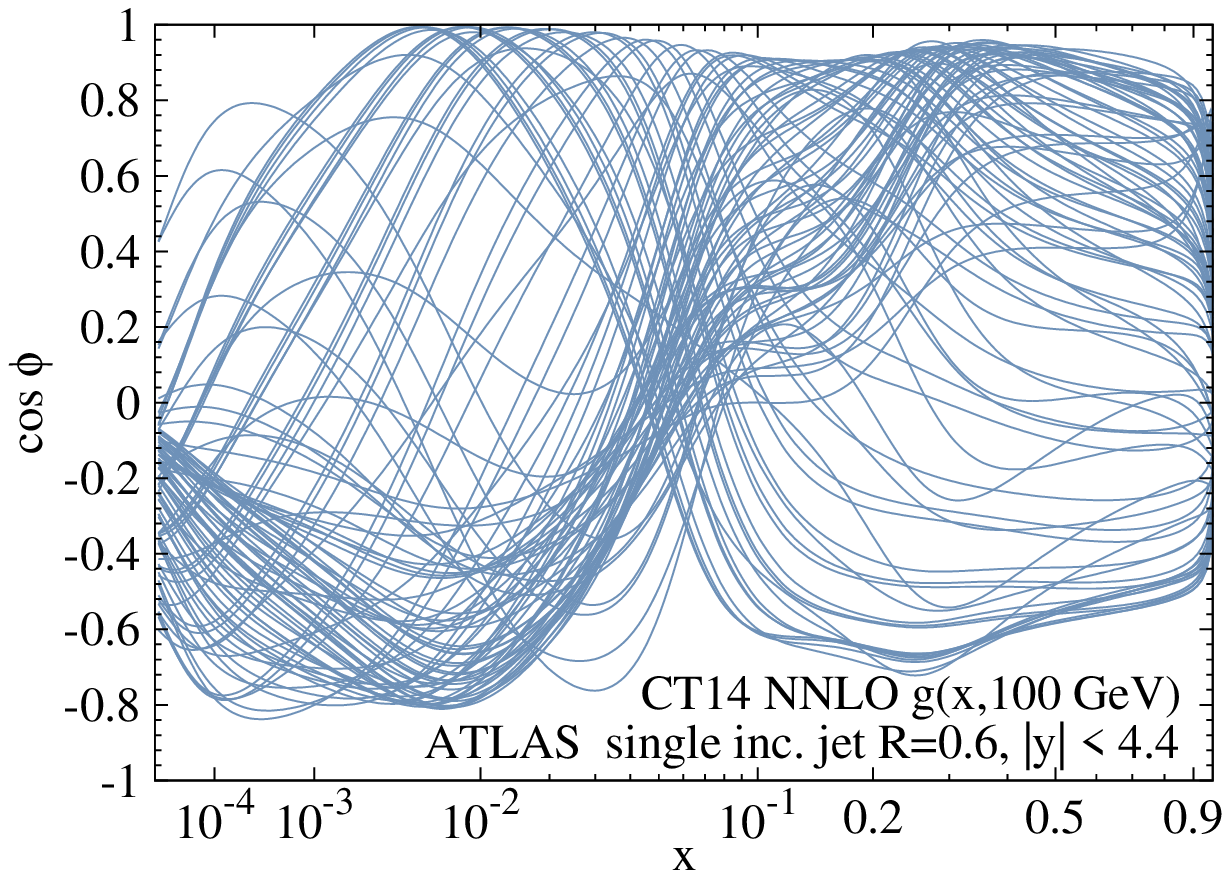}
\includegraphics[width=0.49\textwidth]{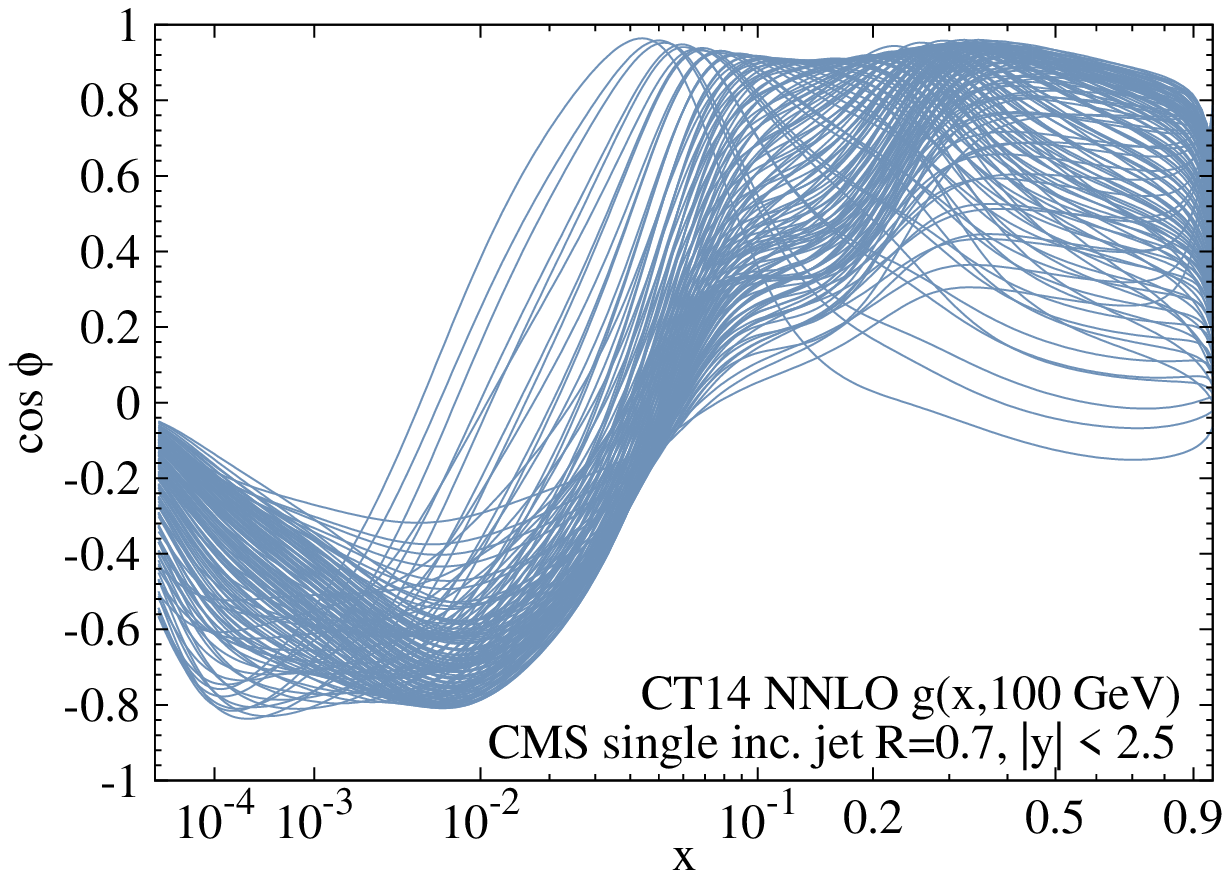}
\caption{The correlation cosine $\cos\phi$ \cite{Nadolsky:2008zw}
between the g-PDF at the specified $x$ value on the horizontal axis and
NLO predictions for the CDF \cite{Aaltonen:2008eq} (upper left panel), D\O~ \cite{Abazov:2008ae} (upper right panel),
 ATLAS \cite{Aad:2011fc} (lower left panel) and
CMS \cite{Chatrchyan:2012bja} (lower right panel) inclusive jet cross
sections at $Q=100$ GeV.
\label{fig:jetcosphi_g}}
\end{figure}

\begin{figure}[p]
\includegraphics[width=0.49\textwidth]{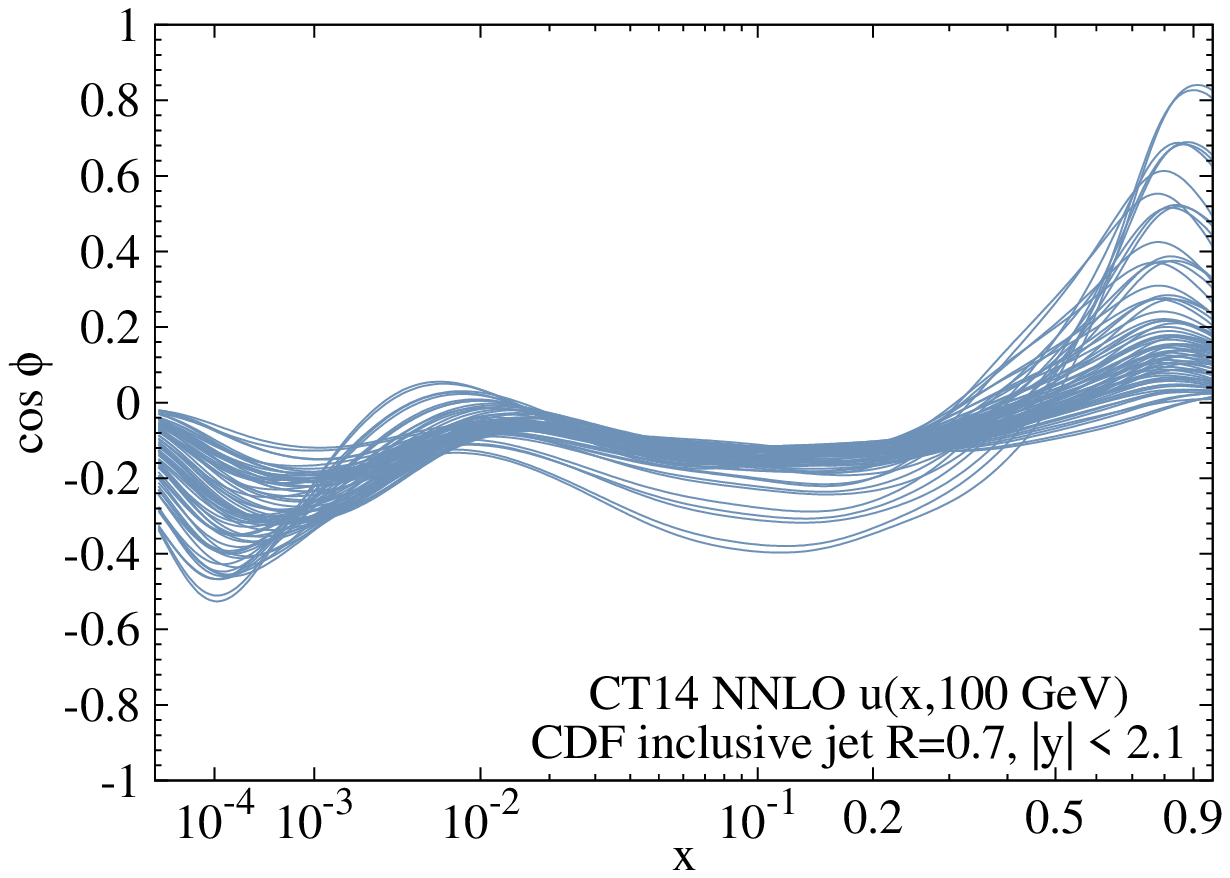}
\includegraphics[width=0.49\textwidth]{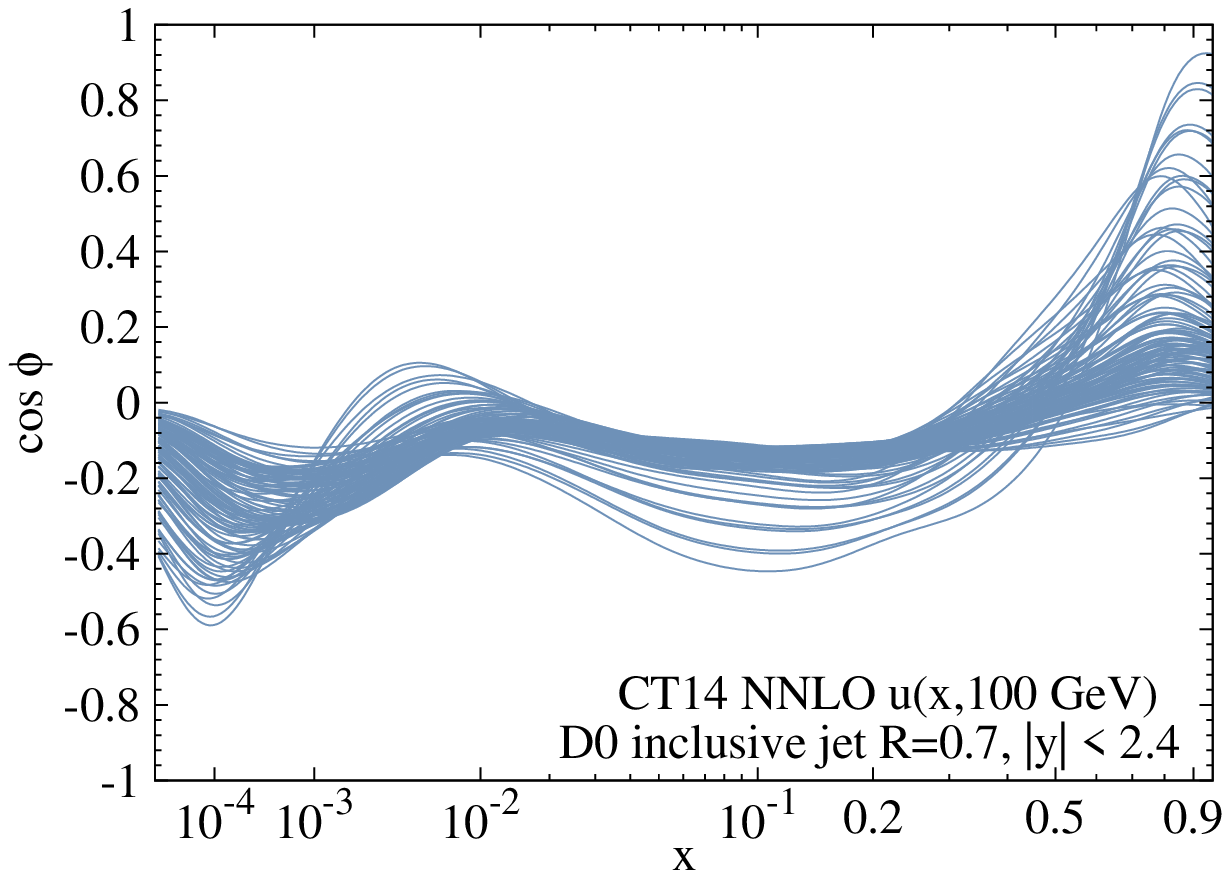}
\includegraphics[width=0.49\textwidth]{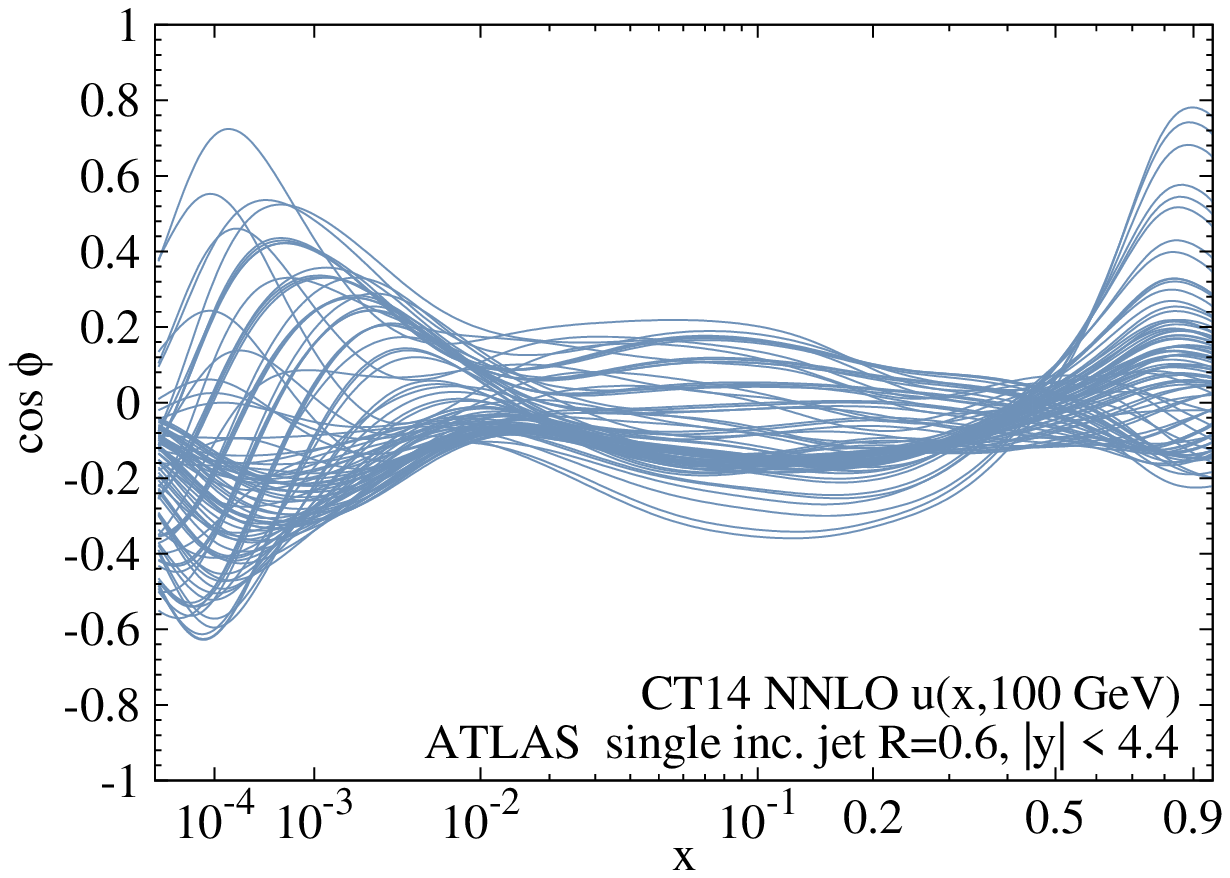}
\includegraphics[width=0.49\textwidth]{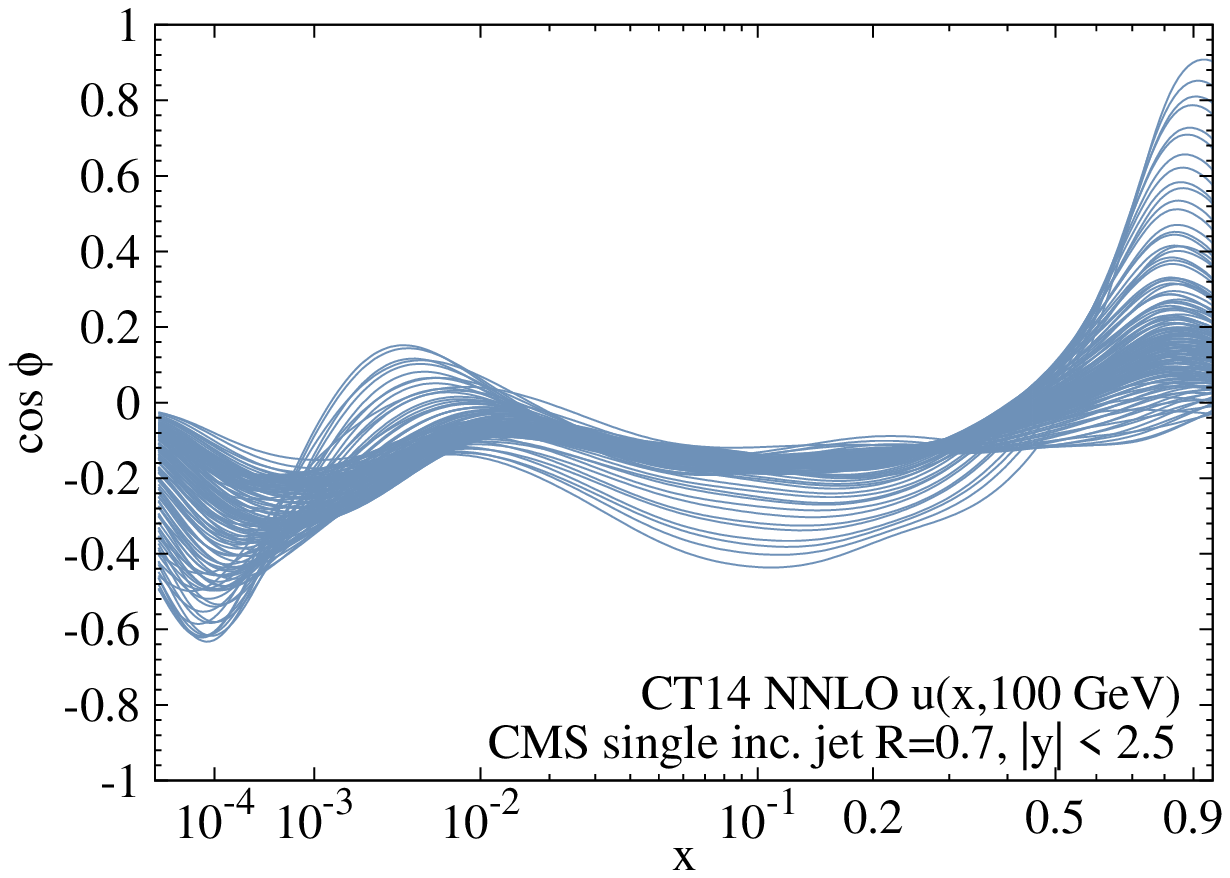}
\caption{The correlation cosine $\cos\phi$ \cite{Nadolsky:2008zw}
between the $u$-PDF at the specified $x$ value on the horizontal axis and
NLO predictions for the CDF \cite{Aaltonen:2008eq} (upper left panel), D\O~ \cite{Abazov:2008ae}  (upper right panel),
 ATLAS \cite{Aad:2011fc} (lower left panel) and
CMS \cite{Chatrchyan:2012bja}  (lower right panel) inclusive jet cross
sections at $Q=100$ GeV.
\label{fig:jetcosphi_u}}
\end{figure}

For the ATLAS \cite{Aad:2011fc}, CMS \cite{Chatrchyan:2012bja}, CDF \cite{Aaltonen:2008eq} and D\O~ \cite{Abazov:2008ae} inclusive jet
data sets, the correlation cosine, $\cos\phi$, for gluon PDF
is plotted in Fig.~\ref{fig:jetcosphi_g} using NLO QCD
theory to evaluate the theoretical cross section.
Again, the lines
correspond to individual $p_{Tj}$ bins of the data.
We observe that the CDF and D\O~  jet cross sections
are highly correlated with the
gluon PDF $g(x,Q)$ at $x\gtrsim 0.05$, and anticorrelated at small
$x$ as a consequence of the momentum sum rule. The ATLAS and CMS jet cross sections are highly correlated with $g(x,Q)$ in a much wider
range, $x>0.005$. In contrast, the PDF-induced correlation of the
jet cross sections with the quark PDFs, such as $u(x,Q)$ in the
Fig.~\ref{fig:jetcosphi_u}, is at most moderate.
The ATLAS and CMS jet data therefore
have the potential to reduce the gluon uncertainty, but
significant reduction will require the data from Run 2.

\subsubsection{High-luminosity lepton charge asymmetry from the Tevatron \label{sec:D0Wasy}}
Forward-backward asymmetry ($A_{ch}$)  distributions of charged leptons from
inclusive weak boson production at the Tevatron are uniquely sensitive
to the average slope of the ratio $d(x,Q)/u(x,Q)$ at large $x$, of
order 0.1 and above. In the CT14 analysis, we include several data sets of
$A_{ch}$ measured at $\sqrt{s} = $ 1.8 and 1.96 TeV by the CDF and
D\O~ Collaborations. The CDF Run-1 data set on
$A_{ch}$  \cite{Abe:1994rj,Abe:1996us}, which was instrumental in resolving
conflicting information on the large-$x$ behavior of $u(x,Q)$ and
$d(x,Q)$ from contemporary
fixed-target DIS experiments \cite{Berger:1988tu,Martin:1988aj,Lai:1994bb,Martin:1994kn}, is supplemented by the CDF Run-2 data set at $170\mbox{
pb}^{-1}$ \cite{Acosta:2005ud}. $A_{ch}$ data at $\sqrt{s}=1.96\mbox{
TeV}$ from D\O~ in the electron \cite{D0:2014kma}
and muon \cite{Abazov:2007pm}  decay channels, for $9.7\mbox{
fb}^{-1}$ and $0.3 \mbox{ fb}^{-1}$, are also included.
In all $A_{ch}$ data sets, we include subsamples with
the cuts on the transverse momentum $p_{T\ell}$ of the final-state
lepton specified in Table~\ref{tab:EXP_2}.

The electron data set ($9.7\mbox{ fb}^{-1}$) from D\O~ that we now
include replaces the $0.75\mbox{ fb}^{-1}$ counterpart
set \cite{Abazov:2008qv}, first included in CT10. This replacement has
an important impact on the determination of the large-$x$ quark PDFs;
thus, these new $A_{ch}$ data sets are perhaps the most challenging
and valuable among all that were added in CT14.

The D\O~ $A_{ch}$
data have small experimental errors, and hence push the
limits of the available theoretical calculations.
Relatively small differences in the average slope
(with respect to $x$) of the $d/u$ ratio in the
probed region can produce large variations in
$\chi^2_{n}$ for the Tevatron charge asymmetry \cite{Berger:1988tu,Martin:1988aj,Lai:1994bb}.
By varying the minimal selection cuts on $p_{T\ell}$ of the lepton, it
is possible to probe subtle features of the large-$x$ PDFs. For that,
understanding of the transverse momentum dependence in both experiment and
theory is necessary, which demands evaluation of transverse momentum resummation effects.

When the first Tevatron Run-2 $A_{ch}$ data sets were implemented in CT fits,
significant tensions were discovered between the electron and muon
channels, and even between different $p_{T\ell}$ bins within one decay channel.
The tensions prompted a detailed study in the CT10 analysis
\cite{Lai:2010vv}. The study found that various $p_{T\ell}$ bins of the electron and muon
asymmetries from D\O~ disagree with DIS experiments and among themselves.

In light of these unresolved tensions,
we published a CT10 PDF ensemble at NLO, which did
not include the D\O~ Run-2 $A_{ch}$ data and yielded a $d/u$ ratio
that was close to that ratio in CTEQ6.6 NLO.  An alternative
CT10W NLO ensemble was also constructed. It
included four $p_{T\ell}$ bins of that data and predicted a
harder $d/u$ behavior at $x\rightarrow 1$. When constructing the  counterpart
CT10 NNLO PDFs in \cite{Gao:2013xoa}, we
took an in-between path and included only
the two most inclusive $p_{T\ell}$ bins, one from the
electron \cite{Abazov:2008qv}, and one from the
muon \cite{Abazov:2007pm} samples. This choice still resulted in a
larger $d/u$ asymptotic value in CT10 NNLO than in CTEQ6.6.

The new $A_{ch}$ data for $9.7\mbox{ fb}^{-1}$ in the
 electron channel is
more compatible with the other global fit in the data that we included.
Therefore, CT14 includes the D\O~ $A_{ch}$ measurement in
 the muon channel with $p_{T\ell} > 20$ GeV \cite{Abazov:2007pm} and
 in the electron channel with $p_{T\ell} > 25$\,
 GeV \cite{D0:2014kma}. The replacement does not affect the general
 behavior of the PDFs, except that the CT14 $d/u$ ratio at high $x$
follows the trends of CTEQ6.6 NLO and CT10 NLO,
rather than of CT10W NLO and CT10 NNLO.

\subsubsection{New HERA data}
CT14 includes a combined HERA-1 data set of
reduced cross sections for semi-inclusive DIS production of open charm
\cite{Abramowicz:1900rp}, and measurements of the longitudinal
structure function $F_L(x,Q)$ in neutral-current DIS \cite{Collaboration:2010ry}.
The former replaces independent data sets of charm structure
functions and reduced cross sections from H1 and ZEUS \cite{Aaron:2011gp,
  Breitweg:1999ad, Chekanov:2003rb, Aktas:2005iw}. Using the combined
HERA charm data set, we obtain a slightly smaller uncertainty on the
gluon at $x < 0.01$ and better constraints on charm mass than with
independent sets \cite{Gao:2013wwa}.
The latter HERA
data set, on $F_L$, is not independent from the combined HERA set on
inclusive DIS \cite{Aaron:2009aa}, but has only nine data points and
does not significantly change the global $\chi^2$. Its utility is
primarily to prevent unphysical solutions for the gluon PDF at small
$x$ at the stage of the PDF error analysis.

\subsubsection{Other LHC results}

One class of LHC data that could potentially play a large role \cite{Nadolsky:2008zw}
in the determination of the gluon distribution, especially at high
$x$, is the differential distributions of $t\bar{t}$ production, now
available from ATLAS~\cite{Aad:2014zka} and CMS~\cite{Chatrchyan:2012saa,Khachatryan:2015oqa}.
However, these data are not included into our fit, as the
differential NNLO $t\bar{t}$ cross section predictions for the LHC are
not yet complete and the total cross section measurements lack statistical power.~\cite{Mitov:DIS2015}. In addition, constraints on the
PDFs from $t\bar t$ cross sections are mutually correlated with the
values of QCD coupling and top quark mass. NLO electroweak
corrections, playing an important
role~\cite{Kuhn:2013zoa,Kuhn:2006vh} for
these data, are still unavailable for some $t\bar t$ kinematic
distributions. Once these calculations are completed, they will be
incorporated in future versions of CT PDFs. For now, we simply
show predictions from CT14 for the $t\bar{t}$
distributions using the approximate NNLO calculations
in Section~\ref{sec:LHCPredictions}.

\subsection{Summary of theoretical calculations}
\subsubsection{QCD cross sections}
The CT14 global analysis prioritizes the selection of
 published data for which NNLO predictions are available,
and theoretical uncertainties of various kinds
are well understood. Theoretical calculations for neutral-current DIS are based on the NNLO
implementation \cite{Guzzi:2011ew} of the S-ACOT-$\chi$ factorization
scheme \cite{Aivazis:1993pi,Collins:1998rz,Tung:2001mv} with massive quarks.
For inclusive distributions in
the low-mass Drell-Yan process, NNLO predictions are obtained with
the program VRAP \cite{Anastasiou:2003yy,Anastasiou:2003ds}.
Predictions for $W/Z$ production and
weak boson charge asymmetries with $p_{T\ell}$ cuts
are obtained with the NNLL-(approx. NNLO) program
ResBos \cite{Balazs:1995nz,Balazs:1997xd,Landry:2002ix,Guzzi:2013aja}, as in the previous
analyses.

As already mentioned in the introduction,
two exceptions from this general rule concern
 charged-current DIS and collider jet production. Both have
 unique sensitivities to crucial PDF combinations, but are still known only to
 NLO. The CCFR and NuTeV data on inclusive
 and semi-inclusive charge-current DIS are indispensable for
 constraining the strangeness PDF; single-inclusive jet
 production at the Tevatron and now at the LHC are essential for
 constraining the gluon distribution.
Yet, in both categories, the experimental uncertainties
 are fairly large and arguably diminish the impact of missing
 NNLO effects. Given the importance of these measurements,
our approach is then to include these data in our NNLO global PDF
fits, but evaluate their matrix elements at NLO.

According to this choice, we do not rely on the use of threshold
resummmation
techniques~\cite{Kidonakis:2000gi,Kumar:2013hia,Kumar:2013hia}
 to approximate the NNLO
corrections in jet production. Nor do we remove the LHC jet data due to the
kinematic limitations of such resummation techniques \cite{Ball:2014uwa}. A large
effort was invested in the CT10 and CT14 analyses
to estimate the possibility of biases
in the NNLO PDFs due to using NLO cross sections
for jet production \cite{Gao:2012he,Ball:2012wy}.
The sensitivity of the central PDFs and their uncertainty to plausible
NNLO corrections was estimated with a variation of Cacciari-Houdeau's
method \cite{Cacciari:2011ze}, by
introducing additional correlated systematic errors in jet production
associated with the residual dependence on QCD scales and a potential
missing contribution of a typical magnitude expected from an NNLO
correction.
These exercises produced two conclusions. First, the scale
variation in the NLO jet cross section
is reduced if the central renormalization and factorization scales are
set equal to the transverse momentum $p_T$
of the individual jet in the data bin. This
choice is adopted both for the LHC and Tevatron
jet cross sections. In the recently completed partial NNLO calculation
for jets produced via $gg$ scattering
\cite{Currie:2013dwa,Ridder:2013mf},
this scale choice leads to an NNLO/NLO K-factor that is both smaller than
for the alternative scale equal to the leading jet's $p_T$, and
is relatively constant over the range
of the LHC jet measurements \cite{Carrazza:2014hra}.
Second, the plausible effect of the residual QCD scale
dependence at NLO can be estimated as a correlated uncertainty in the
CT10 NNLO fit. Currently it has marginal effect
on the central PDF fits and the PDF uncertainty.

The CT14 analysis computes NLO cross sections for inclusive jet
production with the help of
{\sc FastNLO}~\cite{Wobisch:2011ij} and {\sc ApplGrid}~\cite{Carli:2010rw} interfaces to {\sc NLOJET++} \cite{Nagy:2001fj,Nagy:2003tz} .
A series of benchmarking exercises  that we had completed
\cite{Gao:2012he,Ball:2012wy} verified that the fast interfaces are in
good agreement among themselves and with an independent NLO calculation
in the program MEKS \cite{Gao:2012he}. Both ATLAS and CMS have measured the inclusive jet cross sections for two jet sizes. We use the larger of the two sizes (0.6 for ATLAS and 0.7 for CMS) to further reduce the importance of NNLO corrections.

\subsubsection{Figure-of-merit function \label{sec:FigureOfMerit}}
In accord with the general procedure summarized in Ref.~\cite{Gao:2013xoa},
the most probable solutions for CT14 PDFs are found by a minimization of the
function
\begin{equation}
\chi^2_{global} = \sum_{n=1}^{N_{exp}} \chi^2_n + \chi^2_{th}.
\end{equation}
This function
sums contributions $\chi^2_n$ from $N_{exp}$ fitted experiments and includes a
contribution $\chi^2_{th}$ specifying theoretical conditions
(``Lagrange Multiplier
constraints'') imposed on some PDF parameters.
In turn, the $\chi^2_n$ are constructed as in Eq. (14) of \cite{Gao:2013xoa} and account for both uncorrelated and correlated experimental errors.
Section 3 of that paper includes a detailed review
of the statistical procedure that we continue to follow.
Instead of repeating that review,
we shall briefly remind the reader about the usage of
the tolerance and quasi-Gaussian $S$ variables when constructing
the error PDFs.

The minimum of the $\chi_{global}^2$ function is found iteratively
by the method of steepest descent using the program {\tt MINUIT}.
The boundaries of the 90\% C.L. region around the minimum of $\chi^2_{global}$,
and the eigenvector PDF sets quantifying the associated uncertainty,
are found by iterative diagonalization of the Hessian matrix
\cite{Stump:2001gu,Pumplin:2001ct}. The 90\% C.L. boundary in CT14 and CT10
analyses is determined according to two tiers of
criteria, based on
the increase in the global $\chi^2_{global}$ summed over all experiments,
and on the agreement with individual experimental data sets \cite{Lai:2010vv,Gao:2013xoa,Dulat:2013hea}.
The first type of condition demands that the global $\chi^2$ does not increase
above the best-fit value by more than $\Delta\chi^2=T^2$, where
the 90\% C.L. region corresponds to $T \approx 10$.
The second condition introduces a penalty term $P$,
called Tier-2 penalty, in $\chi^2$
when establishing the confidence region,
which quickly grows when the fit ceases to agree with
any specific experiment within the 90\% C.L. for that experiment.
The effective function $\chi^2_{\mathrm eff}=\chi^2_{global}+P$
is scanned along each eigenvector direction until $\chi^2_{eff}$
increases above the tolerance bound,
or rapid $\chi^2_{eff}$ growth due to the penalty $P$ is triggered.

The penalty term is constructed as
\begin{equation}
  P = \sum_{n=1}^{N_{exp}} (S_n)^k \theta(S_n)
\end{equation}
from the equivalent Gaussian variables $S_n$ that
obey an approximate standard normal distribution
independently of the number of data points $N_{pt,n}$ in the experiment.
Every $S_{n}$ is a monotonically increasing
function of the respective $\chi^2_n$ given in \cite{Dulat:2013hea,Lewis:1988}.
The power $k=16$ is chosen so that $(S_n)^k$ sharply increases
from zero when $S_n$ approaches 1.3, the value corresponding
to the 90\% C.L. cutoff. The implementation of
$S_n$ is fully documented in the appendix of Ref.~\cite{Dulat:2013hea}.

\subsubsection{Correlated systematic errors \label{sec:CorrErr}}
In many of the data sets included in the CT14 analysis, the reported correlated systematic errors
from experimental sources dominate over the statistical errors. Care
must therefore be taken in the treatment of these
systematic errors to avoid artificial biases in the best-fit outcomes, such as the
bias described by D'Agostini in \cite{D'Agostini:1993uj,D'Agostini:1999cm}.

Our procedure for handling the systematic errors is reviewed in Secs. 3C and 6D of
\cite{Gao:2013xoa}; see also a related discussion in the appendices
of \cite{Pumplin:2002vw} and \cite{Ball:2012wy}. The
correlated errors for a given experiment, and effective shifts in the theory or data that they cause, are estimated in a linearized approximation by
including a contribution in the figure-of-merit function $\chi^2$
proportional to the correlation matrix.
A practical
implementation of this approach runs into a dilemma of distinguishing
between the additive and multiplicative correlated
errors, which are often not separated in the experimental publications, but must
follow different prescriptions to prevent the bias.  It is the matrix
$\beta_{i,\alpha}$ of {\it relative} correlated errors that is typically
published; the {\it absolute} correlated errors
must be reconstructed from $\beta_{i,\alpha}$ by
following the prescription for either the additive or multiplicative
type.

In inclusive jet production, the choice between the additive and multiplicative treatments
modifies the large-$x$ behavior of the gluon PDF. This has been
studied in the CT10 NNLO analysis, cf. Sec. 6D of
\cite{Gao:2013xoa}. In general, the dominant sources of
systematic error, especially at the Tevatron and LHC, should be treated as multiplicative rather than
additive; that is, by assuming that the relative systematic error
corresponds to a fixed
fraction of the theoretical value, and not of the central data value.
The final CT14 PDFs were derived under this assumption, by
treating the systematic errors as multiplicative in all
experiments.\footnote{According to terminology adopted in
  Refs.~\cite{Ball:2012wy,Gao:2013xoa}, CT14 implements the correlated
  errors according to the ``extended $T$'' prescription for all experiments,
i.e., by normalizing the relative correlated errors by the current
theoretical value in each iteration of the fit.} Of course, this is just one option on the table: alternative candidate fits
of the CT14 family were also performed, by treating some correlated
errors as additive. They produced the PDFs that generally lie within the
quoted uncertainty ranges, as in the previous exercise
documented in \cite{Gao:2013xoa}.

\section{Overview of CT14 PDFs as functions of $x$ and $Q$ \label{sec:OverviewCT14}}

\begin{figure}[htbp]
\includegraphics[width=0.49\textwidth]{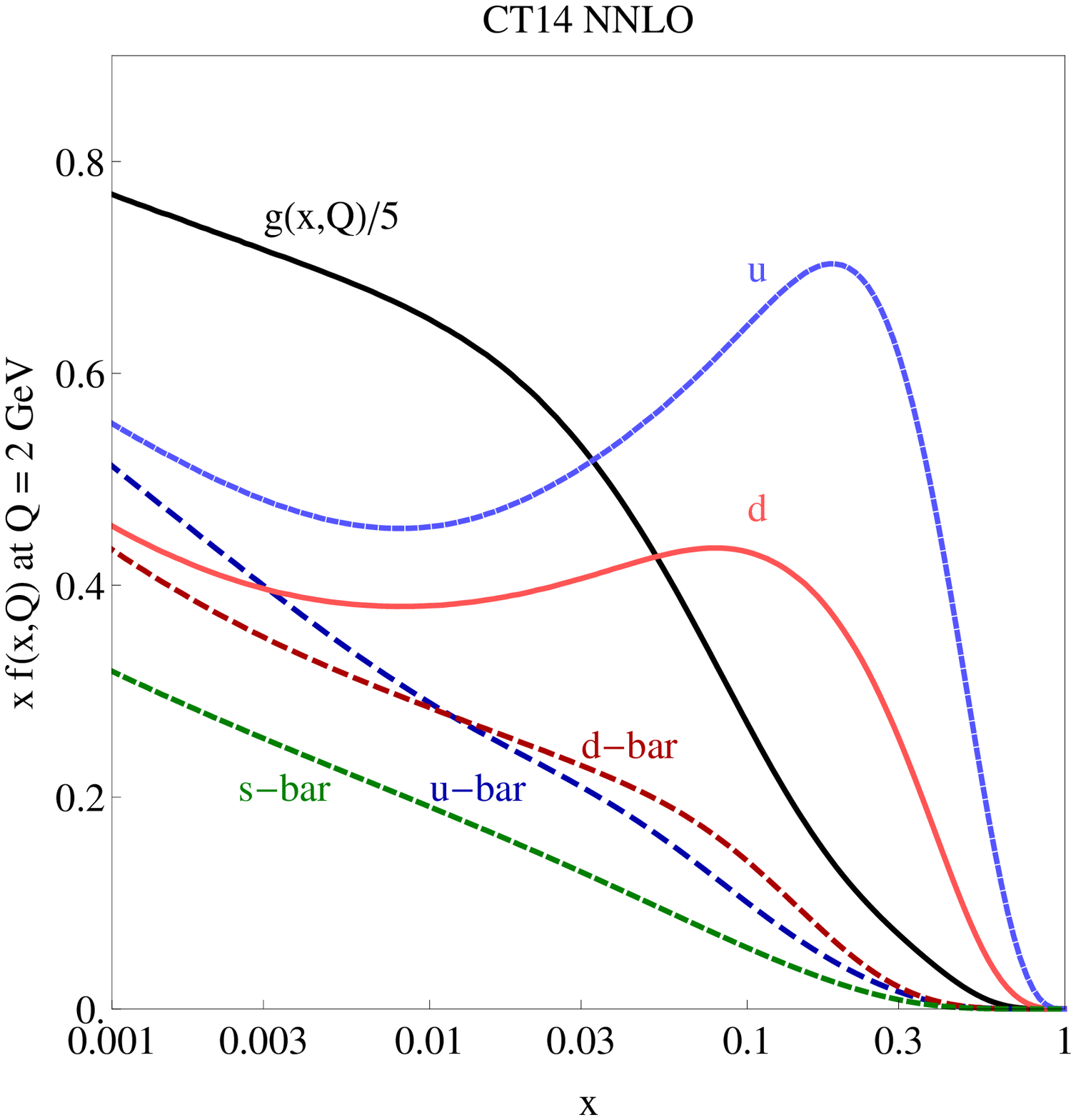}
\includegraphics[width=0.49\textwidth]{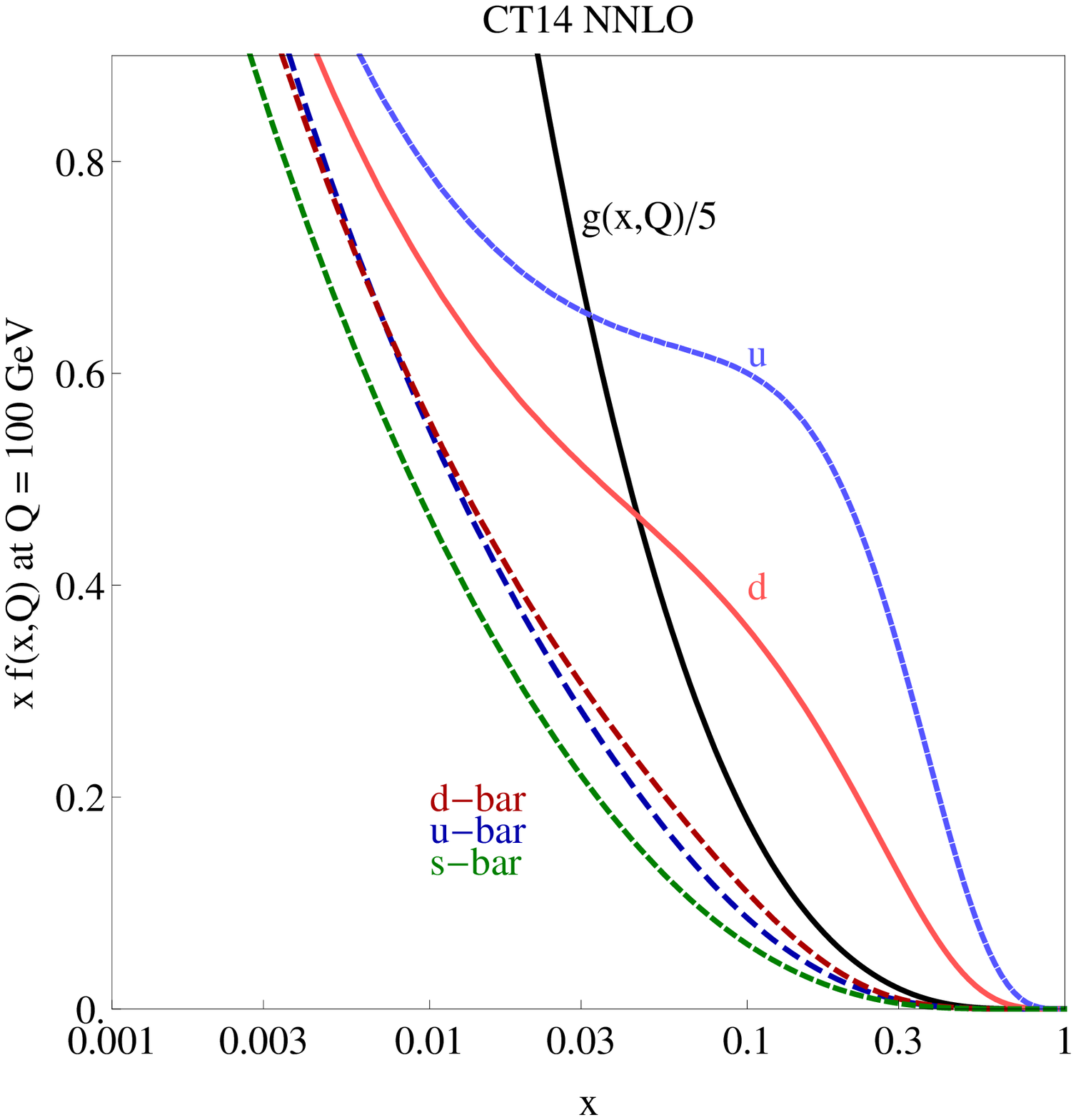}
\caption{The CT14 parton distribution functions at  $Q = 2$ GeV and $Q
  = 100$ GeV for $u, \overline{u}, d, \overline{d}, s = \overline{s}$,
  and $g$.
\label{fig:ct14pdf}}
\end{figure}

Figure \ref{fig:ct14pdf} shows an overview of the CT14 parton
distribution functions, for $Q = 2$ and $100$ GeV.
The function $x f(x,Q)$ is plotted versus $x$, for flavors $u,
\overline{u}, d, \overline{d}, s = \overline{s}$, and $g$.
We assume $s(x,Q_0)=\bar s(x,Q_0)$, since their difference is
consistent with zero and has large uncertainty \cite{Lai:2007dq}.
The plots show the {\em central fit} to the {\em global data} listed
in Tables~\ref{tab:EXP_1} and \ref{tab:EXP_2}, corresponding to the
lowest total $\chi^2$ for our choice of PDF parametrizations.

\begin{figure}[tb]
\center
\includegraphics[width=0.43\textwidth]{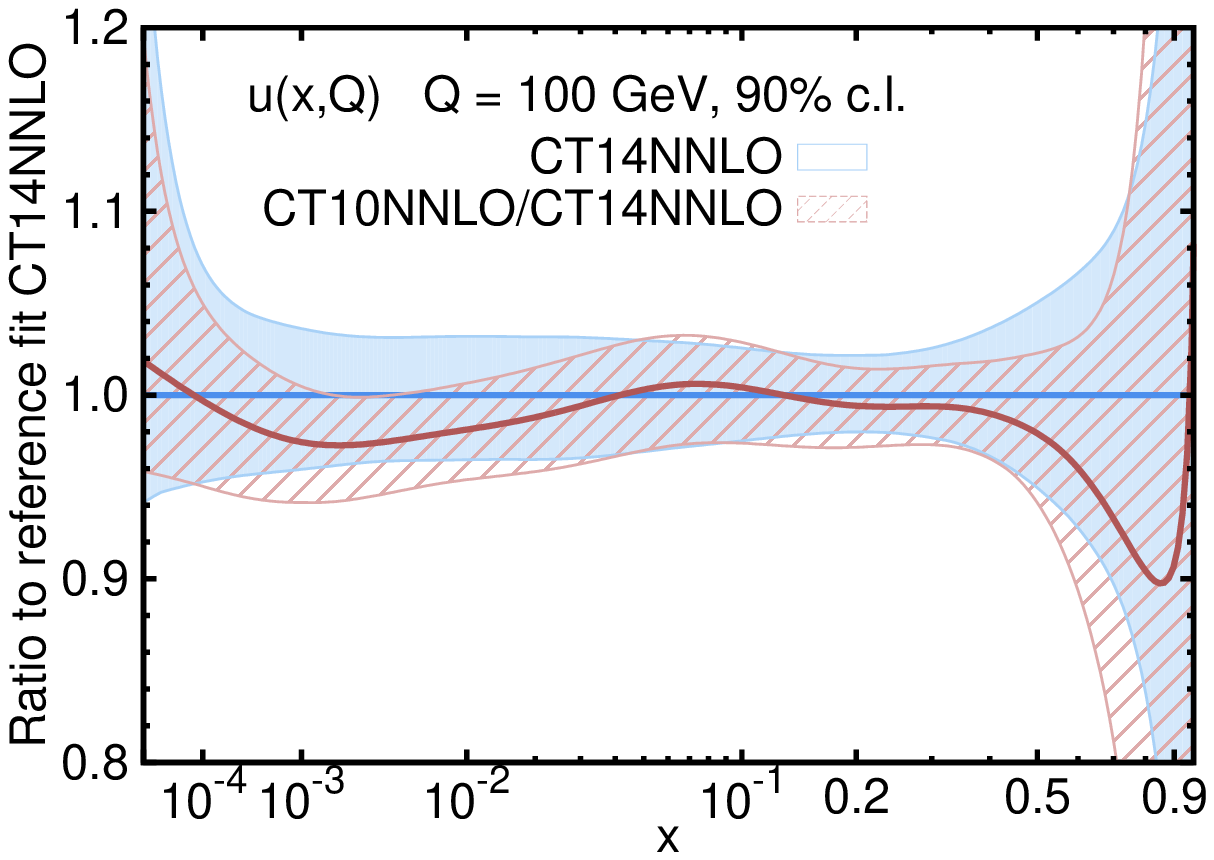}
\includegraphics[width=0.43\textwidth]{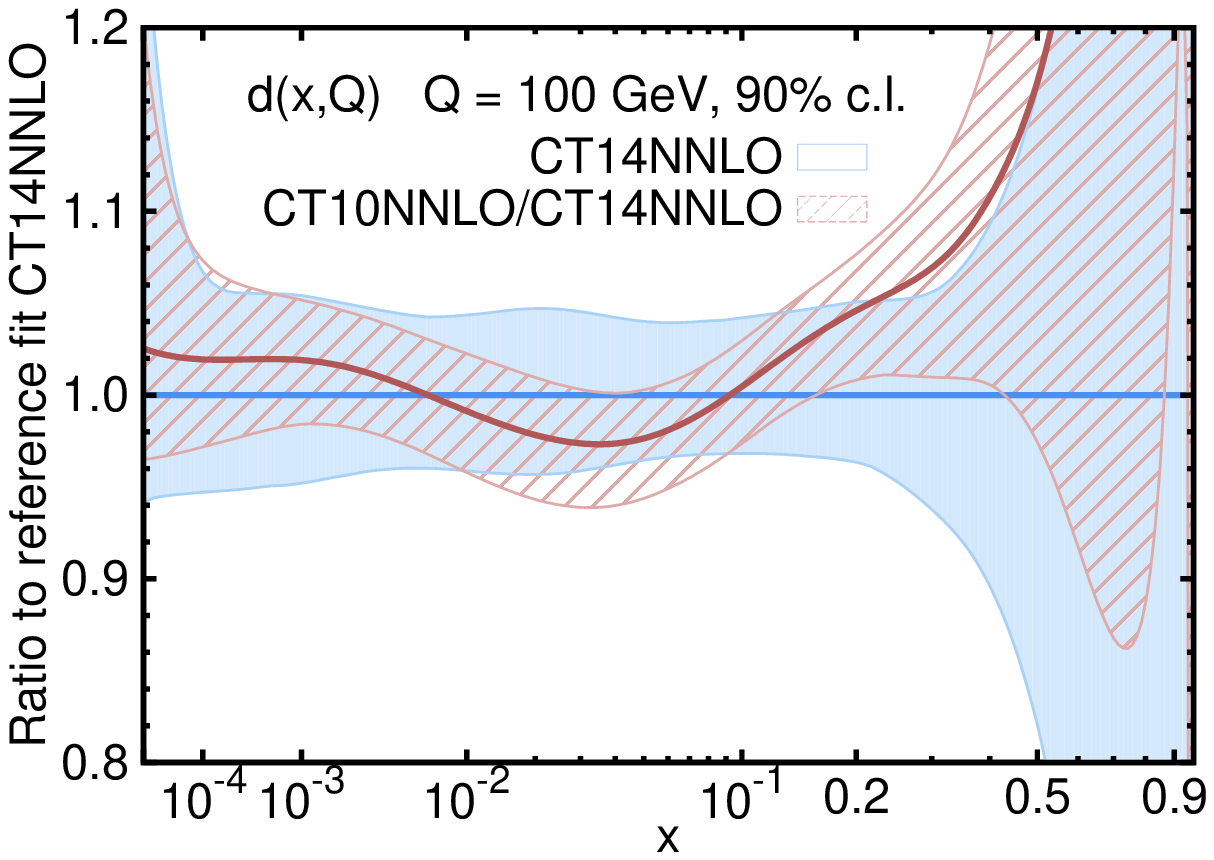}
\includegraphics[width=0.43\textwidth]{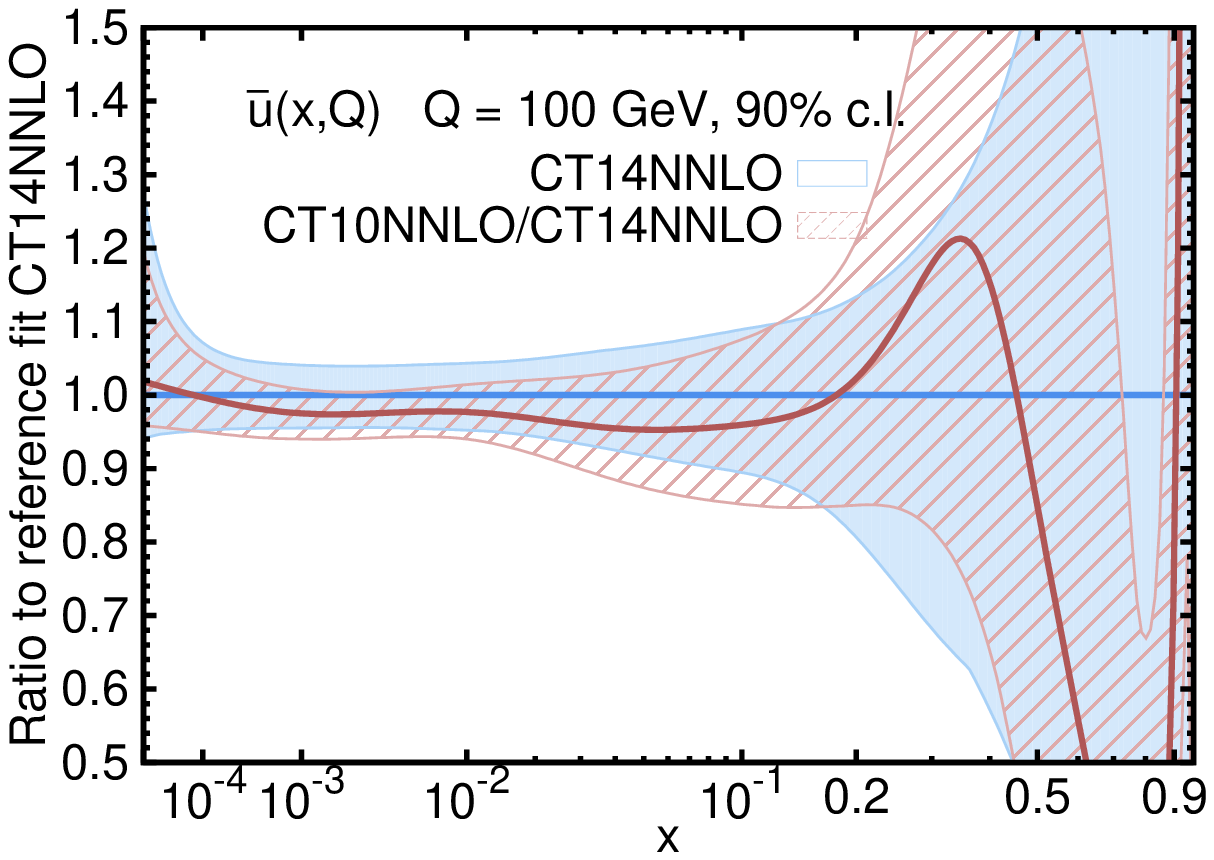}
\includegraphics[width=0.43\textwidth]{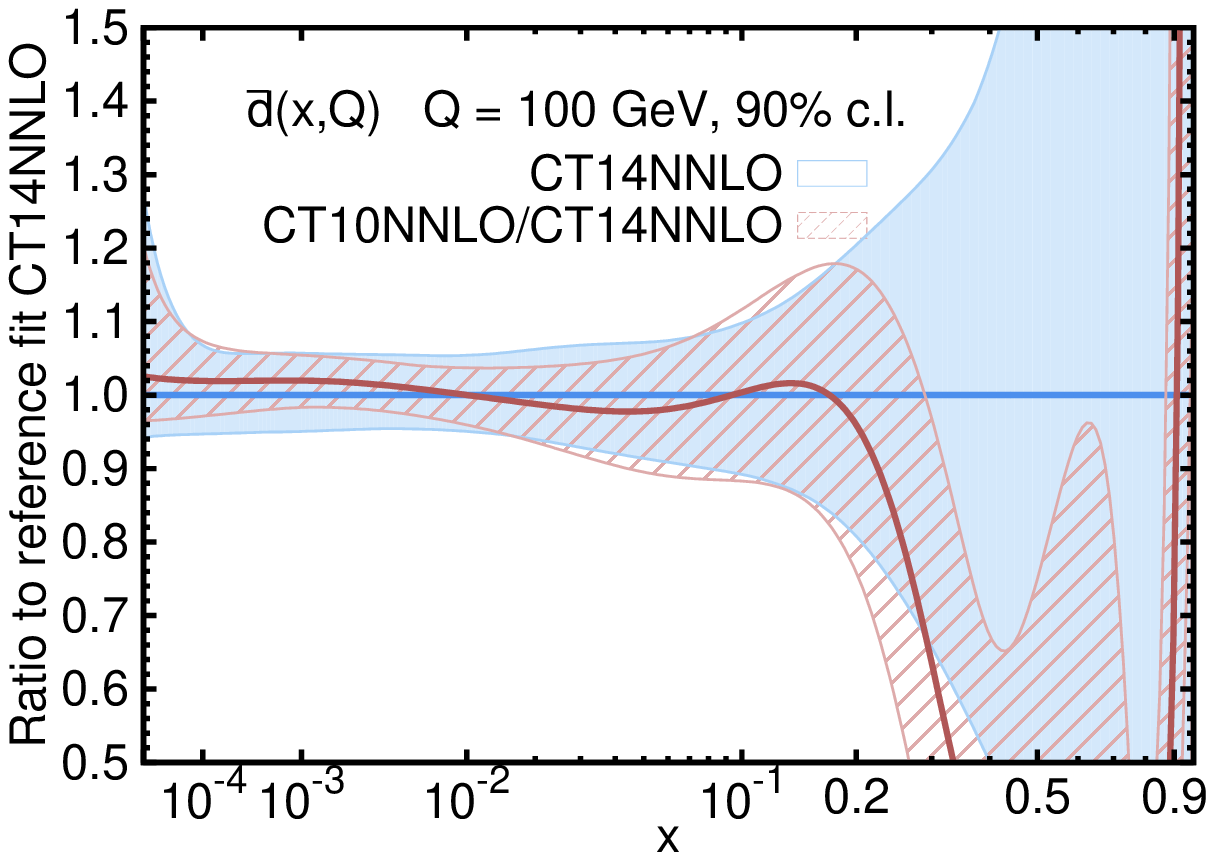}
\includegraphics[width=0.43\textwidth]{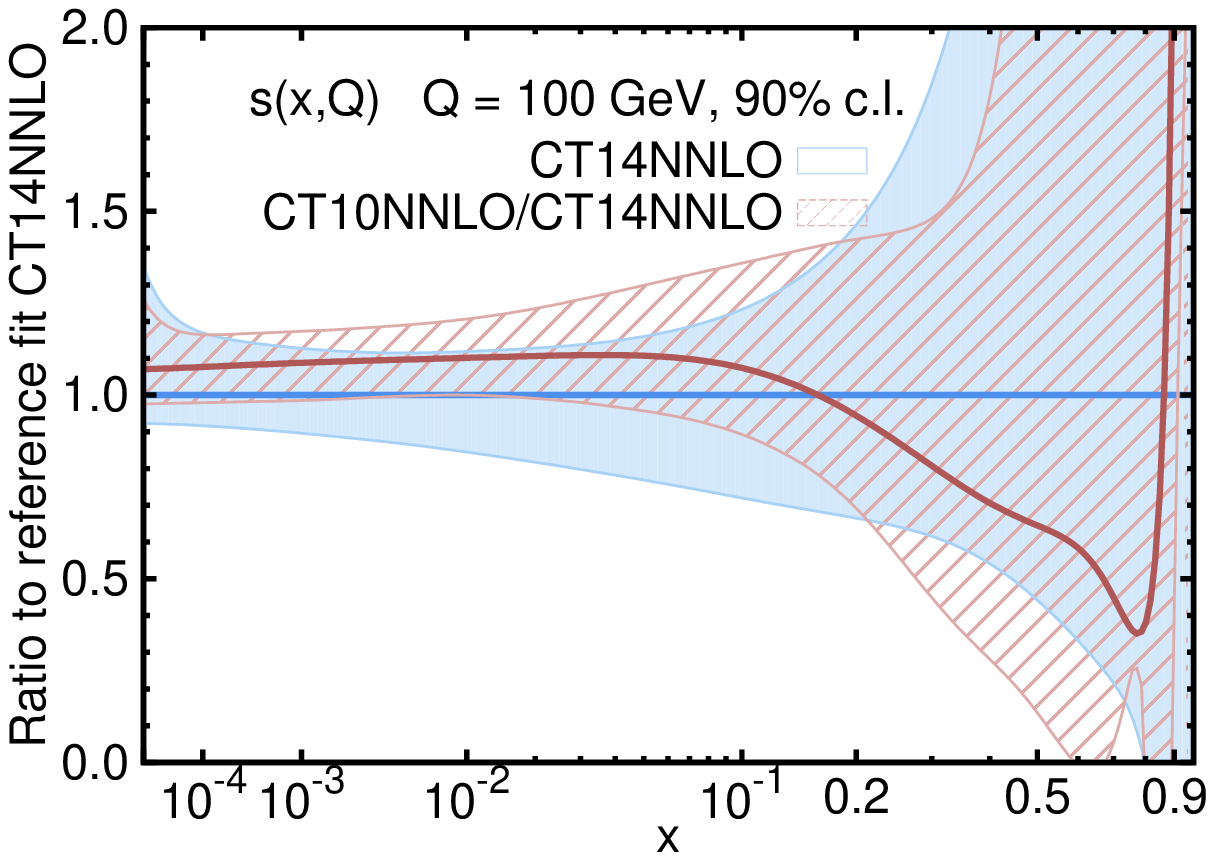}
\includegraphics[width=0.43\textwidth]{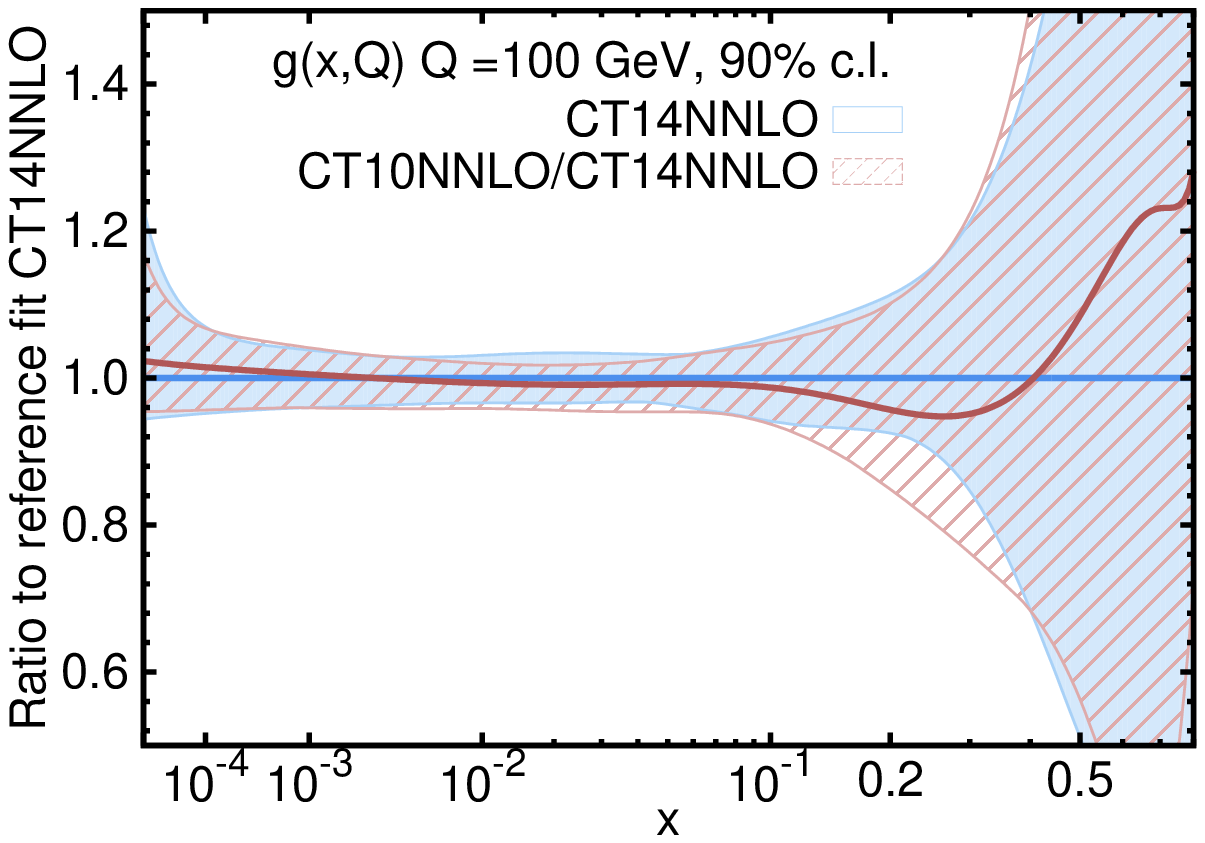}
\caption{A comparison of 90\% C.L. PDF uncertainties from CT14 NNLO (solid blue)
  and CT10 NNLO (red dashed) error sets.
  Both error bands are
  normalized to the respective central CT14 NNLO PDFs.
\label{fig:PDFbands1}}
\end{figure}

The relative changes between the CT10
NNLO and CT14 NNLO ensembles are best visualized by comparing
their PDF uncertainties. Fig.~\ref{fig:PDFbands1}
compares the PDF error bands at 90\% confidence level
for the key flavors, with each band normalized to the respective
best-fit CT14 NNLO PDF. The blue solid and red dashed error bands are
obtained for CT14 and CT10 NNLO PDFs at $Q=100$ GeV, respectively.

Focusing first on the $u$ and $d$ flavors in the upper four
subfigures,
we observe that the $u$ and $\bar u$ PDFs  have mildly increased
in CT14 at $x < 10^{-2}$,
while the $d$ and $\bar d$ PDFs have become slightly smaller.
These changes can be attributed to a more flexible parametrization
form adopted in CT14, which modifies the $SU(2)$ flavor composition of
the first-generation PDFs at the smallest $x$ values in the fit.

The CT14 $d$-quark PDF has increased by 5\% at $x \approx 0.05$, after the
ATLAS and CMS $W/Z$ production data sets at
7 TeV were included. At $x \gtrsim 0.1$, the update of the D\O~ charge asymmetry
data set in the electron channel, reviewed in Sec.~\ref{sec:D0Wasy},
has reduced the magnitude of the $d$ quark PDFs by a large amount, and
has moderately increased the $u(x,Q)$ distribution.

The $\bar u(x,Q)$ and $\bar d(x,Q)$ distributions
are both slightly larger at $x =
0.01-0.1$ because of several factors. At $x=0.2-0.5$, where there are
only very weak constraints on the sea-quark PDFs,
the new parametrization form of CT14 results in smaller
values of $\bar u(x,Q)$ and larger values $\bar d(x,Q)$, as compared
to CT10, although for the most part within the combined PDF
uncertainties of the two ensembles.

The central strangeness PDF $s(x,Q)$ in the third row of
Fig.~\ref{fig:PDFbands1} has decreased for $0.01< x < 0.15$, but within the limits of the CT10
uncertainty, as a consequence of the more flexible parametrization,
the corrected calculation for massive quarks in charged-current DIS, and  the
inclusion of the LHC data. The extrapolation of $s(x,Q)$ below $x=0.01$,
where no data directly constrain it, also lies somewhat lower than before;
its uncertainty remains large and compatible with that in CT10.
At large $x$, above about 0.2, the strange quark PDF is
essentially unconstrained in CT14, just as in CT10.

The central gluon PDF (last frame of Fig.~\ref{fig:PDFbands1}) has
increased in CT14 by 1-2\% at $x\approx 0.05$ and has been somewhat
modified at $x > 0.1$ by the inclusion of the LHC jet production,
by the multiplicative treatment of correlated errors, and by the other
factors discussed above.
For $x$ between 0.1 to 0.5, the gluon PDF has increased in CT14
as compared to CT10.

\begin{figure}[p]
\center
\includegraphics[width=0.43\textwidth]{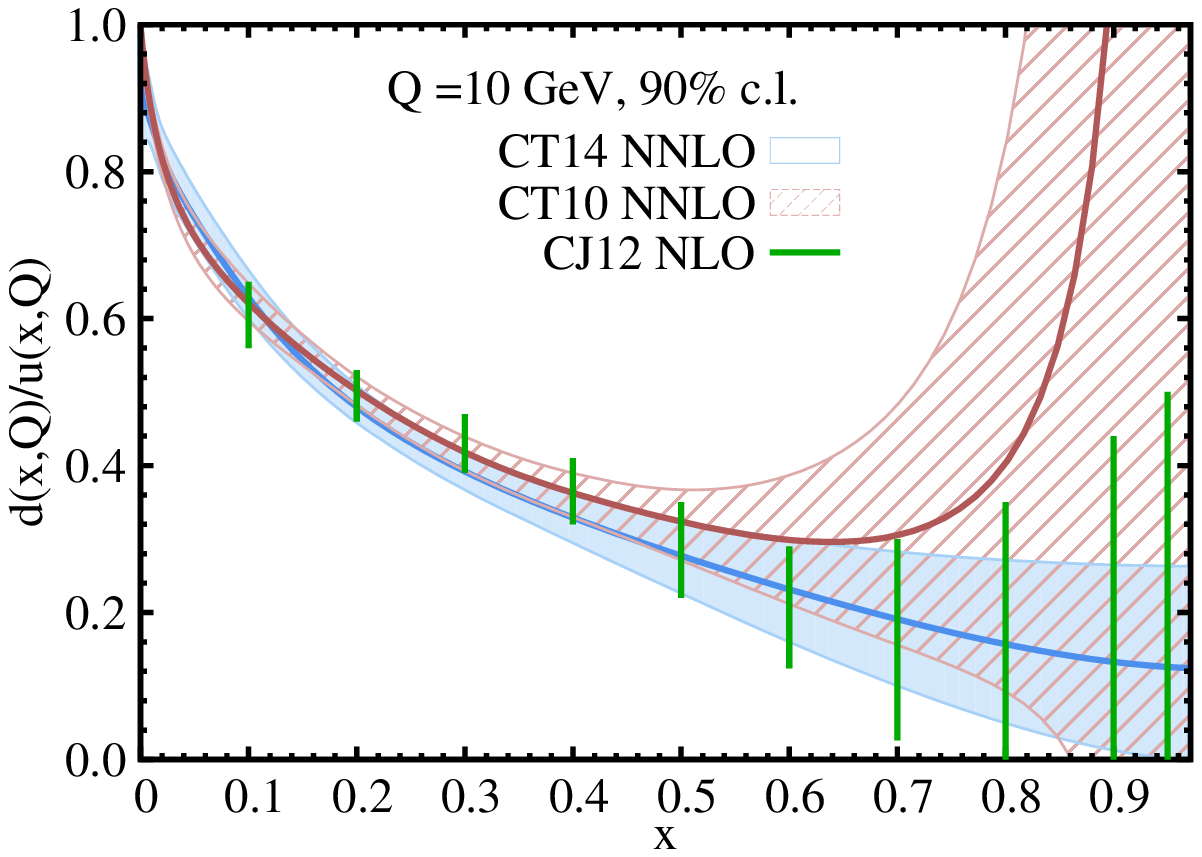}
\includegraphics[width=0.43\textwidth]{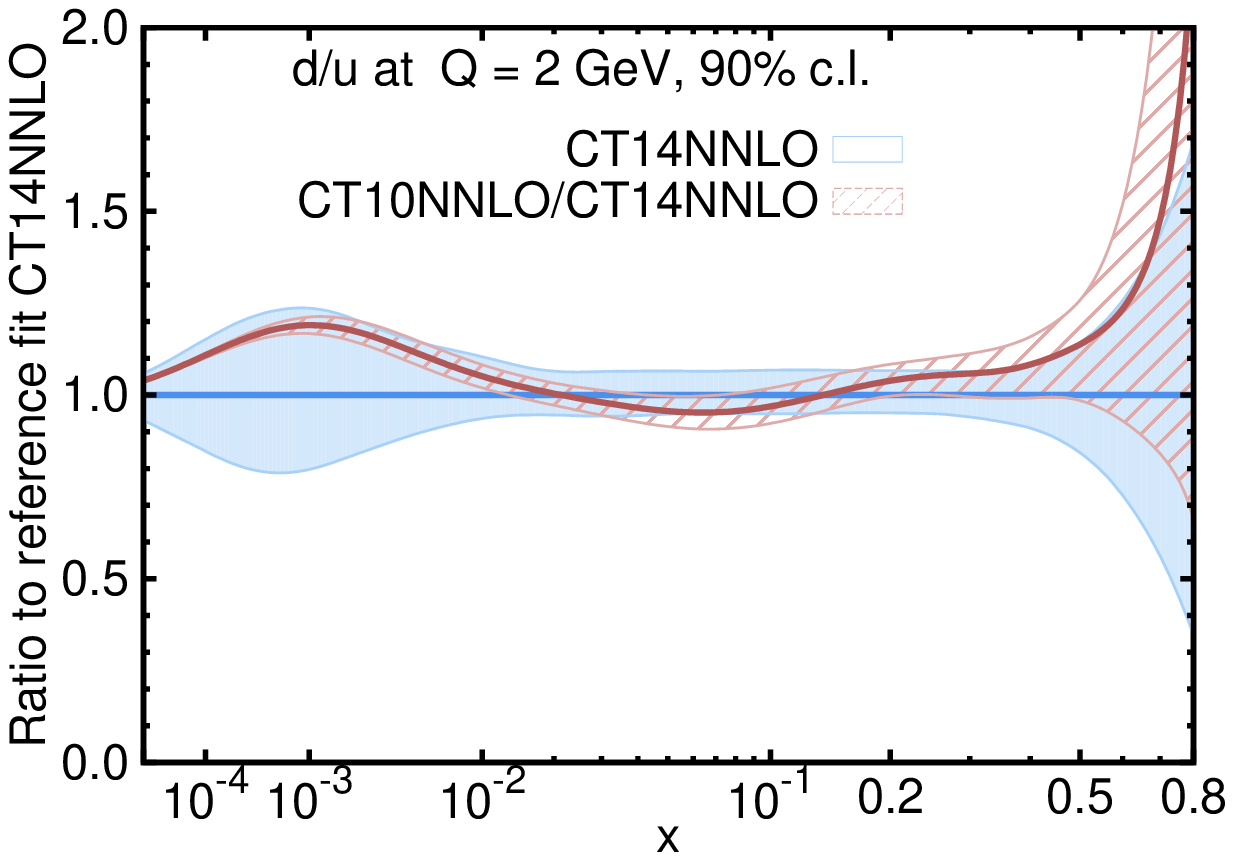}
\includegraphics[width=0.43\textwidth]{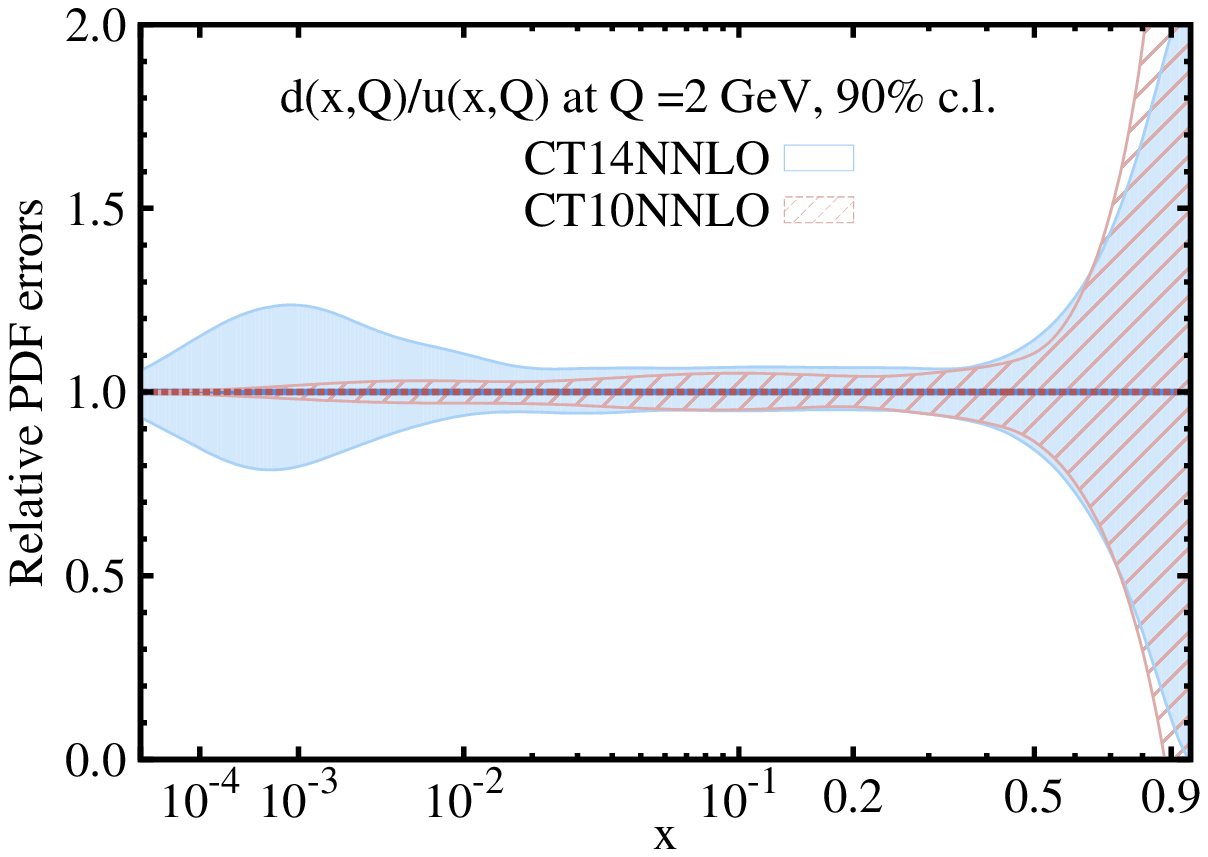}
\caption{A comparison of 90\% C.L. uncertainties on the ratio
  $d(x,Q)/u(x,Q)$ for CT14 NNLO (solid
  blue) and CT10 NNLO (dashed red), and CJ12 NLO (green lines) error ensembles.
\label{fig:DOUband}}
\end{figure}

\begin{figure}[p]
\center
\includegraphics[width=0.43\textwidth]{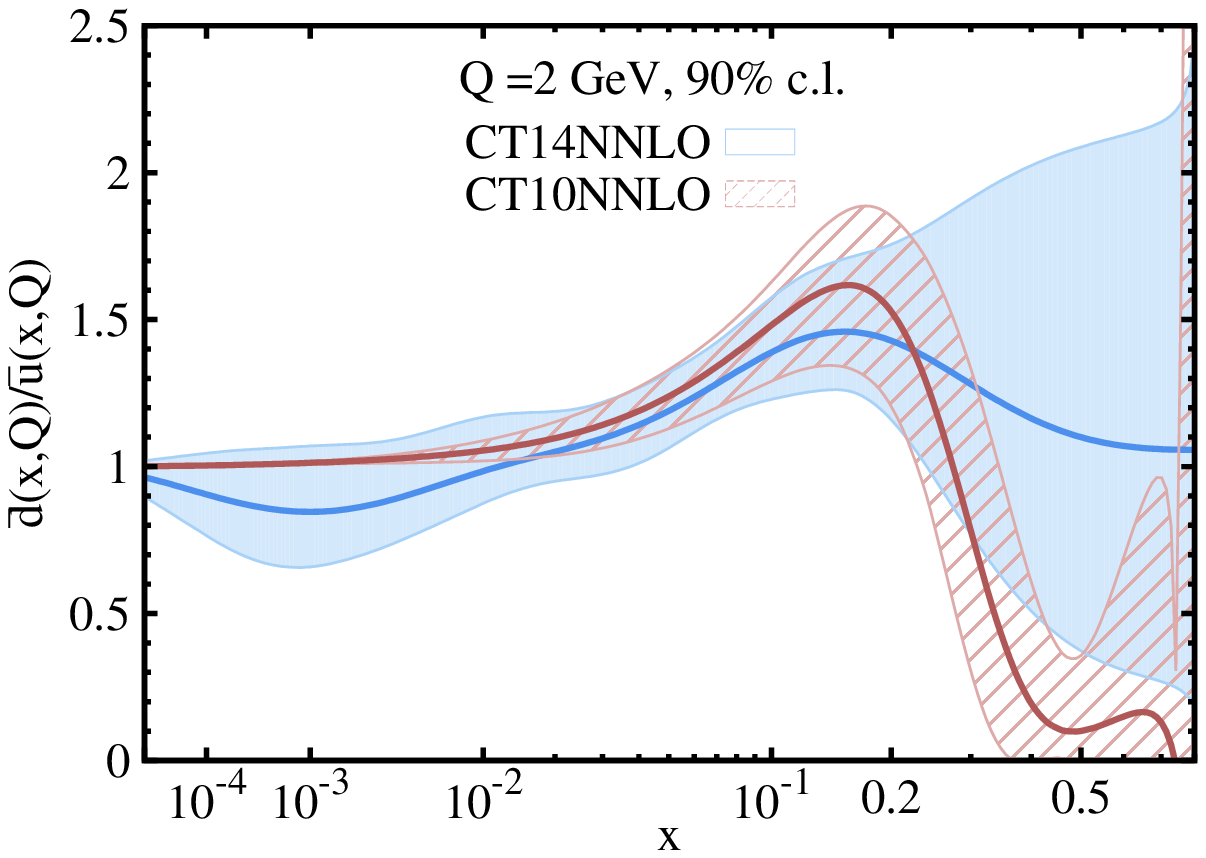}
\includegraphics[width=0.43\textwidth]{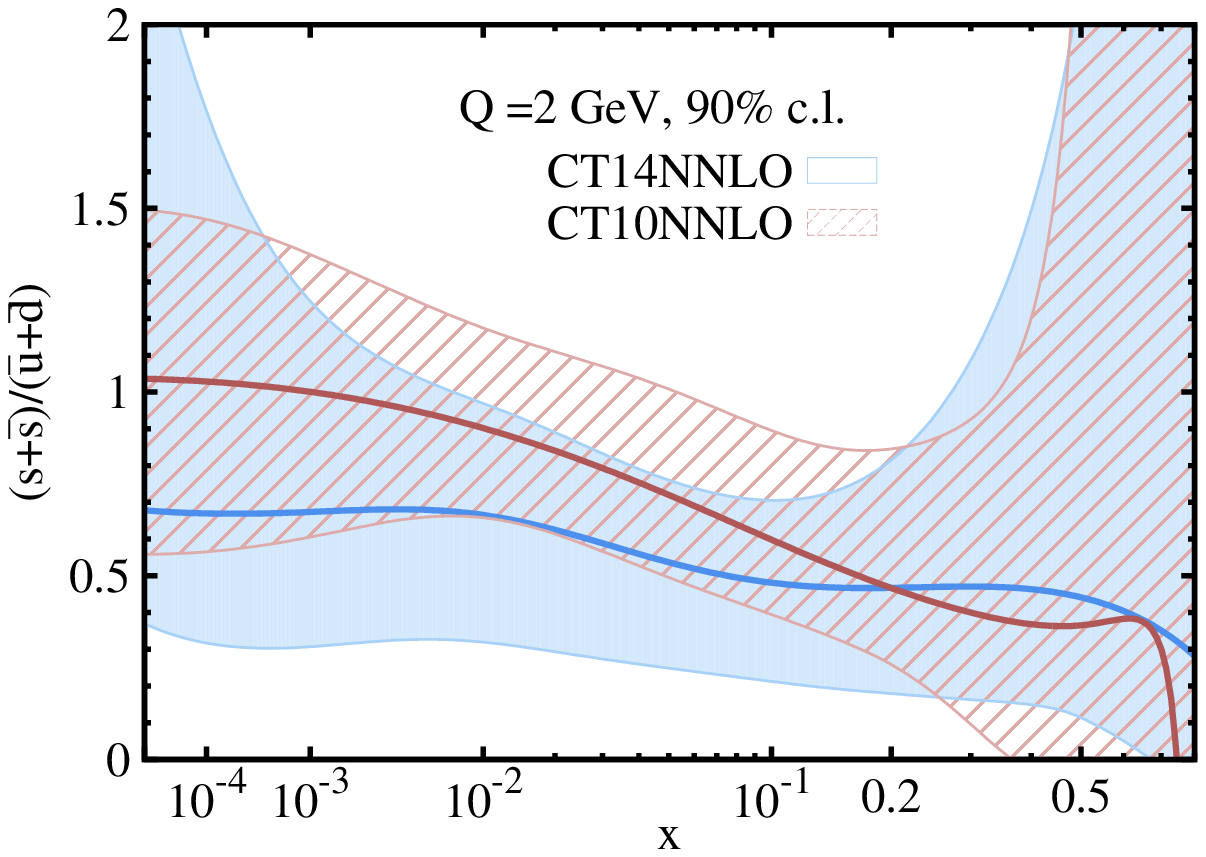}
\caption{A comparison of 90\% C.L. uncertainties on the ratios
  $\bar d(x,Q)/\bar u(x,Q)$ and $\left(s(x,Q)+\bar
  s(x,Q)\right)/\left(\bar u(x,Q) +\bar d(x,Q)\right)$, for CT14 NNLO (solid
  blue) and CT10 NNLO (red dashed) error ensembles.
\label{fig:DBandSBbands}}
\end{figure}

Let us now review the ratios of various
PDFs, starting with the ratio $d/u$ shown in
Fig.~\ref{fig:DOUband}. The changes in $d/u$ in CT14 NNLO, as compared
to CT10 NNLO, can be summarized as a reduction of the central ratio at $x > 0.1$,
caused by the $9.7\mbox{ fb}^{-1}$ D\O~ charge asymmetry data, and an
increased uncertainty at $x < 0.05$ allowed by the new
parametrization form. At $x>0.2$, the central CT14 NNLO ratio is
lower than that of CT10 NNLO, while their relative PDF uncertainties
remain about the same. This can be better seen from a direct comparison
of the relative PDF uncertainties (normalized to their respective
central PDFs) in the third inset.
The collider charge asymmetry data constrains $d/u$ at
$x$ up to about 0.4. At even higher $x$, outside of the experimental
reach, the behavior of the CT14 PDFs reflects
the parametrization form, which now allows $d/u$ to
approach any constant value at $x\rightarrow 1$.

At such high $x$, the CTEQ-JLab analysis (CJ12) \cite{Owens:2012bv}
has independently determined the ratio $d/u$ at NLO, by
including the fixed-target DIS data at lower $W$ and higher $x$
that is excluded by a selection cut $W > 3.5 \mbox{ GeV}$ in
CT14,  and by considering higher-twist and nuclear effects
that can be neglected in the kinematic range of CT14 data. The CT14
uncertainty band on $d/u$ at NNLO lies for the most part between the
CJmin and CJmax predictions at NLO that demarcate the CJ12 uncertainty,
cf. the first inset of Fig.~\ref{fig:DOUband}. We see that the CT14
predictions on $d/u$ at $x > 0.1$, which were derived
from high-energy measurements that are not affected by
nuclear effects, fall within the CJ12 uncertainty range
obtained from low-energy DIS with an estimate of various effects beyond
leading-twist perturbative QCD. The ratio should be  stable to inclusion of
NNLO effects; thus, the two ensembles  predict a
similar trend for collider observables sensitive to $d/u$.

Turning now to the ratios of sea quark PDFs in
Fig.~\ref{fig:DBandSBbands}, we observe that the uncertainty on $\bar
d(x,Q)/\bar u(x,Q)$ in the left inset has also increased at small $x$
in CT14 NNLO. At $x > 0.1$, we assume that both $\bar u(x,Q_0)$ and
$\bar d(x,Q_0)$ are proportional to $(1-x)^{a_2}$ with the same power
$a_2$; the ratio
$\bar d(x,Q_0)/\bar u(x,Q_0)$ can thus approach a constant value that
comes out to be close to 1 in the central fit, while the
parametrization in CT10 forced it to vanish. The uncertainty on
$\bar d/\bar u$ has also increased across most of the $x$ range.

The overall reduction in the strangeness PDF at $x > 0.01$ leads to a
smaller ratio of the strange-to-nonstrange sea quark PDFs, $\left(s(x,Q)+\bar
  s(x,Q)\right)/\left(\bar u(x,Q) +\bar d(x,Q)\right)$, presented in
  the right inset of Fig.~\ref{fig:DBandSBbands}. At $x < 0.01$, this
  ratio is determined entirely by parametrization form and was found
  in CT10 to be consistent with the exact $SU(3)$ symmetry of PDF
  flavors, $\left(s(x,Q)+\bar
  s(x,Q)\right)/\left(\bar u(x,Q) +\bar d(x,Q)\right) \rightarrow 1$
  at $x\rightarrow 0$, albeit with a large uncertainty. The
  $SU(3)$-symmetric asymptotic solution at $x\rightarrow 0$
is still allowed in CT14 as a possibility, even though the asymptotic
limit of the central CT14 NNLO has been reduced and is now at about
0.6 at $x=10^{-5}$. The uncertainty of
strangeness has increased at such small $x$ and now allows $\left(s(x,Q)+\bar
  s(x,Q)\right)/\left(\bar u(x,Q) +\bar d(x,Q)\right)$ between 0.35
  and 2.5 at $x=10^{-5}$.

\section{Comparisons with hadronic experiments \label{sec:TheoryVsData}}

\subsection{Electroweak total cross sections at the LHC}

Measurements of total cross sections for production of massive
electroweak particles at hadron colliders provide cornerstone
benchmark tests of the Standard Model. These relatively simple
observables  can be both measured with high precision and
predicted in NNLO QCD theory with small uncertainties.
In this subsection, we collect NNLO theory predictions based on CT14 and CT10
NNLO PDFs for inclusive $W$ and $Z$ boson production, top-quark pair
production, Higgs-boson production (through gluon-gluon fusion), at the
LHC with center-of-mass energies of 8 and 13 TeV.
These theoretical predictions can be compared to the corresponding experimental
measurements. We also examine correlations between PDF uncertainties
of the total cross sections in the context of the Hessian formalism,
following the approach summarized in Ref.~\cite{Nadolsky:2008zw}.
PDF-driven correlations reveal relations between PDF uncertainties of
QCD observables through their shared PDF parameters.

The masses of the top quark and Higgs boson
are set to $m_t^{pole}=173.3$ GeV and $m_H=125$ GeV, respectively,
in this work.
The $W$ and $Z$ inclusive cross sections
(multiplied by branching ratios for the decay into one charged lepton flavor),
are calculated by using the \textsc{Vrap} v0.9 program~\cite{Anastasiou:2003ds,Anastasiou:2003yy} at NNLO in QCD, with the renormalization
and factorization ($\mu_R$ and $\mu_F$) scales set equal
to the invariant mass of the vector boson.
The total inclusive top-quark pair cross sections are calculated with
the help of the program \textsc{Top++}
v2.0~\cite{Czakon:2013goa,Top++} at NNLO+NNLL accuracy, with QCD
scales set to the mass of the top quark.
The Higgs boson cross sections via gluon-gluon fusion
are calculated at NNLO in QCD by using  the \textsc{iHixs} v1.3
program~\cite{Anastasiou:2011pi},
in the heavy-quark effective theory (HQET) with finite top quark mass correction,
and with the QCD scales set equal
to the invariant mass of the Higgs boson.

\begin{figure}[tb]
  \begin{center}
  \includegraphics[width=0.385\textwidth]{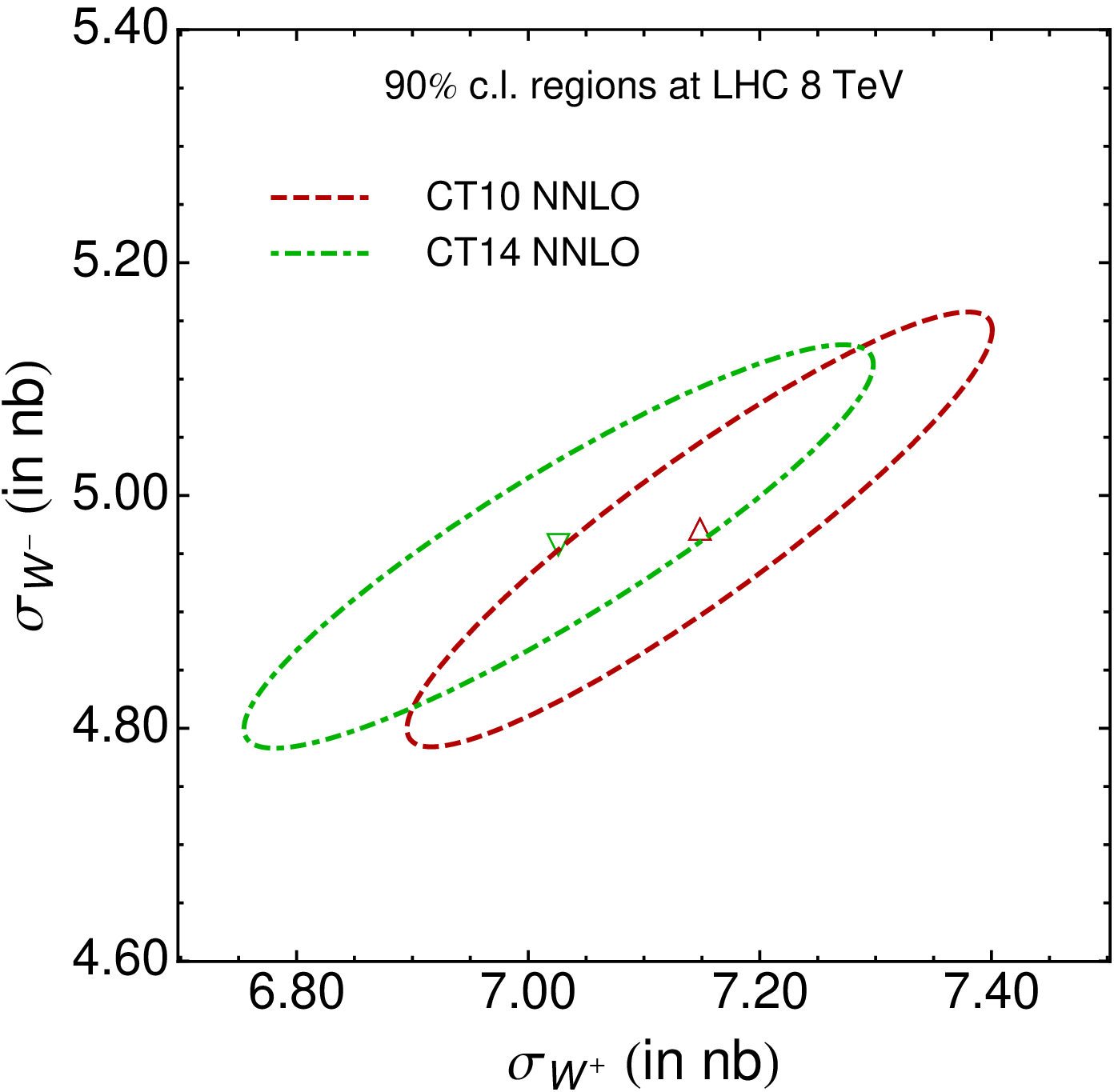}\hspace{0.2in}
  \includegraphics[width=0.40\textwidth]{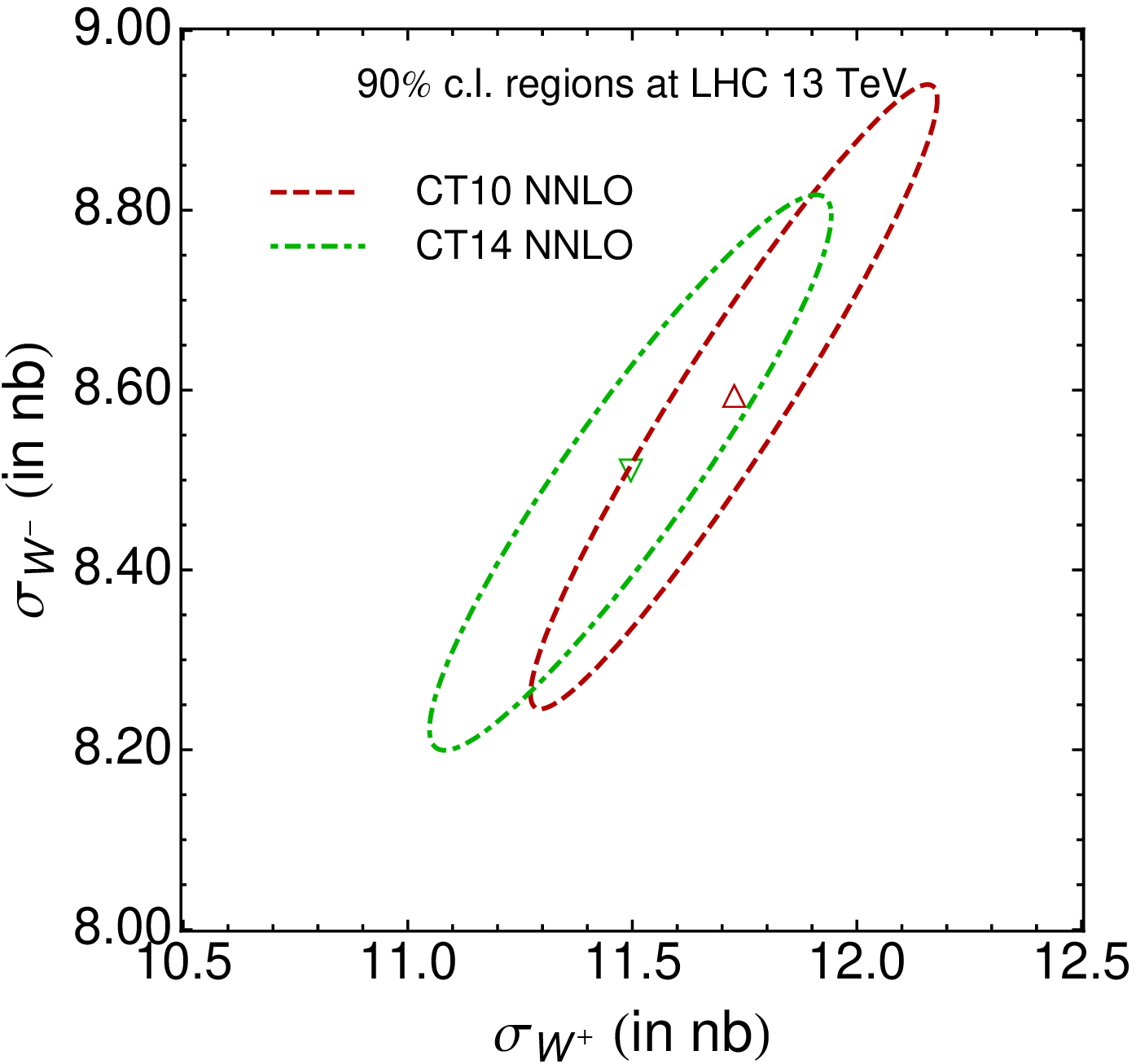}
  \end{center}
  \vspace{-1ex}
  \caption{\label{fig:inc1}
The CT14 and CT10 NNLO 90\% C.L. error ellipses for the $W^{-}$  and $W^{+}$ cross
sections, at the LHC 8 and 13 TeV.
}
\end{figure}

\begin{figure}[tb]
  \begin{center}
  \includegraphics[width=0.395\textwidth]{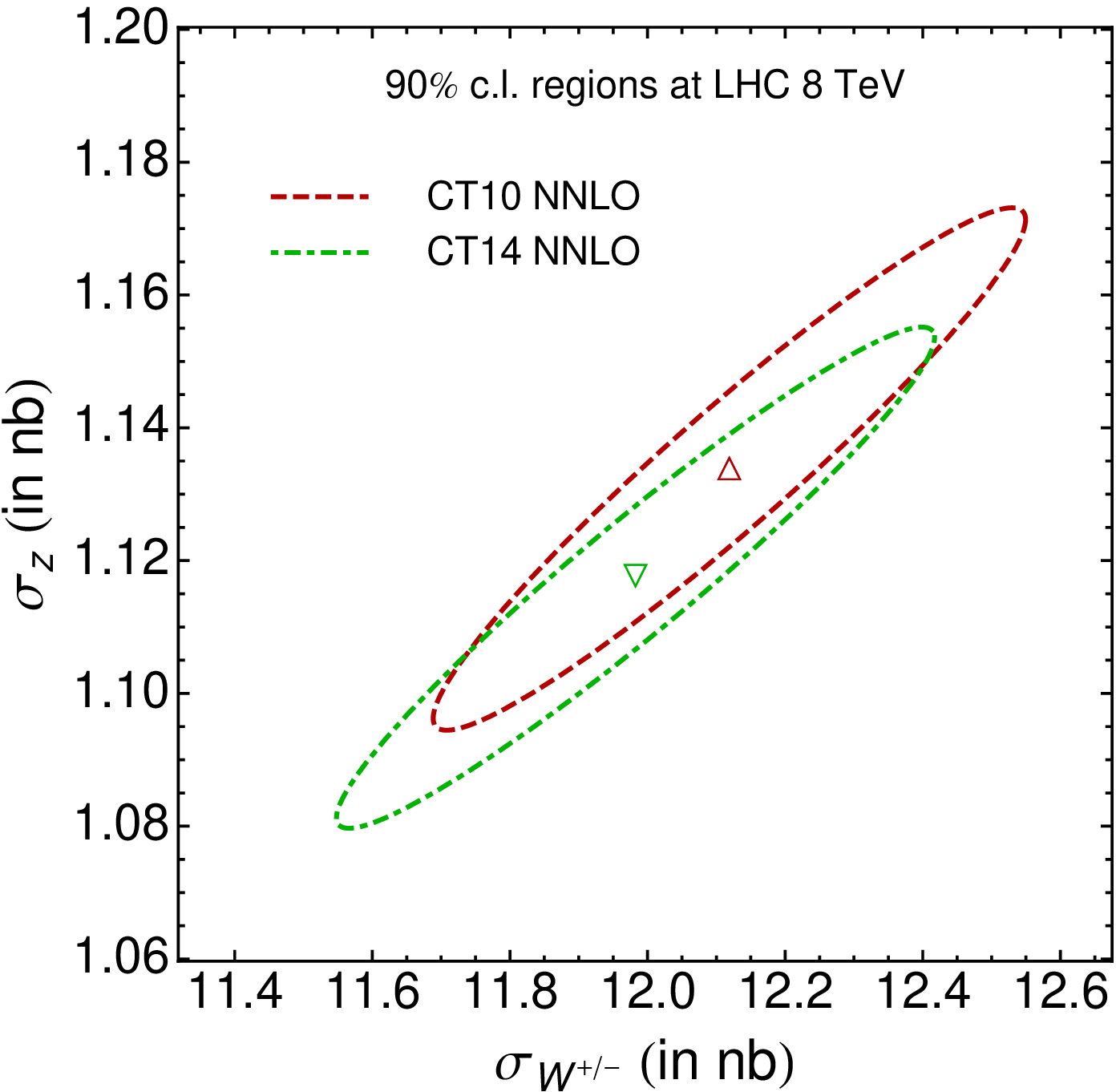}\hspace{0.2in}
  \includegraphics[width=0.40\textwidth]{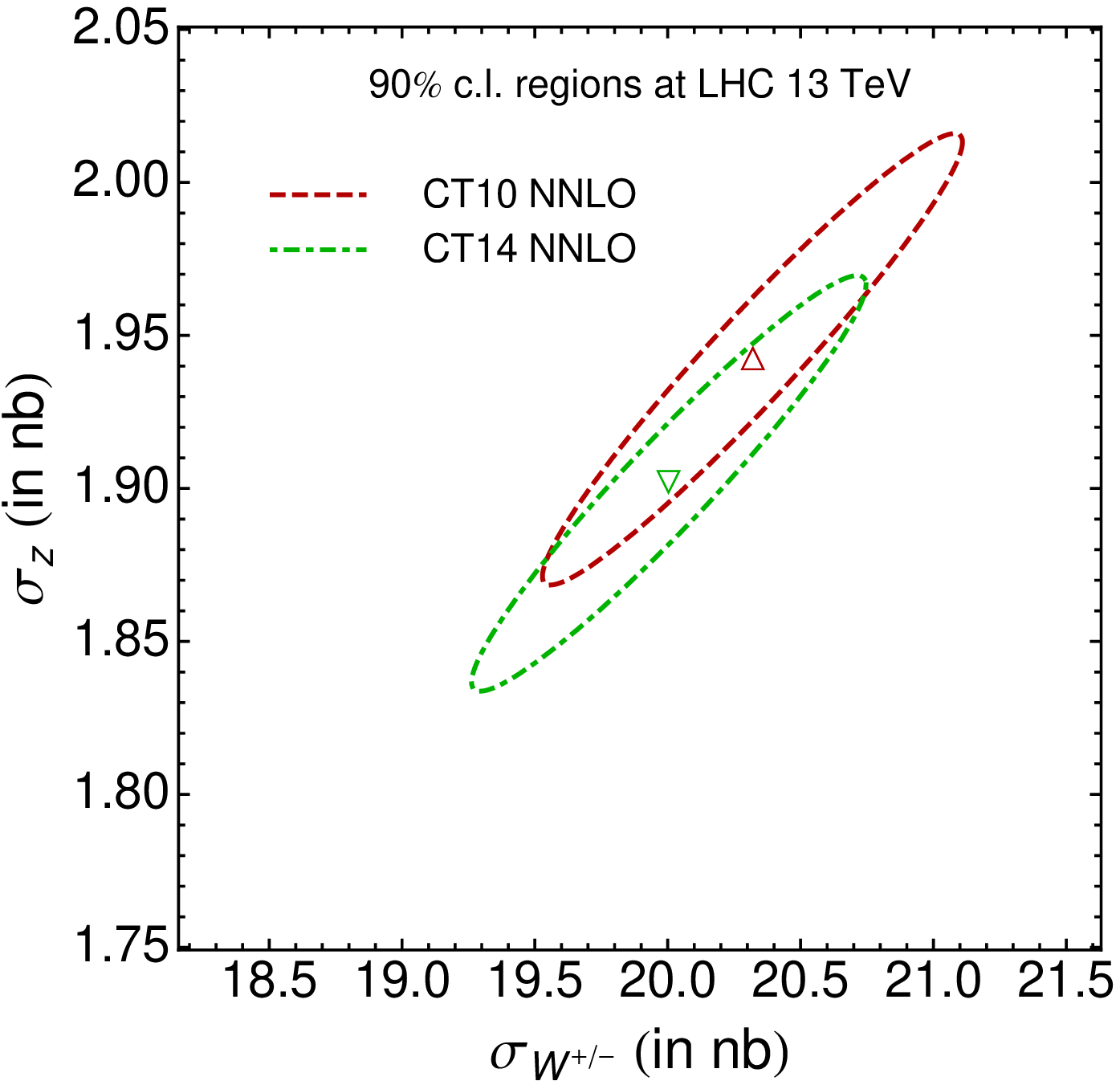}
  \end{center}
  \vspace{-1ex}
  \caption{\label{fig:inc2}
CT14 and CT10 NNLO 90\% C.L. error ellipses for $Z$ and  $W^{\pm}$ cross
sections, at the LHC 8 and 13 TeV.
}
\end{figure}

\begin{figure}[tb]
  \begin{center}
  \includegraphics[width=0.4\textwidth]{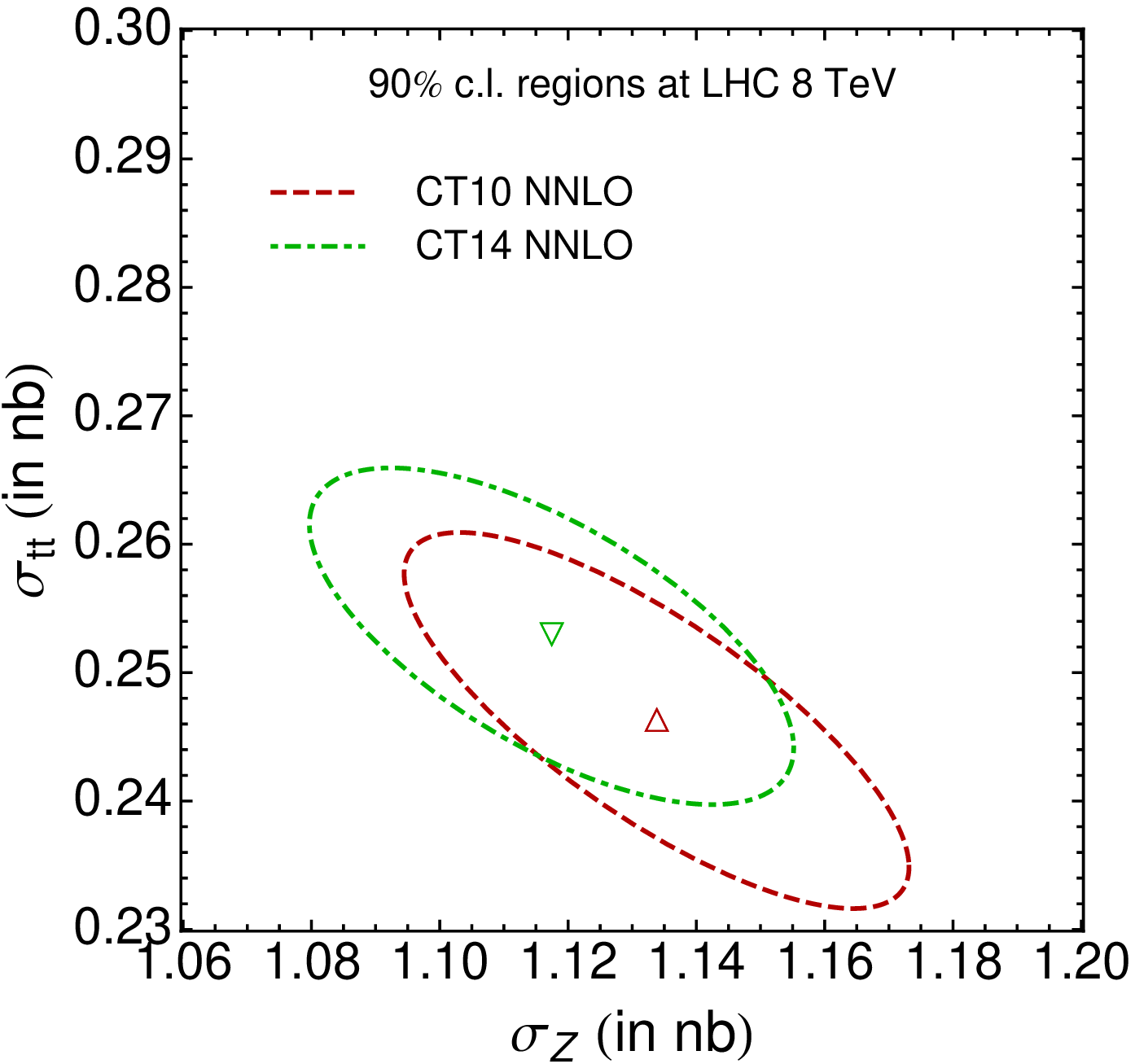}\hspace{0.2in}
  \includegraphics[width=0.4\textwidth]{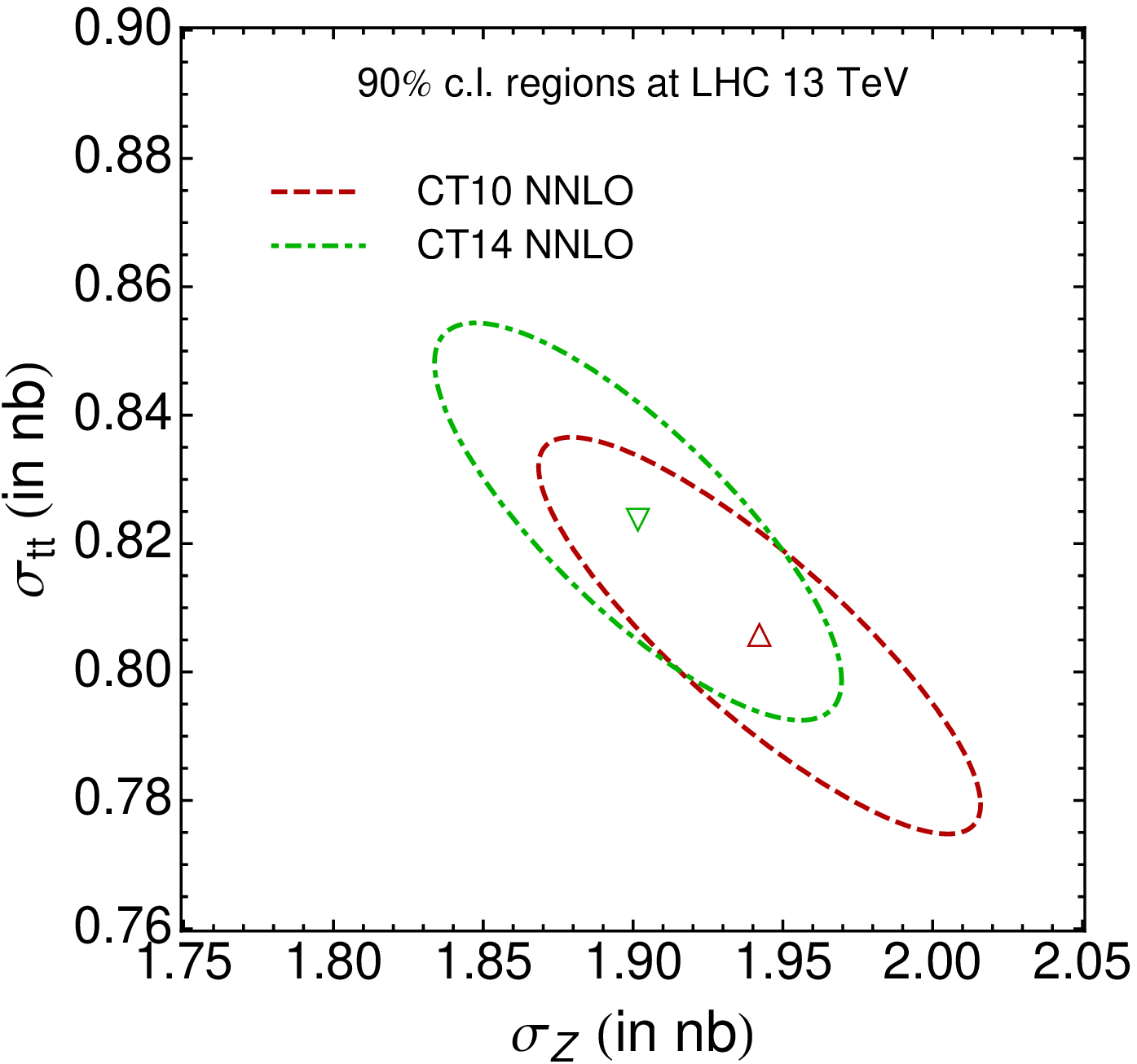}
  \end{center}
  \vspace{-1ex}
  \caption{\label{fig:inc3}
CT14 and CT10 NNLO 90\% C.L. error ellipses for $t\bar{t}$ and $Z$
cross sections, at the LHC 8 and 13 TeV.
  }
\end{figure}

\begin{figure}[tb]
  \begin{center}
  \includegraphics[width=0.4\textwidth]{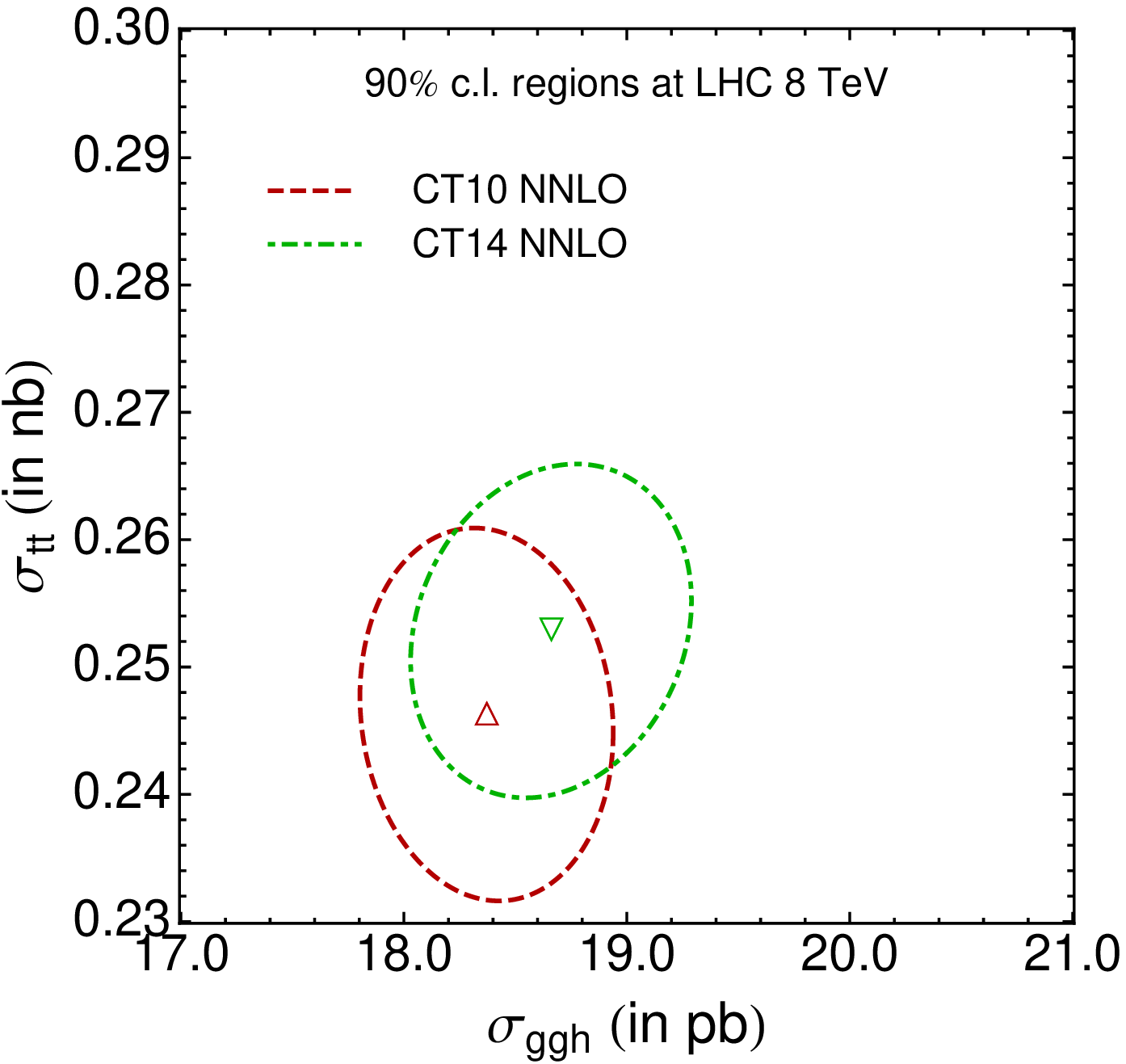}\hspace{0.2in}
  \includegraphics[width=0.4\textwidth]{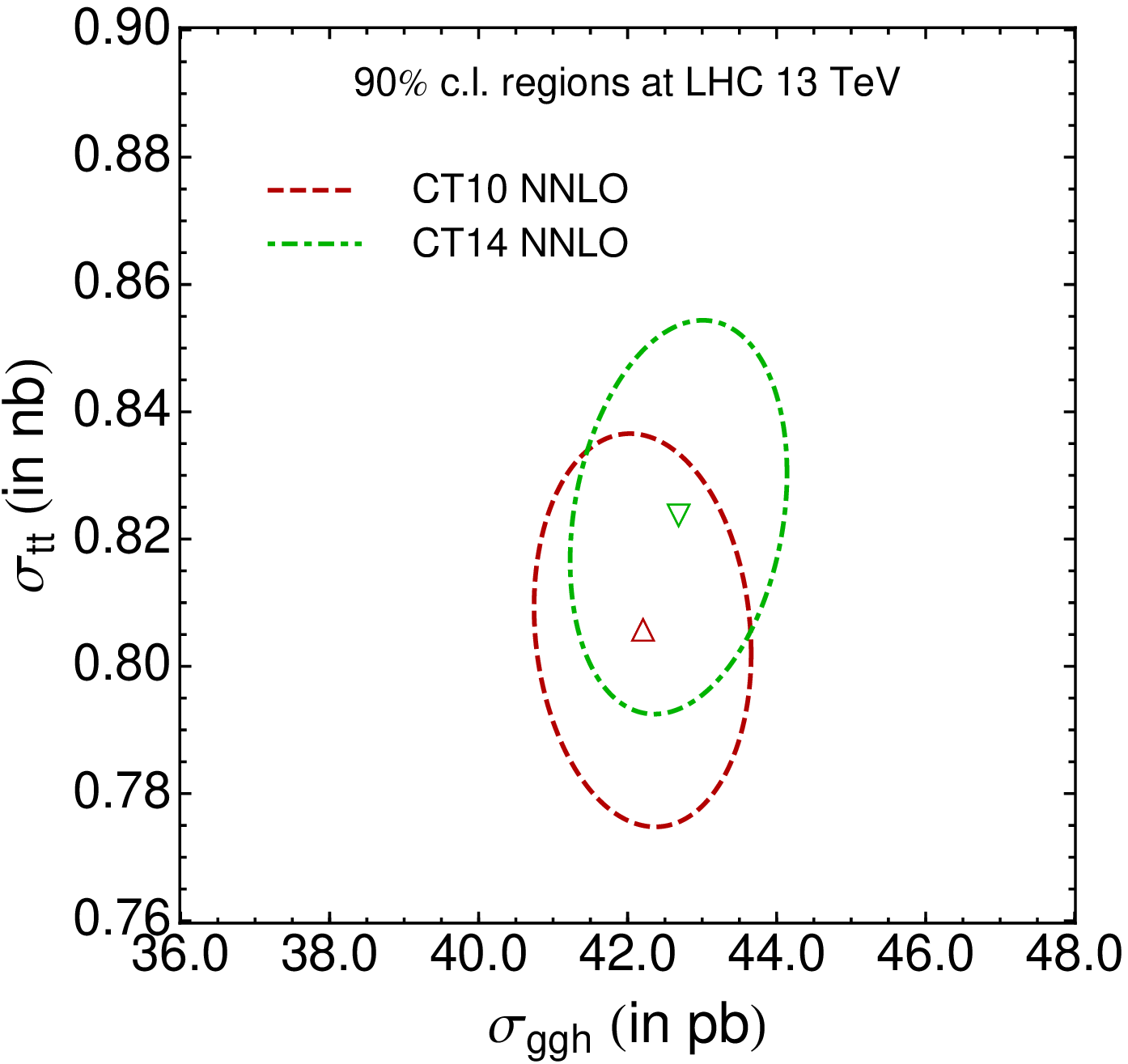}
  \end{center}
  \vspace{-1ex}
  \caption{\label{fig:inc4}
CT14 and CT10 NNLO 90\% C.L. error ellipses for $t\bar{t}$ and $ggH$ cross sections, at the LHC 8 and 13 TeV.
}
\end{figure}

Figs.~\ref{fig:inc1} -- \ref{fig:inc4} show central predictions and
90\% C.L. regions for ($W^+$, $W^-$), ($Z$,$W^{\pm}$),
($t\bar{t}$,$Z$) and ($t\bar{t}$,$ggH$) pairs of inclusive cross sections at
the LHC 8 and 13 TeV. In each figure, two elliptical confidence
regions are shown, obtained with either
 CT14 or CT10 NNLO PDFs. These can be used to read off PDF
 uncertainties and correlations for each pair of cross sections.
For example, Figs.~\ref{fig:inc1} and \ref{fig:inc2}
indicate that the PDF induced uncertainties, at the 90\% C.L., are about 3.9\%, 3.7\%,
and 3.7\% for $W^+$, $W^-$, and $Z$ boson production at the LHC 13 TeV,
respectively, with CT14 NNLO PDFs.
As compared to the results using CT10 NNLO PDFs, the ratio of the
total inclusive cross sections of $W^+$ to $W^-$ productions 
at the LHC 13 TeV
is smaller by about one percent
when using CT14 NNLO PDFs which also provide a slightly larger error
(by about half percent) in that ratio.
Specifically, the CT14 NNLO predictions of that ratio at the 68\% C.L. are
$1.42_{-1\%}^{+1.2\%}$ at LHC 8 TeV, and
$1.35_{-1\%}^{+1\%}$ at LHC 13 TeV,
respectively.
The central predictions at 8 TeV are in agreement
with the recent CMS measurements~\cite{Chatrchyan:2014mua}.
They also show that the electroweak gauge boson cross sections
are highly correlated with each other; in fact, much of the
uncertainty is driven in this case by the small-$x$ gluon
\cite{Nadolsky:2008zw}.

In Fig.~\ref{fig:inc3}, we observe a moderate anti-correlation between
the top-quark pair and the $Z$ boson production cross sections. This
is a consequence of the proton momentum sum rule mediated by the gluon
PDF \cite{Nadolsky:2008zw}.
In Fig.~\ref{fig:inc4}, the Higgs boson cross section through
gluon-gluon fusion does not have a pronounced correlation or anti-correlation
with the top-quark cross section, because they are dominated by the gluon
PDF in different $x$ regions. The Higgs boson and $t\bar{t}$ cross
section predictions are further examined in Section~\ref{sec:Uncert}.
As a result of the changes in PDFs from CT10 to CT14,
both the calculated Higgs boson and top-quark pair production cross
sections have increased slightly,
while the electroweak gauge boson cross sections have decreased.
However, the changes of the central predictions
are within the error ellipses of either CT14 or CT10.

\subsection{LHC and Tevatron inclusive jet cross sections}

We now turn to the comparisons of CT14 PDFs with new LHC cross sections on
inclusive jet production. We argued in Section~\ref{sec:Setup}
that PDF uncertainty of inclusive jet production at the LHC is
strongly correlated with the gluon PDF in a wider range of $x$ than in the
counterpart measurements at the
Tevatron. The true potential of LHC jets for constraining the gluon PDF also
depends on experimental uncertainties, which we can now explore for
the first time using the CMS and ATLAS data on inclusive jet cross
sections at 7 TeV.

We first note that, in the context of our analysis, the
single-inclusive jet measurements at the LHC are
found to be in reasonable consistency with the other global data, including
Tevatron Run-2 single-inclusive jet cross sections measured by
the CDF and D\O~ collaborations. The values of $\chi^2$ for the four jet
experiments (ID=504, 514, 535, and 538) are listed at the end of
Table~\ref{tab:EXP_2}. We obtain very good fits ($\chi^2/N_{pt}=$1.09
and 0.55) to the D\O~ and ATLAS
jet data sets, and moderately worse fits ($\chi^2/N_{pt}=$1.45
and 1.33) to the CDF and CMS data sets. The description of the
Tevatron jet data sets has been examined as a part of the
CT10 NNLO study \cite{Gao:2013xoa},
where it was pointed out that the $\chi^2$ for the CDF
Run-2 measurement tends to be increased by
random, rather than systematic, fluctuations of the
data. In regards to describing the Tevatron jet data,
the CT14 NNLO PDFs follow similar trends as CT10 NNLO.

\subsubsection{CMS single-inclusive jet cross sections}

\begin{figure}[tb]
\center
\includegraphics[width=0.50\textwidth]{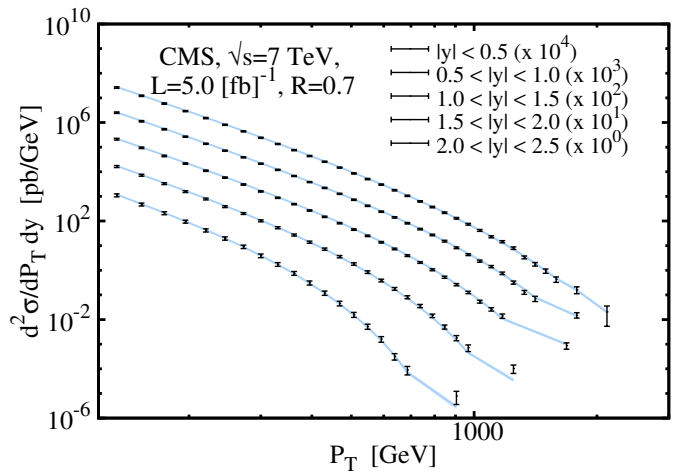}
\caption{Comparison of data and theory for the CMS 7 TeV inclusive jet
  production, for CT14 NNLO PDFs.
Measurements of $d^2\sigma / dp_{T}dy$ for 5 rapidity bins
are plotted as functions of jet $p_{T}$.
The points are data with total experimental errors, obtained by adding
the statistical and systematic errors in quadrature. The bands are theoretical calculations with 68\% C.L.\ PDF uncertainties.
\label{fig:ds5381}
}
\end{figure}

\begin{figure}[tb]
\center
\includegraphics[width=0.43\textwidth]{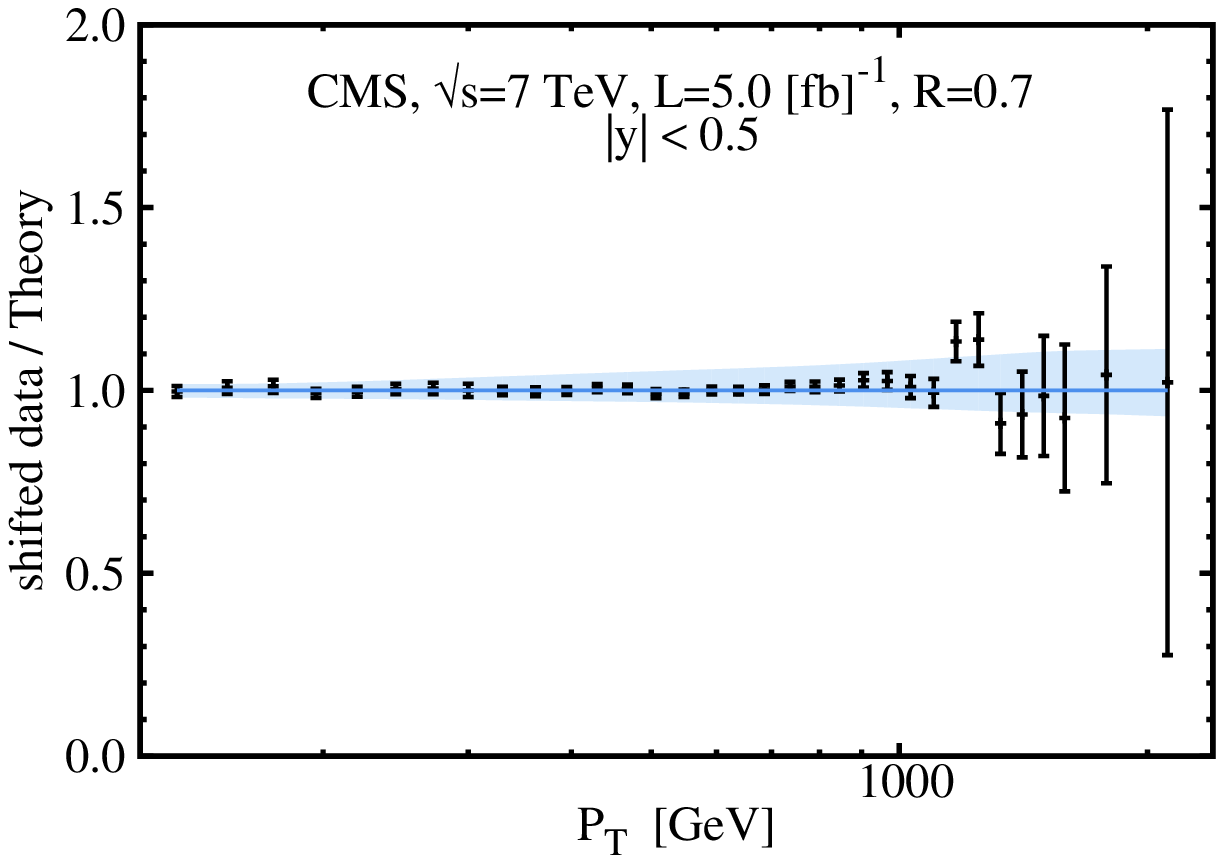}
\includegraphics[width=0.43\textwidth]{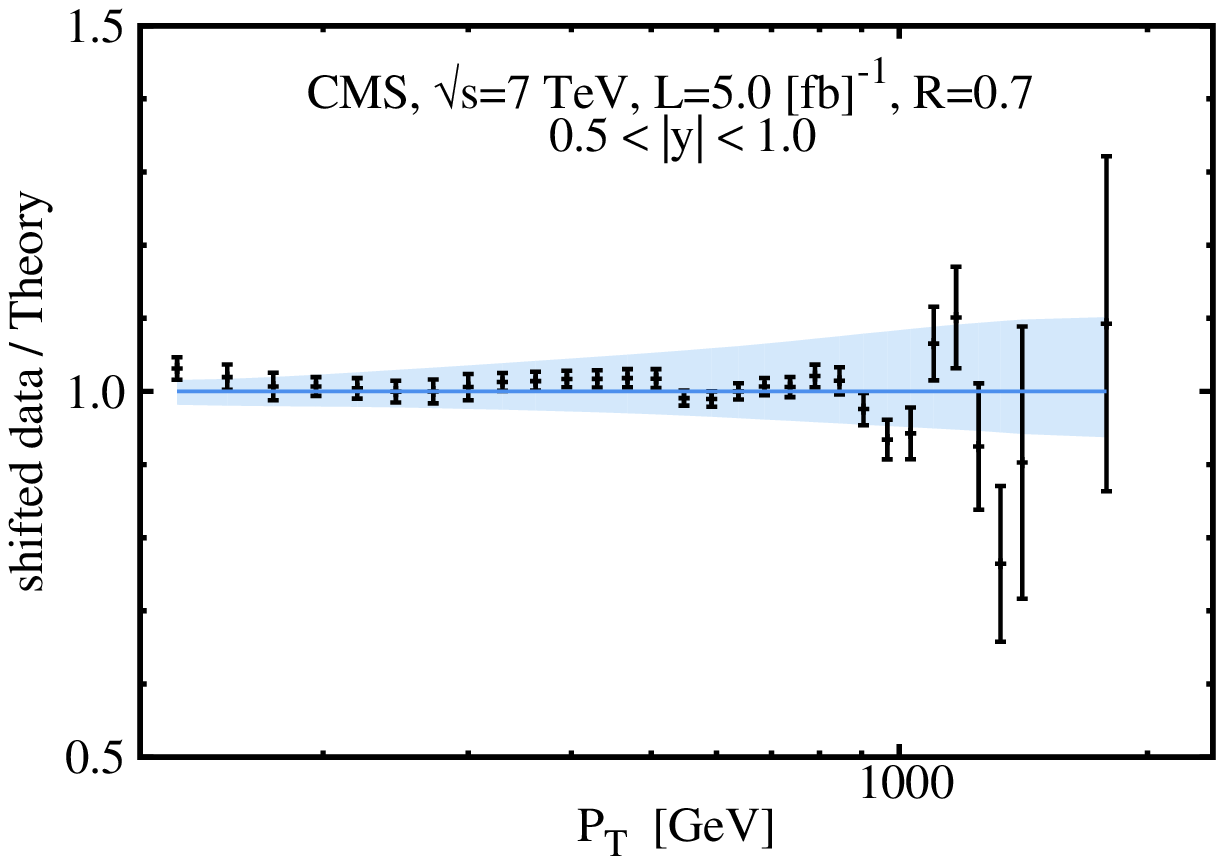}
\includegraphics[width=0.43\textwidth]{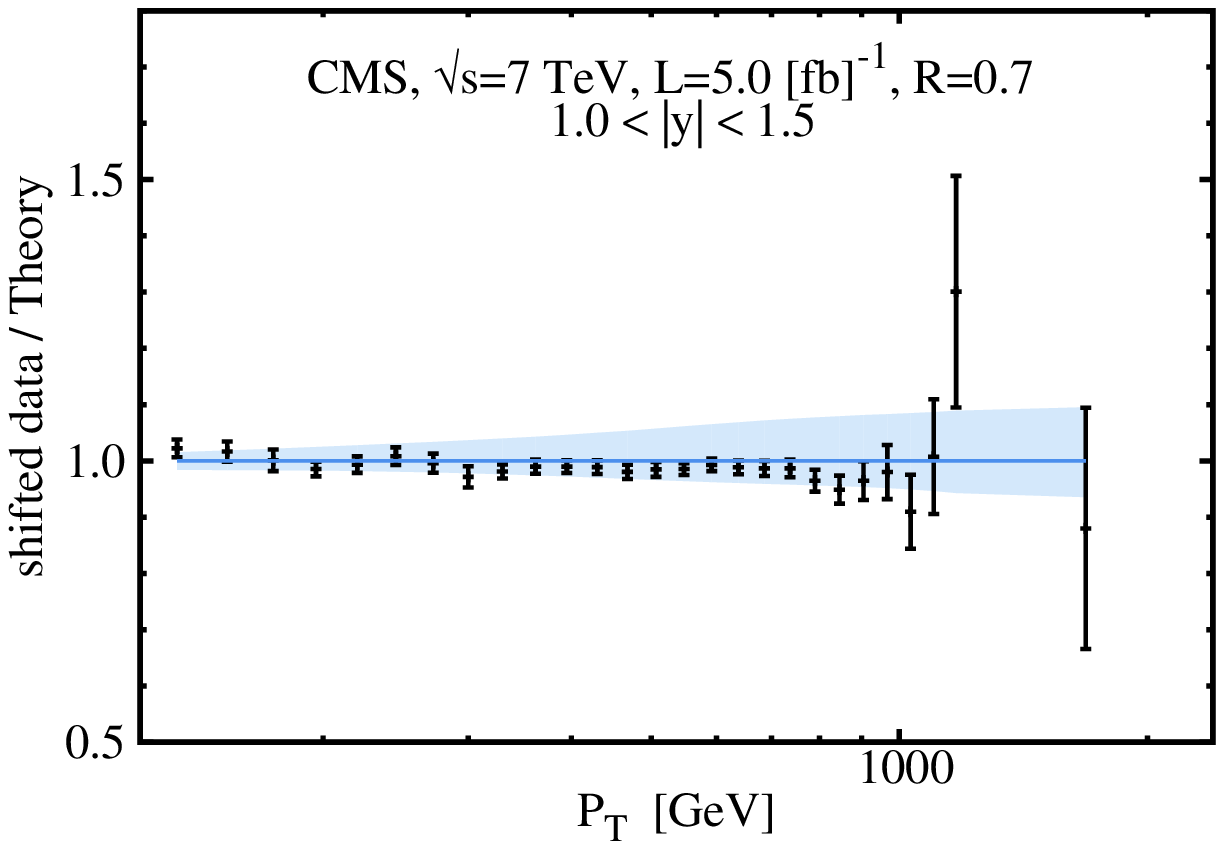}
\includegraphics[width=0.43\textwidth]{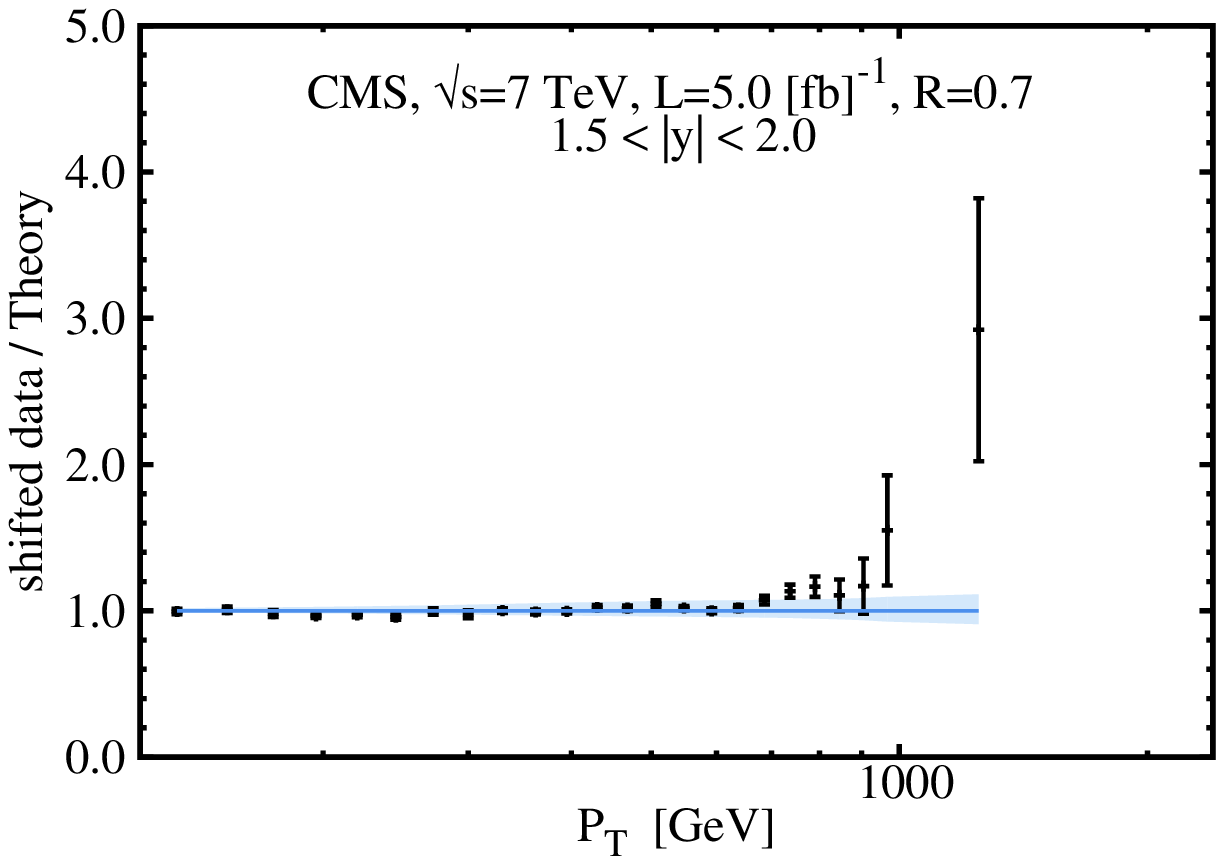}
\includegraphics[width=0.43\textwidth]{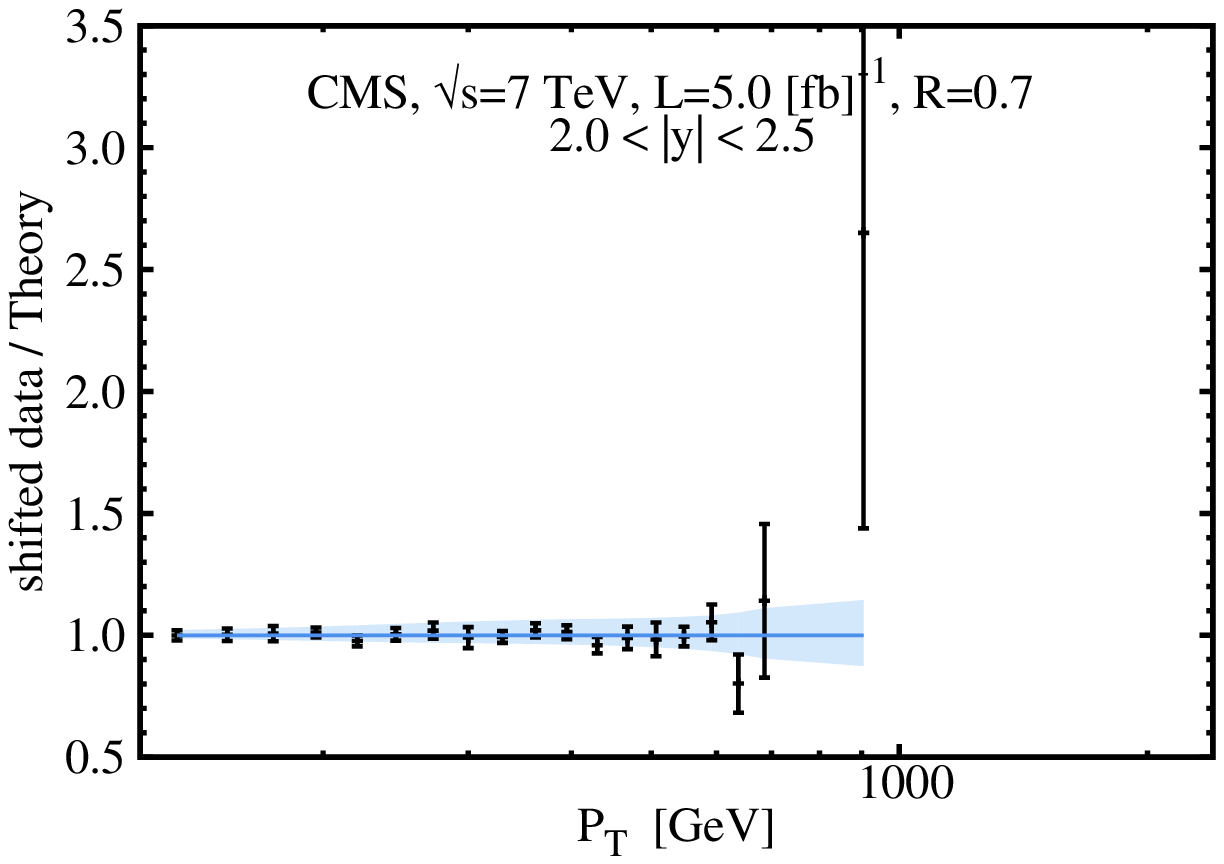}
\caption{Same as Fig.~\ref{fig:ds5381},
shown as the ratio of shifted data for CMS 7 TeV
divided by theory. The error bars
correspond to total uncorrelated errors. The shaded region
shows the 68\% C.L. PDF uncertainties.
\label{fig:ds5382}}
\end{figure}

\begin{figure}[tb]
\includegraphics[width=0.43\textwidth]{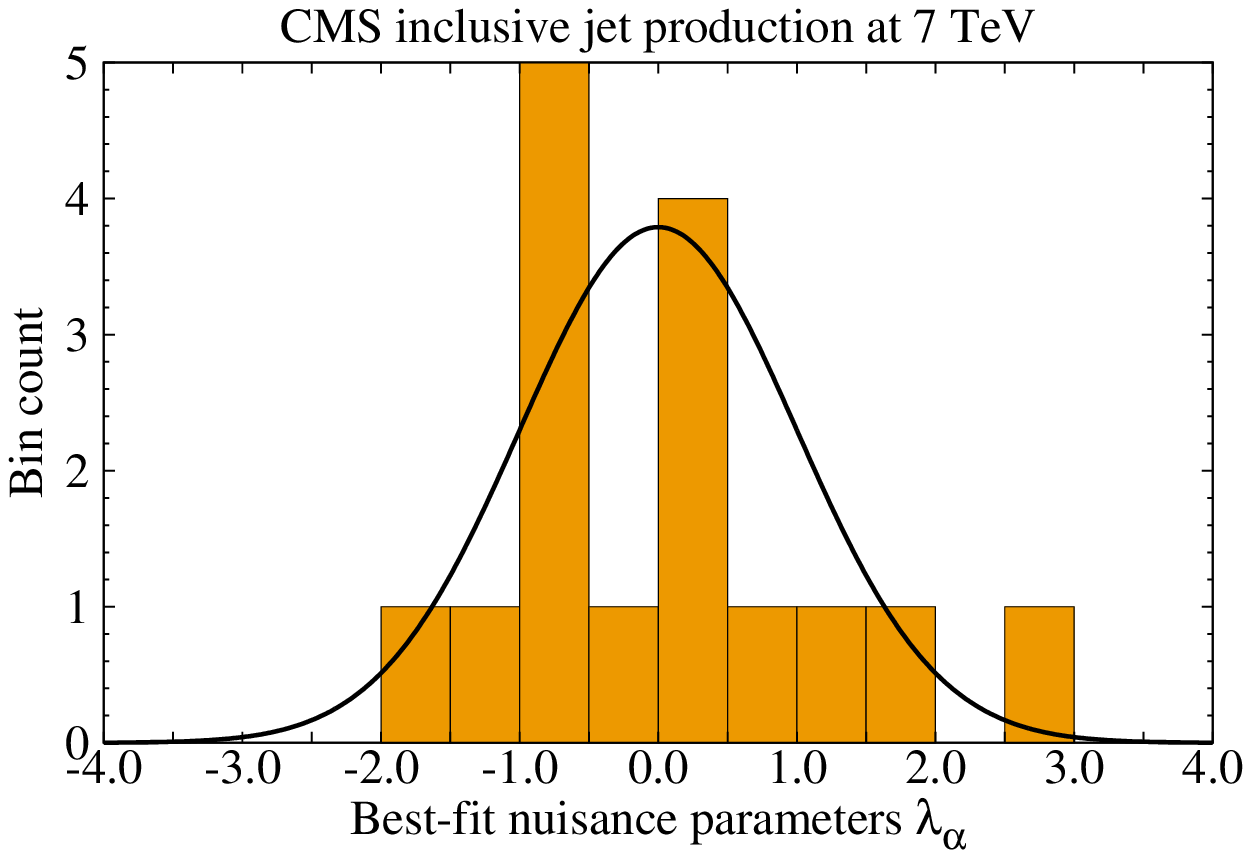}
\caption{Histogram of optimized nuisance parameters $\lambda_\alpha$
for the sources of correlated systematic errors
of the CMS 7 TeV inclusive jet production. The curve is the
standard normal distribution expected in the ideal case.
\label{fig:ds5384}}
\end{figure}

Figure ~\ref{fig:ds5381} shows a comparison between the measurements
for the CMS inclusive jet data at 7 TeV~\cite{Chatrchyan:2012bja} and
NLO theory prediction~\cite{Ellis:1992en,Kunszt:1992tn,Wobisch:2011ij} utilizing
CT14 NNLO PDFs. We  discussed earlier in the paper that
the missing NNLO contributions to the hard-scattering
cross section can be anticipated to be small under our QCD scale
choices, compared to the experimental uncertainty.

The CMS data, with 5 ${\rm fb}^{-1}$ of integrated luminosity, employ
the anti-$k_T$ jet algorithm~\cite{Cacciari:2011ma} with jet radius
$R=0.7$. The measurements are divided into 5 bins of rapidity and
presented as a function of the $p_T$ of the jet, with a total
of 133 data points. The theoretical prediction based on the CT14 NNLO PDFs
reproduces the behavior of experimental cross sections across thirteen
orders of magnitude.

Fig.~\ref{fig:ds5382} provides a more detailed look at these
distributions, by plotting the shifted central data values divided by
the theory. The data are shifted by optimal amounts based on the
treatment of the systematic errors as nuisance parameters,
cf. Ref.~\cite{Gao:2013xoa}. The error bars for the shifted data
include only uncorrelated errors,
i.e. statistical and uncorrelated systematic errors added in quadrature.
Here we notice moderate differences (up to a few tens of percent of the
central prediction) between theory and shifted data, which
elevate $\chi^2$ for this data set by about 2.5 standard deviations
for the central CT14 PDF set, or less for the error PDF sets.

Although they are not statistically significant, the origin of these mild
discrepancies can be further explored by studying the correlated shifts
allowed by the systematic uncertainties. In our implementation of
systematic errors \cite{Gao:2013xoa},
each correlated uncertainty $\alpha$ is associated with a normally distributed
random nuisance parameter $\lambda_\alpha$. When $\lambda_\alpha \neq 0$,
it may effectively shift the
central value of the data point $i$ in the fit by
\[
\beta_{i,\alpha} \lambda_\alpha= \sigma_{i,\alpha} X_i \lambda_{\alpha} \, ,
\]
where $\sigma_{i,\alpha}$ is the published fractional 1-$\sigma$ uncertainty of
data point $i$ due to systematic error $\alpha$. $X_i$ is the cross
section value that normalizes the fractional systematic uncertainty
\cite{Gao:2013xoa}, set equal to the
theoretical value $T_i$ in the procedure of the current analysis.

Each $\lambda_\alpha$ is adjusted to optimize the
agreement between theory and data. Fig.~\ref{fig:ds5384} shows a
histogram of the best-fit $\lambda_\alpha$ for the 19 sources of the
systematic errors published by CMS~\cite{Chatrchyan:2012bja}.
In an ideal situation,  the optimized
$\{\lambda_{\alpha}: \alpha = 1 ... 19\}$ would be normally distributed with a
mean value of 0 and standard deviation of 1.
The actual distribution of the $\lambda_\alpha$ values in
Fig.~\ref{fig:ds5384} appears to be somewhat narrower than the
standard normal one. This and relatively high $\chi^2/N_{pt}=1.33$
may indicate that either
uncorrelated systematic uncertainties are underestimated, or
higher-order theoretical calculations are needed to describe the data.

\subsubsection{ATLAS single-inclusive jet cross sections}
\begin{figure}[tb]
\vspace{10pt}
\includegraphics[width=0.50\textwidth]{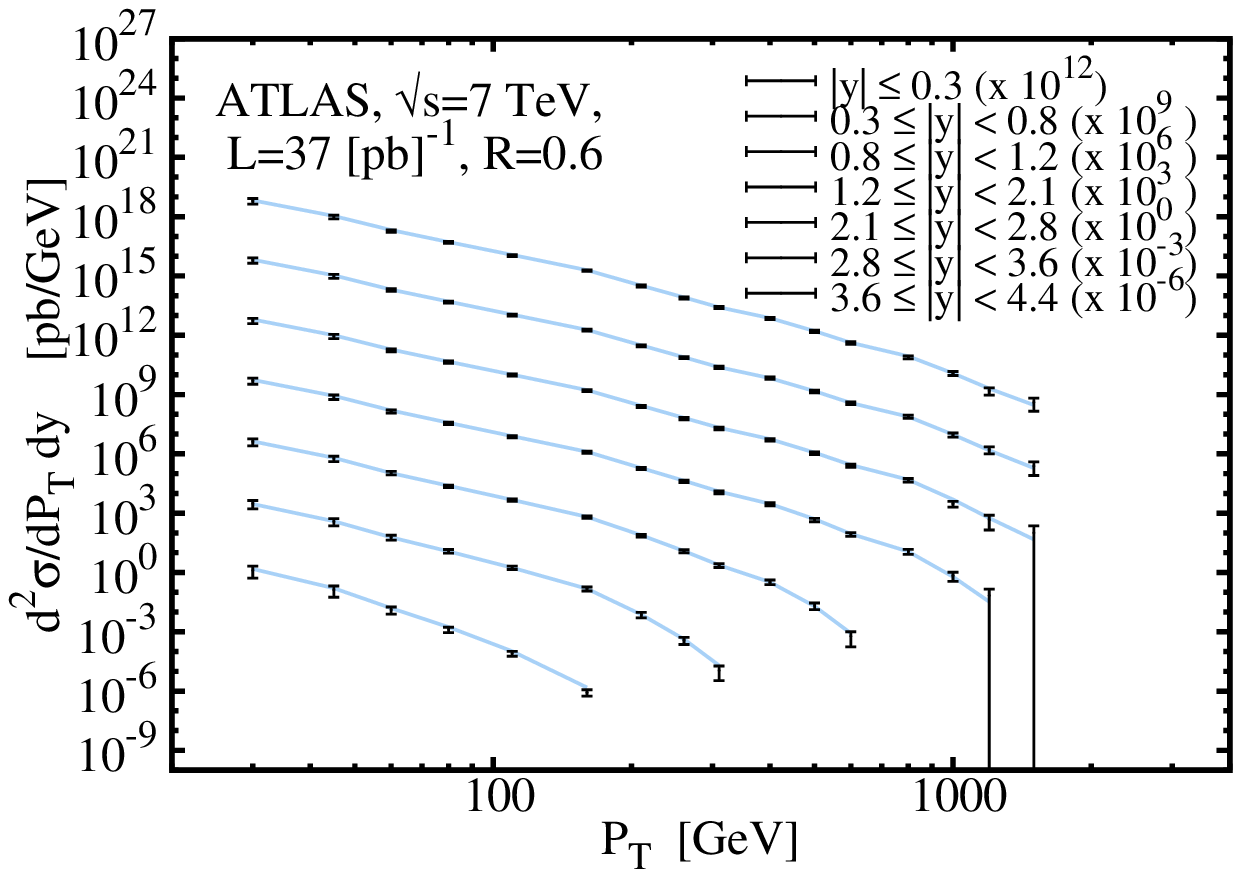}
\caption{
Comparison of data on $d^2\sigma / dp_{T}dy$
and NLO theory for the ATLAS 7 TeV inclusive jet
  production, using CT14 NNLO PDFs.
\label{fig:ds5351}}
\end{figure}

\begin{figure}[p]
\includegraphics[width=0.30\textwidth]{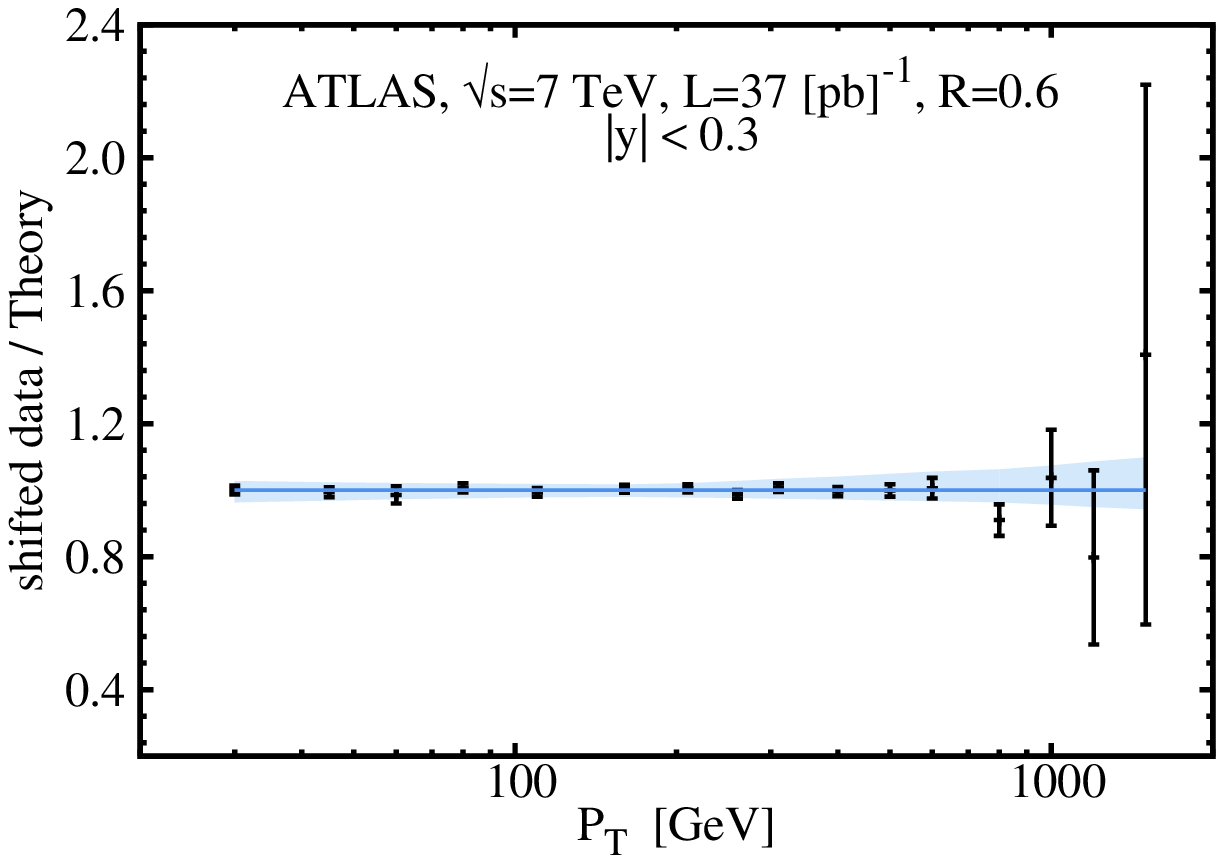}
\includegraphics[width=0.30\textwidth]{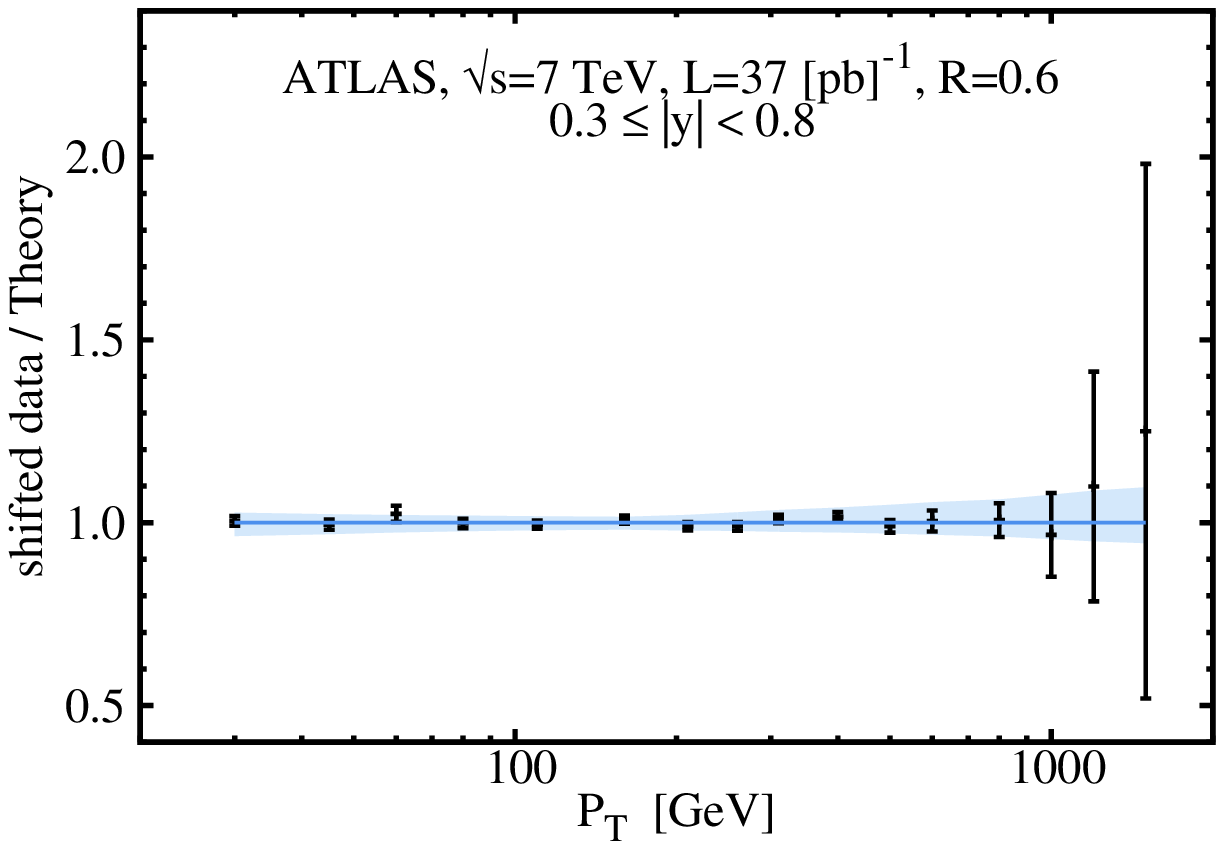}
\includegraphics[width=0.30\textwidth]{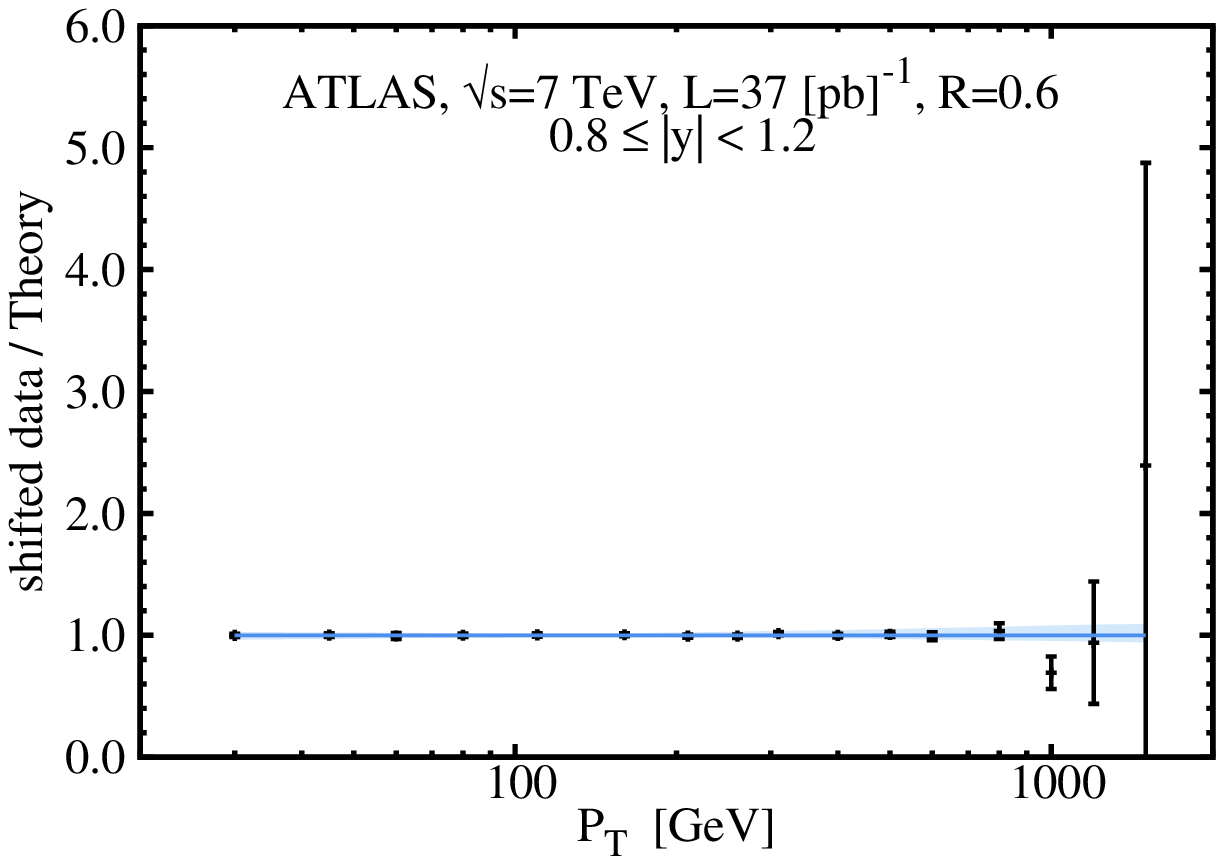}
\includegraphics[width=0.30\textwidth]{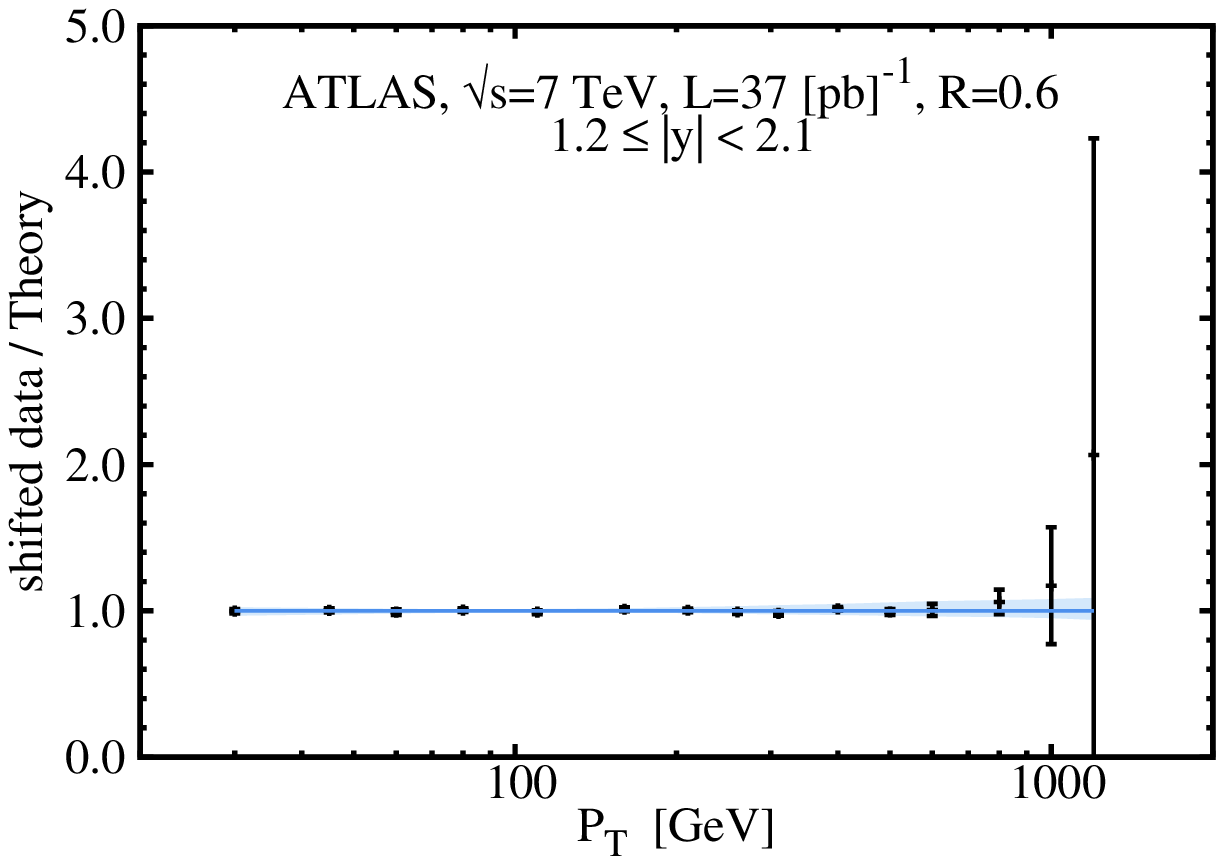}
\includegraphics[width=0.30\textwidth]{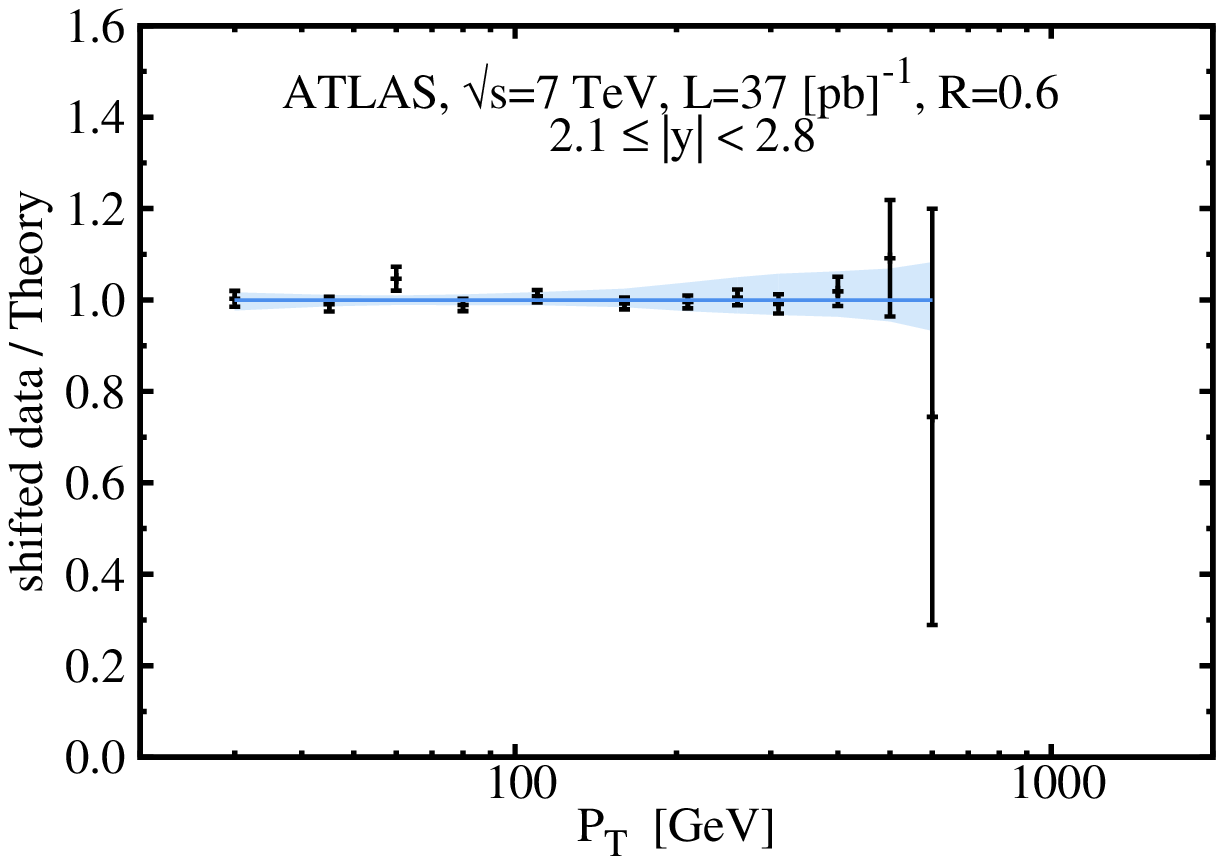}
\includegraphics[width=0.30\textwidth]{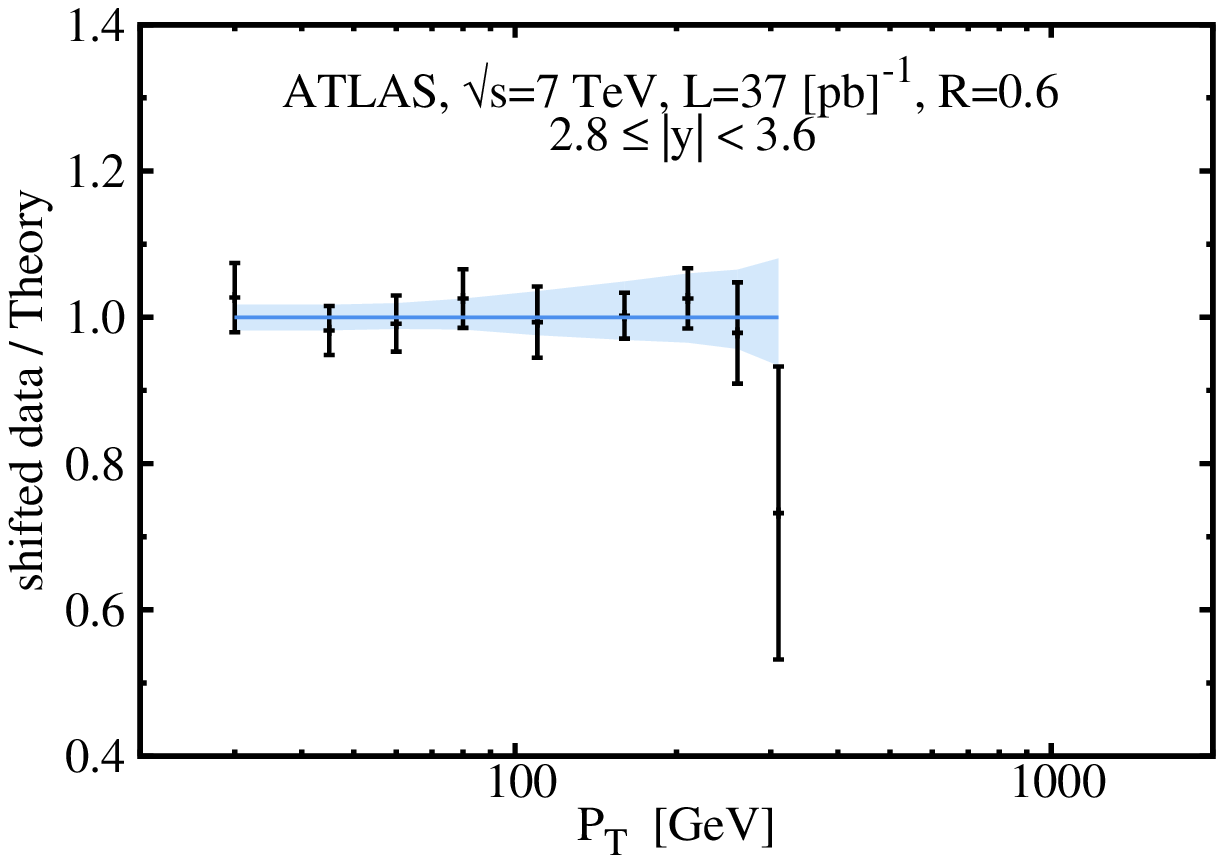}
\includegraphics[width=0.30\textwidth]{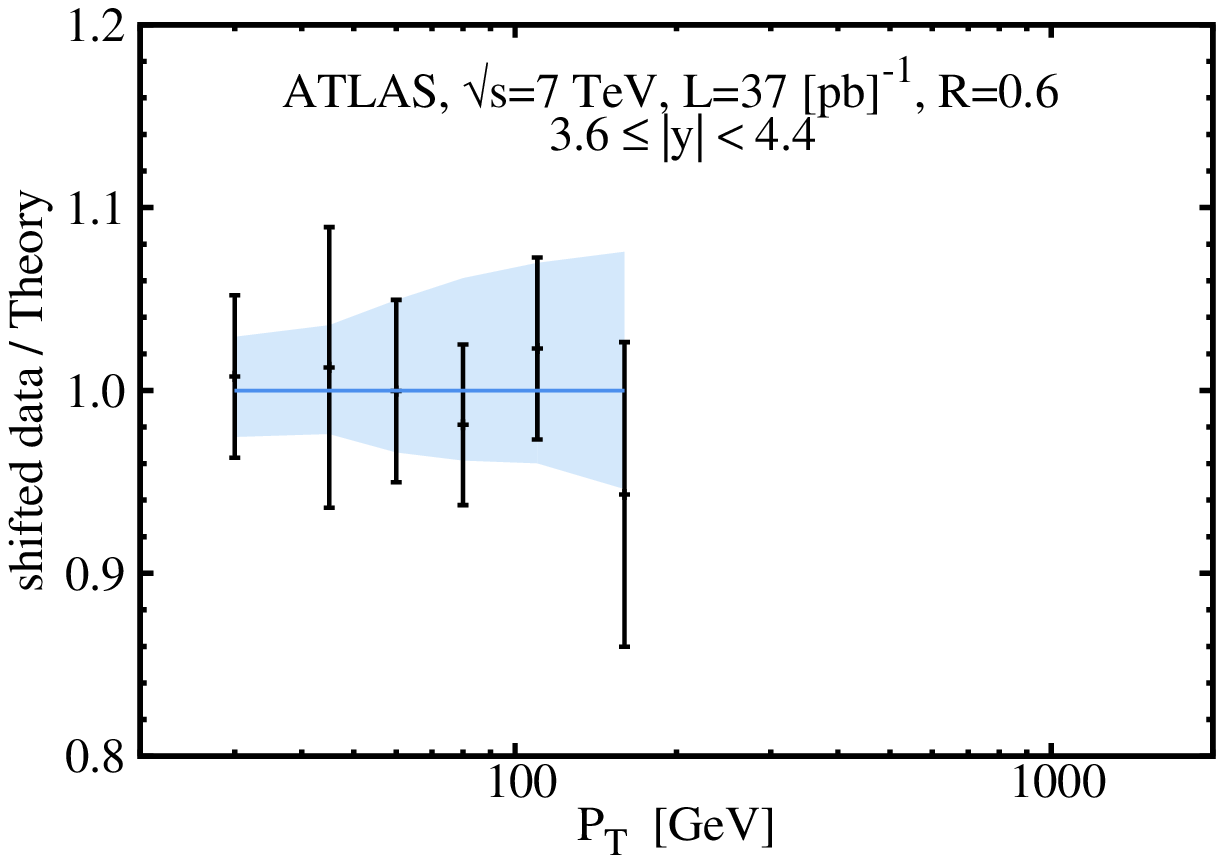}
\caption{Same as Fig.~\ref{fig:ds5351}, shown
as the ratio of shifted data for ATLAS 7 TeV divided by
the NLO theory.
The error bars
correspond to total uncorrelated errors. The shaded region
shows the 68\% C.L. PDF uncertainties.
\label{fig:ds5353}}
\end{figure}

\begin{figure}[p]
\includegraphics[width=0.50\textwidth]{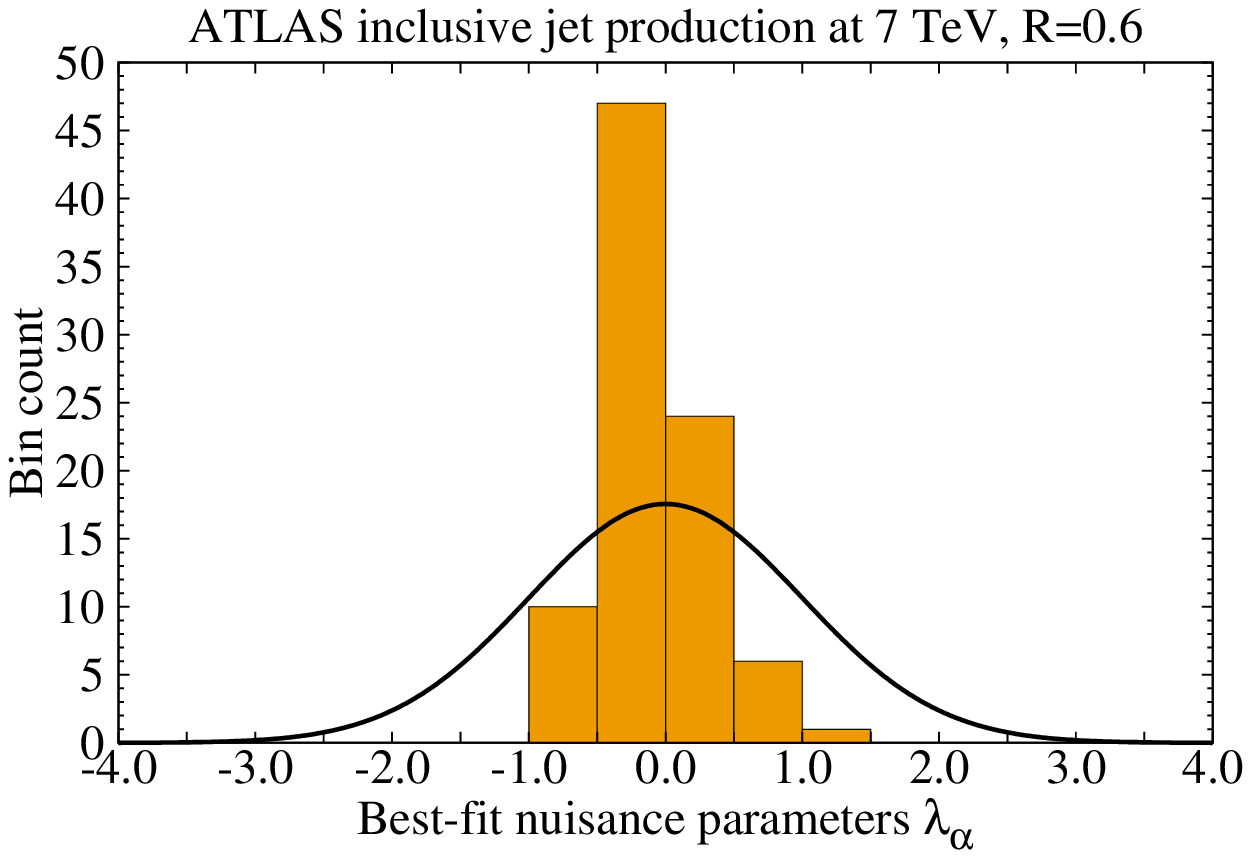}
\caption{Histogram of optimized nuisance parameters $\lambda_\alpha$
for the sources of correlated systematic errors
of the ATLAS 7 TeV inclusive jet production.
\label{fig:ds5354}}
\end{figure}

Equivalent comparisons for the ATLAS 7 TeV inclusive jet
production with $37\mbox{ pb}^{-1}$ of integrated
luminosity~\cite{Aad:2011fc}
are presented in Figs.~\ref{fig:ds5351} -- \ref{fig:ds5354}.
In this case, we compare to data in 7 bins of rapidity for
the anti-$k_T$ jet algorithm~\cite{Cacciari:2011ma} with jet radius $R=0.6$.
The agreement is excellent in all figures, not the least because both
statistical and systematic errors are still large in this early data set.
Among  119 sources of experimental errors that were identified, many
have little impact on the best fit. The resulting distribution
of the nuisance parameters in Fig.~\ref{fig:ds5354}
at the best fit is much narrower than the ideal Gaussian distribution,
indicating that most of the correlated sources need not deviate
from their nominal values when the PDFs are fitted.

To summarize, Figs.~\ref{fig:ds5381}-\ref{fig:ds5354} demonstrate
that CT14 PDFs agree with both sets of
CMS and ATLAS single-inclusive jet cross sections.
The ATLAS collaboration also measured inclusive jet production at
center of mass energy $\sqrt{s}=$ 2.76 TeV and published ratios
between the 2.76 and 7 TeV measurements in Ref.~\cite{Aad:2013lpa}.
These two measurements are well described by the theory prediction using CT14,
with a $\chi^2/N_{pt}\approx 1$. Furthermore, the ATLAS collaboration published
the inclusive jet measurements using another choice of jet radius of 0.4~\cite{Aad:2011fc}.
Both ATLAS and CMS collaborations measured cross sections for dijet production~\cite{Aad:2011fc,Chatrchyan:2012bja}
based on the same data sample of the single-inclusive jet measurements.
These measurements are not included in the CT14 global analysis because of
the correlations between the two (di-jet and single-inclusive jet) data sets.
However, it has been verified that the CT14 analysis gives a good
description for all these data sets as well.

\subsection{Differential cross sections for lepton pair production at the LHC}

\subsubsection{Charged lepton pseudorapidity distributions in
  $W/Z$ boson production \label{sec:LHCWZrap}}

\begin{figure}[tb]
\includegraphics[width=0.4\textwidth]{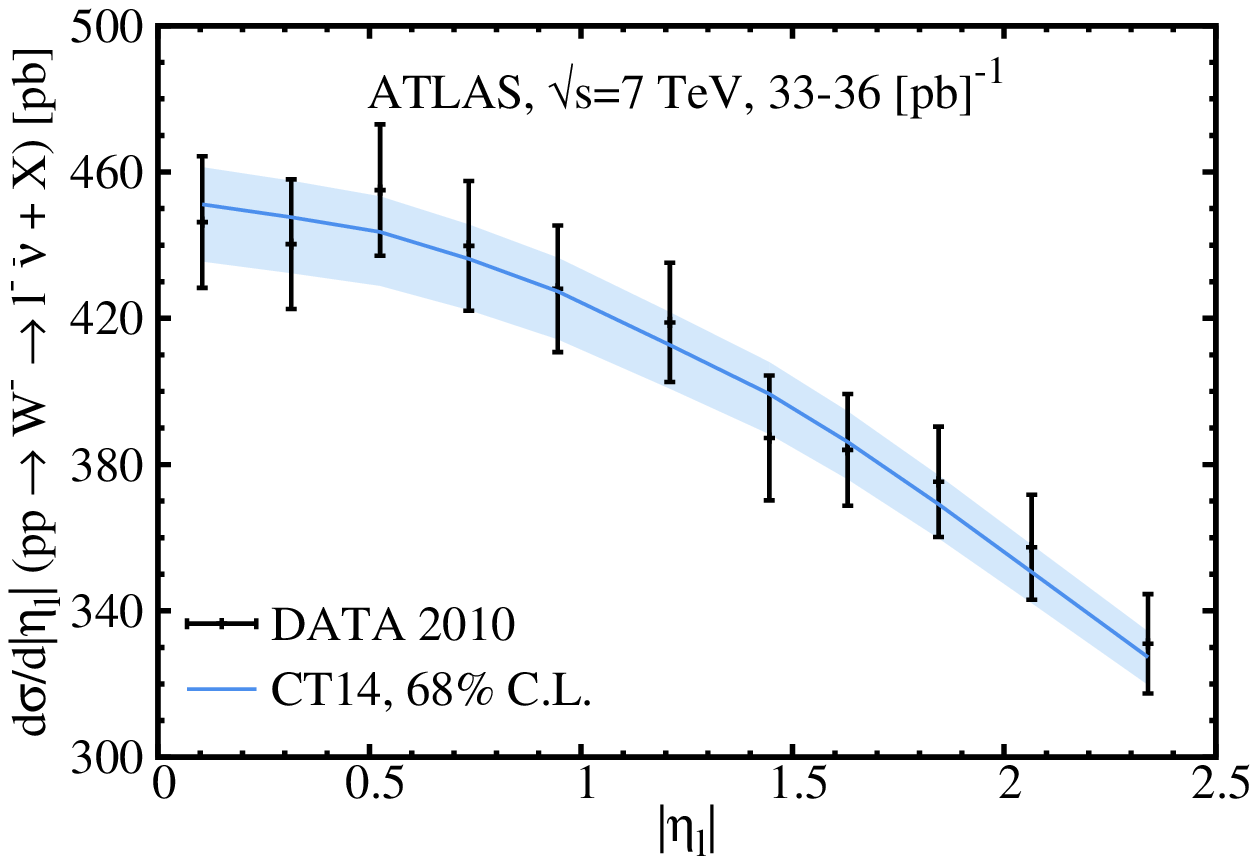}
\includegraphics[width=0.4\textwidth]{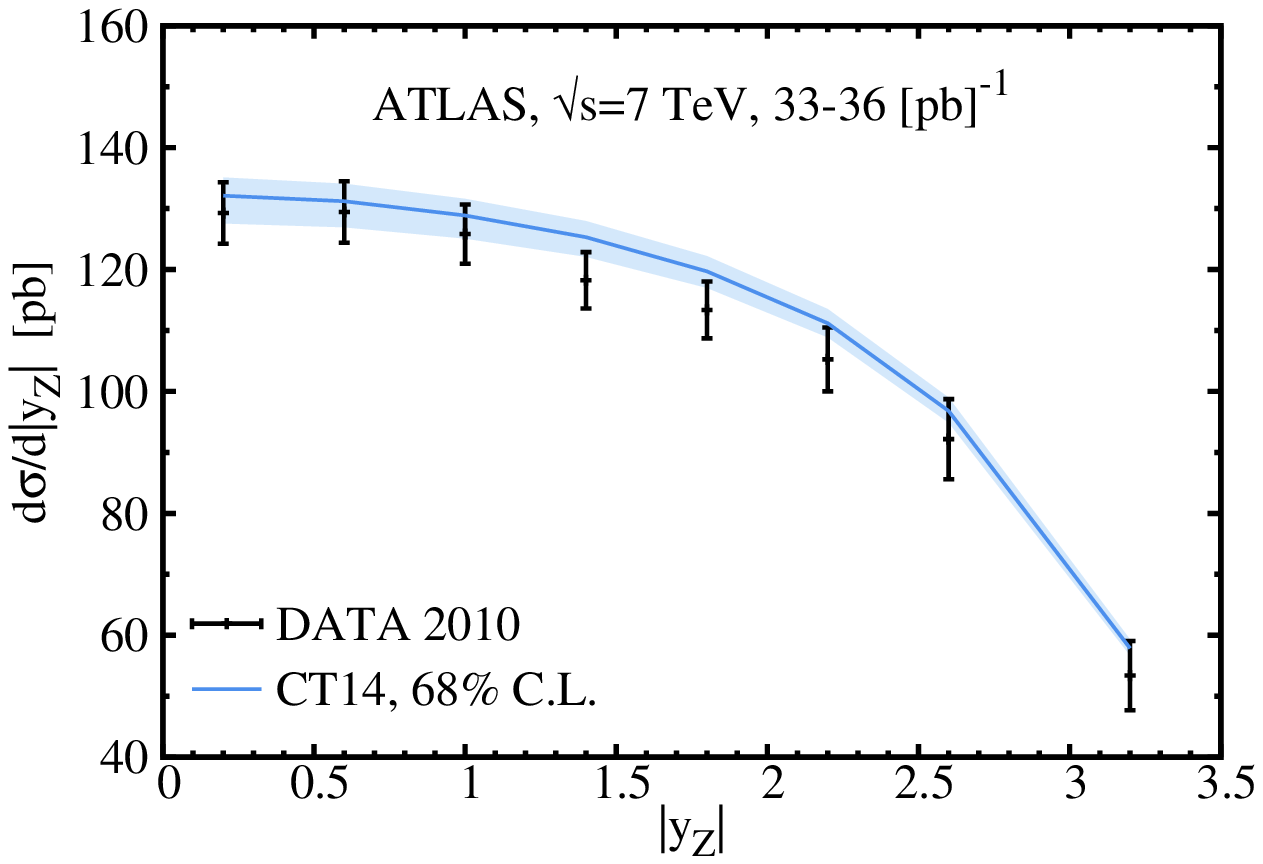}
\includegraphics[width=0.4\textwidth]{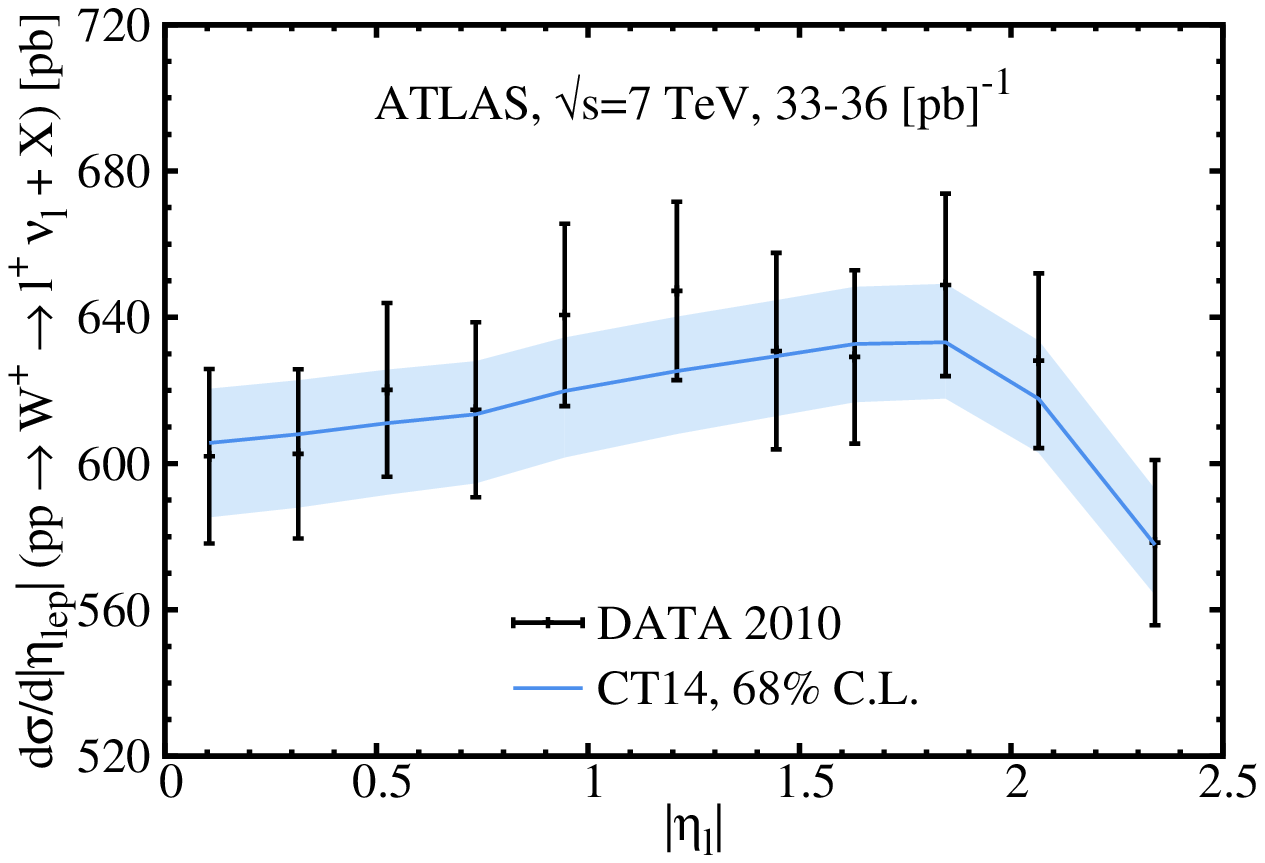}
\caption{
Comparison between the 2010 ATLAS measurements~\cite{Aad:2011dm} of the
$W^{\pm}$ charged-lepton pseudorapidity and $Z$ boson rapidity distributions
at $\sqrt{s}=$ 7 TeV, and the \textsc{ResBos} theory using CT14 NNLO PDFs.
\label{fig:ds268-123}}
\end{figure}

Differential cross sections for production of massive vector bosons
set important constraints on the flavor composition of the proton, notably
on the $u$ and $d$ quarks, anti-quarks and their ratios.
Figure~\ref{fig:ds268-123} compares  CT14 NNLO theoretical
predictions  with pseudorapidity ($|\eta_\ell|$)
distributions of charged leptons from
inclusive $W^\pm$ and $Z^0$ production and decay in the
2010 ATLAS 7 TeV data sample with 33-36 ${\rm pb}^{-1}$ of integrated
luminosity~\cite{Aad:2011dm}.
Theoretical predictions are computed using the program
\textsc{ResBos}.
The black data points represent the unshifted central values of the
data. The error bars indicate the total (statistical+systematic)
experimental error.
The blue band is the CT14 PDF uncertainty evaluated at the 68\% C.L.
These three measurements share correlated systematic errors.
From the figures, we see that the data are described well by theory
over the entire rapidity range, even in the absence of correlated
systematic shifts. The PDF uncertainties are similar in their size
to those of the experimental measurements and, overall, the
theory predictions are within one standard deviation of the data.

\subsubsection{
Influence of $W$ boson charge asymmetry measurements at the LHC}

\begin{figure}[tb]
\includegraphics[width=0.4\textwidth]{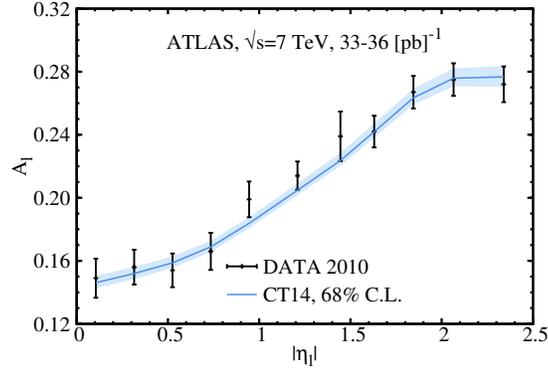}
\caption{
 $W^{\pm}$ charge asymmetry as a function of lepton pseudorapidity
measured by the ATLAS Collaboration, compared to the 68\% C.L. CT14
NNLO uncertainty band. The kinematic requirements are $p_{T\ell} > 20$
GeV, $p_{T\nu_\ell} > 25$ GeV and $M_T^{\ell\nu_\ell} > 40$ GeV.
\label{fig:ds268asy}}
\end{figure}

\begin{figure}[ht!]
\includegraphics[width=0.4\textwidth]{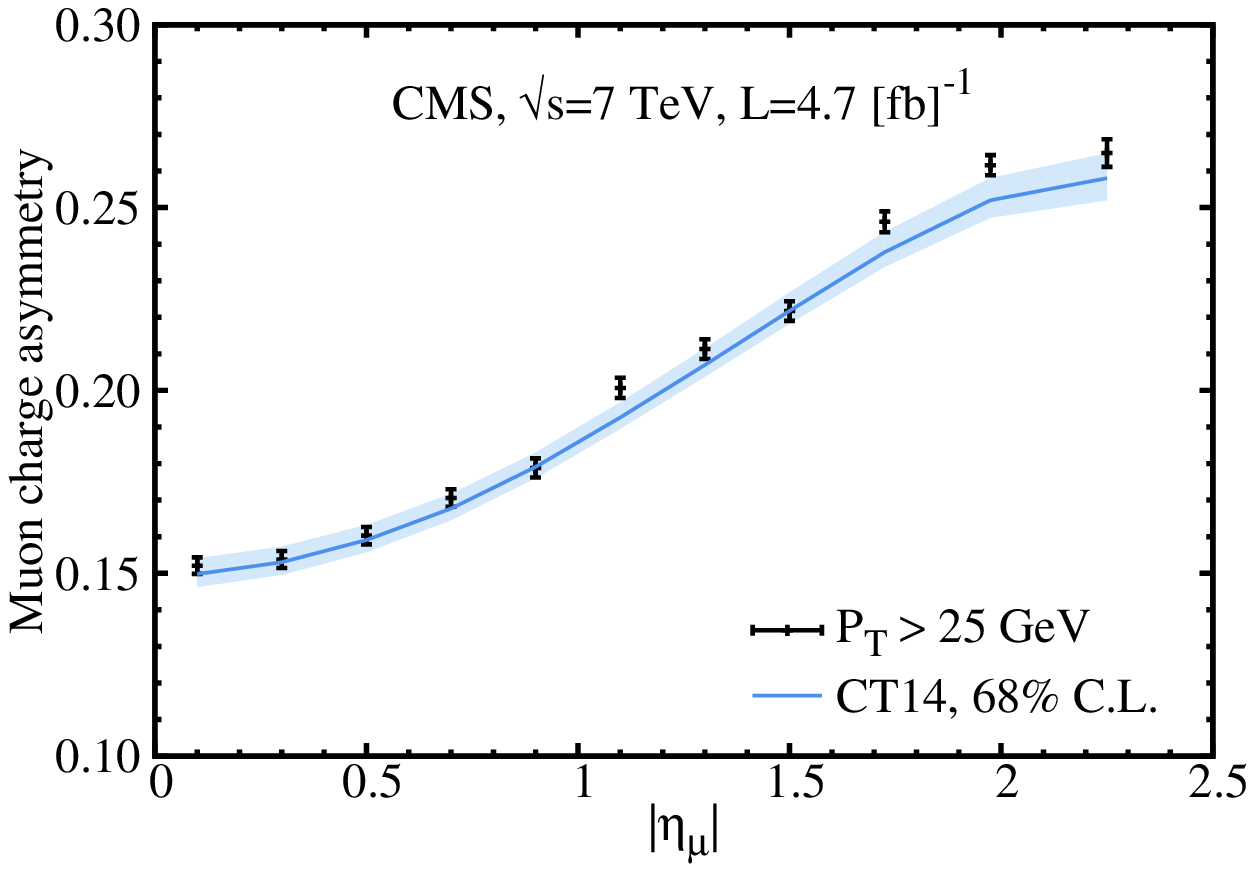}
\includegraphics[width=0.4\textwidth]{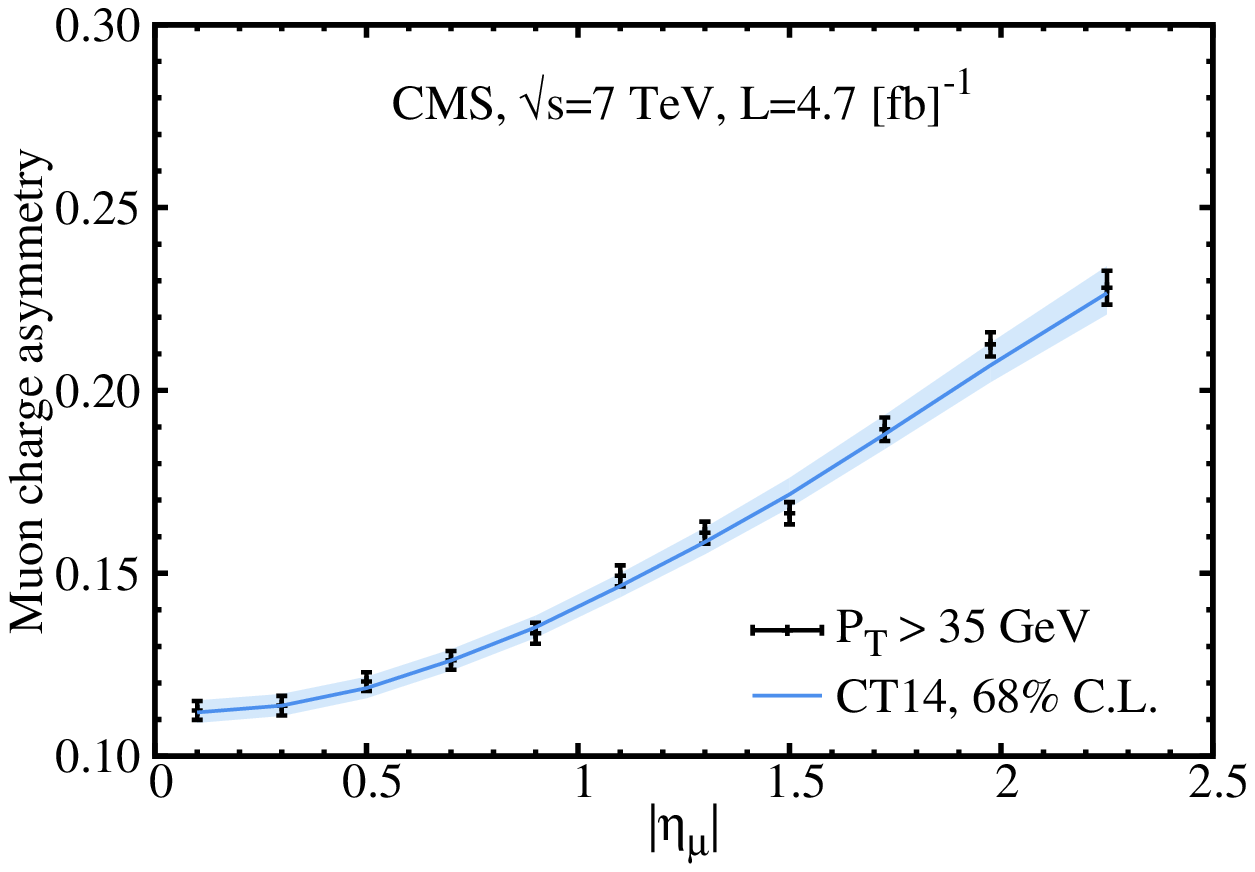}
\includegraphics[width=0.4\textwidth]{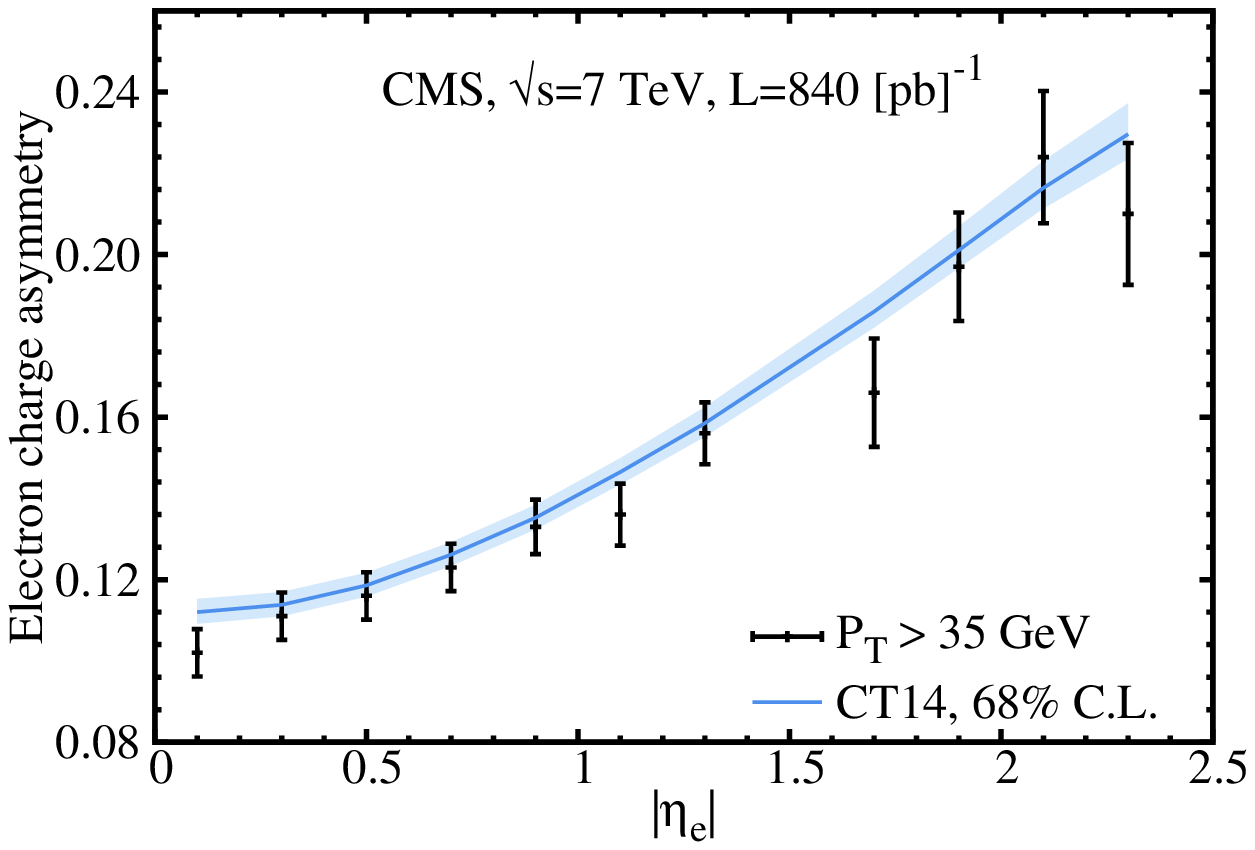}
\caption{Charge asymmetry of decay muons and electrons
from $W^{\pm}$ production measured by the CMS experiment.
The data values have $p_{T\ell} > 25$ or $35$ GeV for the muon data and
$p_{T\ell} > 35$ GeV for the electron data.
The vertical error bars on the data points indicate total (statistical and systematic) uncertainties.
The curve shows the CT14 theoretical calculation;
the shaded region is the PDF uncertainty at 68\% C.L.
\label{fig:ds266}}
\end{figure}

Another handy observable for determining the parton distribution functions
is the {\em charge asymmetry}
for $W^{+}$ and $W^{-}$ bosons produced in $pp$ or $p\bar p$ collisions.
This process has been measured
both at the Tevatron and at the LHC. As the asymmetry involves a ratio of
the cross
sections, many experimental systematic errors cancel, leading to very precise
results. Without these collider data, the main information about the
difference between the light flavors, $d, \overline{d}$
and $u, \overline{u}$,
would come from the BCDMS and NMC experiments, which are
measurements of muon deep-inelastic scattering on proton and deuteron
targets. Under the assumption of charge symmetry between the nucleons,
the difference of the proton and deuteron cross sections distinguishes
between the $u$ and $d$ PDFs in a nucleon.
However, the deuteron measurements are subject to nuclear binding
corrections, which have been estimated by introducing nuclear
models~\cite{Accardi:2011fa,Owens:2012bv,Martin:2012da}, but are not calculable
from first principles. In contrast, the $W^{\pm}$ charge asymmetry
data from the Tevatron and LHC colliders directly provide information
about the difference between $d$ and $u$ flavors, without the need for
nuclear corrections. By including the ATLAS and CMS charge asymmetry
data, we are able to obtain, for the first time, direct experimental
constraints on the differences of the quark and antiquark PDFs for $u$
and $d$ flavors at $x\approx 0.02$ typical for the 7 TeV kinematics.
\begin{figure}[tb]
\includegraphics[width=0.50\textwidth]{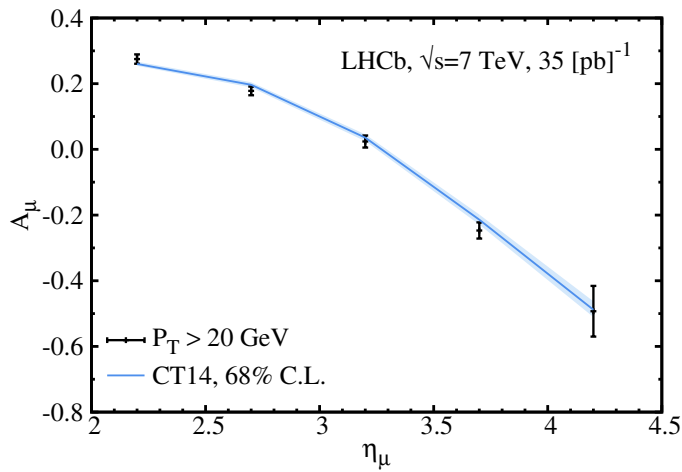}
\caption{Charge asymmetry of decay muons
from $W^{\pm}$ production measured by the LHCb experiment.
\label{fig:ds241asy}}
\end{figure}

Figure~\ref{fig:ds268asy} shows a comparison of data and theory,
for the lepton charge asymmetry of inclusive $W^{\pm}$  production,
from the ATLAS experiment at the LHC 7 TeV~\cite{Aad:2011dm}.
These asymmetry data are correlated with the
$W/Z$ rapidity measurements discussed in the previous subsection; all
four $W/Z$ data sets are included in the CT14 global analysis
using a {\em shared} correlation matrix from
the ATLAS publication~\cite{Aad:2011dm}.
The measurement was carried out with several kinematic cuts.
The lepton transverse momentum was required to be greater than 20 GeV,
the missing transverse energy to be greater than 25 GeV,
and the lepton-neutrino transverse mass to be greater than 40 GeV.
The shaded region is the PDF uncertainty of CT14 NNLO at 68\% C.L.
Again the points with error bars represent the unshifted data with
the experimental errors added in quadrature.
The data fluctuate around the CT14 predictions and are described well
by the CT14 error band.

Figure~\ref{fig:ds266} presents a similar comparison of the unshifted
data and CT14 NNLO theory
for the charge asymmetry of decay muons~\cite{Chatrchyan:2013mza}
and electrons~\cite{Chatrchyan:2012xt} from inclusive $W^{\pm}$ production
from the CMS experiment at the LHC 7 TeV.
The asymmetry for muons is measured
with 4.7 ${\rm fb}^{-1}$ of integrated luminosity,
with $p_{T\ell} > 25$ and $35$ GeV; the asymmetry for electrons is measured
with 840 ${\rm pb}^{-1}$ and $p_{T\ell} > 35$ GeV.
Here we note that the CMS measurement
does not apply a missing $E_{T}$ cut to $A_{ch}$, contrary to the
counterpart ATLAS $A_{ch}$ measurement.
Theory predictions are the same for both the
muon and electron channels with the same cuts.
The muon and electron data are consistent with one another, but
the muon data have smaller statistical and systematic uncertainties,
as is apparent in Fig.~\ref{fig:ds266}.
All three subsets of CMS $A_{ch}$ agree with predictions using CT14;
their $\chi^2$ is further improved by optimizing the correlated shifts.
The electron data and the muon data with the $p_{T\ell}$  cut
of 35 GeV are included in the CT14 global analysis.
The muon data with a $p_{T\ell}$ cut of 25 GeV are not included
in the CT14 analysis, but nevertheless are well described.

In the LHCb measurement of the charged lepton asymmetry at 7
TeV~\cite{Aaij:2012vn}, the muons are required to have a transverse
momentum greater than 20 GeV. The corresponding comparison of the CT14
NNLO predictions to the LHCb $A_{ch}$ data is shown in Fig.~\ref{fig:ds241asy}.
The LHCb case is especially interesting,
as the LHCb acceptance for charged leptons extends beyond
the rapidity range measured by ATLAS and CMS.
Thus, the LHCb results are sensitive to the $u$ and $d$ quark PDFs
at larger $x$ values than at the ATLAS or CMS.
Good agreement between data and theory is again observed.

\begin{figure}[ht]
\includegraphics[width=0.4\textwidth]{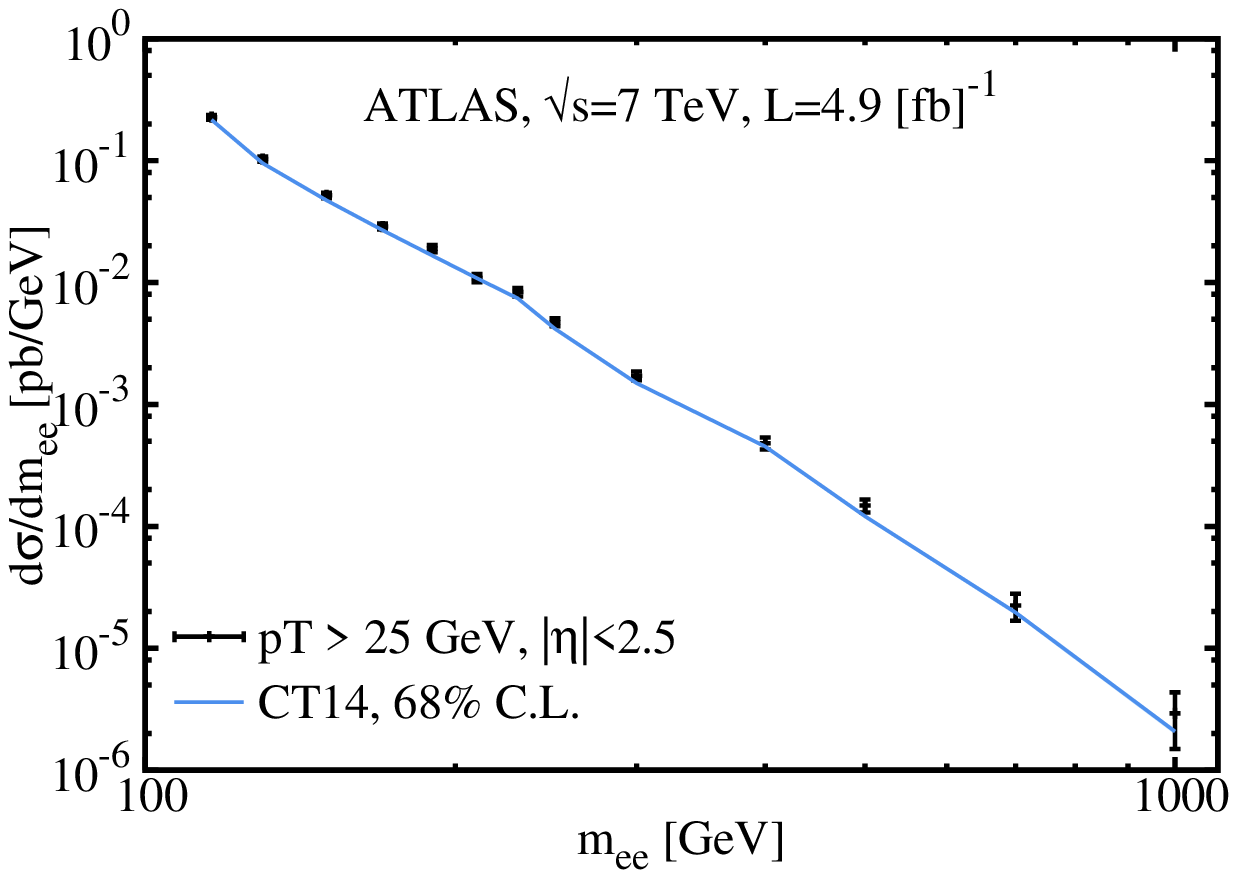}
\includegraphics[width=0.4\textwidth]{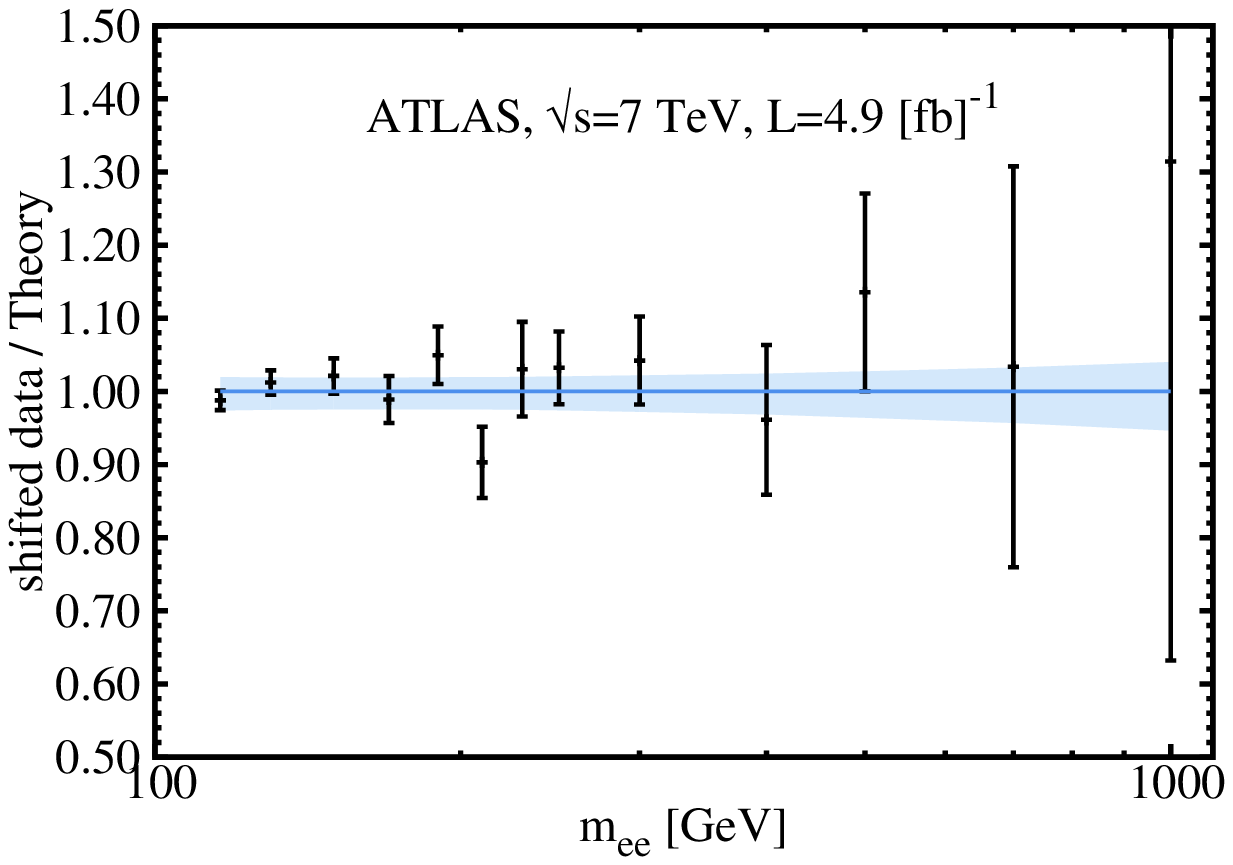}
\caption{Invariant mass distributions of Drell-Yan pairs in the
  high-mass region by ATLAS 7 TeV \cite{Aad:2013iua},
with superimposed NNLO predictions  based on CT14 NNLO PDFs. The left
subfigure shows the differential cross sections as a function of the
dilepton mass $m_{ee}$. The right subfigure shows the ratio of ATLAS shifted
data to CT14 theory predictions.
\label{fig:ds275}}
\end{figure}

\begin{figure}[ht]
\includegraphics[width=0.4\textwidth]{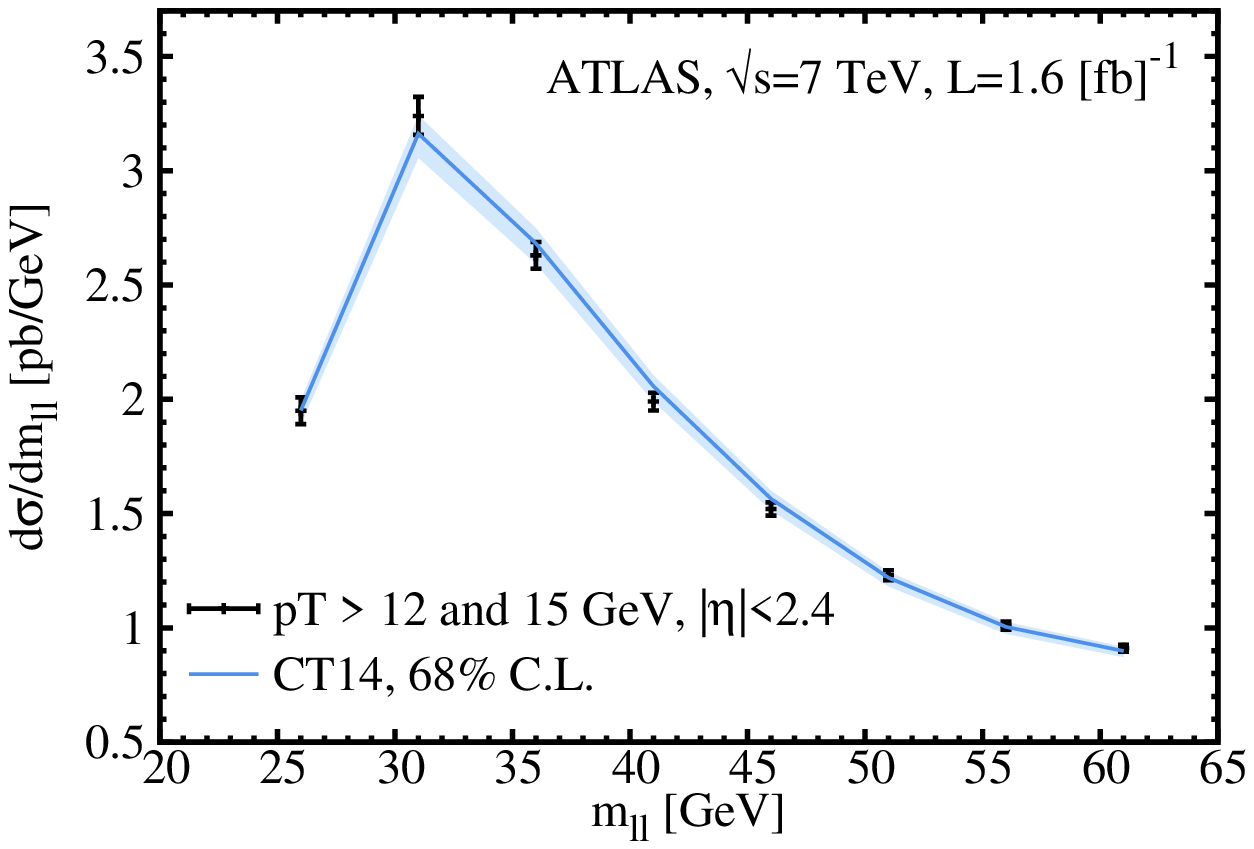}
\includegraphics[width=0.4\textwidth]{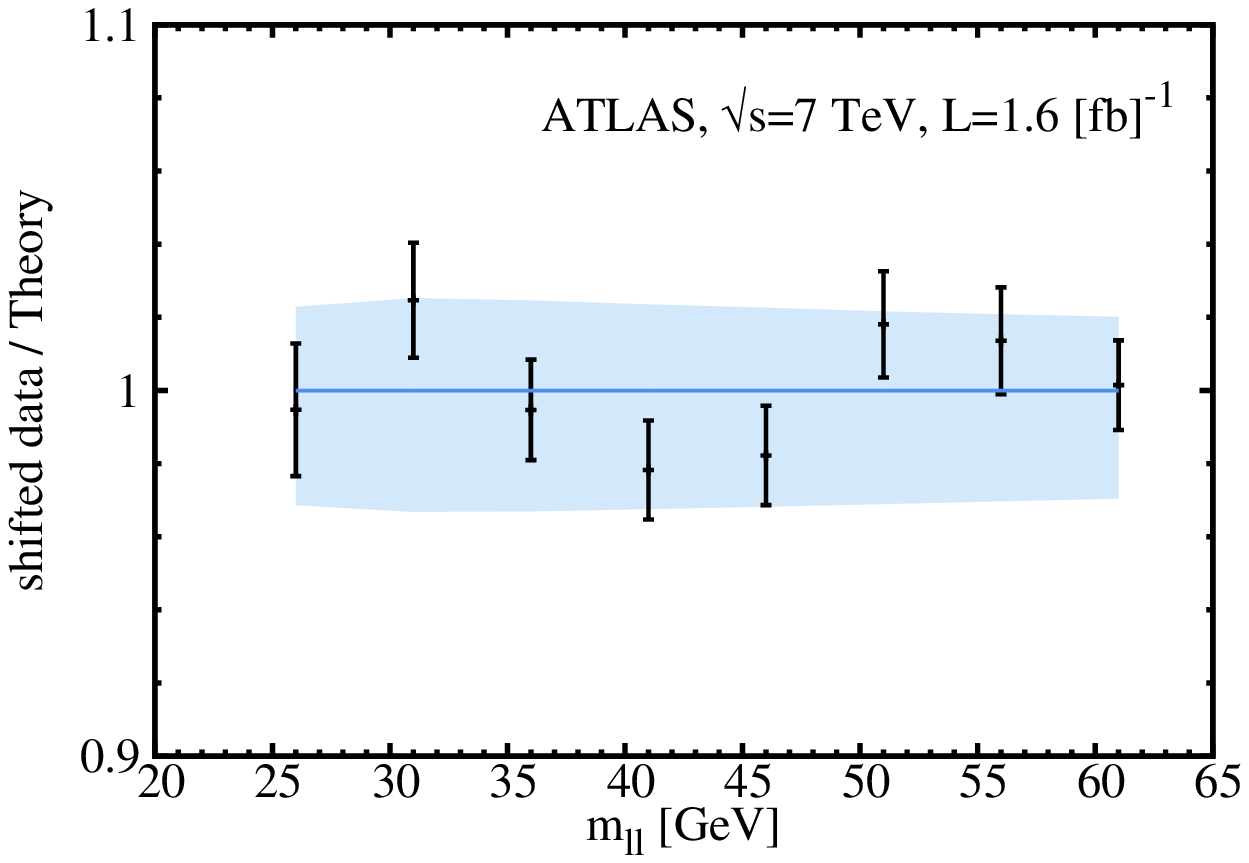}\\
\includegraphics[width=0.4\textwidth]{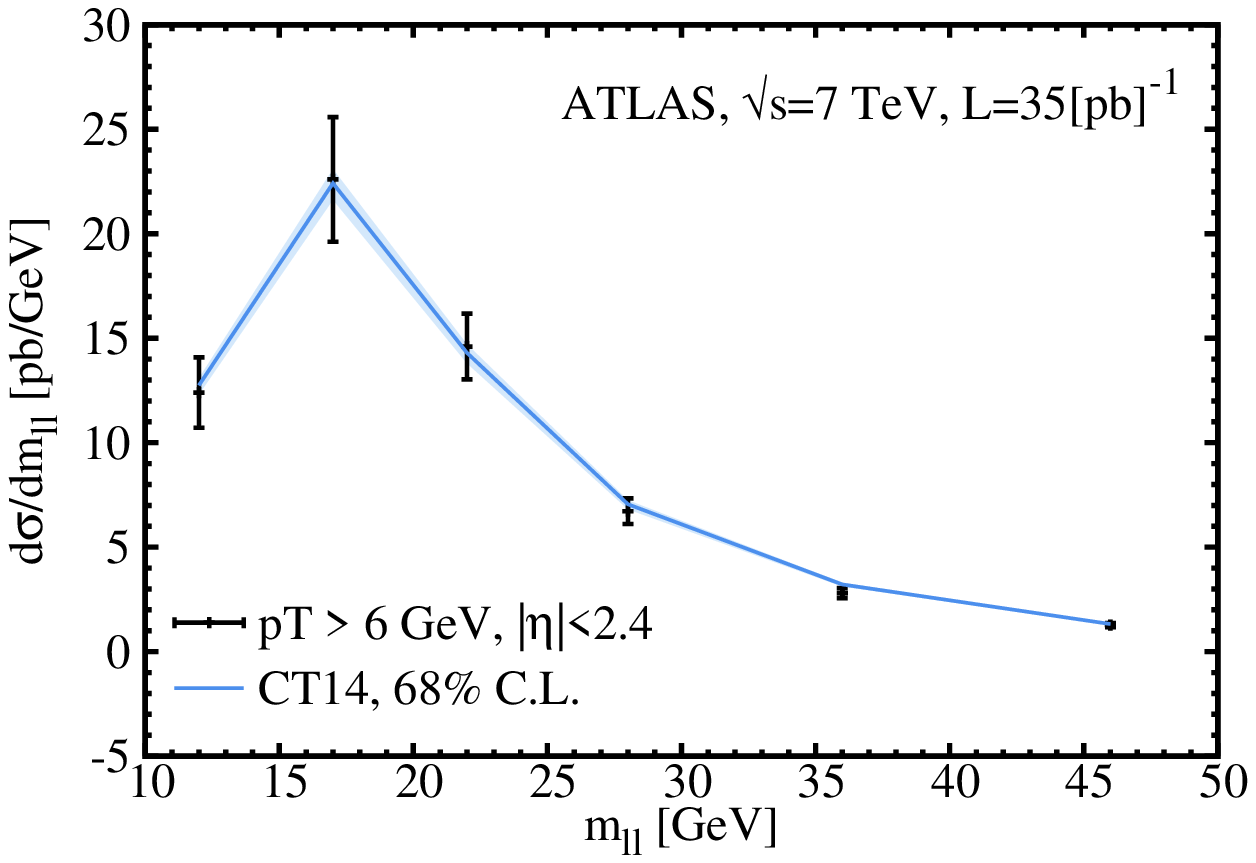}
\includegraphics[width=0.4\textwidth]{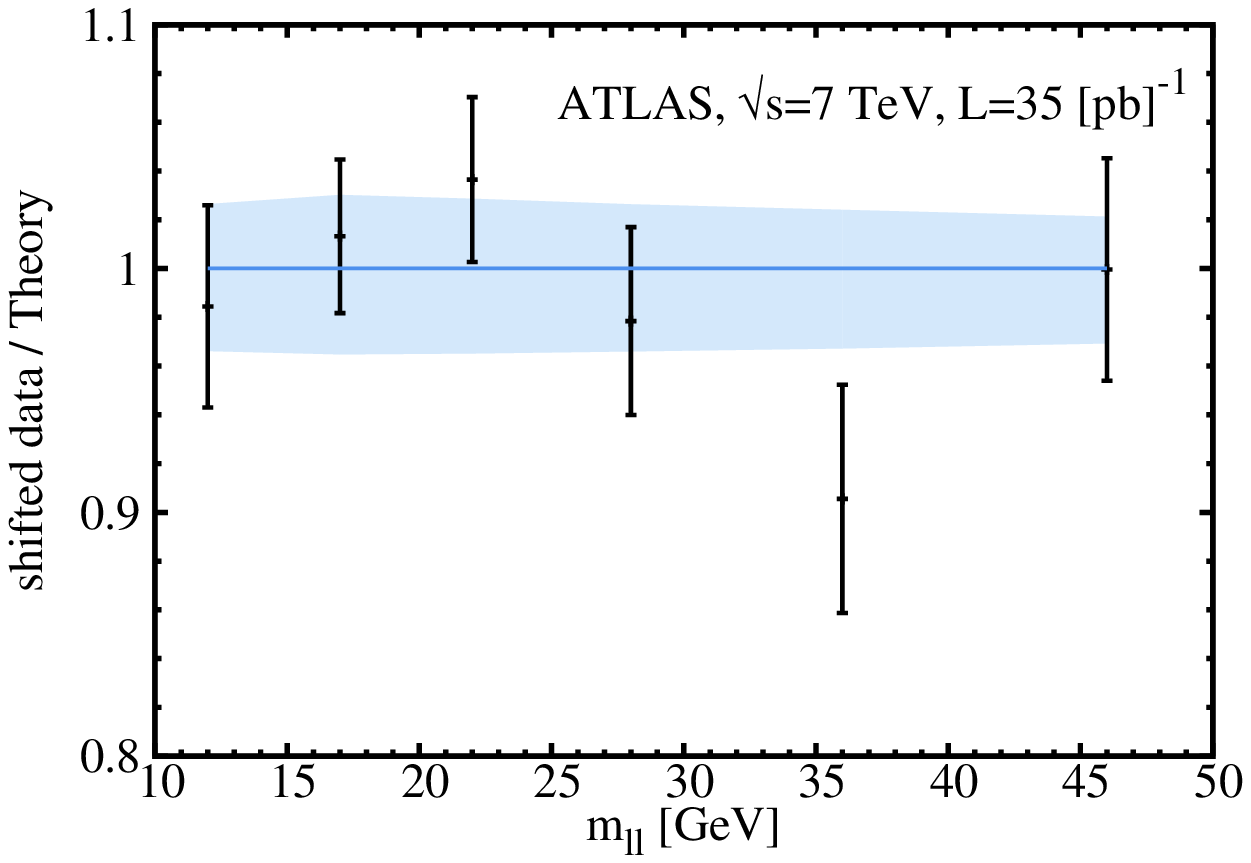}
\caption{
Same as in Fig.~\ref{fig:ds275}, for ATLAS 7 TeV differential distributions
of Drell-Yan pairs in the low-mass and extended low-mass regions \cite{Aad:2014qja}.
\label{fig:ds2767}}
\end{figure}

\subsubsection{Production of Drell-Yan pairs at ATLAS}
 In Figs.~\ref{fig:ds275} and \ref{fig:ds2767},
we compare CT14 NNLO predictions to ATLAS 7 TeV
measurements of differential cross sections for production of
high-mass \cite{Aad:2013iua}
and low-mass \cite{Aad:2014qja} Drell-Yan pairs, plotted
as a function of dilepton invariant mass $m_{\ell \ell}$.
The experimental cross sections correspond to the ``electroweak Born
level'', unfolded from the raw data by correcting for electroweak
final-state radiation.
The high-mass data sample corresponds to $116 < m_{\ell \ell} < 1000 $ GeV.
At low dilepton masses, we compare to the combined electron+muon
sample at $26 < m_{\ell \ell} < 66$ GeV for
${\sc L} = 1.6\mbox{  fb}^{-1}$ in the upper row, as well as to
the muon sample at $12 < m_{\ell \ell} < 66$ GeV for
${\sc L} = 35\mbox{  pb}^{-1}$ in the lower row. Fiducial
acceptance cuts on the decay leptons are specified inside the
figures. Correlated experimental uncertainties are included in the
comparison.

On the theory side, the cross sections are calculated
at NNLO in QCD with {\sc ApplGrid}~ interface \cite{Carli:2010rw} to FEWZ
\cite{Melnikov:2006kv,Gavin:2010az,Gavin:2012sy,Li:2012wna}, and
including photon-scattering contributions.
Experimental uncertainties in these cross sections tend to be larger
than the PDF uncertainties, as illustrated by the figures,
hence we only compare these data to
the CT14 predictions {\it a posteriori}, without actually including
them in the CT14 fit.

It can be observed in the figures that CT14 NNLO PDFs agree well
with the high-mass and low-mass data samples both in terms of the
cross sections (in the left subfigures) and ratios of the shifted data
to theoretical predictions (right subfigures). The PDF uncertainty
bands, indicated by light-blue color, approximate the average
behavior of the experimental data without systematic discrepancies.

\subsection{$W^{\pm}$ charge asymmetry from the D\O~ experiment at the Tevatron}

We reviewed above that, historically,
measurements of $W^{\pm}$ charge asymmetry at the Tevatron
have been important in the CTEQ-TEA global analysis.
For example, the CTEQ6 PDFs (circa 2002) and CT10 PDFs (circa 2010-2012)
included the $W^{\pm}$ asymmetry data from the CDF and D\O~ experiments
to supplement the constraints on $u$ and $d$ quark PDFs at $x > 0.1$
from fixed-target DIS experiments.
The charge asymmetry at the Tevatron probes the differences of the
{\em slope in $x$} of the PDFs for $u$ and $d$ flavors.

\begin{figure}[ht!]
\includegraphics[width=0.43\textwidth]{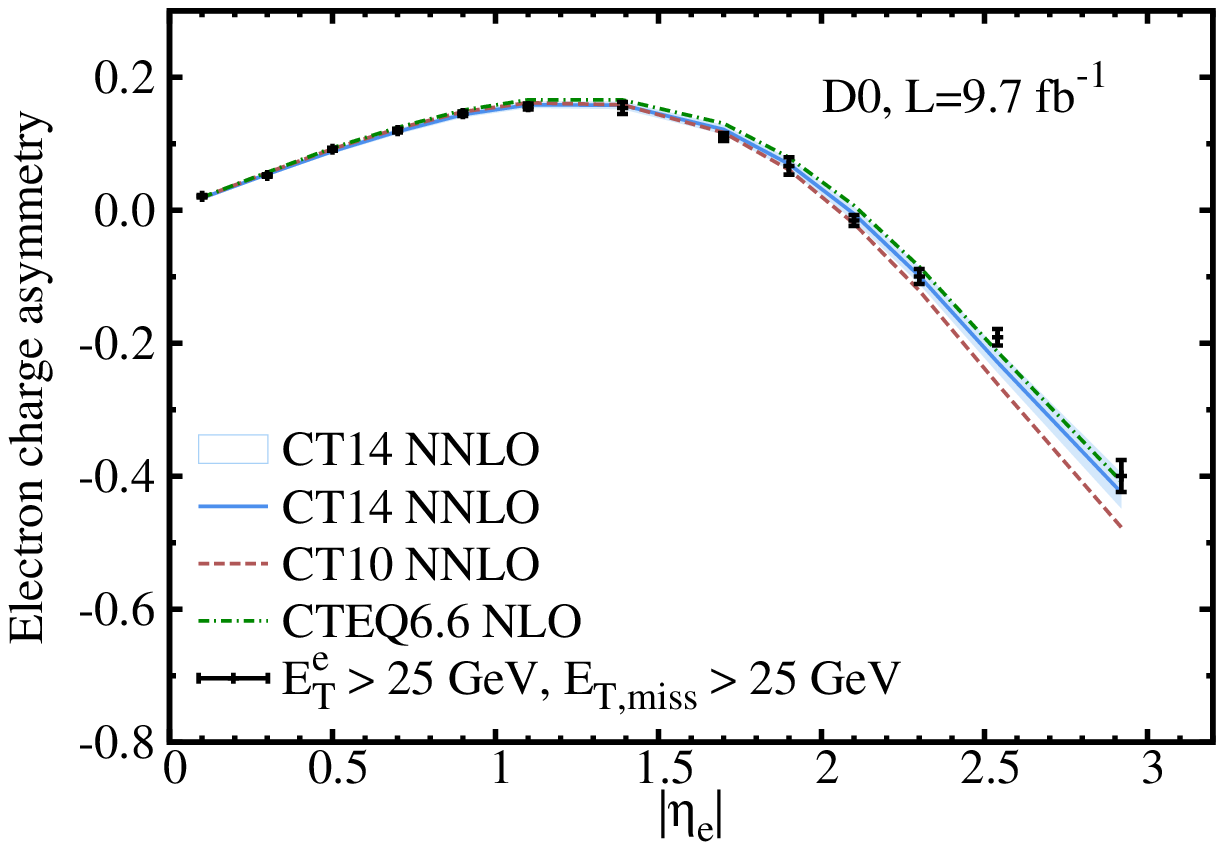}
\includegraphics[width=0.43\textwidth]{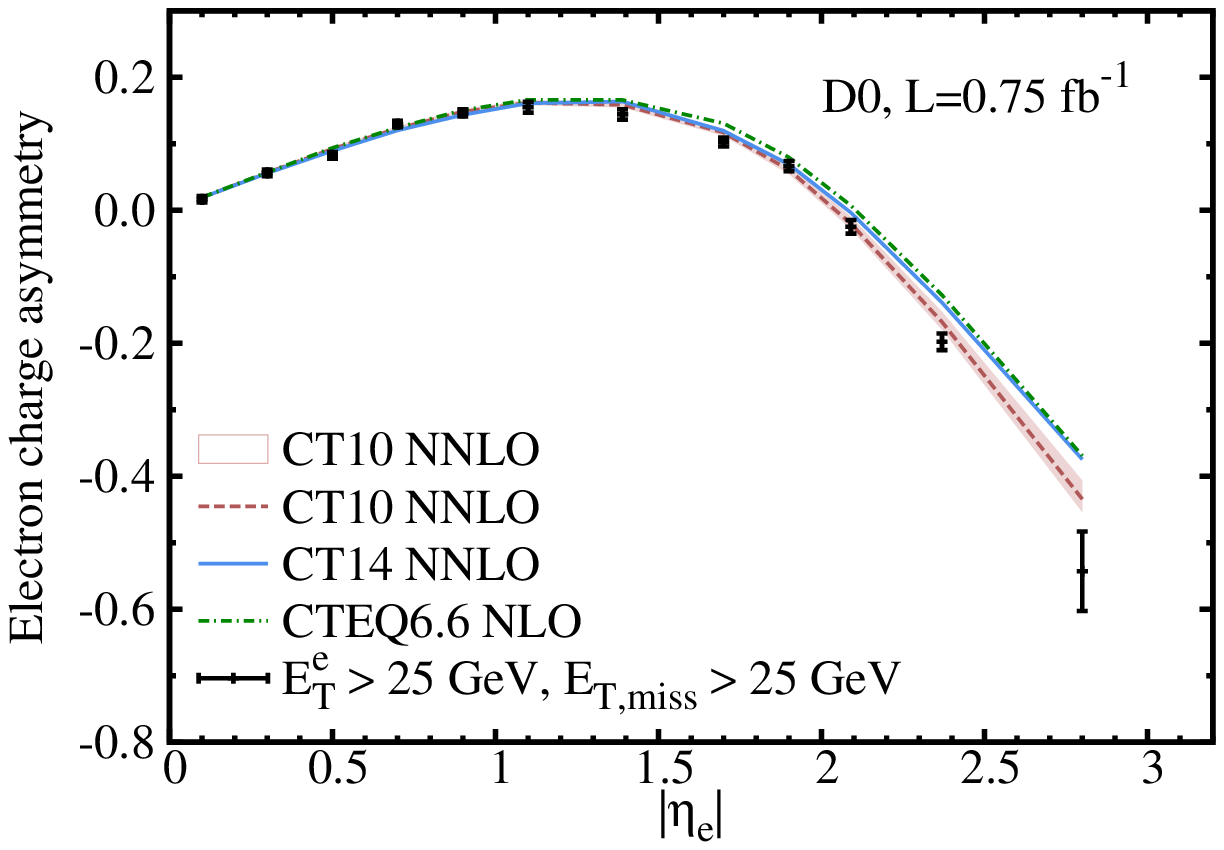}
\caption{
Charge asymmetry of decay electrons from $W^{\pm}$ production
measured by the D\O~ experiment in Run-2 at the Tevatron with high (left) and low (right) luminosities, compared to several generations of CTEQ-TEA PDFs.
\label{fig:ds234asya}}
\end{figure}

\begin{figure}[ht!]
\includegraphics[width=0.43\textwidth]{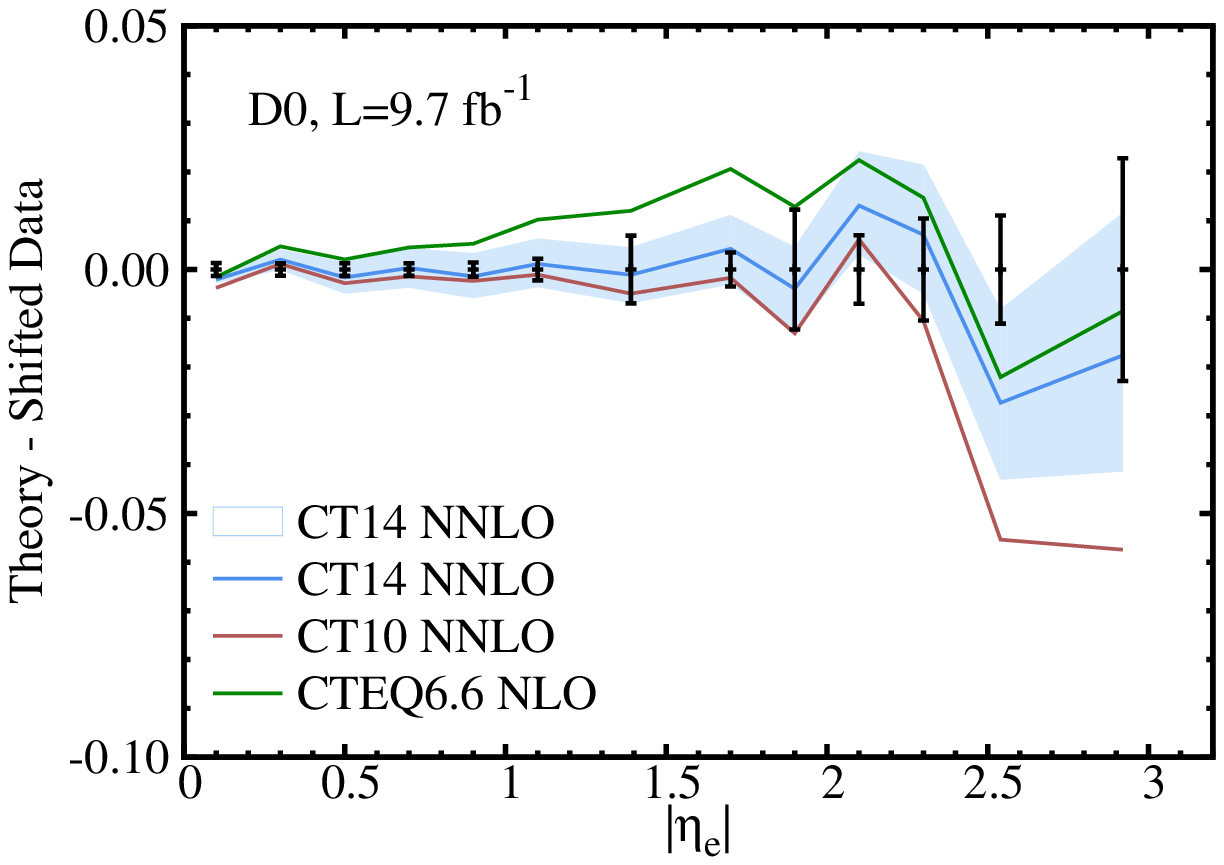}
\includegraphics[width=0.43\textwidth]{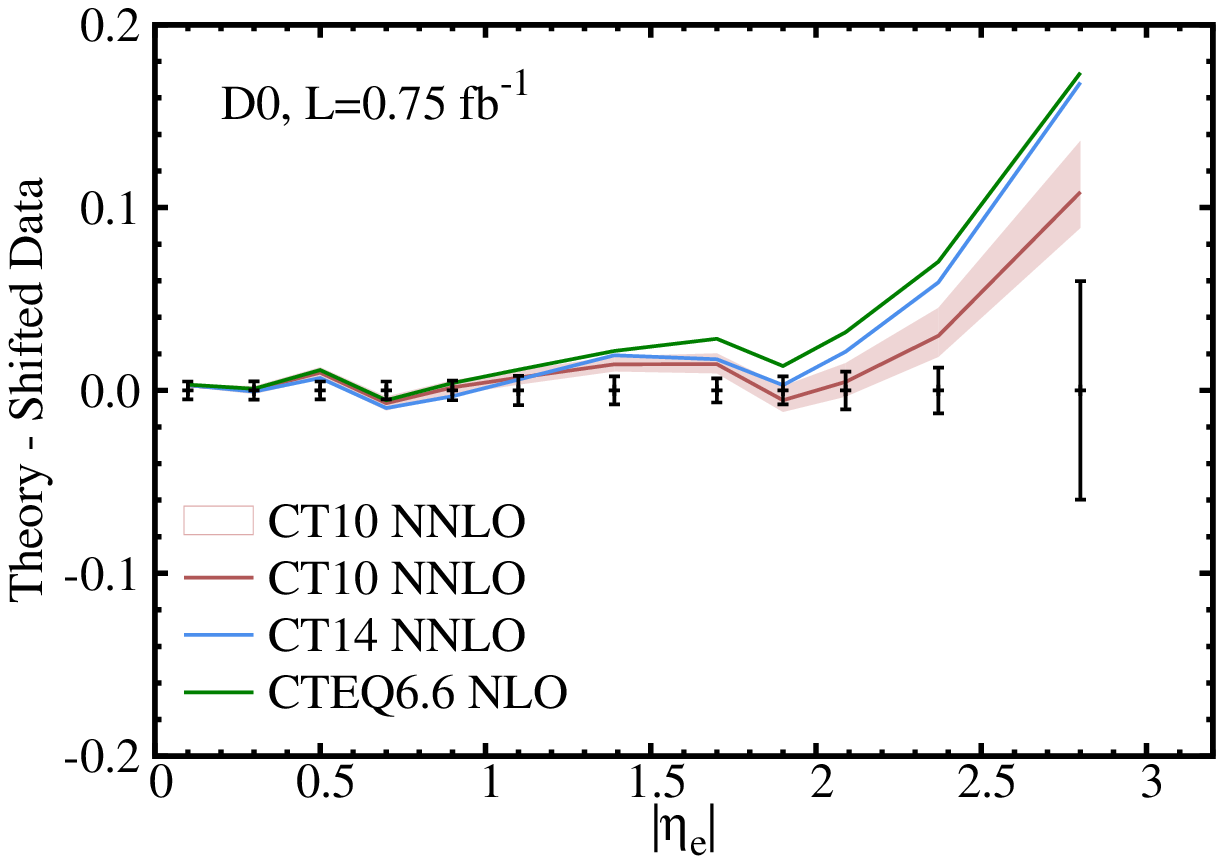}
\caption{
Same as Fig.~\ref{fig:ds234asya}, plotted
as the difference between theory and shifted data for $A_{ch}$ from
D\O~ Run-2 ($9.7\mbox{ fb}^{-1}$).
\label{fig:ds234asyb}}
\end{figure}

A new $W^{\pm}$ charge asymmetry measurement from the D\O~ experiment
at the Tevatron has recently been published,
using the full integrated luminosity ($9.7\mbox{ fb}^{-1}$)
from Run-2~\cite{D0:2014kma}.
The experimental uncertainties, both statistical and systematic,
are smaller than in the previous $A_{ch}$ measurement~\cite{Abazov:2008qv}.
Figure \ref{fig:ds234asya} compares the D\O~ Run 2 data and various
theoretical predictions at NNLO for both the latest (left) and the
previous D\O~ data set (right). We show the unshifted data with the total
experimental errors as error bars, and the 68\% C.L.\ PDF
uncertainties as the shaded regions. As an alternative representation,
Figure \ref{fig:ds234asyb} shows the differences between theory and
shifted data, where the error bars represent the uncorrelated experimental errors.
From the two figures, we conclude that it is difficult to fit
both data sets well,
given the smallness of the systematic shifts associated with $A_{ch}$.
While the $9.7\mbox{ fb}^{-1}$ electron data set is in better
agreement with the global data, including
the D\O~ muon \cite{Abazov:2007pm} and CDF  \cite{Acosta:2005ud}
$A_{ch}$ measurements, the best-fit $\chi^2/N_{pt}$ for the $9.7\mbox{
  fb}^{-1}$ sample remains relatively high (about 2) and is sensitive
to detailed implementation of NNLO corrections.
 In-depth studies on the D\O~ asymmetry data will be presented
in a forthcoming paper. When the high-luminosity D\O~ $A_{ch}$ measurement was
substituted for the low-luminosity one, we observed reduction in the
$d/u$ ratio at $x > 0.1$ compared to CT10W NLO and CT10 NNLO sets.

In total, constraints from the LHC and Tevatron
$W/Z$ differential cross sections and asymmetries
lead to important changes in the quark sector PDFs, as  documented in
Sec.~\ref{sec:OverviewCT14}. At $x\lesssim 0.02$, we obtain more
realistic error bands for the $u$, $\bar u$, $d$, $\bar d$ PDFs upon
including the ATLAS and CMS data sets. At $x > 0.1$, the
high-luminosity D\O~ charge asymmetry and other compatible experiments
predict a softer behavior of $d(x,Q)/u(x,Q)$ than in CT10W.

\subsection{Constraints on strangeness PDF
from CCFR, NuTeV, and LHC experiments}

Let us now turn to the strangeness PDF $s(x,Q)$, which has become smaller at
$x >0.05$ in CT14 compared to our previous
analyses, CT10 and CTEQ6.6. Although the CT14 central $s(x,Q)$ lies
within the error bands of either earlier PDF set,
it is important to verify that it is consistent
with the four fixed-target measurements that are
known to be sensitive to $s(x,Q)$: namely,
measurements of dimuon production in neutrino and antineutrino
collisions with iron targets, from the CCFR~\cite{Goncharov:2001qe}
and NuTeV~\cite{Mason:2006qa} collaborations (ID=124-127).

\begin{figure}[p]
\includegraphics[width=0.70\textwidth]{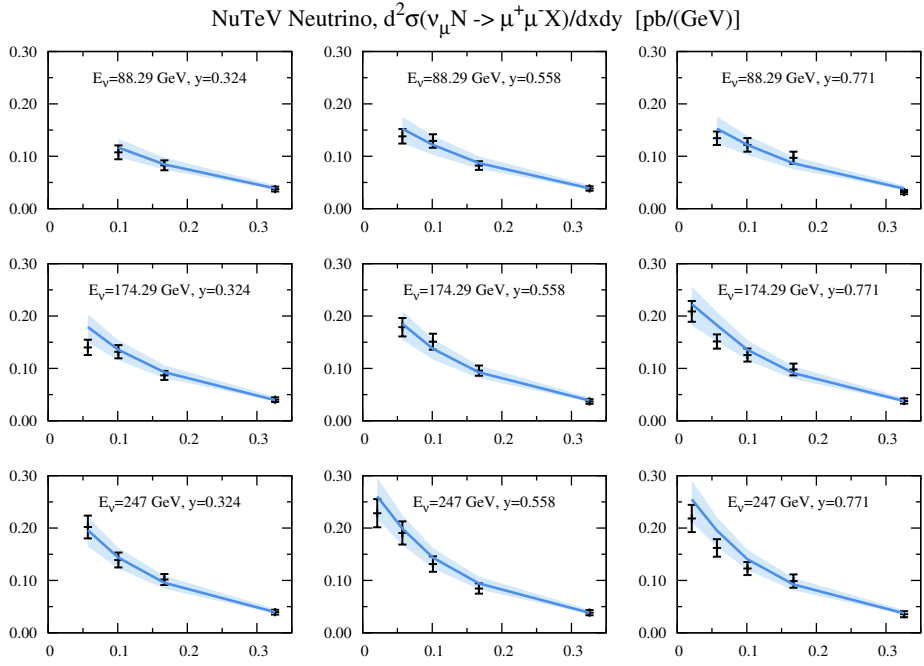}
\caption{Comparison of data and theory for the NuTeV measurements
of dimuon production in neutrino-iron collisions.
The data are
expressed in the form of $d^2\sigma/dxdy$ and
shown as a function of $x$
for a certain $y$ and neutrino energy.
\label{fig:DS124}}
\end{figure}

\begin{figure}[p]
\includegraphics[width=0.70\textwidth]{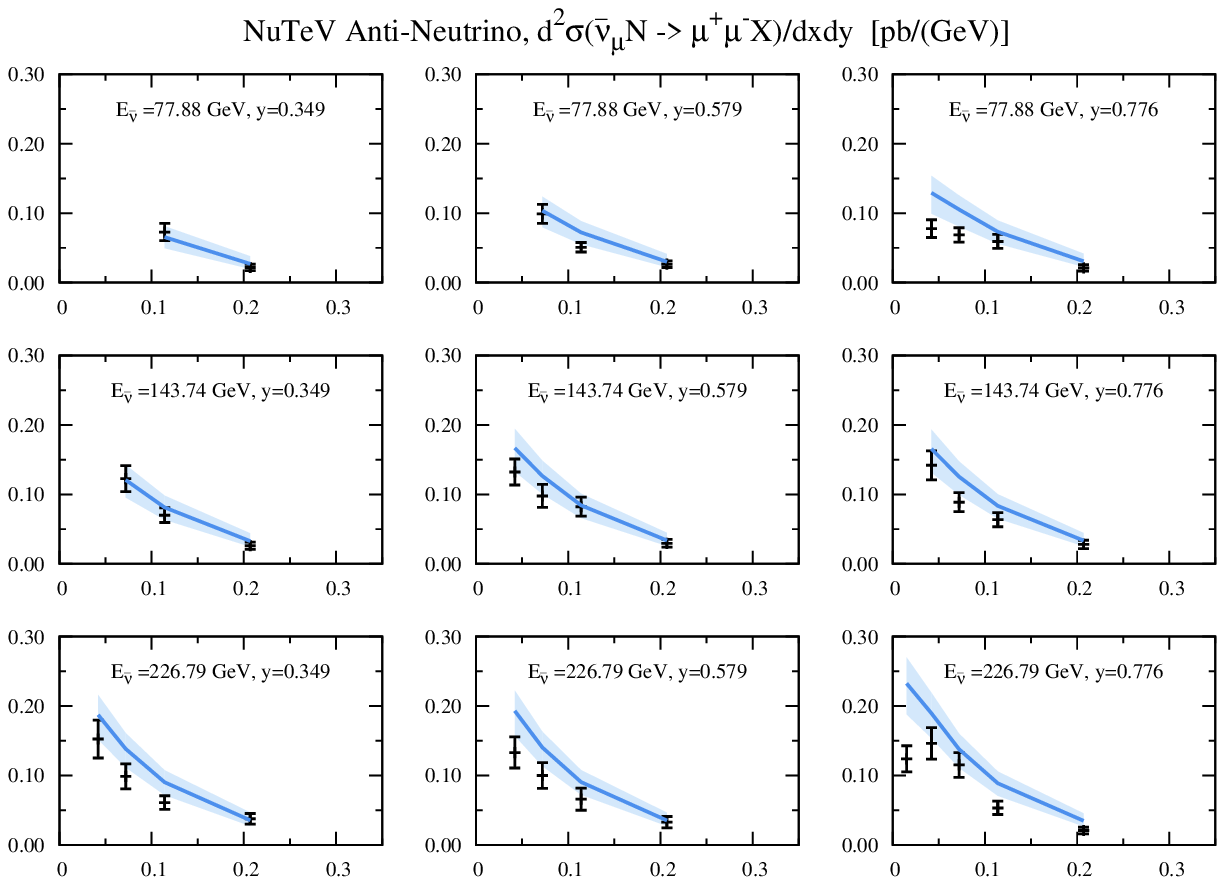}
\caption{Same as Fig.~\ref{fig:DS124}, for the NuTeV measurements
of dimuon production in antineutrino-iron collisions.
\label{fig:DS125}}
\end{figure}

\begin{figure}[p]
\includegraphics[width=0.70\textwidth]{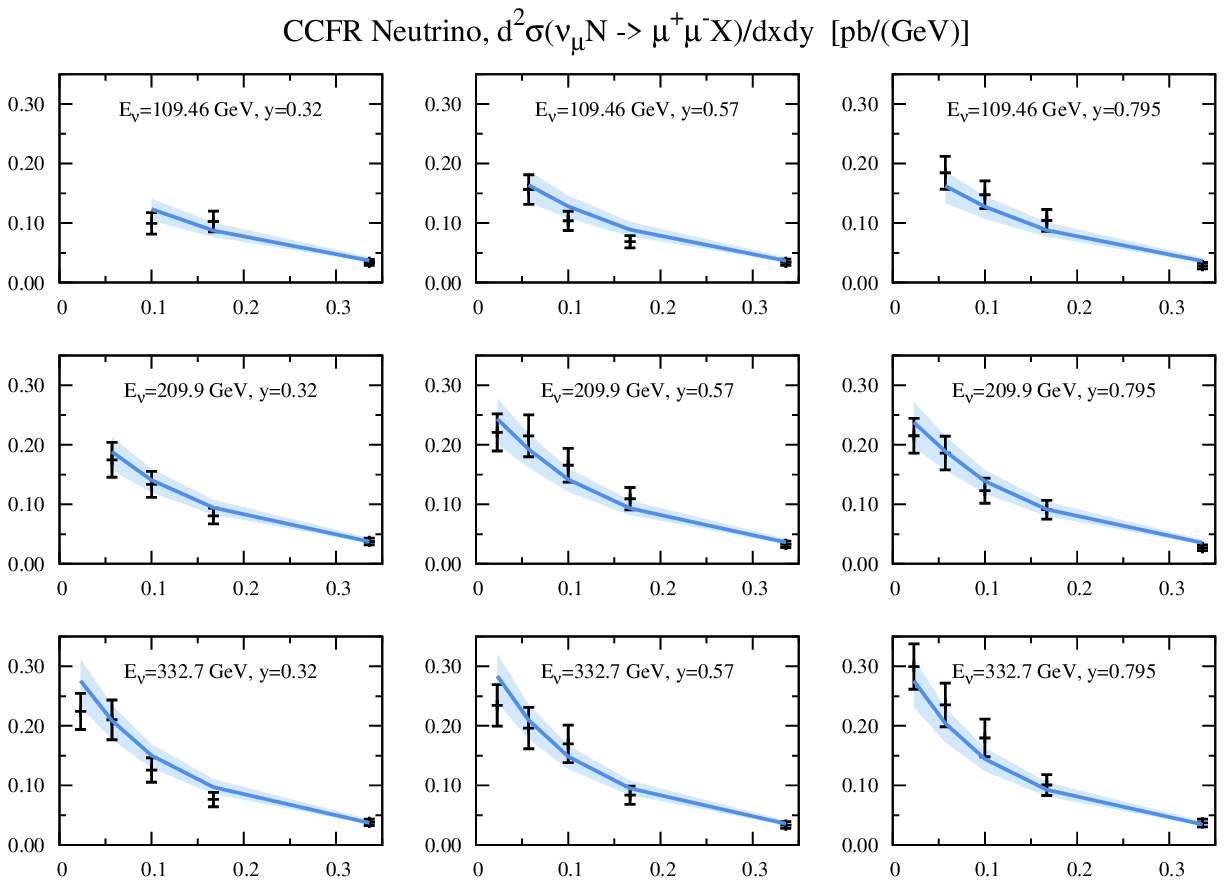}
\caption{Same as Fig.~\ref{fig:DS124}, for the CCFR measurements
of dimuon production in neutrino-iron collisions.
\label{fig:DS126}}
\end{figure}

\begin{figure}[p]
\includegraphics[width=0.70\textwidth]{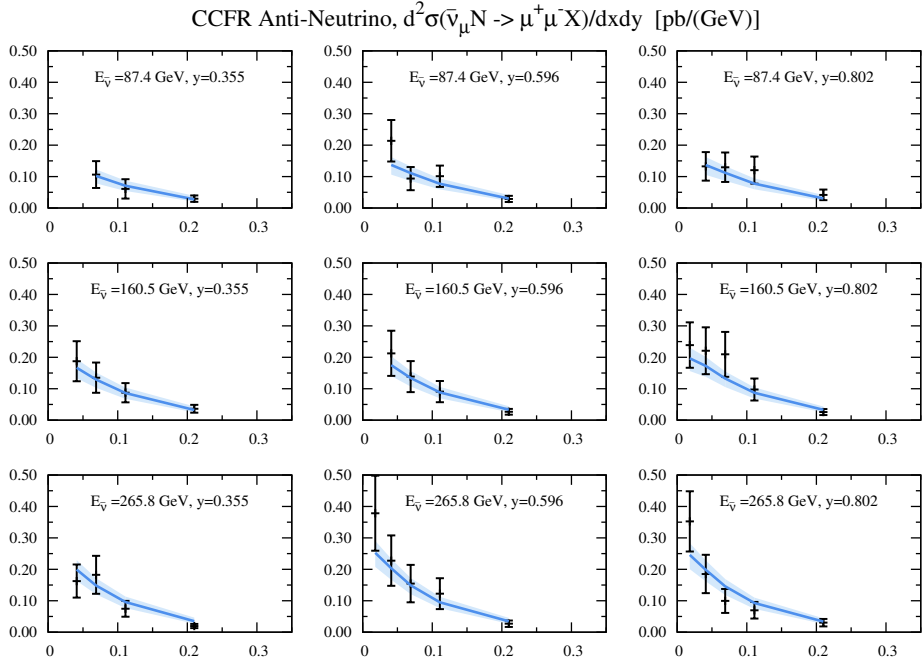}
\caption{Same as Fig.~\ref{fig:DS124}, for the CCFR measurements
of dimuon production in antineutrino-iron collisions.
\label{fig:DS127}}
\end{figure}

Predictions using previous
CTEQ PDFs were in agreement with these four experiments.
In Table~\ref{tab:EXP_1} for CT14, the four corresponding $\chi^2$ values
are also good.
Supporting evidence comes from the point-by-point comparisons in
Figs.~\ref{fig:DS124} and \ref{fig:DS125}, between the
theoretical cross sections for CT14 NNLO PDFs
and the dimuon data from the NuTeV experiment in neutrino
and antineutrino scattering. The analogous comparisons for
the CCFR experiment are in Figs.~\ref{fig:DS126} and
\ref{fig:DS127}. Given the size of the measurement errors
and of the PDF uncertainty, it is clear that CT14 central predictions provide a
good description of the dimuon cross sections.
Also, our estimate for the {\em uncertainty} of the strange PDF
looks reasonable: it is comparable to the measurement errors
for these cross sections, which are known to be sensitive mostly
to the strange quark PDF.

Nevertheless, the CT14 central strangeness PDF lies on the lower side
of the CT10 PDF uncertainty in some kinematic ranges.
As mentioned in the introduction, the reduction
is in part attributable to elimination of a computational
error (wrong sign of a term) in the treatment of heavy-quark mass effects
in charged-current DIS in post-CTEQ6.5 analyses, and in part
from other sources, especially introduction of the LHC $W/Z$ data,
and more flexible parameterizations for all PDF flavors.

The ATLAS and CMS experimental collaborations have recently
published studies on the strangeness content of the proton and have come
to somewhat discrepant conclusions. On the ATLAS side, two papers were published, one in 2012~\cite{Aad:2012sb}, and one in
2014~\cite{Aad:2014xca}. In the 2012 study, the inclusive DIS and inclusive $W^{\pm}$ and $Z$ boson production measurements~\cite{Aad:2011dm}
were employed to determine the strangeness fraction of the proton for one
value of $(x,Q)$. In the 2014 study, the  ATLAS 7 TeV $W +c\mbox{-jet}$, $W+D^{(\star)}$ \cite{Aad:2014xca}, and inclusive $W^{\pm}/Z$
cross sections were used. These two analyses determined the
ratio ($r^s$) of strange to down-sea quark PDF,
\begin{equation}
r^s \equiv 0.5 \frac{(s+ \overline{s})}{\overline{d}}
  ~~~~\textrm{at} ~~~~x=0.023, ~~~~  Q=1.4~\textrm{GeV}.
\end{equation}
They find
\begin{eqnarray}
&&r^s = 1.00^{+0.25}_{-0.28},   ~~~~\textrm{ATLAS} ~ (2012),
\nonumber\\
&&r^s = 0.96^{+0.26}_{-0.30}   ~~~~\textrm{ATLAS} ~ (2014),
\end{eqnarray}
which imply a rather large strangeness density.

In 2014, the CMS collaboration~\cite{Chatrchyan:2013mza} determined
the ratios
\begin{equation}
R^s \equiv \frac{(s+ \overline{s})}{\overline{u} + \overline{d}},  ~~~~\textrm{at} ~~~~x=0.023, ~~~~  Q=1.4~\textrm{GeV}
\end{equation}
and
\begin{equation}
\kappa^s(Q^2)= \frac{\int_0^1 x \left[s(x,Q^2) + \overline{s}(x,Q^2)\right]dx}{\int_0^1 x \left[\overline{u}(x,Q^2) + \overline{d}(x,Q^2)\right]dx} ,
\end{equation}
by using inclusive DIS, the
charge asymmetry of decay muons from $W^{\pm}$ production~\cite{Chatrchyan:2013mza},
and $W$ + charm production differential cross sections~\cite{Chatrchyan:2013uja} at 7 TeV. They obtain
\begin{eqnarray}
&&R^s = 0.65^{+0.19}_{-0.17} ,
\nonumber\\
&&\kappa^s(Q^2=20 \textrm{ GeV}^2)= 0.52^{+0.18}_{-0.15} ~~~~\textrm{CMS} ~ (2014).
\end{eqnarray}
Notice that ATLAS and CMS use two different definitions, $r^s$ and
$R^s$, for the strangeness fraction, which
are supposed to coincide at the initial scale $Q_0=1.4$ GeV, if $\bar
u(x,Q_0) = \bar d(x,Q_0)$.\footnote{Both ATLAS and CMS studies are
  performed in the HERAFitter framework~\cite{Alekhin:2014irh}
and assume SU(2)-symmetric sea quark PDF parametrizations
at the initial scale $Q_0=$ 1.4 GeV.}

For comparison, at the factorization scale $Q=1.4$ GeV and $x=0.0234$,
the CT14 and CT10 predictions are
\begin{eqnarray}
&&r^s_{\rm CT14\, NNLO} = \frac{\bar{s}(x,Q)}{\bar{d}(x,Q)} = 0.53\pm 0.20,
\nonumber\\
&&r^s_{\rm CT10\, NNLO} = \frac{\bar{s}(x,Q)}{\bar{d}(x,Q)} = 0.76\pm 0.17.
\label{CT14rs}
\end{eqnarray}
Both CT14 and CT10 indicate a smaller strangeness than the ATLAS
result and are compatible with CMS; the $r^s$ ratio is smaller for
CT14 than  for CT10.

The NOMAD Collaboration has also completed a study of the strange quark PDF,
relying on  $\nu+Fe\rightarrow \mu^+ + \mu^- +X$
measurements~\cite{Samoylov:2013xoa} at lower energies than NuTeV and CCFR.
They find that the strangeness suppression factor is
\begin{equation}
 \kappa^s (20 \textrm{ GeV}^2) = 0.591 \pm 0.019 ,
\end{equation}
also yielding a smaller strangeness density than the ATLAS result.
In another recent study by S. Alekhin and collaborators~\cite{Alekhin:2014sya},
the strange quark distribution
and the ratios $r^s$ and $\kappa^s$ were determined in a QCD
analysis including the NuTeV, CCFR, NOMAD and CHORUS measurements.
The study uses the fixed-flavor-number (FFN) scheme for the
heavy-flavor treatment.  Their main result is
 $\kappa^s(20 \textrm{ GeV}^2) = 0.654 \pm 0.030$.
The CT14 and CT10 predictions for this quantity are
 \begin{eqnarray}
\kappa^s_{\rm CT14\, NNLO} &=& 0.62 \pm  0.14, \nonumber\\
\kappa^s_{\rm CT10\, NNLO} &=& 0.73 \pm 0.11.
\label{CT14kappas}
\end{eqnarray}
The CT14 calculation is consistent with the NOMAD central value.
However, the CT14 PDF uncertainty is considerably larger than
the uncertainty quoted in the NOMAD paper, partly because of a
different convention for the PDF uncertainty.

\subsection{The CMS $W+c$ production measurement}

Another experimental measurement that
has direct access to the strange quark distribution is
the associated production of $W$ boson and charm quark at the LHC.
Such a measurement
was reported by the CMS collaboration for $\sqrt s=7\, {\rm TeV}$ and a total integrated luminosity of 5 ${\rm
  fb}^{-1}$~\cite{Chatrchyan:2013uja}.
Cross sections and ratios of cross sections with the observed  $W^+$
and $W^-$ bosons were measured
differentially with respect to the absolute value of the pseudorapidity
of the charged lepton from the $W$ boson decay.
As the theoretical cross section is not yet known at NNLO,
the data were not directly included in the global fit, but are
compared here to NLO calculations based on MCFM
6.0~\cite{Campbell:2010ff}, assuming
a non-zero charm quark mass, and excluding contributions
from gluon splitting into a $c\bar c$ pair.
The renormalization and factorization scales are set to the virtuality
of the $W$ boson. The transverse momentum of the
charged lepton is required to be at least $25\,{\rm GeV}$.
The theoretical calculation applies the same kinematical
cuts as in the experimental analysis, but at the parton level.

\begin{figure}[h!]
  \begin{center}
  \vspace{20pt}
  \includegraphics[width=0.4\textwidth]{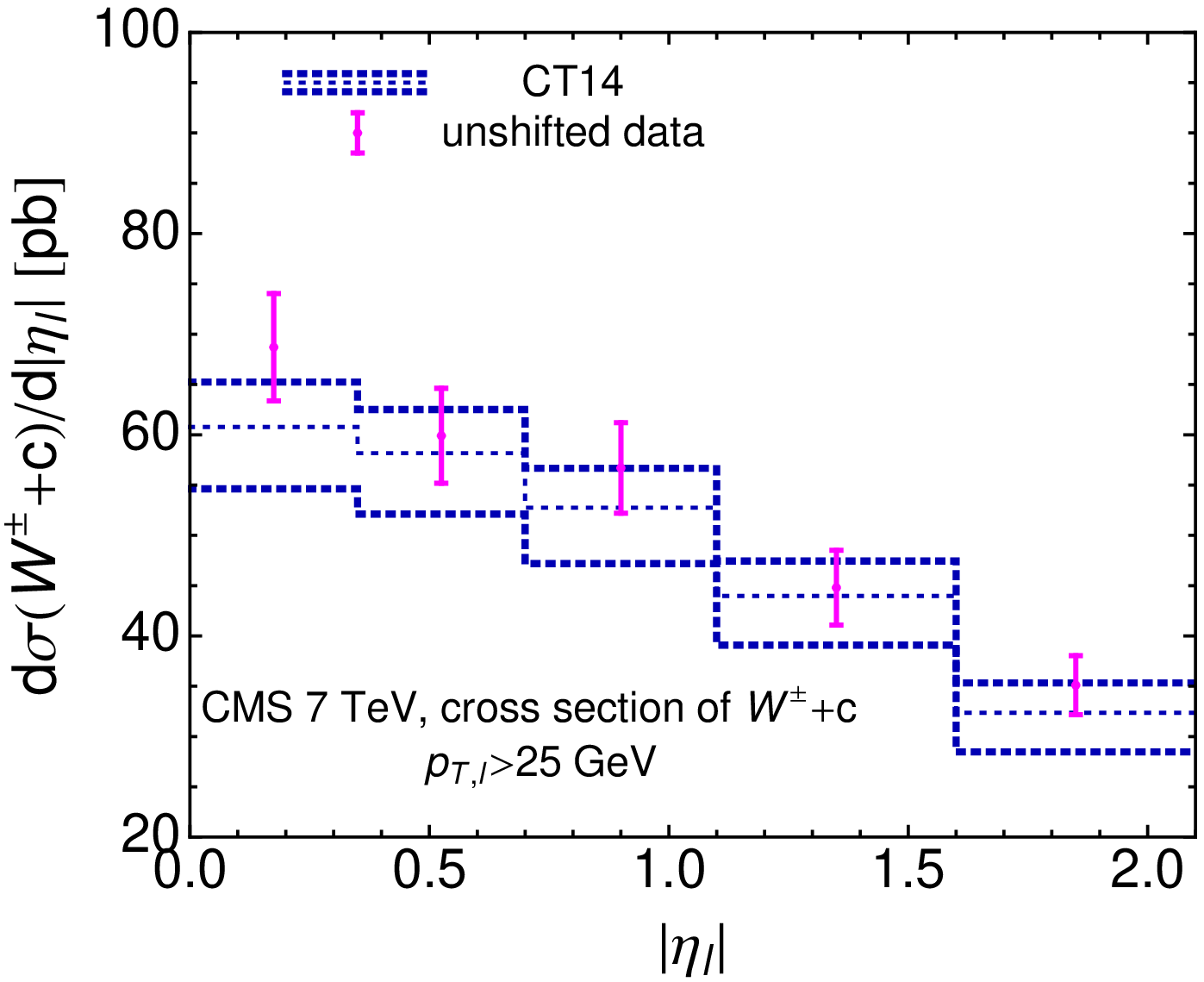}\hspace{0.2in}
  \includegraphics[width=0.4\textwidth]{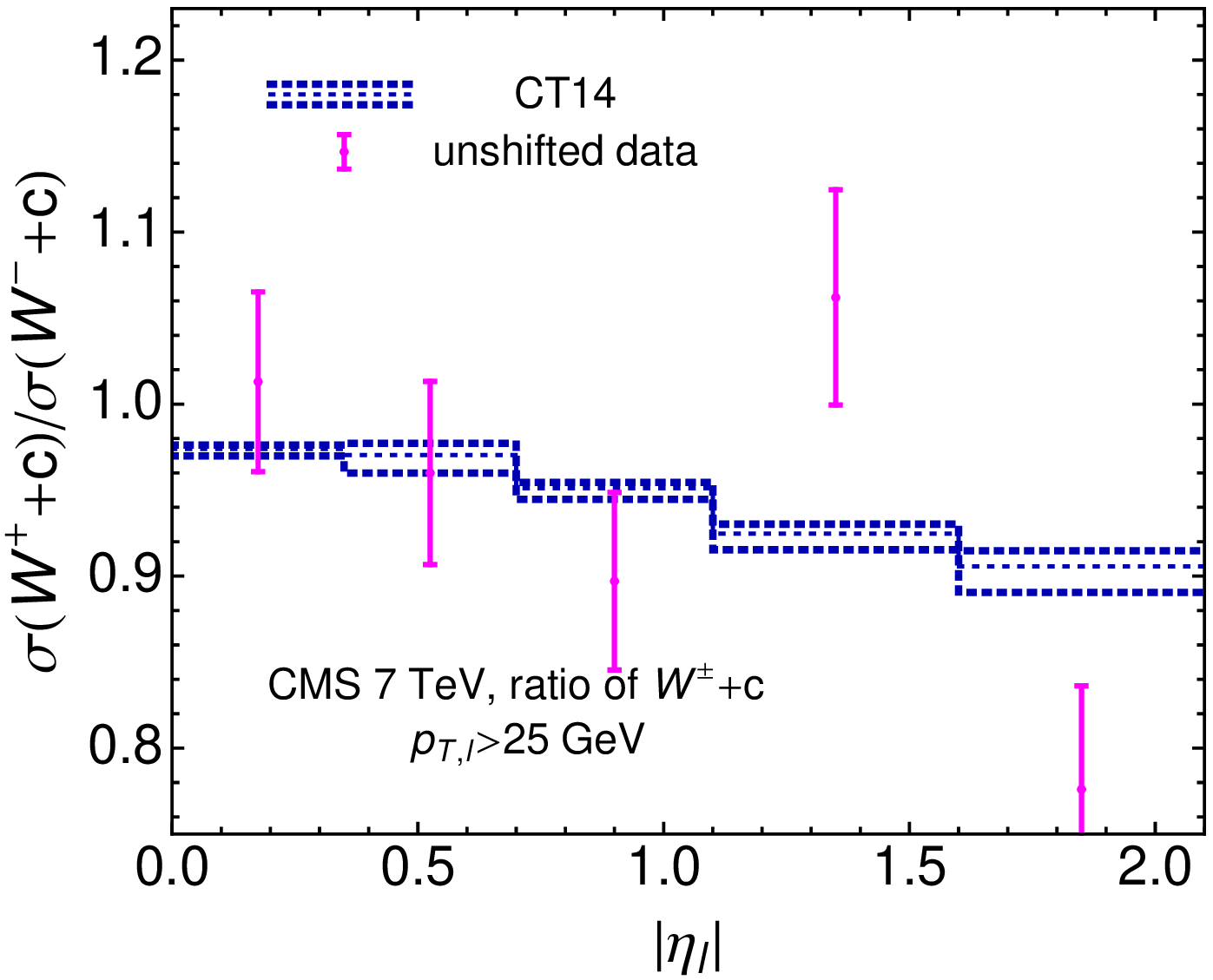}
  \end{center}
  \vspace{-1ex}
  \caption{\label{fig:wc1}
 Comparison of the CT14 predictions to $W^{\pm}+c$ differential cross
 sections (left) and to the ratio of $W^{+}+\bar c$ to $W^{-}+c$
 cross sections (right) from the CMS measurement at 7 TeV.}
\end{figure}

The left panel of Fig.~\ref{fig:wc1} shows the pseudorapidity distribution
of the decay charged lepton from $W$ boson decay
in $W^\pm + c$ production at 7 TeV.
The format of the figures is the same as in the previous comparisons.
The total experimental errors in the figures are reasonably close to
the 68\% C.L. PDF uncertainties. With further experimental and theoretical improvements,
the process may contribute to the reduction of the PDF uncertainty.

The right panel shows the {\em ratio} of charged lepton rapidity distributions
in $W^+ + {\bar c}$ and $W^- + c$ production, which
provides a handle on the strangeness asymmetry, $s - \bar s$.
The CT14 parametrization allows for no intrinsic $s$-asymmetry
at the initial scale $Q_0$. (At higher scales, a tiny asymmetry is
generated by 3-loop DGLAP evolution.)
Our prediction reproduces the average trend of the data, however, the experimental errors are larger than the PDF uncertainties.

\begin{figure}[tb]
  \begin{center}
  \includegraphics[width=0.4\textwidth]{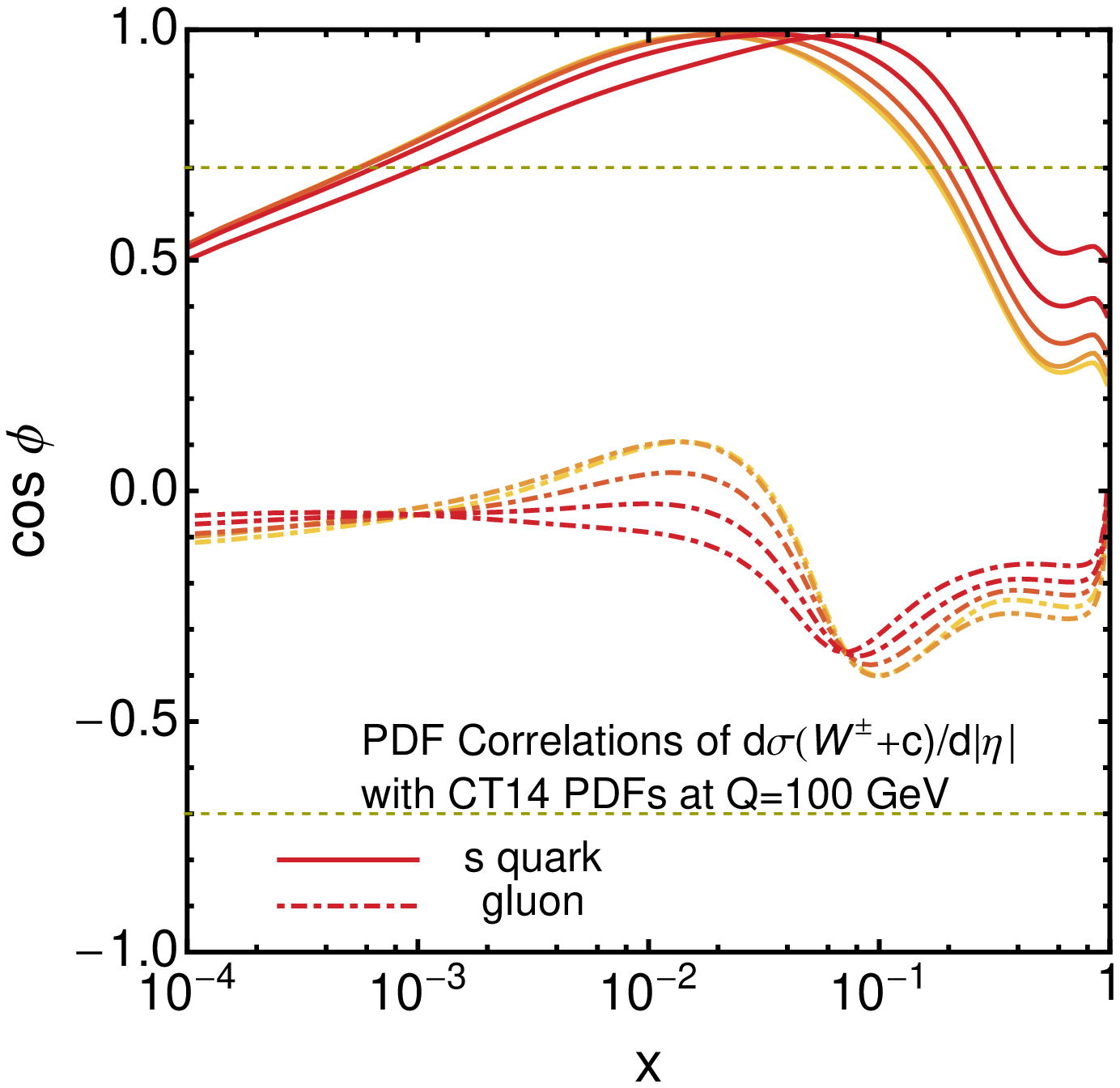}\hspace{0.2in}
  \includegraphics[width=0.4\textwidth]{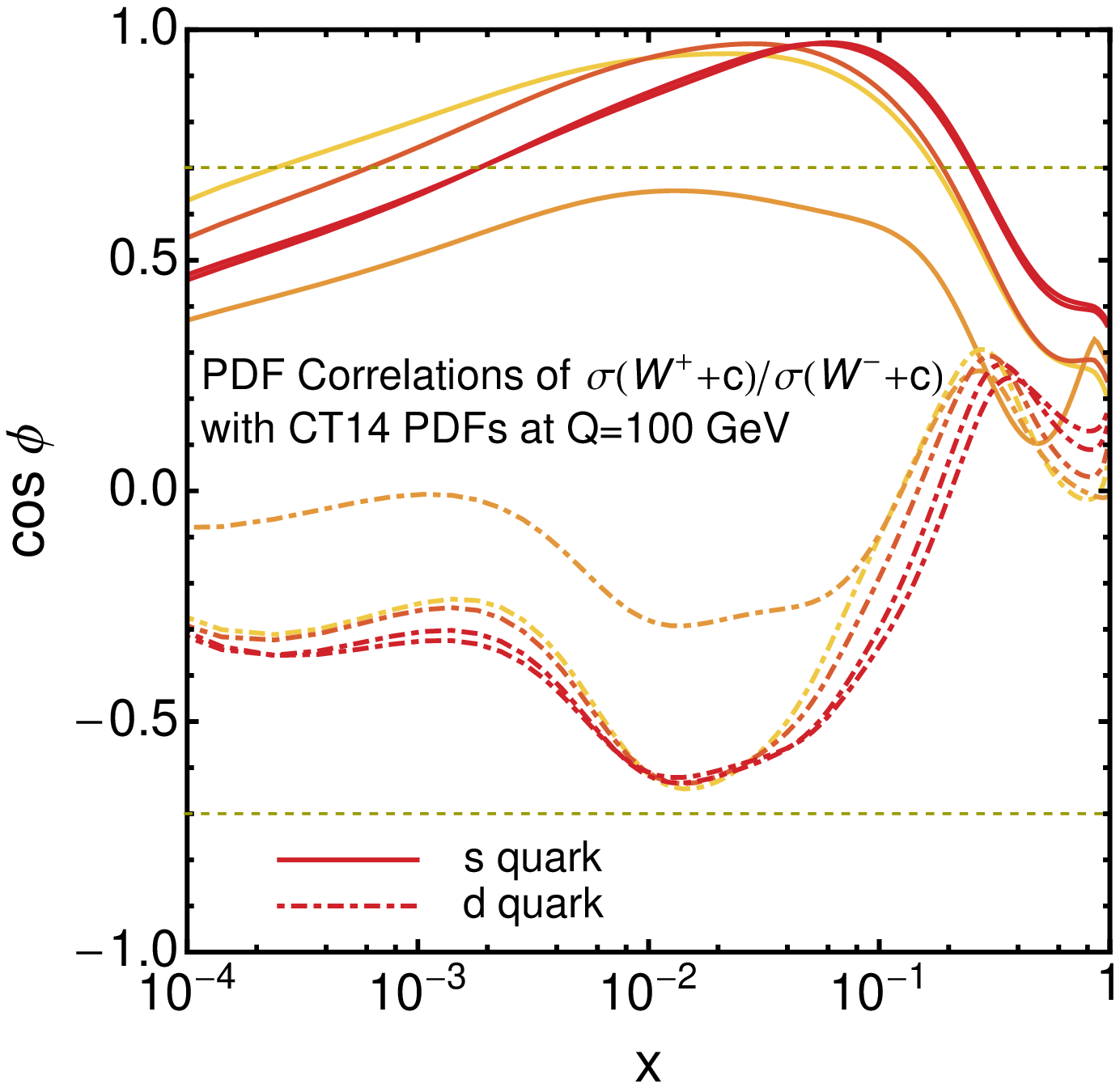}
  \end{center}
  \vspace{-1ex}
  \caption{\label{fig:wc2}
 Correlation cosines between the PDFs of select flavors,
 the $W^{\pm}+c$ cross section, and the $W^+/W^-$ cross section ratio,
 as a function of $x$ in the PDF.}
\end{figure}

\begin{figure}[tb]
  \vspace{2ex}
  \begin{center}
  \includegraphics[width=0.4\textwidth]{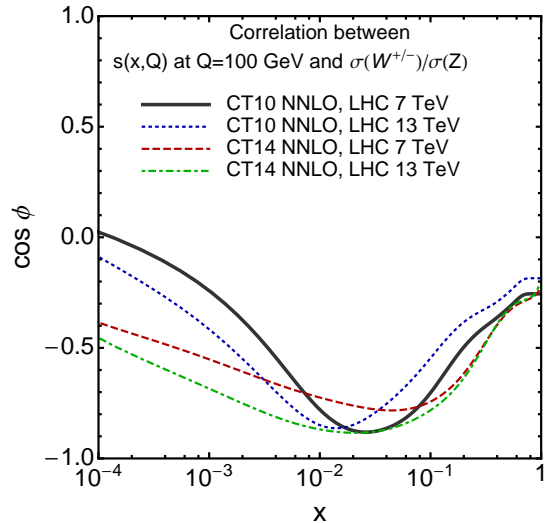}
  \end{center}
  \vspace{-2ex}
  \caption{\label{fig:wc3}
 Correlation cosines between the ratio of total cross sections
 for $W^{\pm}$ production and $Z$ boson production,
 and $s$ quark PDF from CT14 and CT10 NNLO sets, for LHC 7 and 13 TeV.}
\end{figure}

The specific $x$ ranges that are probed by CMS $W^\pm+c$ cross sections can be
identified by plotting correlation cosines \cite{Nadolsky:2008zw} between
the PDFs of various flavors, the $W^\pm+c$ cross section, or cross
section ratios.
Fig.~\ref{fig:wc2} shows such correlation cosines
for the $s$ quark, gluon, and $d$ quark PDFs at the factorization
scale of 100 GeV.
Lines in darker colors correspond to bins with larger rapidities.
In the case of the differential cross section,
the PDF correlations are most significant
for the strange quark distributions at $x=0.01-0.1$,
as indicated by their strong correlations with $\cos\phi \sim 1$.
The gluon does not play a significant role,
due to its relatively smaller uncertainty in the same $x$ region.
In the case of the cross section ratio,
the correlation with the strangeness is still dominant.
But also, at large rapidity, the $d$ quark contribution
to the $W^-$ cross sections is mildly anti-correlated, indicating
that the ratio has marginal sensitivity to $d(x,Q)$
at $x$ around $0.01$.

Another well-known probe of the strangeness
content of the proton is provided by the ratio
of total cross sections for LHC $W^{\pm}$ and
$Z$ boson production~\cite{Nadolsky:2008zw}. The correlation cosine
between $\sigma(W^\pm)/\sigma(Z)$ and $s(x,Q)$ can be viewed in Fig.~\ref{fig:wc3}.
As expected, we observe strong anti-correlation in a
certain $x$ range at all LHC center-of-mass energies.
Compared to CT10, the $x$ region of the strongest sensitivity shifts
to higher $x$, and the $x$ dependence gets flatter in CT14.

\section{Impact on Higgs boson and
$t\bar t$ cross sections at the LHC \label{sec:LHCPredictions}}
\label{sec:Uncert}

Gluon fusion provides the largest cross section for production of a
Higgs boson. It was the most important process for the discovery of
the Higgs boson in 2012, and it continues to be essential for detailed
studies of Higgs boson properties. A great deal of
benchmarking of hard-cross sections and PDFs
for the $gg$ initial state was carried out both before and after the
discovery~\cite{Dittmaier:2011ti,Alekhin:2011sk,Gao:2012he,Dittmaier:2012vm,Ball:2012wy,Heinemeyer:2013tqa}.
This was motivated in part by the fact that the PDF uncertainty for the $gg$ initial state was comparable to the renormalization and
factorization scale uncertainties in the theoretical cross section at NNLO for producing a Higgs boson through gluon fusion.
The recent calculation of the gluon fusion process at NNNLO~\cite{Anastasiou:2015ema} has reduced the scale uncertainty
in the hard cross section still further, making the PDF uncertainty even more critical.

Similarly, production of a $t\bar t$ final state is crucial to many
analyses at the LHC, as both a standard model signal and as a
background to new physics. By far the dominant subprocess for $t\bar
t$ production at the LHC is $gg\rightarrow t\bar t$, making $t\bar t$
production an important benchmark for understanding the $gg$ PDF
luminosity~\cite{Nadolsky:2008zw}, especially with the current
calculation of the $t\bar t$ total inclusive cross section
now available at NNLO~\cite{Czakon:2013goa,Top++}.

Using CT10 PDFs, we have recently
performed detailed analyses of the predictions for
$gg\rightarrow H$ and $t\bar t$ cross sections,
as well as their uncertainties from both the PDFs
and the strong coupling $\alpha_s$~\cite{Dulat:2013kqa,Schmidt:2014gda}.
In this section we update these studies and
review CT14 predictions for $gg\rightarrow H$ and $t\bar t$ total and
differential cross sections.

\subsection{Higgs boson from gluon fusion at the LHC} \label{sec:Higgs}

We begin with an analysis of the PDF and $\alpha_s$ uncertainties
for $gg\rightarrow H^0$. For this,
we have utilized the NNLO code iHixs 1.3~\cite{Anastasiou:2011pi},
choosing the Higgs boson mass to be $M_H = 125\ {\rm GeV}$, and
with both the renormalization and factorization scales fixed at $\mu=M_H$.
Here, we have included the finite top quark mass correction (about 7\%)
to the fixed-order NNLO result obtained using the HQET (with
infinite top quark mass approximation).

To calculate the 90\% C.L. PDF and $\alpha_s$
uncertainties of an arbitrary cross section $X$ according
to the most conventional (Hessian) method
\cite{Pumplin:2001ct}, we provide error PDFs (56 in the
case of CT14) to probe independent combinations of the PDF parameters for the central $\alpha_s(M_Z)=0.118$,
plus two additional PDFs
obtained from the best fits with $\alpha_s(M_Z)=0.116$ and 0.120. Using
the error sets, the combined PDF+$\alpha_s$ uncertainty on $X$
is estimated by adding the PDF and $\alpha_s$ uncertainties in quadrature
\cite{Lai:2010nw}. The quadrature-based combination is exact if
$\chi^2$ has a quadratic dependence,
and $X$ has a linear dependence on the PDF fitting parameters in the
vicinity of the best fit.
To account for some mild nonlinearities, asymmetric errors are allowed in
the positive and negative directions of each eigenvector in the
fitting parameter space.

Another method for estimating the PDF and $\alpha_s$ uncertainties on $X$
introduces Lagrange multipliers (LM) \cite{Stump:2001gu}. It does not
rely on any assumptions about the functional dependence of $X$
on the PDF parameters.  Instead, the PDFs are refitted a
number of times, while fixing $X$ to take some user-selected value in each fit.
Then the uncertainty in $X$ can be estimated by looking at
how $\chi^2$ in the series of fits varies depending on
the input value of $X$.  The
downside of the LM method is that it requires to repeat the PDF fit many
times in order to calculate the uncertainty of each given observable.
It is clearly impractical for general-purpose experimental analyses;
however, it can be straightforwardly performed for a few selected
observables.  As a side
benefit, the LM method also provides an easy way to see which
experimental data sets in the PDF global analysis have the most impact on
the PDF dependence of $X$.
Thus, in this section we will perform both the LM and Hessian analyses
of the uncertainties for the Higgs boson and $t\bar t$ cross sections
at the LHC.

\begin{figure}[h]
\vspace{10pt}
\begin{center}
\includegraphics[width=0.49\textwidth]{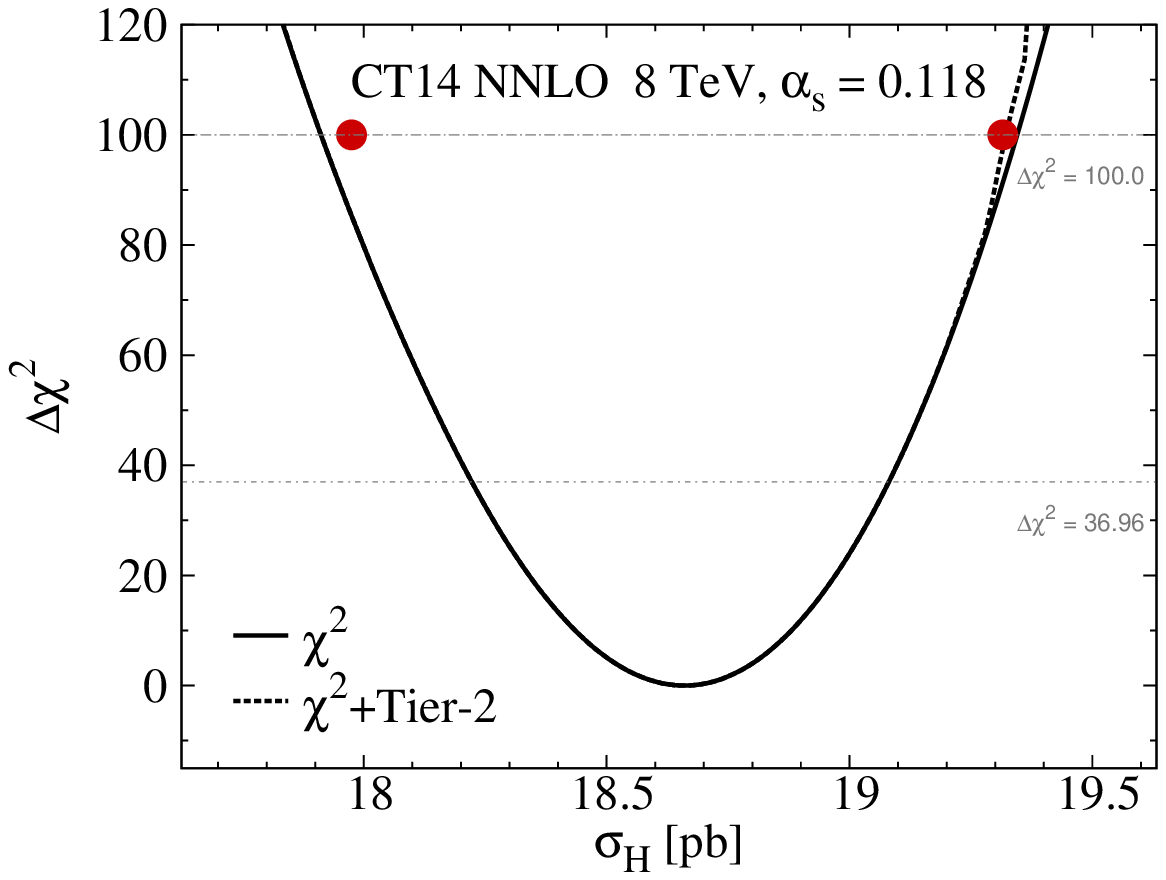}
\includegraphics[width=0.49\textwidth]{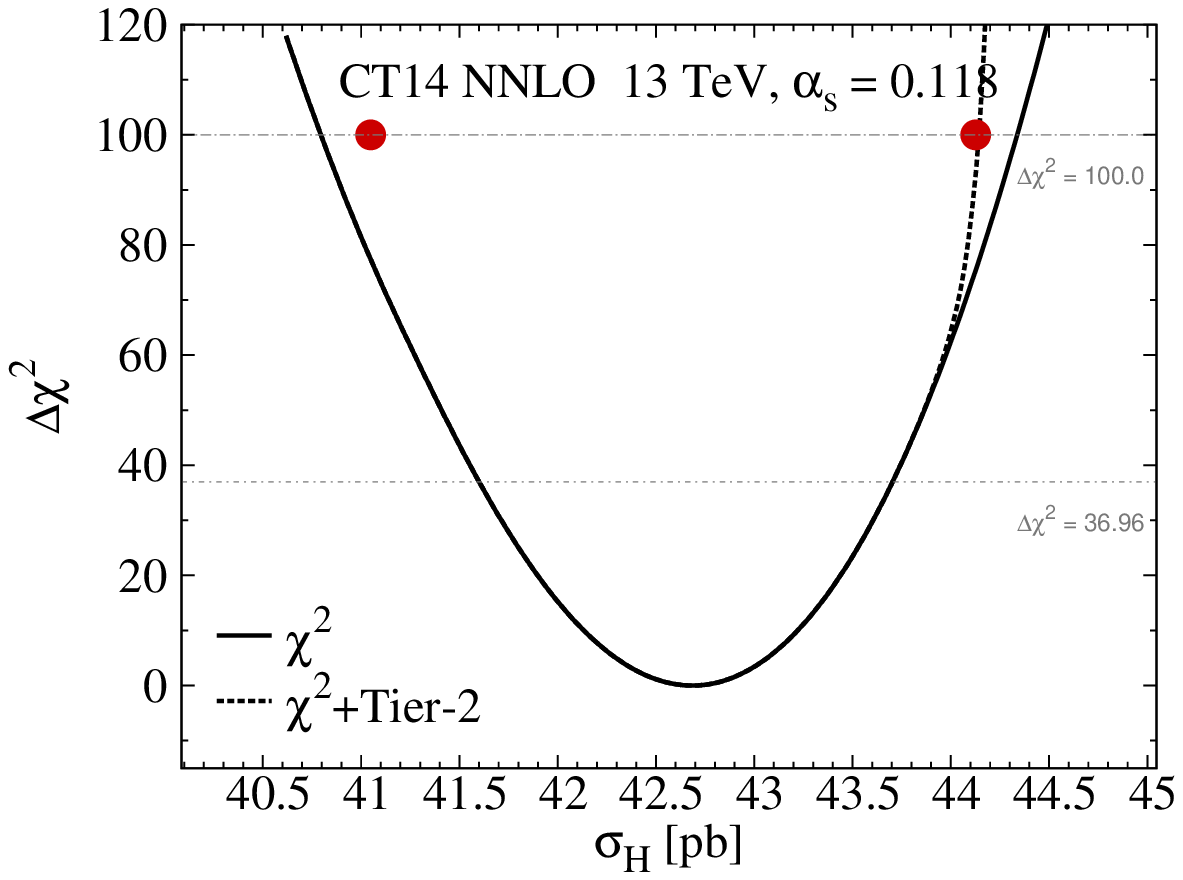}
\end{center}
\vspace{-10pt}
\caption{
Dependence of the increase in $\chi^2$ in the constrained CT14 fit
on the expected cross section $\sigma_{H}$ at the LHC 8 and 13 TeV,
for $\alpha_{s}(M_{Z}) = 0.118$. The solid and dashed curves are for
the constrained fits without and with the Tier-2 penalties,
respectively.   The red dots correspond to the upper
and lower 90\% C.L. limits calculated by the Hessian method.
\label{fig:LMparab}}
\end{figure}

We first do these calculations while keeping the strong coupling
fixed at its central value of $\alpha_s(M_Z)=0.118$ recommended by
the PDF4LHC group.
The uncertainties obtained this way are purely due to the PDFs.
The results of the LM analysis are illustrated by
Fig.~\ref{fig:LMparab},  where we plot the change $\Delta\chi^2$
in $\chi^2$ as a function of the tentative cross section $\sigma_H$
for Higgs boson production via gluon fusion in $pp$
collisions at energies $\sqrt{s} = $ 8 and 13 TeV.
$\Delta \chi^2=0$ corresponds to the best-fit PDFs to the CT14
experimental data set, so that the minimum of the approximately parabolic
curves is at our best-fit prediction for $\sigma_H$.
Non-zero $\Delta \chi^2$ are obtained with an extra constraint that
enforces $\sigma_H$
to take the values on the horizontal axis that deviate from the best-fit ones.
We have plotted the changes of
both the simple $\chi^2$ (solid) and  the $\chi^2 + $Tier-2 penalty (dashed),
in order to see the effects
of requiring that no particular data set is too badly fit
in the global analysis.
(As defined in the Appendix of Ref.~\cite{Dulat:2013hea}
and in \cite{Lai:2010vv},
the Tier-2 penalty makes use of the variable $S_{n}$,
which gives a measure of the goodness-of-fit
for each individual data set.
A large $S_n$ means that the experiment is not consistent
with the theory.)
We see that the two curves are almost identical over much of the
range plotted, only beginning to diverge
when $\sigma_{H}$ is far from the best-fit value,
and one or more experimental data sets can no longer be satisfactorily fit.

We can estimate asymmetric errors $(\delta\sigma_{H})_{\pm}$ at the $90\%$
C.L. by allowing a tolerance $\Delta\chi^2=T^2$, with
$T$ of about 10.  Given the nearly parabolic nature of these plots, we see that
the $68\%$ C.L. errors can be consistently defined using a range
corresponding to $\Delta\chi^2=(T/1.645)^2$.   The $90\%$ C.L. and $68\%$
C.L. tolerance values are indicated by the upper and lower horizontal
lines, respectively, in each of the plots.
 Finally, the red dots are the upper and lower 90\% C.L. limits from the
Hessian method analysis. They agree quite well with the LM analysis using the
$\chi^2+$Tier-2 penalty at both 8 and 13 TeV.
The effect of the Tier-2 penalty is modest, the deviations from
the parabolic behavior are small.

\begin{figure}[h]
\vspace{10pt}
\includegraphics[width=0.49\textwidth]{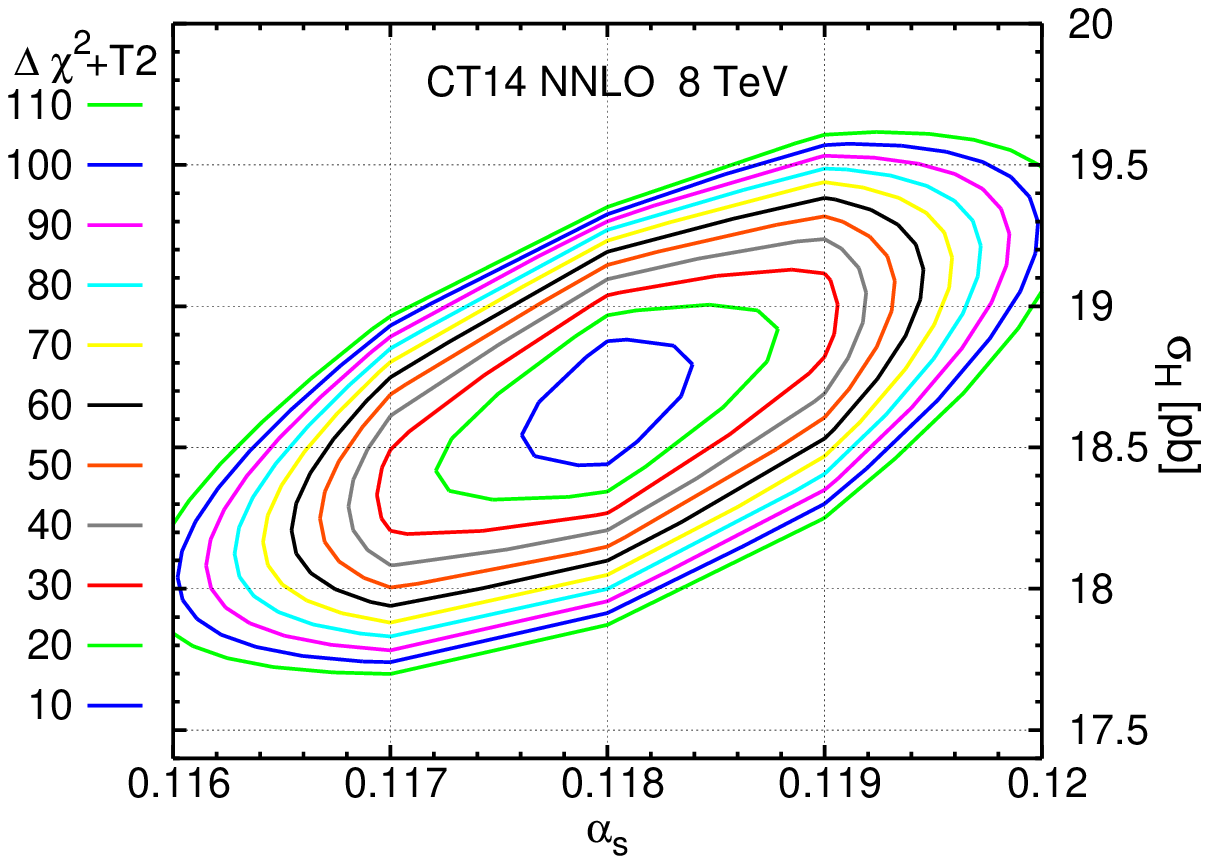}
\includegraphics[width=0.49\textwidth]{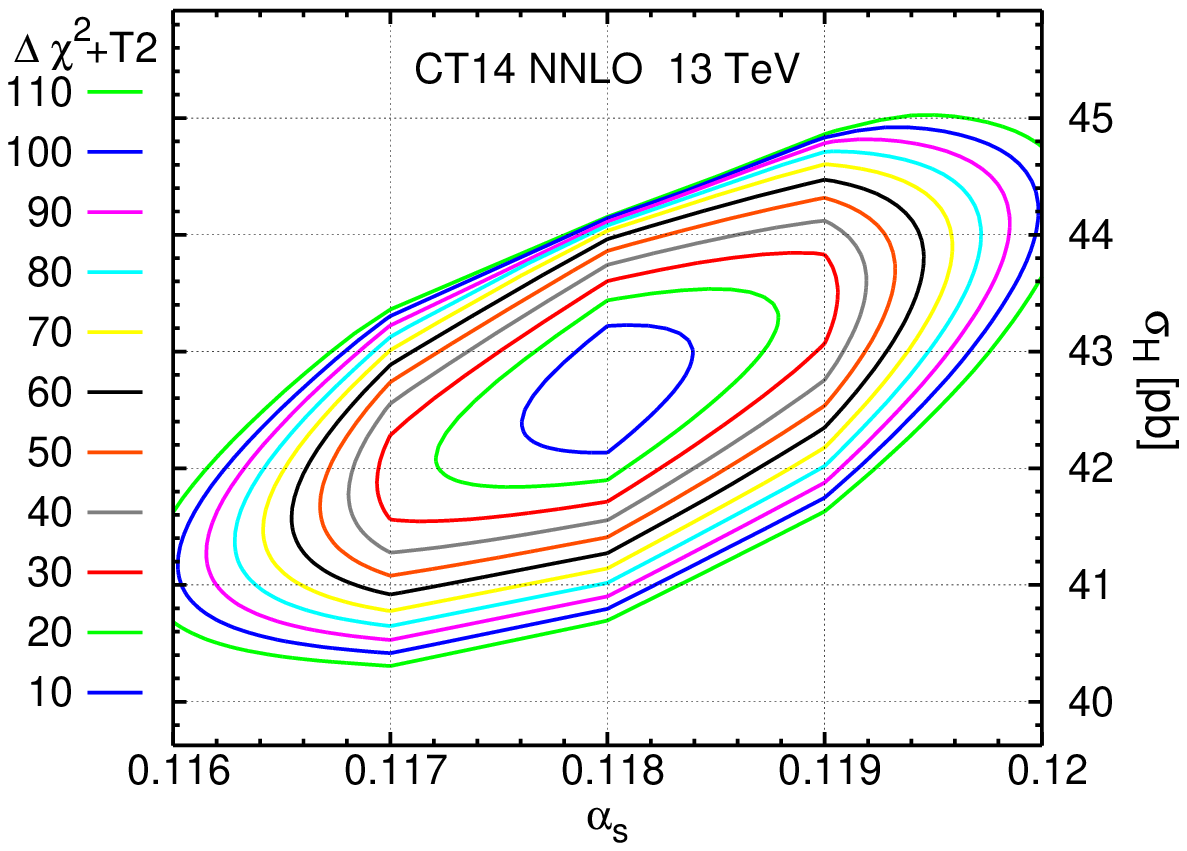}
\caption{Contour plots of $\Delta \chi^{2}\left(\alpha_s(M_Z),\sigma_H\right)$ plus Tier-2 penalty
in the $(\alpha_{s}(M_Z),\sigma_{H})$ plane,
for $\sigma_H$ at the LHC 8 and 13 TeV.
\label{fig:contours}}
\end{figure}

We next perform a LM scan by allowing both the $\sigma_H$ cross
section and $\alpha_s(M_Z)$ to vary as ``fitting
parameters'',  and by including the
world-average constraints on $\alpha_s(M_Z)$ directly into the
$\chi^2$ function.  (Details can be obtained in
Ref.~\cite{Dulat:2013kqa}.)  We examine $\chi^2$ as a function of
$(\alpha_{s}(M_Z),\sigma_{H})$ and trace out contours of constant
$\chi^{2}$+Tier-2 penalty in the $(\alpha_{s},\sigma_{H})$ plane in
Fig.~\ref{fig:contours}, for $\sqrt{s}=8$ and 13 TeV.

A contour here is the locus of points in the $(\alpha_{s}, \sigma_{H})$ plane
along which the constrained value of $\chi^{2}+$Tier-2 is constant.
We see from Fig.~\ref{fig:contours} that the values of
$\sigma_{H}$ and $\alpha_{s}(M_{Z})$ are strongly correlated,
as expected, since the $gg$ fusion cross
section is proportional to $\alpha_s(M_Z)^2$.
Larger values of $\alpha_{s}(M_{Z})$ correspond to
larger values of $\sigma_{H}$ for the same goodness-of-fit
to the global data, even though there is a partially compensating
decrease of the $gg$ luminosity.
The effect of the Tier-2 penalty is very small, being most noticeable
for values of $\alpha_s$ around its global average
of 0.118, which results in a squeezing of the ellipses in that region.

Table \ref{tbl:xsecs} recapitulates the results from Figs.~\ref{fig:LMparab} and
\ref{fig:contours} by listing the central values of $\sigma_H$,
the PDF uncertainties, and combined PDF $+ \ \alpha_s$ uncertainties
as obtained by the Hessian and LM methods. Here, the PDF $+ \ \alpha_s$ uncertainty at 68\% C.L. is  obtained from the result at 90\% C.L. by a scaling factor of $1/1.645$.
\begin{table}[h!]
\begin{center}
\begin{tabular}{l|l|l}
\hline
$gg \to H$ (pb), PDF unc., $\alpha_s = 0.118$\ \ &\qquad\quad\ 8 TeV  &\qquad\quad\ 13 TeV \\
\hline
 68\% C.L. (Hessian)&
 $18.7 +2.1\% -2.3\%$      &
 $42.7 +2.0\% -2.4\%$      \\
\hline
 68\% C.L. (LM)&
 $ +2.3\% -2.3\%$          &
 $ +2.4\% -2.5\%$          \\
\hline
\hline
$gg \to H$ (pb), PDF+$\alpha_s$ unc.&\qquad\quad\ 8 TeV   &\qquad\quad\ 13 TeV \\
\hline
 68\% C.L. (Hessian)&
 $18.7 +2.9\% -3.0\%$      &
 $42.7 +3.0\% -3.2\%$      \\
\hline
 68 \% C.L. (LM)&
 $ +3.0\% -2.9\%$          &
 $ +3.2\% -3.1\%$          \\
\hline
\end{tabular}
\end{center}
\caption{\label{tbl:xsecs}
Uncertainties of $\sigma_{H}(gg \to H)$ computed by the Hessian and LM methods, with Tier-2 penalty included.
The $68\%$ C.L. errors are given as percentages of the central values.
The PDF-only uncertainties are for $\alpha_s(M_Z)=0.118$.
}
\end{table}

\begin{figure}[h]

  \begin{center}
  \includegraphics[width=0.7\textwidth]{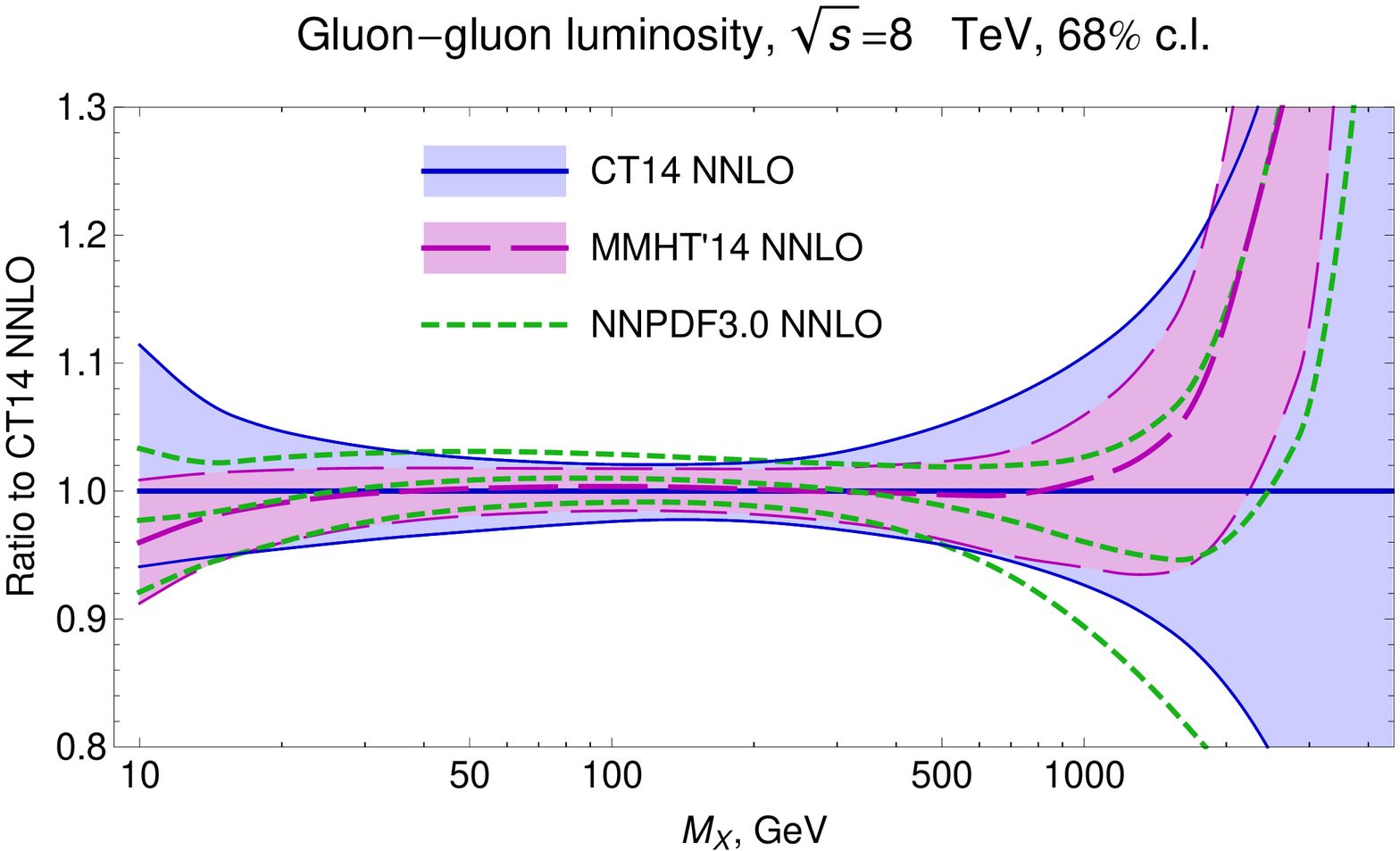} 
  \includegraphics[width=0.7\textwidth]{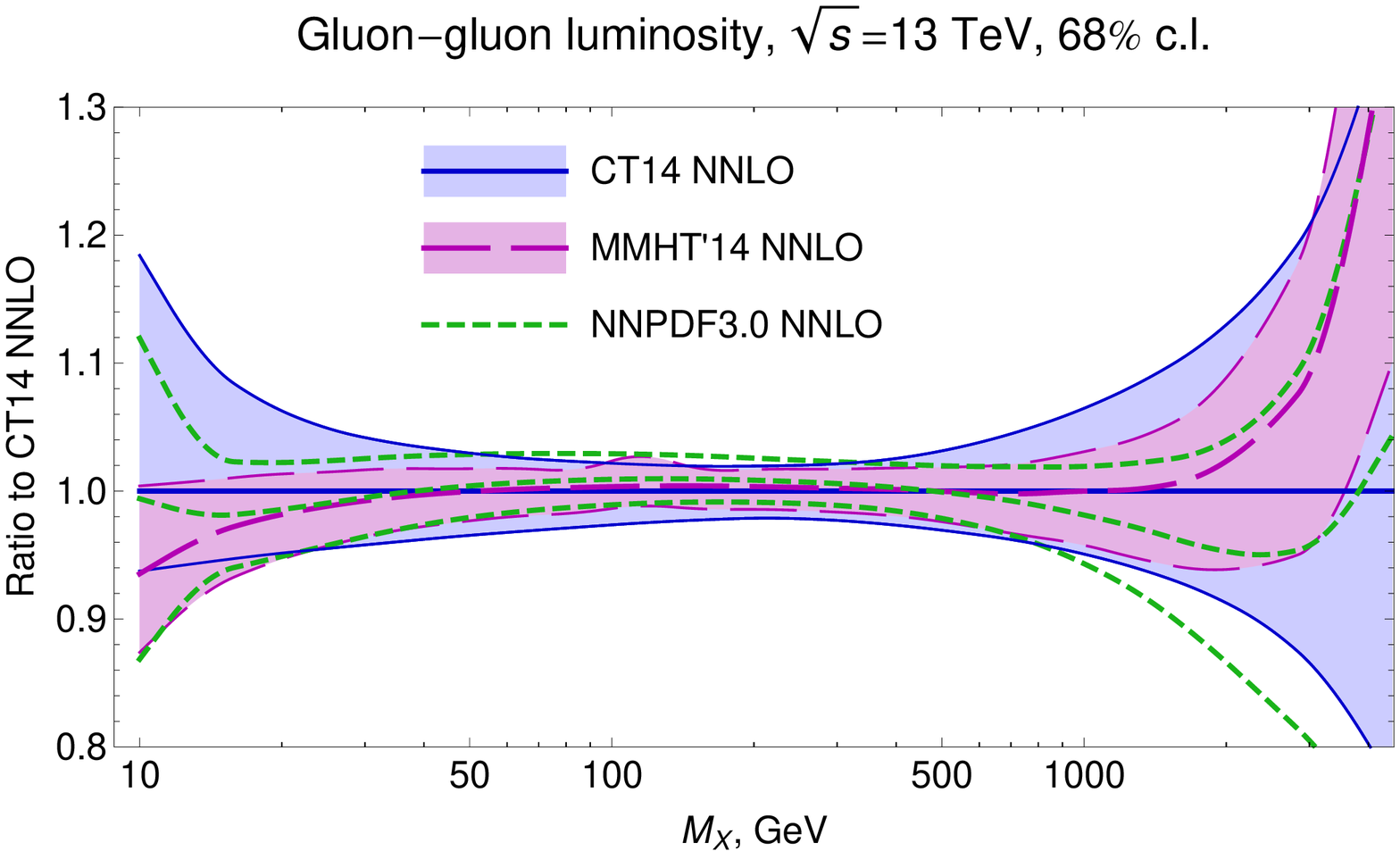}
  \end{center}
\caption{ The $gg$ PDF luminosities
for CT14, MMHT2014 and NNPDF3.0 PDFs
at the LHC with $\sqrt{s}=8$ and 13 TeV, with $\alpha_s(M_Z)=0.118$.
\label{fig:luminosityratio}}
\end{figure}

The $gg$ PDF luminosities for CT14, MMHT2014~\cite{Harland-Lang:2014zoa} and NNPDF3.0~\cite{Ball:2014uwa} PDFs at 13 TeV are shown in Fig.~\ref{fig:luminosityratio}. The parton luminosity is defined as in Ref.~\cite{Campbell:2006wx}.
All central values and uncertainty bands agree very well among the three global PDFs, in the $x$ range sensitive to
Higgs production.
In Table~\ref{tbl:HiggsXStable}, we compare the predictions
for $\sigma_H$ from CT14 with those from MMHT2014, NNPDF3.0, and CT10.
Compared to CT10, predicted $\sigma_H$ values for CT14 NNLO have
increased by 1-1.5\%. Along with the changes also present in the
updated PDFs from the two other PDF groups, the modest increase in the CT14 gluon brings $\sigma_H$ from the global PDF groups into a remarkably good agreement. The projected spread due to the latest NNLO PDFs in the total cross section $\sigma_H$ at 13 TeV will be about the same in magnitude as the scale uncertainty in its NNNLO prediction.

\begin{table}
\vspace{2ex}
\begin{center}
\begin{tabular}{c|c|c|c|c}
\hline \hline
  & CT14 & MMHT2014  & NNPDF3.0 & CT10\\
\hline
8 TeV &
 $18.66^{+2.1\%}_{-2.3\%}$ &
 $18.65^{+1.4\%}_{-1.9\%}$ &
 $18.77^{+1.8\%}_{-1.8\%}$ &
 $18.37^{+1.7\%}_{-2.1\%}$\\
\hline
13 TeV &
 $42.68^{+2.0\%}_{-2.4\%}$ &
 $42.70^{+1.3\%}_{-1.8\%}$ &
 $42.97^{+1.9\%}_{-1.9\%}$ &
 $42.20^{+1.9\%}_{-2.5\%}$\\
\hline
\hline
\end{tabular}
\end{center}
\caption{\label{tbl:HiggsXStable}
Higgs boson production cross sections (in pb)
for the gluon fusion channel  at the LHC, at 8 and 13 TeV
center-of-mass energies, obtained using the CT14, MMHT2014, NNPDF3.0, and CT10
PDFs, with a common value of $\alpha_s(M_Z)=0.118$. The errors
given are due to the PDFs at the 68\% C.L.
}
\end{table}

\begin{figure}[h]
\begin{center}
\includegraphics[width=0.49\textwidth]{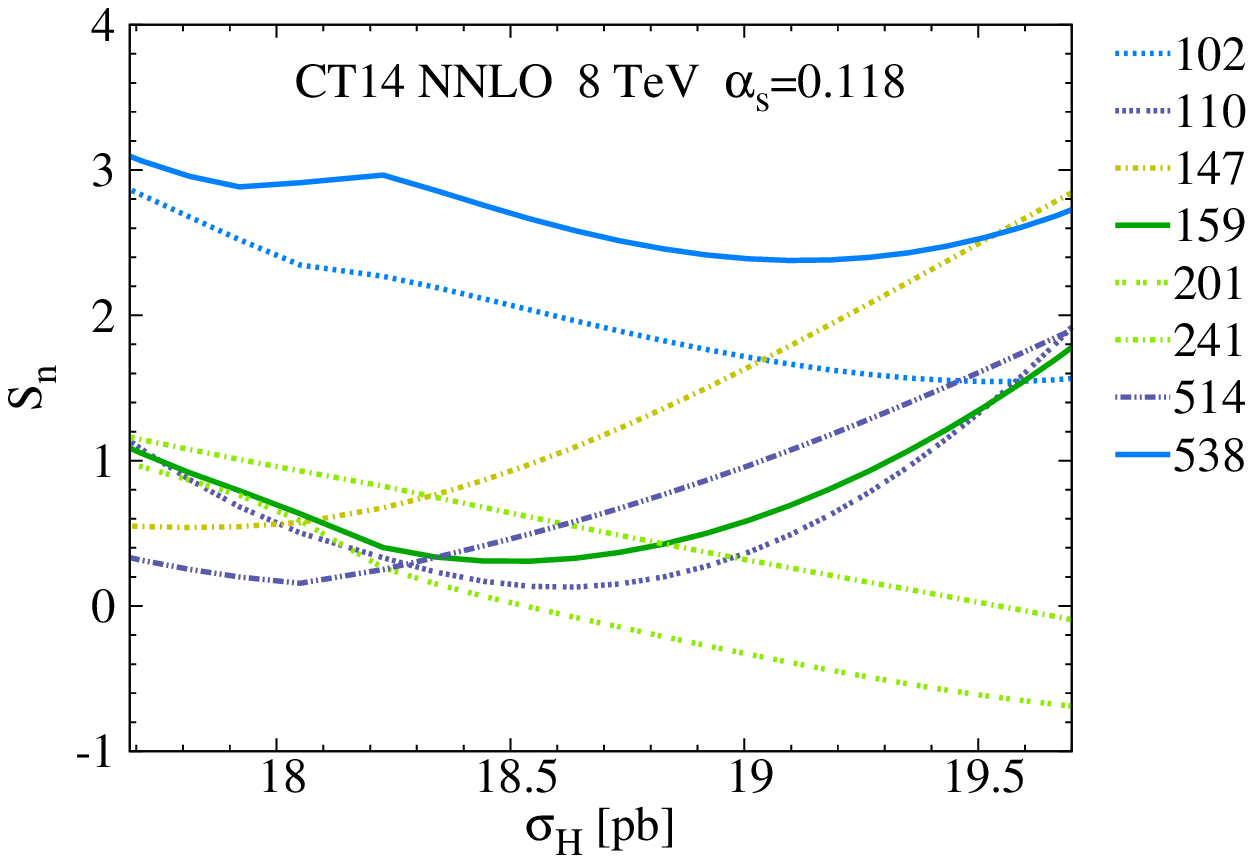}
\includegraphics[width=0.49\textwidth]{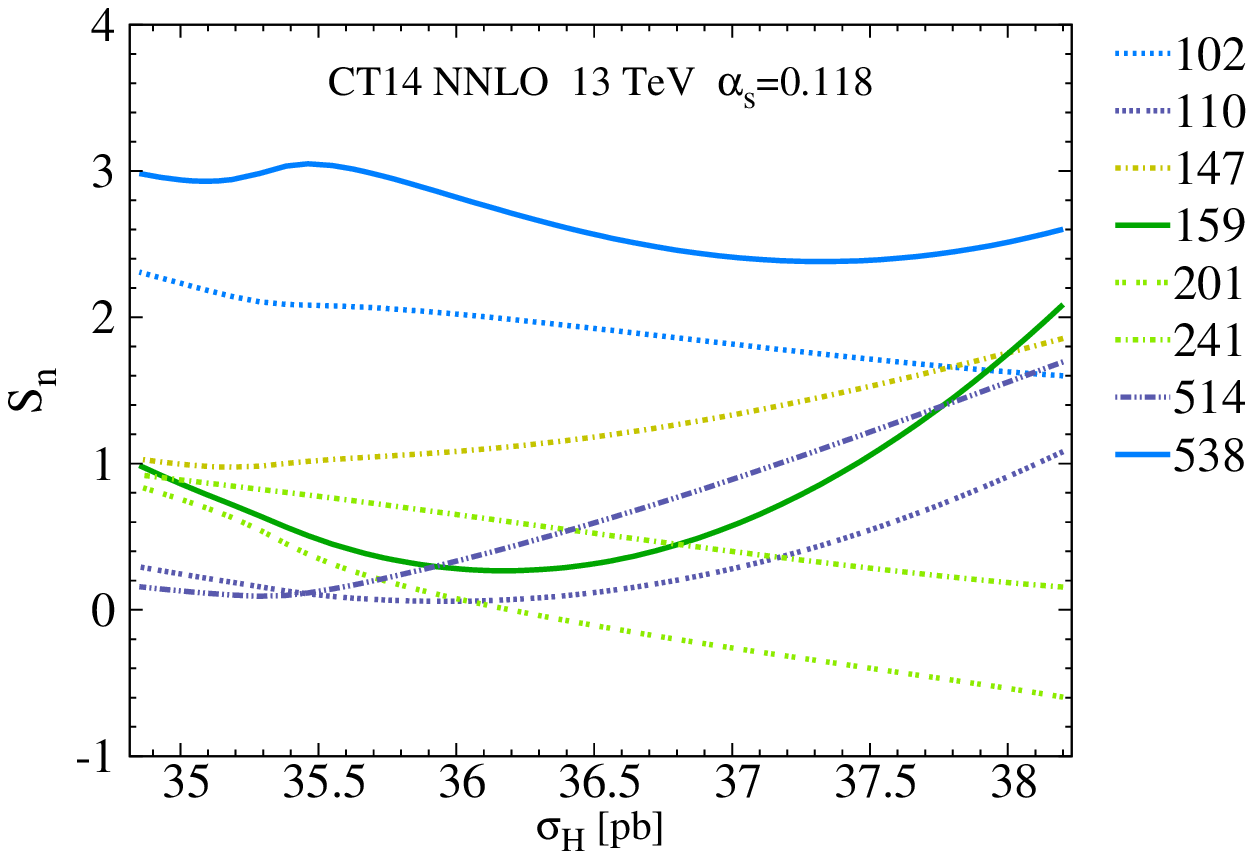}
\end{center}
\vspace{-10pt}
\caption{ The equivalent Gaussian variable $S_n$ versus $\sigma_H$
at the LHC with $\sqrt{s}=8$ and 13 TeV.
\label{fig:spartan}}
\end{figure}

Besides providing an estimate of the PDF uncertainty,
the LM analysis allows us to identify the experimental data sets that are
most sensitive to variations of $\sigma_H$. In the LM scan of
$\sigma_H$, we monitor the changes of the
equivalent Gaussian variable $S_n$ for each included experimental data set.
In the plots of $S_n$ values vs. $\sigma_H$, of the type presented in
Fig.~\ref{fig:spartan}, we select the experiments whose $S_n$
(closely related to $\chi^2_n$)
depends strongly on $\sigma_H$. Such experiments typically impose
the tightest constraints on $\sigma_H$, when their $S_n$ quickly grows with $\sigma_H$.

We see that, although the CMS 7 TeV inclusive jet data (538) is
relatively poorly fit by CT14 NNLO, it is also not very sensitive to
the expected Higgs cross section.
The data sets most relevant to the Higgs cross section
are the HERA inclusive data set (159) at both larger and smaller
values of $\sigma_H$, as well as combined charm production cross
sections from HERA (147); D\O~ Run 2
inclusive jet (514); and CCFR $F_2^p$ (110) at larger  $\sigma_H$.  At
small $\sigma_H$, the most sensitive data set is BCDMS $F_2^d$ (102),
with some sensitivity also from E605 Drell-Yan (201) and LHCb 7 TeV
charge asymmetry (241). Sensitivity of $\sigma_H$ to CCFR dimuon data
observed with CT10 \cite{Dulat:2013kqa}
is no longer present.

\subsection{$t\bar{t}$ production cross section at the LHC}
\label{sec:ttbar}

Next, we consider theoretical predictions and their uncertainties for
the total inclusive cross section for $t\bar{t}$ production at the
LHC, and also present some differential cross sections.

In the $t\bar t$ case, the comparison between the Hessian and Lagrange multiplier methods for finding uncertainties is very
similar to that found for the Higgs cross section.  Therefore, we just present our final estimates
for the total inclusive cross section from the \textsc{Top++} code~\cite{Top++},
given in Table~\ref{topXsectable}.
Recent experimental measurements
of the total inclusive cross section for top-quark pair production
at the LHC are given in Table~\ref{topXsectable2}, together with ATLAS and CMS combined determinations at $\sqrt{s}$ = 7 and 8 TeV.

\begin{table}[h!]
\begin{center}
\begin{tabular}{l|l|l|l}
\hline
$pp \to t\bar{t}$ (pb),\ PDF unc., $\alpha_s=0.118$\ \  &\qquad\quad\  7 TeV &\qquad\quad\  8 TeV  &\qquad\quad\  13 TeV \\
\hline
 68\% C.L. (Hessian)&
 $177 +4.4\% -3.7\%$       &  
 $253 +3.9\% -3.5\%$       &
 $823 +2.6\% -2.7\%$       \\
\hline
 68\% C.L. (LM)&
                           &
 $ +4.8\% -4.6\%$          &
 $ +2.9\% -2.9\%$          \\
\hline
\hline
$pp \to t\bar{t}$ (pb),\ PDF$+\alpha_s$ &\qquad\quad\  7 TeV &\qquad\quad\  8 TeV  &\qquad\quad\  13 TeV \\
\hline
 68\% C.L. (Hessian)&
 $ +5.5\% -4.6\%$          &  
 $ +5.2\% -4.4\%$          &
 $ +3.6\% -3.5\%$          \\
\hline
 68\% C.L. (LM)&
                           &
 $ +5.1\% -4.7\%$          &
 $ +3.6\% -3.5\%$          \\
\hline
\end{tabular}
\end{center}
\vspace{-2ex}
\caption{\label{topXsectable}
CT14 NNLO total inclusive cross sections for top-quark pair production
at LHC center-of-mass energies of 7, 8, and 13 TeV,
for an assumed top-quark mass of 173.3 GeV.
}
\end{table}

\begin{table}[h!]
\begin{center}
\begin{tabular}{l|l|l}
\hline
$\sigma_{t\bar{t}}^\textrm{exp} \textrm{(pb)}$ \ \  &\qquad\quad\  7 TeV (dilepton channel) &\qquad\quad\  8 TeV  (lep+jets) \\
\hline
ATLAS~\cite{Aad:2011yb},\cite{Aad:2015pga}    &  $177\pm 20^\textrm{(stat)}\pm 14^\textrm{(syst)}\pm 7^\textrm{(lumi.)}$ & $260\pm 1^\textrm{(stat)} {^{+22} _{-23}}^\textrm{(syst)}\pm 8^\textrm{(lumi.)}\pm 4^\textrm{(beam)}$  \\
\hline
CMS~\cite{Chatrchyan:2012bra},\cite{CMS:2012gza}    &  $161.9\pm 2.5^\textrm{(stat)} {^{+5.1} _{-5.0}}^\textrm{(syst)}\pm 3.6^\textrm{(lumi.)}$ & $228.4\pm 9.0^\textrm{(stat)}{^{+29} _{-26}}^\textrm{(syst)}\pm 10^\textrm{(lumi.)}$ \\
\hline
\hline
       \ \  &\qquad\quad\  7 TeV (lepton+jets, di-lepton, all-jets) &\qquad\quad\  8 TeV  (dilepton channel) \\
\hline
ATLAS and CMS & \\
Combined~\cite{ATLAS:2012dpa},\cite{CMS:2014gta}   &  $173.3\pm 2.3^\textrm{(stat)}\pm 7.6^\textrm{(syst)}\pm 6.3^\textrm{(lumi.)}$ & $241.5\pm 1.4^\textrm{(stat)}+5.7^\textrm{(syst)}\pm 6.2^\textrm{(lumi.)}$   \\
\hline
\end{tabular}
\end{center}
\vspace{-2ex}
\caption{\label{topXsectable2}
Measurements of total inclusive cross sections for top-quark pair production
at LHC center-of-mass energies of 7, 8, and 13 TeV,
for an assumed top-quark mass of 172.5 GeV.}
\end{table}

For comparison, predictions and PDF-only errors using CT10NNLO PDFs give $\sigma_{t\bar{t}}=246^{+4.1\%}_{-3.4\%}$ pb at 8 TeV
and $\sigma_{t\bar{t}}=806^{+2.5\%}_{-2.2\%}$ pb at 13 TeV at 68\% C.L.
Here we find that the Hessian and the LM methods
are in very good agreement in CT14 at $\sqrt{s}=13$ TeV, and agree
slightly worse at $\sqrt{s}=8$ TeV.
Measurements of $t\bar{t}$ pair production can potentially
constrain the gluon PDF at large $x$, if correlations between
the gluon, $\alpha_s(M_Z)$ and the top-quark mass are
accounted for. Given the current experimental precision of $t\bar{t}$
measurements, the impact of such data in a global PDF fit is expected
to be moderate; related exploratory studies can be found in
Refs.~\cite{Guzzi:2014wia,Czakon:2013tha}.

\begin{figure}[tb]
\includegraphics[width=0.70\textwidth]{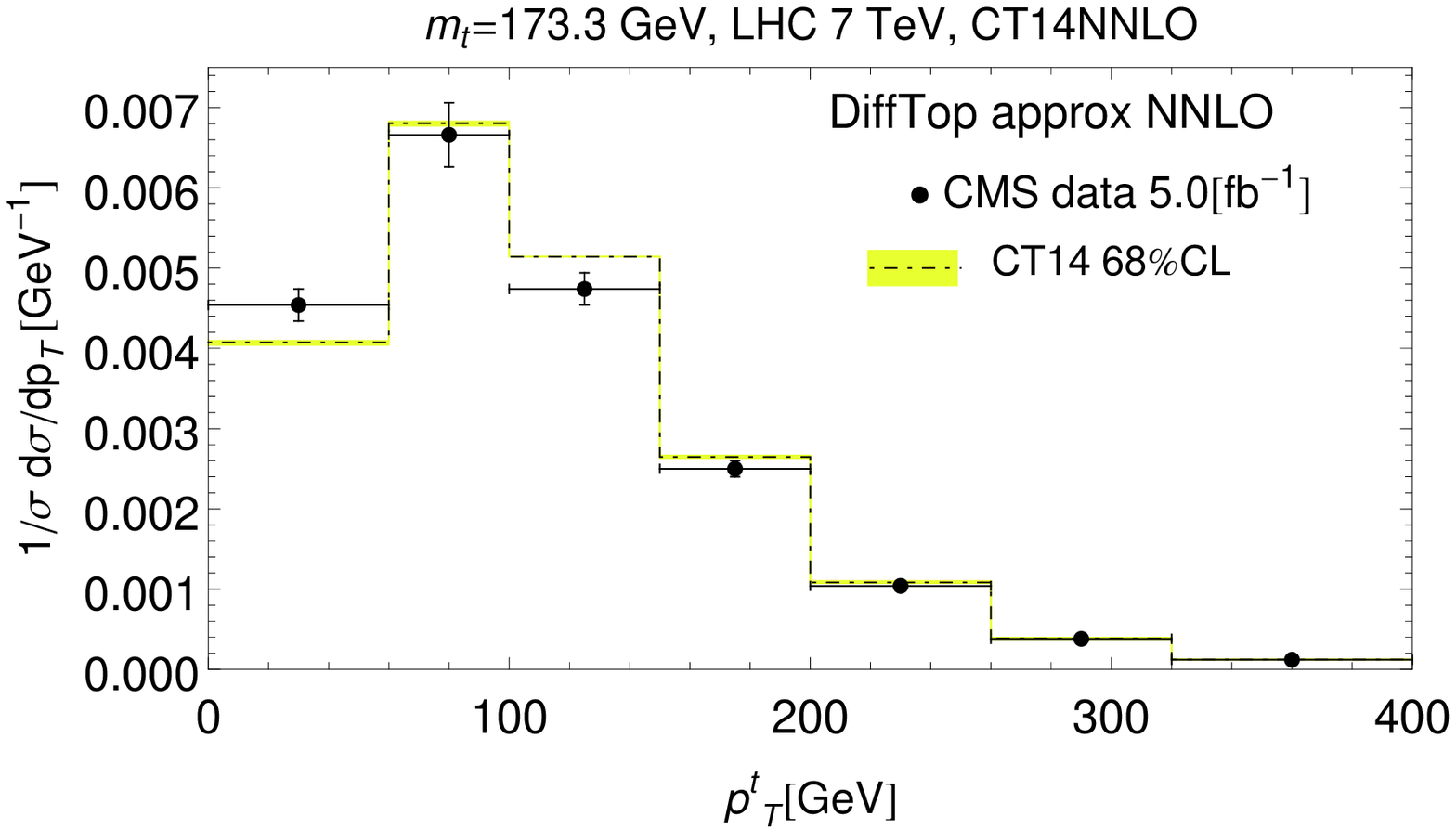}\\
\includegraphics[width=0.70\textwidth]{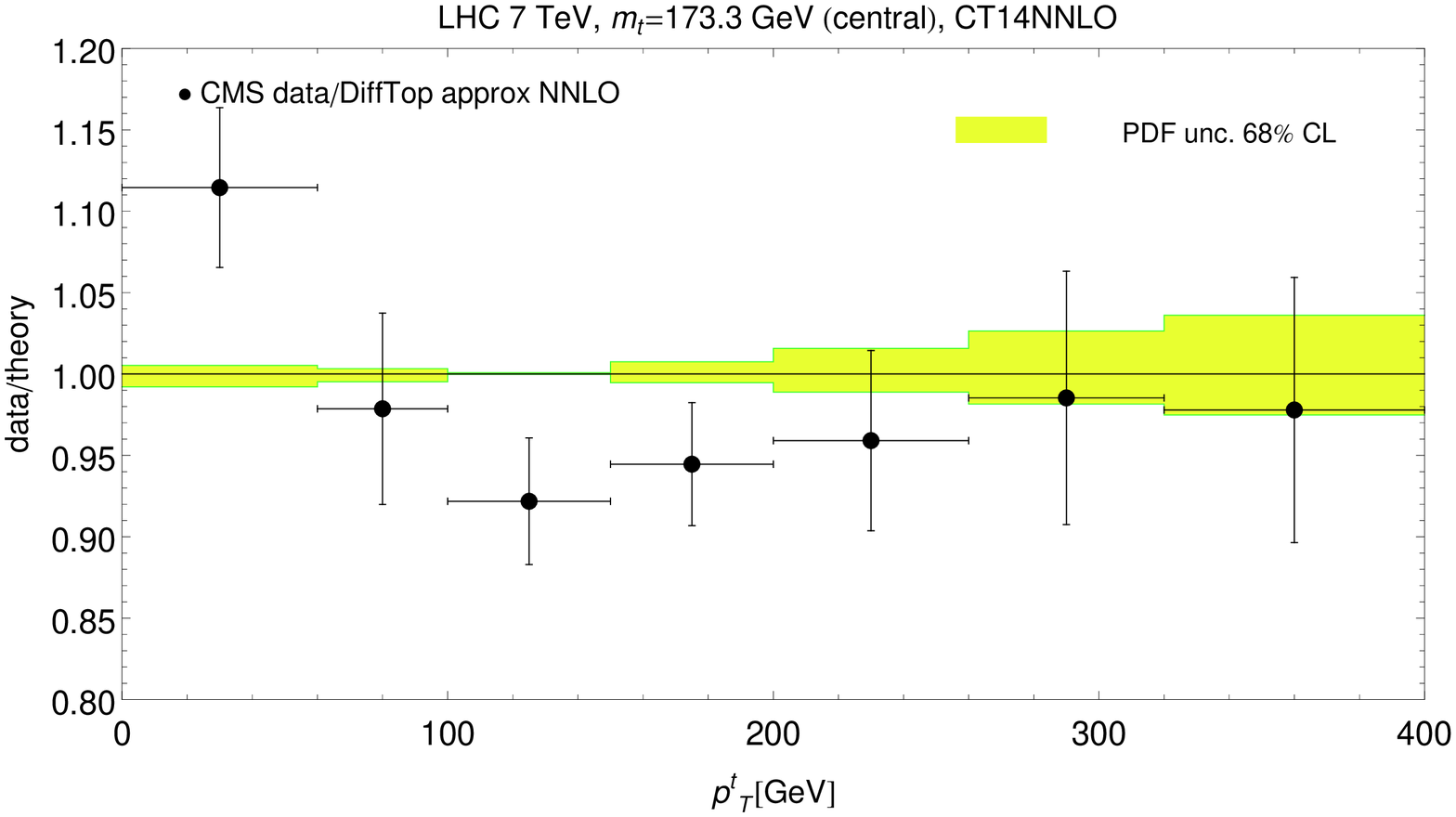}
\caption{Normalized final-state top-quark $p_T$ differential distribution at CMS 7 TeV.
\label{fig:4F-fig1}}
\end{figure}

\begin{figure}[tb]
\includegraphics[width=0.70\textwidth]{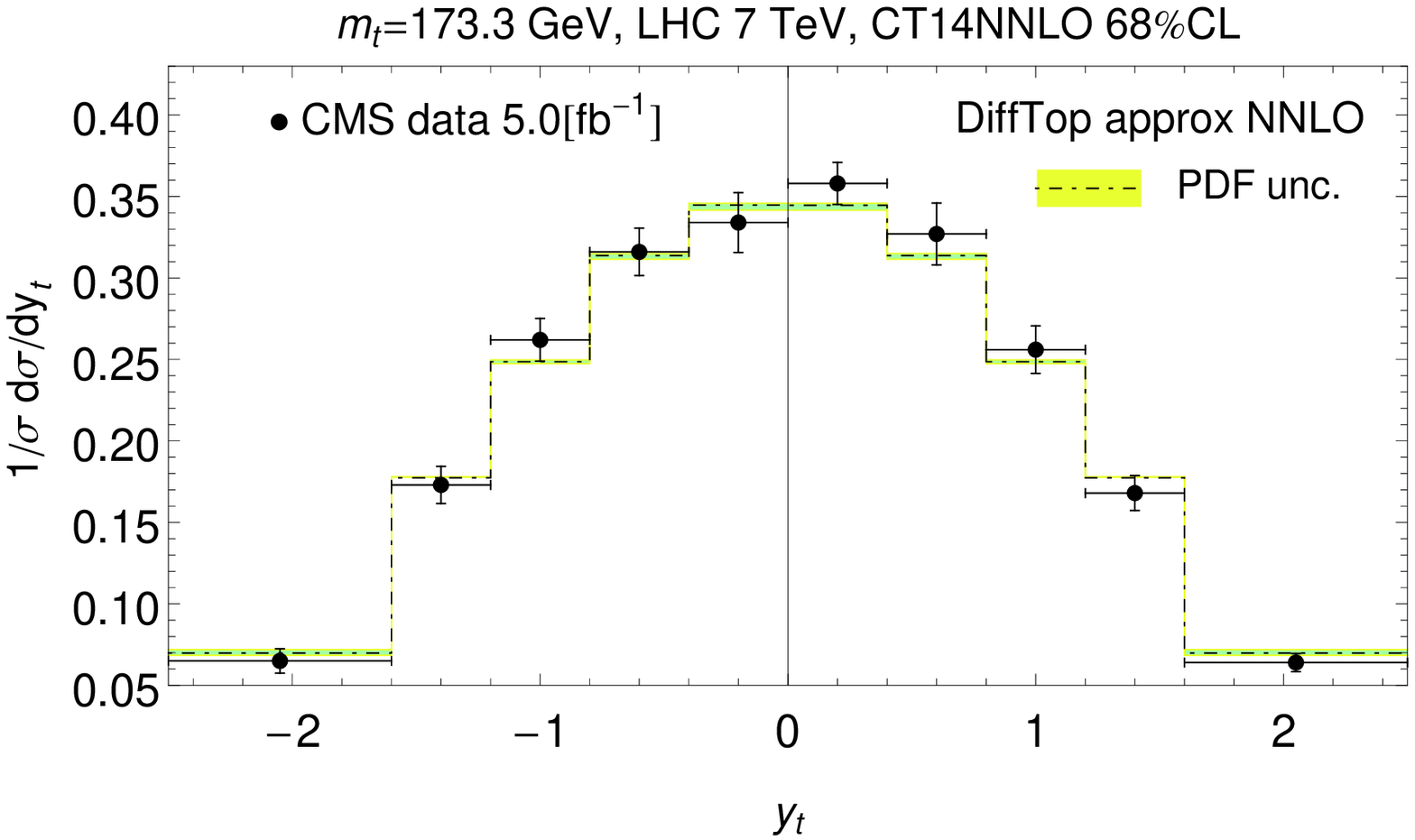}\\
\includegraphics[width=0.70\textwidth]{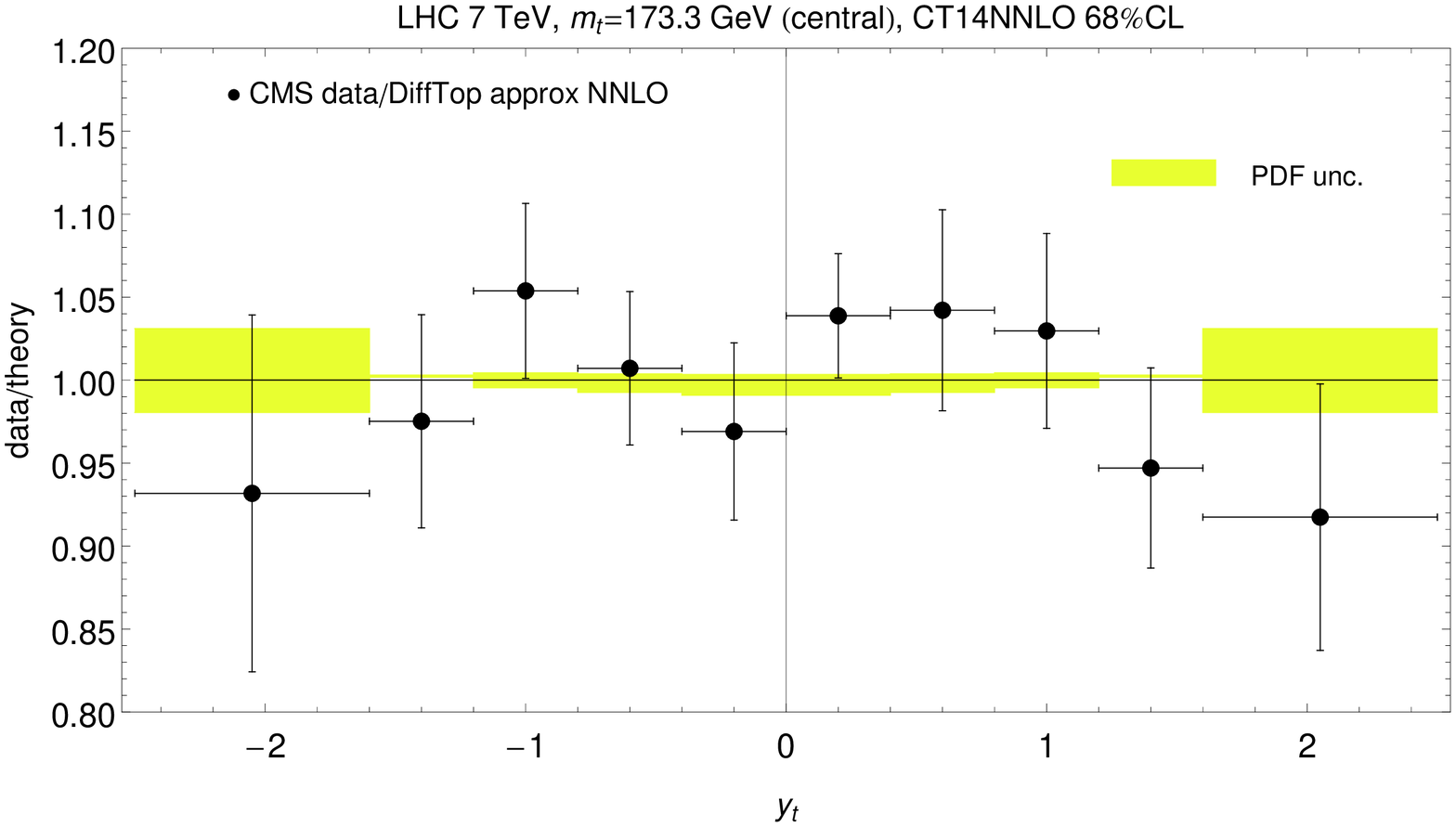}
\caption{Normalized final-state top-quark rapidity distribution at CMS 7 TeV.
\label{fig:4F-fig1y}}
\end{figure}

\begin{figure}[tb]
\includegraphics[width=0.70\textwidth]{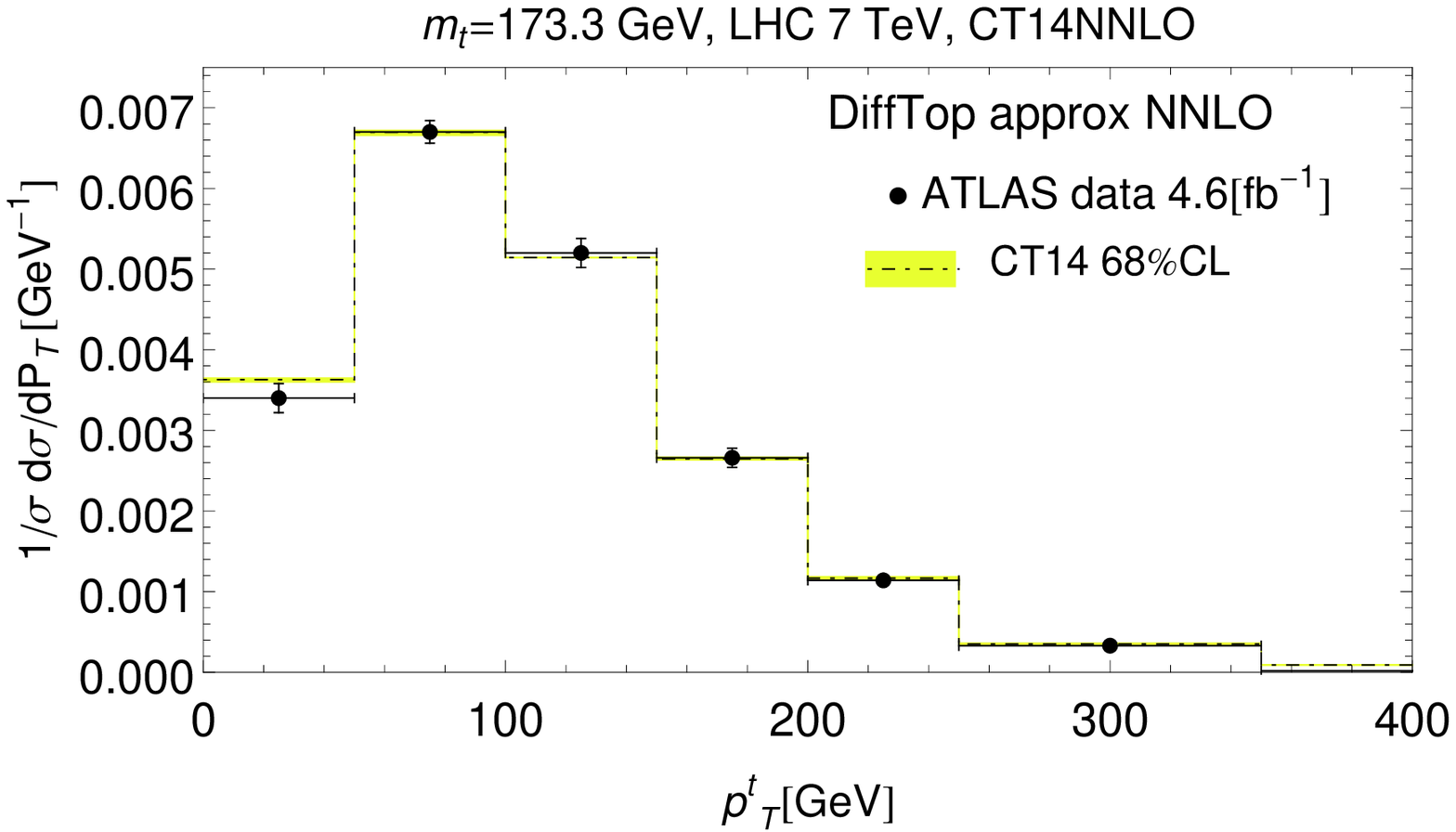}\\
\includegraphics[width=0.70\textwidth]{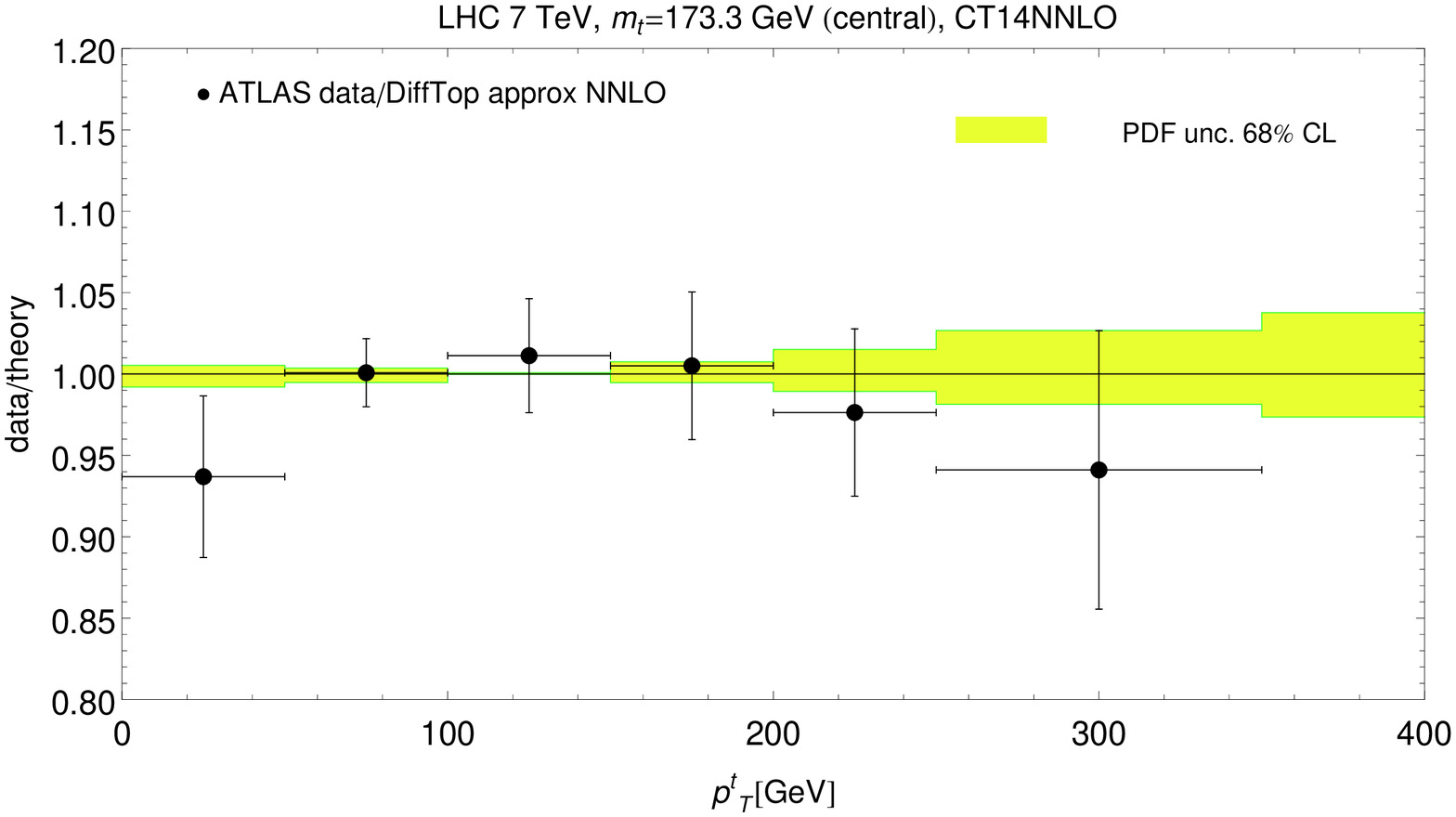}
\caption{Normalized final-state top-quark $p_T$ differential distribution at ATLAS 7 TeV.
\label{fig:4F-fig2}}
\end{figure}

\begin{figure}[tb]
\includegraphics[width=0.43 \textwidth]{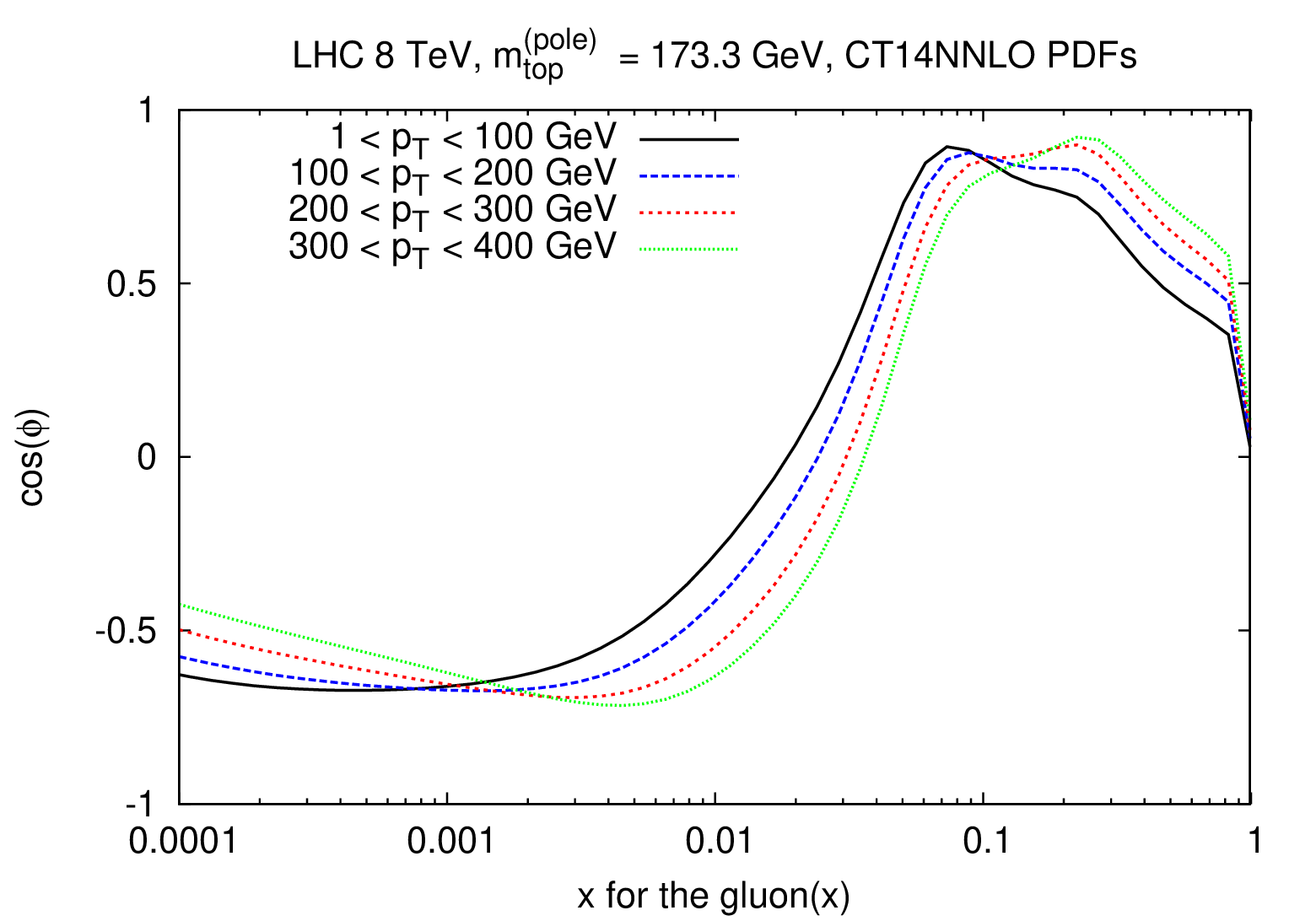}
\includegraphics[width=0.43 \textwidth]{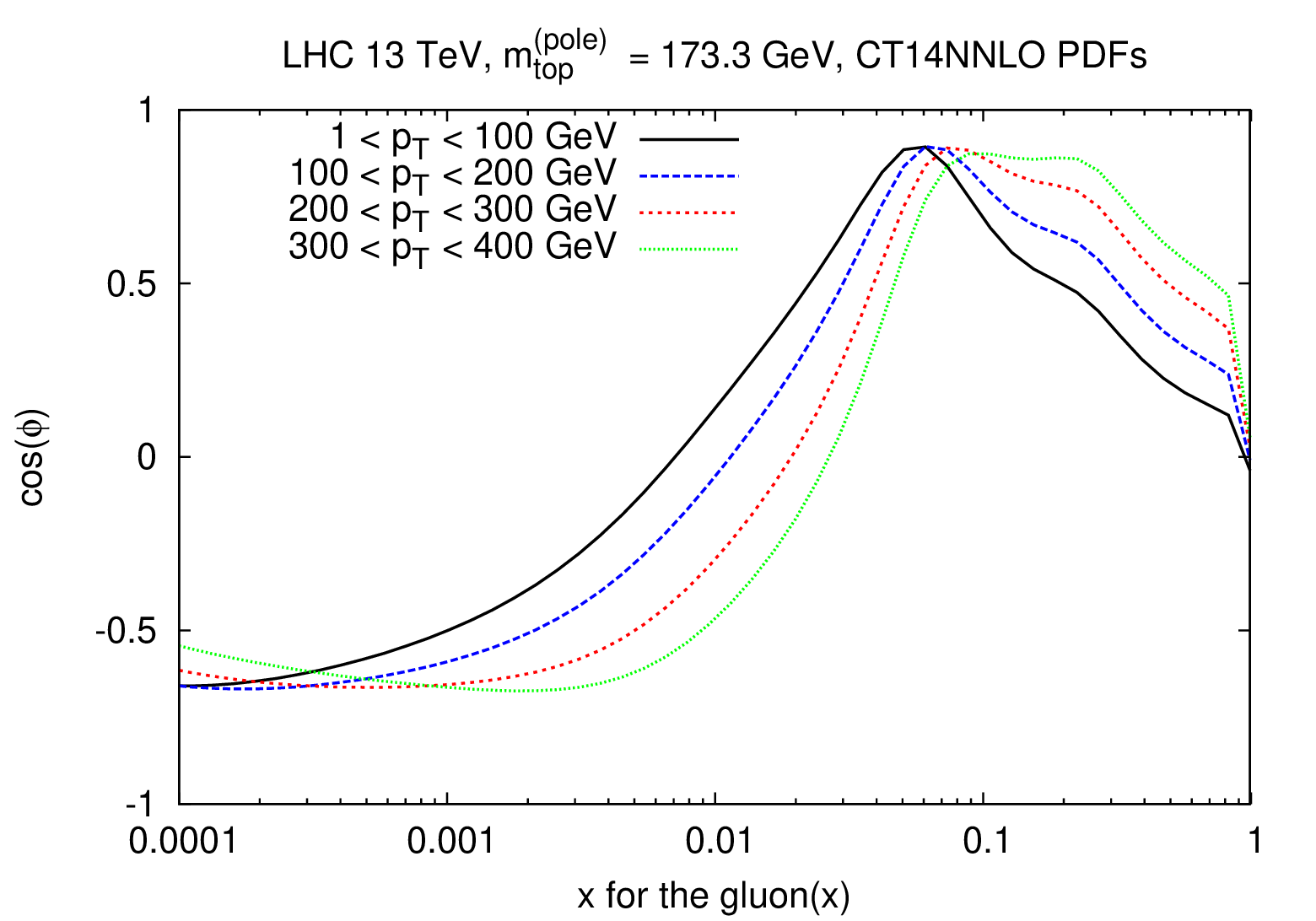}
\caption{The correlation cosine as a function of $x$-gluon for the top-quark $p_T$ distribution
in $t\bar{t}$ production at the LHC at $\sqrt{s}$ = 8 and 13 TeV.
\label{fig:4F-fig3}}
\end{figure}

\begin{figure}[ht!]
\includegraphics[width=0.43\textwidth]{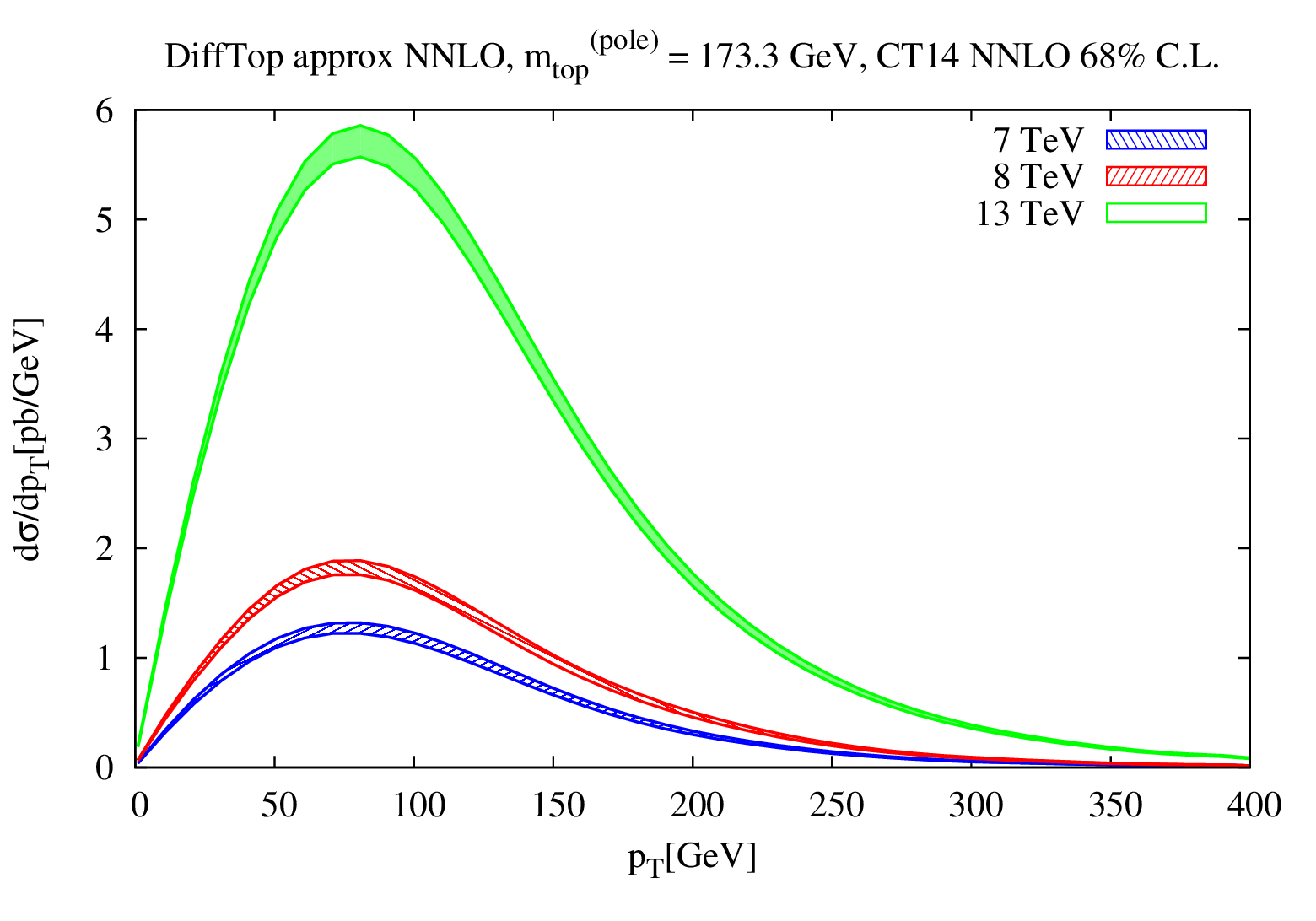}
\includegraphics[width=0.43\textwidth]{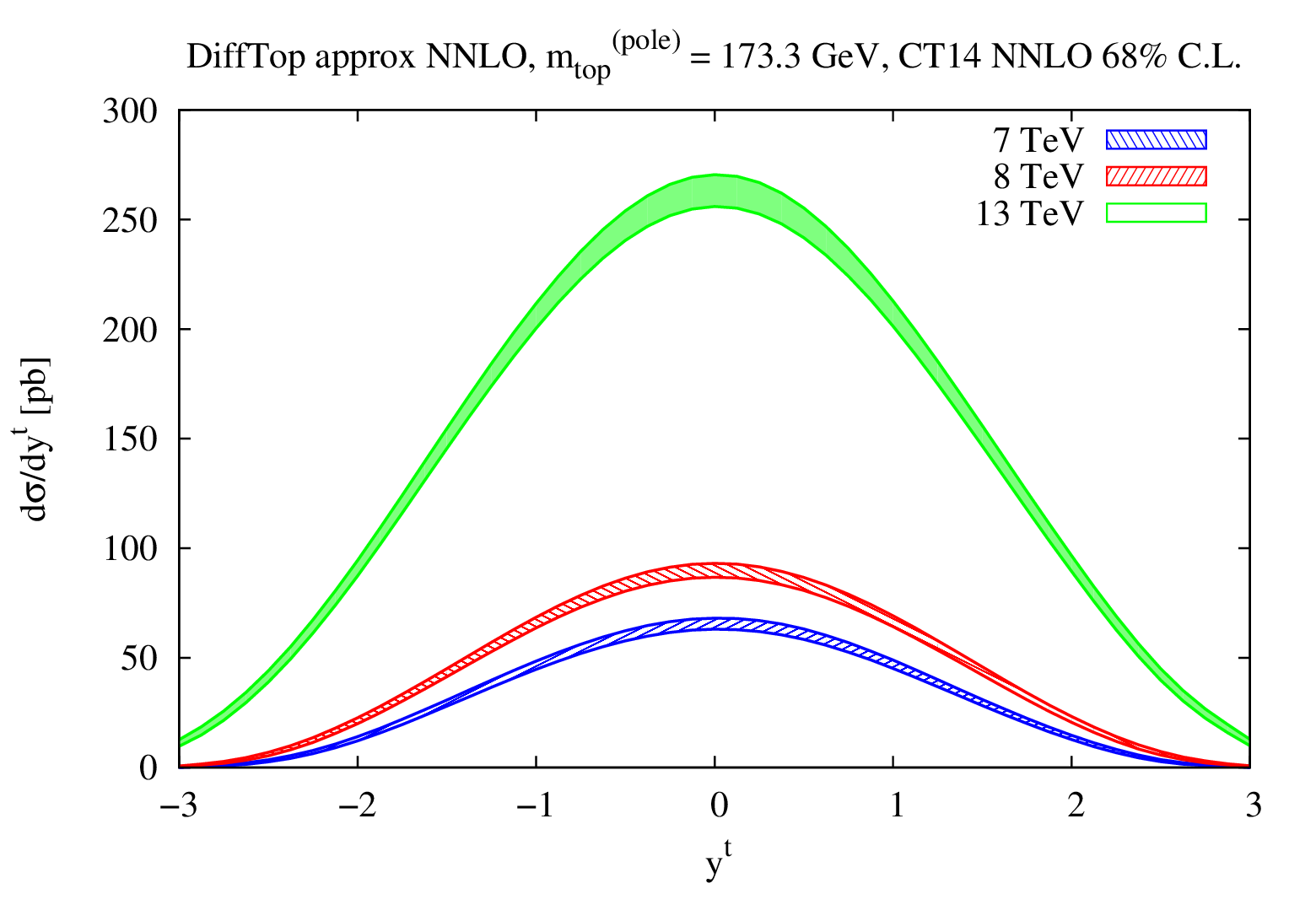}
\caption{Absolute differential $p_T$ and $y$ distributions for the
  final-state top-quark in $t\bar{t}$ production at the LHC at
  $\sqrt{s}$ = 7, 8, and 13 TeV.
\label{fig:4F-fig4}}
\end{figure}

In Figs.~\ref{fig:4F-fig1},~\ref{fig:4F-fig1y} and~\ref{fig:4F-fig2}, the normalized top-quark transverse momentum $p_T$ and rapidity $y$ distributions
at approximate NNLO $({\cal O}(\alpha_s^4))$ are compared to the CMS~\cite{Chatrchyan:2012saa} and ATLAS~\cite{Aad:2014zka} measurements,
at a center of mass energy $\sqrt{s}$ = 7 TeV.
The yellow bands represent the CT14 PDF uncertainty evaluated at
the 68\%C.L. with the program \textsc{DiffTop}
\cite{Guzzi:2014wia} based on QCD threshold expansions
beyond the leading logarithmic approximation, for one-particle
inclusive kinematics. The value of the top-quark mass here
is $m_t=173.3$ GeV in the ``pole mass'' definition.
In Fig.~\ref{fig:4F-fig3} the
correlation cosine between the differential top-quark $p_T$ distribution
and the momentum fraction $x$ carried by the gluon is shown, in four different $p_T$ bins at the LHC $\sqrt{s}$ = 8 and 13 TeV.
The cosine correlation at $\sqrt{s}$ = 7 TeV exhibits identical
features to that of $\sqrt{s}$ = 8 TeV. It is therefore omitted.
A strong correlation between the $p_T$ distribution and large
$x$-gluon ($x\approx 0.1$) is observed for both LHC energies,
although the cosines exhibit different patterns of $x$ dependence.
Finally, in Fig.~\ref{fig:4F-fig4} we present
the absolute, rather than normalized,
differential $p_T$ and $y$ distributions
for top-quark production, together with the relative PDF uncertainties,
at the LHC with $\sqrt{s}$ = 7, 8 and 13 TeV.


\section{Discussion and Conclusion}\label{sec:conclude}

In this paper, we have presented CT14, the next generation of NNLO (as well as LO and NLO) parton distributions from a global analysis by the CTEQ-TEA
group. With rapid improvements in LHC measurements, the focus of the
global analysis has shifted toward providing accurate predictions in
the wide range of $x$ and $Q$ covered by the LHC data. This
development requires a long-term multi-prong effort in
theoretical, experimental, and statistical areas.

In the current study, we have added enhancements that open the door
for long-term developments in CT14 methodology geared toward the goals
of LHC physics. This is the first CT analysis that includes
measurements of inclusive production of vector bosons
\cite{Aaij:2012vn, Chatrchyan:2013mza, Chatrchyan:2012xt, Aad:2011dm}
and jets \cite{Aad:2011fc, Chatrchyan:2012bja} from
the LHC at 7 and 8 TeV as input for the fits. We also include new
data on charm production from DIS at HERA \cite{Abramowicz:1900rp}
and precise measurements of the electron charge asymmetry from D\O~ at $9.7\mbox{ fb}^{-1}$
\cite{D0:2014kma}.
These measurements allow us to probe new combinations of quark
flavors that were not resolved by the previous data sets. As most of
these measurements contain substantial correlated systematic
uncertainties, we have implemented these correlated errors and have examined their
impact on the PDFs.

On the theory side, we have introduced a more flexible parametrization to better capture variations in the PDF dependence.  A series of benchmark
tests of NNLO cross sections, carried out in the run-up for the CT14
fit for all key fitted processes,
has resulted in better agreement with most experiments
and brought accuracy of most predictions to the truly NNLO level. We
examined the PDF errors for the important LHC
processes and have tested the consistency of the Hessian and Lagrange
Multiplier approaches. Compared to CT10, the new inputs
and theoretical advancements resulted in a softer $d/u$ ratio at large
$x$, a lower strangeness PDF at $x>0.01$, a slight increase in the
large-$x$ gluon (of order 1\%), and wider uncertainty bands on $d/u$,
$\bar d/\bar u$, and $q-\bar q$ combinations at $x$ of order $0.001$
(probed by LHC $W/Z$ production). Despite these changes in central
predictions, the CT14 NNLO PDFs remain consistent with CT10 NNLO within the
respective error bands.

Some implications of CT14 predictions for phenomenological observables
were reviewed in Sections \ref{sec:TheoryVsData} and \ref{sec:LHCPredictions}.
Compared to calculations with CT10 NNLO, the $gg \rightarrow H$ total
cross section has increased slightly in CT14: by 1.6\% at the LHC 8 TeV
and by 1.1 \% at 13 TeV. The $t\bar t$ production cross sections have also increased
in CT14 by 2.7\% at 8 TeV and by 1.4\% at 13 TeV. The $W$ and $Z$
cross sections, while still consistent with CT10, have slightly
changed as a result of reduced strangeness. Common ratios of
strangeness and non-strangeness PDFs for CT14 NNLO, shown in
Eqs.~(\ref{CT14rs}) and (\ref{CT14kappas}), are consistent with
the independent
ATLAS, CMS, and NOMAD determinations within the PDF uncertainties.

The final CT14 PDFs are presented in the form of 1 central and 56
Hessian eigenvector sets at NLO and NNLO. The 90\% C.L. PDF
uncertainties for physical observables can be estimated from these
sets using the symmetric \cite{Pumplin:2002vw} or asymmetric
\cite{Lai:2010vv,Nadolsky:2001yg} master formulas by adding
contributions from each pair of sets in quadrature. These PDFs are
determined for the central QCD coupling of $\alpha_s(M_Z)=0.118$,
consistent with the world-average $\alpha_s$ value. For estimation of
the combined PDF+$\alpha_s$ uncertainty, we provide two additional
best-fit sets for $\alpha_s(M_Z)=0.116$ and 0.120. The
90\% C.L. variation due to $\alpha_s(M_Z)$ can be estimated as a half of the
difference in predictions from the two $\alpha_s$ sets. The
PDF+$\alpha_s$ uncertainty, at 90\% C.L., and including correlations,
can also be determined by adding the PDF uncertainty and $\alpha_s$
uncertainty in quadrature \cite{Lai:2010nw}.

At leading order, we provide two PDF sets,
obtained assuming 1-loop evolution of $\alpha_s$ and
$\alpha_s(M_Z)=0.130$; and 2-loop evolution of $\alpha_s$
and $\alpha_s(M_Z)=0.118$.
Besides these general-purpose PDF sets, we provide a series of (N)NLO
sets for $\alpha_s(M_Z)=0.111-0.123$
and additional sets in heavy-quark
schemes with up to 3, 4, and 6 active flavors.
Phenomenological applications of the CT14 series
and the special CT14 PDFs
(such as allowing for nonperturbative intrinsic charm contribution)
will be discussed in a follow-up study \cite{CT14paper2}.

Parametrizations for the CT14 PDF sets are distributed in a standalone
form via the CTEQ-TEA website \cite{CT14website}, or as a part of
the LHAPDF6 library \cite{LHAPDF6}. For backward compatibility with
version 5.9.X of LHAPDF, our website also provides CT14 grids in the
LHAPDF5 format, as well as an update for the CTEQ-TEA module
of the LHAPDF5 library, which must be included during compilation
to support calls of all eigenvector sets included with CT14 \cite{LHAPDF5}.

\begin{acknowledgments}

This work was supported by the U.S. DOE Early Career Research Award
DE-SC0003870; by the U.S. Department
of Energy under Grant No. DE-FG02-96ER40969, DE-SC0013681, and DE-AC02-06CH11357;
by the U.S. National
Science Foundation under Grant No. PHY-0855561 and PHY-1417326; by Lightner-Sams Foundation;
and by the National Natural Science Foundation of China under Grant
No. 11165014 and 11465018.

\end{acknowledgments}

\section*{Appendix: Parametrizations in CT14}

Parton distribution functions are measured by parameterizing their
$x$-dependence at a low scale $Q_0$.  For each choice
of parameters, the PDFs are computed at higher scales by
DGLAP evolution; and the parameters
at $Q_0$ are adjusted to optimize the fit to a
wide variety of experimental data.
Traditional parametrizations for each flavor are of the form
\begin{eqnarray}
x \, f_a(x,Q_0) \, = \, x^{a_1} \, (1-x)^{a_2} \,P_a(x) \label{xfa}
\end{eqnarray}
where the $x^{a_1}$ behavior at $x \to 0$ is guided by Regge theory, and
the  $(1-x)^{a_2}$ behavior at $x \to 1$ is guided by spectator counting
rules.  The remaining factor $P_a(x)$ is assumed to be slowly varying,
because there is no reason to expect fine structure in it even at scales
below $Q_0$, and evolution from those scales up to $Q_0$ provides
additional smoothing.

In the previous CTEQ analyses, $P_a(x)$ in Eq.~(\ref{xfa}) for each
flavor was chosen as an exponential of a polynomial in $x$ or $\sqrt{x}\,$;
e.g.,
\begin{eqnarray}
P(x) = \exp(a_0 + a_3\sqrt{x} + a_4 x + a_5 x^2)
\label{eq:ct10valence}
\end{eqnarray}
for $\mathrm{u}_v(x)$ or $\mathrm{d}_v(x)$ in CT10 \cite{Gao:2013xoa}.
The exponential form conveniently enforces the desired
positive-definite behavior for the PDFs, and it suppresses
non-leading behavior in the limit $x \to 0$ by a factor $\sqrt{x}$,
which is similar to what would be expected from a secondary
Regge trajectory.  However, this parametrization has two undesirable features.
First, because the exponential function
can vary rapidly, the power laws $x^{a_1}$ and $(1-x)^{a_2}$,
which formally control the $x \to 0$ and $x \to 1$ limits, need
not actually dominate in practical regions of small $x$
(say $x \lesssim 0.001$) or large $x$ (say $x \gtrsim 0.6$).
Second, the qualitative similarity of $\exp(a_3 \, \sqrt{x})$, $\exp(a_4 \, x)$,
and $\exp(a_5 \, x^2)$ to each other causes the parameters
$a_3$, $a_4$, $a_5$ to be strongly correlated
with each other in the fit.  This correlation
may destabilize the $\chi^2$ minimization
and compromise the Hessian approach to uncertainty analysis, since
that approach is based on a quadratic dependence of $\chi^2$ on
the fitting parameters, which is only guaranteed close to the
minimum.

We introduce a better style of parametrization in CT14.
We begin by replacing $P_a(x)$ by a polynomial in
$\sqrt{x}$,  which avoids the rapid variations invited by
an exponential form.  Low-order polynomials have been used previously
by many other groups; however, polynomials with higher powers
were less widespread. We add them now to provide more
flexibility in the parametrization.  In particular,
for the best-constrained flavor combination
$\mathrm{u}_v(x) \equiv \mathrm{u}(x) \, - \, \mathrm{\bar{u}}(x)$
we use a fourth-order polynomial
\begin{eqnarray}
P_{\mathrm{u}_v} \, = \, c_0 \, + \, c_1 \, y \, + \,
c_2 \, y^2 \, + \, c_3 \, y^3 \, + \, c_4 \, y^4,
\label{eq:AppendixPoly}
\end{eqnarray}
where $y = \sqrt{x}$.  But rather than using the
coefficients $c_i$ directly as fitting parameters,
we re-express the polynomial as a linear combination of
\emph{Bernstein polynomials}:
\begin{eqnarray}
P_{\mathrm{u}_v} \, = \, d_0 \, p_0(y) \, + \, d_1 \, p_1(y) \, + \,
d_2 \, p_2(y) \, + \, d_3 \, p_3(y) \, + \, d_4 \, p_4(y),
\end{eqnarray}
where
\begin{eqnarray}
p_0(y) &=& (1 - y)^4, \nonumber \\
p_1(y) &=& 4 \, y \, (1 - y)^3, \nonumber \\
p_2(y) &=& 6 \, y^2 \, (1 - y)^2, \nonumber \\
p_3(y) &=& 4 \, y^3 \, (1 - y), \nonumber \\
p_4(y) &=& y^4.
\label{eq:bernstein}
\end{eqnarray}
This re-expression does not change the functional form
of $P_{\mathrm{u}_v}$: it is still a completely general fourth-order
polynomial in $y = \sqrt{x} \,$.  But the new coefficients
${d_i}$ are less correlated with each other than the old
${c_i}$, because each Bernstein polynomial is strongly peaked
at a different value of $y$.  (The flexibility of the parametrization
can be increased by using higher order polynomials; the generalization
of Eq.~(\ref{eq:bernstein}) to higher orders is obvious---the numerical
factors are just binomial coefficients.)

In practice, we refine this procedure as follows.
First, as a matter of convenience, we set $d_4 = 1$ and
supply in its place an overall constant factor, which is determined
by the number sum rule $\int_0^1 \! \mathrm{u}_v(x) \, dx = 2$.
We then set $d_3 = 1 \, + \, a_1/2$ to suppress
deviations from the $(1-x)^{a_2}$ behavior of $\mathrm{u}_v(x)$ at large $x$
by canceling the first subleading power of $(1-x)$ in $P_{\mathrm{u}_v}$:
\begin{eqnarray}
x \, \mathrm{u}_v(x) \, \to \, \mathrm{const} \times (1\, - \, x)^{a_2} \, \times \,
\left[1 \, + \, {\cal{O}}((1 \, - \, x)^2)\right] \; \;
\mathrm{for} \; \; x \to 1 \; .
\end{eqnarray}
We use the same parametrization for
$\mathrm{d}_v(x) \equiv \mathrm{d}(x) \, - \, \mathrm{\bar{d}}(x)$,
with the same parameter values $a_1$ and $a_2$; but, of
course, independent parameters for the coefficients of the
Bernstein polynomials and the normalization, which is set
by $\int_0^1 \! \mathrm{d}_v(x) \, dx = 1$.

Tying the valence $a_1$ parameters
together is motivated  by Regge theory, and supported by the
observation that the value of $a_1$ obtained in the fit is not
far from the value expected from Regge theory.
(The $a_1$ values for $u$, $d$, $\bar{u}$,
and $\bar{d}$ are expected to be close to $0$ from the
Pomeron trajectory; but that
leading behavior is expected to cancel in $\mathrm{u}_v = \mathrm{u} \, - \, \mathrm{\bar{u}}$
and $\mathrm{d}_v = \mathrm{d} \, - \, \mathrm{\bar{d}}$, revealing the subleading vector
meson Regge trajectory at $a_1 \simeq 0.5\,$.)
Not counting the two normalization parameters that are constrained
by quark-number sum rules, we are left with a total
of 8 fitting parameters for the valence quarks.  This is the same
number of parameters as were used in CT10 NNLO.  As a consistency check,
we find that allowing the $a_1$ and $a_2$ parameters for $\mathrm{d}_v$ to be
independent of those for $\mathrm{u}_v$ would reduce $\chi^2 \approx 3380$ by
less than one unit.  Allowing those parameters to be free would also not
substantially increase the uncertainty range given by the Hessian
procedure, except at very large $x$, where the fractional uncertainty is
already very large.  The additional fractional uncertainty at small $x$
generated by allowing different $a_1$ powers is also not important,
because that uncertainty only appears in the valence quantities
$u(x) - \bar{u}(x)$ and $d(x) - \bar{d}(x)$; while most processes of
interest are governed by the much larger $u(x)$, $d(x)$, $\bar{u}(x)$, $\bar{d}(x)$
themselves.

In addition to theoretical arguments that the power laws $a_2$
should be the same for $\mathrm{u}_v$ and $\mathrm{d}_v$
\cite{BrodskyPrivate},
$\chi^2$ tends to be insensitive to the differences. A large portion of the
data included in the global fit are from electron and muon DIS on
protons, which is more sensitive to $u$ and $\bar{u}$ than to
$\mathrm{d}$ and $\mathrm{\bar{d}}$ because of the squares of their
electric charges. Hence, when similar parametrizations are used for $P_{\mathrm{u}_v}$ and
$P_{\mathrm{d}_v}$, the uncertainties of $a_1(\mathrm{d}_v)$ and
$a_2(\mathrm{d}_v)$ are relatively large.

Our assumption $a_2(\mathrm{u}_v) = a_2(\mathrm{d}_v)$ forces
$\mathrm{u}_v(x)/\mathrm{d}_v(x)$ to approach
a constant in the limit $x \to 1$.  It allows our phenomenological
findings to be relevant for the extensive discussions
of what that constant might be \cite{Accardi:2011fa,Owens:2012bv}.
However, the experimental constraints at
large $x$ are fairly weak:  we can find excellent fits over the range
$-0.5 \, < \, a_2(\mathrm{d}_v) - a_2(\mathrm{u}_v) \, < \, 1.2$
at an increase of only $5$ units in $\chi^2$. Hence both $\mathrm{u}_v(x)/\mathrm{d}_v(x) \to 0$
and $\mathrm{u}_v(x)/\mathrm{d}_v(x) \to \infty$ at $x \to 1\,$
remain fully consistent with the data.  However, our assumption
$a_2(\mathrm{u}_v) = a_2(\mathrm{d}_v)$ does not restrict the calculated
uncertainty range materially in regions where it is not already very large.

By way of comparison, if we use the CT10 NNLO \cite{Gao:2013xoa} form (\ref{eq:ct10valence})
for $\mathrm{u}_v$ and $\mathrm{d}_v$, we obtain a slightly
better fit ($\chi^2$ lower by 8) with an unreasonable $a_2 \approx 0.1 \, $.
Similar behavior led us to fix $a_2 = 0.2$ in CT10 NNLO.

In a different comparison, the MSTW2008 fit \cite{Martin:2009iq, Watt:2012tq}
uses a parametrization for $\mathrm{u}_v$ and $\mathrm{d}_v$ that is
equivalent to Eq.~(\ref{eq:AppendixPoly}) with $c_3 \, = \, c_4 \, = 0$,
with the power-law parameters $a_1$ and $a_2$ allowed to differ
between $\mathrm{u}_v$ and $\mathrm{d}_v$.  If we use this MSTW parametrization
for the valence quarks at our $Q_0 = 1.3 \, \mathrm{GeV}$,
in place of the form we have chosen, the best-fit $\chi^2$ increases
by $64$, even though the total number of fitting parameters is the
same.  This decline in the fit quality comes about because the
freedom to have $a_2(\mathrm{u}_v) \neq a_2(\mathrm{d}_v)$ and
$a_2(\mathrm{u}_v) \neq a_2(\mathrm{d}_v)$ is not actually very helpful,
as noted above; so setting $c_3 \, = \, c_4 \, = 0$ does not leave an
adequate number of free parameters.

The more recent MMHT2014 \cite{Harland-Lang:2014zoa} PDF fit uses full fourth-order
polynomials for $\mathrm{u}_v$ and $\mathrm{d}_v$.  In our fit, however,
we find that no significant improvement in $\chi^2$
would result from treating $d_3({u_v})$ and
$d_3({d_v})$ as free parameters, rather than
choosing them to cancel the first subleading behavior
at $x \to 1$, as we have done.

Meanwhile the HERA PDF fits \cite{Aaron:2009aa,Abramowicz:2015mha,
Zenaiev:2015rfa}
use much more restricted forms, equivalent to
$c_1=c_2=c_3=0$ for $u_v$ and $c_1=c_2=c_3=c_4=0$
for $d_v$.  Those forms
are far too simple to describe
our data set:  using them in place of our choice increases
$\chi^2$ by more than 200.

We made a case in previous work \cite{Pumplin:2009bb} to repackage polynomial
parametrizations like (\ref{eq:AppendixPoly}) as linear combinations of
Chebyshev polynomials of argument $1 \,  -  \, 2\sqrt{x}\,$.  This
method has been adopted in the recent MMHT2014 fit \cite{Harland-Lang:2014zoa}.
However, we now contend that repackaging based on a linear combination
of Bernstein polynomials, as we do in CT14, is much better.
The full functional
forms available in the fit are, of course, the same either way.  But, because
each of the Bernstein polynomials has a single peak, and the peaks occur
at different values of $x$,
the coefficients that multiply those polynomials mainly control distinct
physical regions, and are therefore somewhat independent of each other.

In contrast, every Chebyshev polynomial of argument $1 \,  -  \, 2\sqrt{x}$
has a maximum value $\pm 1$ at both $x=0$ and $x=1$, along with an equal
maximum magnitude at some interior points. All Chebyshev
polynomials are important over the entire range of $x$, so their
coefficients are strongly correlated in the fit.  This causes minor
difficulties in finding the best fit and major difficulties in using the
Hessian method to estimate uncertainties based on orthogonal eigenvectors.
Furthermore, using Bernstein polynomials makes it easy to enforce
the desired positivity of the PDFs in the $x \to 0$ and
$x \to 1$ limits, because each of those limits is controlled by a single
polynomial.

We use a similar parametrization for the gluon, but with a polynomial
of a lower order, because the data provide fewer constraints on the
gluon distribution:
\begin{eqnarray}
P_{g}(y) \, = \, g_0 \, \left[e_0 \, q_0(y) \, + \, e_1 \, q_1(y) \, + \,
q_2(y) \right]
\end{eqnarray}
where
\begin{eqnarray}
q_0(y) &=& (1 - y)^2, \nonumber \\
q_1(y) &=& 2 \, y \, (1 - y), \nonumber \\
q_2(y) &=& y^2.
\end{eqnarray}
However, in place of $y=\sqrt{x}$,  we use the mapping
\begin{eqnarray}
y \, = \, 1 \, - \, (1-\sqrt{x})^2 \, = \, 2\sqrt{x}\, -\, x \;.
\end{eqnarray}
This mapping makes
$y \, = \,  1 \, - \, (1\, - \, x)^2/4 \, + \,
{\cal{O}}((1 \, - \, x)^3)$ and hence
\begin{eqnarray}
P_{g}(y) \, \to \, \mathrm{const} \, + \,
{\cal{O}}\left((1 \, - \, x)^2\right)
\end{eqnarray}
in the limit $x \to 1\,$.  This is an alternative way to
suppress the first subleading power of $(1-x)$ at $x \to 1$.
We have 5 free parameters to describe the gluon distribution,
including $g_0$ which governs the fraction of momentum carried
by the gluons.  The best fit has $a_2 = 3.8$, with the range
$2.6 < a_2 < 5.0$ allowed by an increase of only 5 in $\chi^2$.

In contrast, CT10 NNLO \cite{Gao:2013xoa} again used the form (\ref{eq:ct10valence})
for the gluon distribution, where $a_2$ was frozen at an arbitrary value of
10 because $\chi^2$ was rather insensitive to it.  That left the
same number of free parameters as are used here, but didn't allow
anything to be learned about the behavior at very large $x$.

If we use (\ref{eq:ct10valence}) for the gluon in our present fit,
the resulting $\chi^2$ is nearly as good, but again this choice
yields almost no information about the sixth parameter $a_2$: a range
of $\Delta\chi^2 = 1$ includes $-0.4 < a_2 < 12$. The negative
$a_2$ part of that range corresponds to an integrably singular
gluon probability density at $x \to 1$, which is not actually
forbidden theoretically; but would be totally unexpected.
This older parametrization would bring in unmotivated complexity in the large-x
region that is not indicated by any present data.
To test that our parametrization has adequate flexibility,
we made similar fits using somewhat higher order Bernstein
polynomials, including up to a total of 10 more free parameters.
We calculated the uncertainty for the $gg \to H$ cross section at 8 TeV
using the Lagrange Multiplier method, and found very little variation
in the range of the prediction.  We also calculated the range of
uncertainty in $\alpha_s(m_Z)$ obtained from our fits at 90\% confidence
(including our Tier 2 penalty).  The extra freedom in parametrization
increased the uncertainty range only slightly: $0.111 \, - \, 0.121$ using the
CT14 parametrization; $0.111 \, - \, 0.123$ using the more flexible one.

The sea quark distributions $\bar{d}$ and $\bar{u}$ were parametrized
using fourth-order polynomials in $y$ with the same
mapping $y \, = \, 2\sqrt{x}\, - \, x$ that was used for the gluon.
We assumed $\bar{u}(x)/\bar{d}(x) \to 1$ at $x \to 0$, which implies
$a_1(\bar{u}) = a_1(\bar{d})$.
As the strangeness content is constrained rather poorly,
we used a minimal parametrization $P_{s + \bar{s}} = \mathrm{const}$,
with $a_1$ tied to the common $a_1$ of $\bar{u}$ and $\bar{d}$.
Even fewer experimental constraints apply to the strangeness
asymmetry, so we have assumed $s(x) = \bar{s}(x)$ in this analysis.
Thus, we have just two parameters for strangeness in our Hessian method:
$a_2$ and normalization.  In view of more upcoming data on measuring the asymmetry in the production cross sections of $W+{\bar c}$ and $W+c$
from the LHC, we plan to include $s(x) \ne \bar{s}(x)$ in
our next round of fits.

In all, we have 8 parameters associated with the valence quarks,
5 parameters associated with the gluon, and 13 parameters
associated with sea quarks, for a total of 26 fitting parameters.
Hence there are 52 eigenvector sets generated by the Hessian
method that captures most of the PDF uncertainty.

The Hessian method tends to underestimate the uncertainty
for PDF variations that are poorly constrained, because the
method is based on the assumption that $\chi^2$ is a quadratic
function of the fitting parameters; and that assumption tends
to break down when the parameters can move a long way because
of a lack of experimental constraints.   This can be seen,
for example, for the case of the small-$x$ gluon uncertainty, by
a Lagrange Multiplier scan in which a series of fits are made
with different values of the independent variable
$g(x,Q)$ at $x=0.001$, $Q=Q_0$.

In order to include the wide variation of the gluon distribution
that is allowed at small $x$, we therefore supplement the Hessian
sets with an additional pair of sets that were obtained using the
Lagrange Multiplier method: one with enhanced gluon and one
with suppressed gluon at small $x$, as was already done in CT10.
In CT14, we also include an additional pair of sets with enhanced
or suppressed strangeness at small $x$; although it is possible
that treating $a_1(s)$ as a fitting parameter independent from
$a_1(\bar{u}) = a_1(\bar{d})$ would have worked equally well.

In summary, we have a total of 56 error sets:  $2\times 26$ from the
Hessian method,  supplemented by two extremes of small-$x$ gluon,
and two extremes of small-$x$ strangeness.
Uncertainties from all pairs of error sets are to be summed in
quadrature using the master formulas~\cite{Pumplin:2002vw,Lai:2010vv,Nadolsky:2001yg}.
In comparison, CT10 NNLO had 50 error sets.
The increased flexibility in the CT14 parametrization is warranted
by better experimental constraints and its improved fit to the data.
Indeed, fitting the CT14 data set using the old CT10 parametrizations
yields a best fit that is worse by 60 units in $\chi^2$.

\end{document}